\documentclass[a4paper,10pt]{article}
\usepackage[utf8]{inputenc}
\usepackage{graphicx,tabularx,amsmath,amsthm,amssymb,mathtools,amsfonts,bm,setspace,color,float,caption,subcaption}
\usepackage[pdfpagemode=UseNone,pdfstartview={XYZ null null null}]{hyperref}
\usepackage[longnamesfirst,numbers,square]{natbib}
\usepackage[margin=0.5in]{geometry} 
\usepackage[title]{appendix}

\usepackage{color}

\definecolor{dark-red}{rgb}{0.4,0.15,0.15}
\definecolor{dark-blue}{rgb}{0.15,0.15,0.4}
\definecolor{medium-blue}{rgb}{0,0,0.5}
\hypersetup{colorlinks,linkcolor={dark-red},citecolor={dark-blue},urlcolor={medium-blue}}

\newcommand{\eps}{\varepsilon}

\newcommand{\R}{\mathbb{R}}
\newcommand{\I}{\mathbb{I}}
\newcommand{\C}{\mathbb{C}}
\newcommand{\Z}{\mathbb{Z}}
\newcommand{\M}{\mathbb{M}}
\newcommand{\K}{\mathbb{K}}
\newcommand{\Q}{\mathbb{Q}}
\newcommand{\B}{\mathbb{B}}
\newcommand{\A}{\mathbb{A}}
\newcommand{\D}{\mathbb{D}}
\newcommand{\Laplacian}{\mathbb{L}}

\newcommand{\zero}{\mathbb{O}}

\newcommand{\LL}{\mathcal{L}}
\newcommand{\BB}{\mathcal{B}}
\newcommand{\CC}{\mathcal{C}}
\newcommand{\UU}{\mathcal{U}}
\newcommand{\VV}{\mathcal{V}}
\newcommand{\WW}{\mathcal{W}}

\newcommand{\FF}{\mathcal{F}}
\newcommand{\RR}{\mathcal{R}}
\newcommand{\TT}{\mathcal{T}}

\newcommand{\U}{\bm{u}}
\newcommand{\V}{\bm{v}}

\newcommand{\bU}{\bm{U}}
\newcommand{\bV}{\bm{V}}
\newcommand{\bW}{\bm{W}}
\newcommand{\bF}{\bm{F}}
\newcommand{\bphi}{\bm{\phi}}
\newcommand{\bPhi}{\bm{\Phi}}

\newcommand{\bzero}{\bm{0}}

\newcommand{\orderOne}{\mathcal{O}\left(\varepsilon\right)}
\newcommand{\orderTwo}{\mathcal{O}\left(\varepsilon^2\right)}
\newcommand{\orderThree}{\mathcal{O}\left(\varepsilon^3\right)}
\newcommand{\orderFour}{\mathcal{O}\left(\varepsilon^4\right)}

\newcommand{\clambda}{\overline{\lambda}}

\newtheorem{theorem}{Theorem}[section]

\newtheorem{lemma}[theorem]{Lemma}

\numberwithin{equation}{section}
\allowdisplaybreaks

\title{Pattern Formation and Oscillatory Dynamics in a Two-Dimensional Coupled
  Bulk-Surface Reaction-Diffusion System} \author{Frédéric
  Paquin-Lefebvre \thanks{Dept. of Mathematics, UBC, Vancouver,
    Canada. (corresponding author {\tt paquinl@math.ubc.ca})}
  \enspace, Wayne Nagata \thanks{Dept. of Mathematics, UBC, Vancouver,
    Canada. {\tt nagata@math.ubc.ca}} \enspace, Michael J. Ward
  \thanks{Dept. of Mathematics, UBC, Vancouver, Canada. {\tt
      ward@math.ubc.ca}}}

\begin{document}

\maketitle
\vspace*{-0.5cm}
\begin{abstract}
On a two-dimensional circular domain, we analyze the formation of
    spatio-temporal patterns for a class of coupled bulk-surface
    reaction-diffusion models for which a passive diffusion process
    occurring in the interior bulk domain is linearly coupled to a
    nonlinear reaction-diffusion process on the domain boundary. For
    this coupled PDE system we construct a radially symmetric
    steady state solution and from a linearized stability analysis
    formulate criteria for which this base state can undergo either a
    Hopf bifurcation, a symmetry-breaking pitchfork (or Turing)
    bifurcation, or a codimension-two pitchfork-Hopf bifurcation. For
    each of these three types of bifurcations, a multiple time-scale
    asymptotic analysis is used to derive normal form amplitude
    equations characterizing the local branching behavior of
    spatio-temporal patterns in the weakly nonlinear
    regime. {Among the novel aspects of this weakly nonlinear analysis are the two-dimensionality of the bulk domain, the systematic treatment of arbitrary reaction kinetics restricted to the boundary, the bifurcation parameters which arise in the boundary conditions, and the underlying spectral problem where both the differential operator and the boundary conditions involve the eigenvalue parameter.} The normal form theory is illustrated for
      both Schnakenberg and Brusselator reaction kinetics, and the
      weakly nonlinear results are favorably compared with numerical
      bifurcation results and results from time-dependent PDE
      simulations of the coupled bulk-surface system. Overall, the
      results show the existence of either subcritical or
      supercritical Hopf and symmetry-breaking bifurcations, and
      mixed-mode oscillations characteristic of codimension-two
      bifurcations. Finally, the formation of global structures such
      as large amplitude rotating waves is briefly explored through PDE numerical simulations.
\end{abstract}


\section{Introduction}\label{sec:introduction}

If a passive linear diffusion process in a bounded domain is
  coupled to a nonlinear reaction-diffusion process on the domain
  boundary, spatio-temporal patterns can occur that otherwise would
  not be present without this bulk-surface coupling. Such a pattern
  formation mechanism is relevant in a variety of applications in
  which boundaries play an active role in the overall dynamics. For
  instance, in some biological cell signalling contexts certain
  proteins cycle from an active cellular membrane to a cytoplasmic
  bulk via adsorption and desorption processes. Applications of this
  type include the formation of surface-bound Turing patterns through
  symmetry-breaking instabilities (cf.~\cite{levine2005},
  \cite{ratz2012}, \cite{ratz2014}) as well as the onset of
  Min-protein pole-to-pole oscillations prior to cell division in
  \textit{E. Coli} (cf.~\cite{halatek2012}, \cite{halatek2018}). In
  many prior studies (eg.~\cite{ratz2012}, \cite{ratz2014},
  \cite{levine2005}), the coupled bulk-surface systems have mainly
  been analyzed through either a linear stability analysis, which typically
  involves finding the conditions for a Turing-type diffusion-driven
  instability, or from time-dependent PDE numerical simulations
  (cf.~\cite{madzvamuse2015}, \cite{madzvamuse2016}).

In a simplified geometry consisting of a 1-D spatial bulk
  domain, these models become coupled PDE-ODE models, and were studied
  in \cite{gomez2007}, \cite{gou2015}, \cite{gou2016} and
  \cite{gou2017}. There, dynamically active units, modeled by
  nonlinear ODEs, are spatially segregated and are coupled through a
  linear bulk diffusion field. This setup serves as a modeling
  paradigm for the study of synchrony under diffusion sensing. In
  contrast to the classical types of PDE-ODE models where the coupling
  occurs in all of space (cf.~\cite{korvasova2015}), the type of
  coupling considered here, and in \cite{gomez2007}--\cite{gou2017},
  is restricted to the boundaries, and is expressed in terms of
  Robin-type boundary conditions. In \cite{gomez2007}--\cite{gou2017},
  this class of 1-D coupled PDE-ODE systems was analyzed through a
  combination of linear stability analysis, direct numerical PDE
  simulations, and numerical bifurcation software. The numerical
  bifurcation studies have allowed for the computation of global
  branches of synchronous and asynchronous periodic solutions in terms
  of bulk diffusion coefficients and coupling rates. As an extension
  of the linear stability theory, in \cite{gou2015} a weakly nonlinear
  analysis was developed to study the local branching behavior of
  synchronous oscillations for the idealized case of a single bulk
  species diffusing between two identical membranes, each consisting
  of a single active species.

  To extend this previous work, our goal in this paper is to provide a
  comprehensive weakly nonlinear, or normal form, analysis, to study
  the various bifurcations associated with a class of
  {dimensionless coupled bulk-surface reaction-diffusion
  systems} for which the bulk domain $\Omega$ consists of the disk 
\begin{equation}\label{eq:bulk_domain}
 \Omega = \{ x \in \R^2 \,\, | \,\, \|x\| < R \} \,,
\end{equation}
of radius $R$.  {In the bulk domain we assume that two bulk
  species $U,\,V$ undergo passive diffusion with linear decay in
  $\Omega$. This leads to the following PDEs in the bulk region}
\begin{equation}\label{eq:bulk_RD}
  \frac{\partial U}{\partial t} = D_u \Delta U - \sigma_u U\,, \qquad
  \frac{\partial V}{\partial t} = D_v \Delta V - \sigma_v V\,, \qquad
  x \in \Omega\,, \quad t>0 \,.
\end{equation}
{Here $D_u,\,D_v$ are the constant bulk diffusion
  coefficients, while $\sigma_u,\,\sigma_v$ are the constant bulk decay
  rates. Since $\Omega$ is the disk, the Laplacian $\Delta$ in
  (\ref{eq:bulk_RD}) is conveniently written in terms of polar
  coordinates $(r,\theta)$ as
  $\Delta = \partial_{rr} + r^{-1}\partial_r +
  r^{-2}\partial_{\theta\theta}$. Next, we assume that the flux
  normal to the boundary is proportional to the difference between the
  surface-bound species densities, denoted by $u,\,v$, and the bulk
  species densities evaluated on the boundary. This yields linear
  Robin-type boundary conditions for \eqref{eq:bulk_RD}:}
\begin{equation}\label{eq:BC}
  D_u \left. \frac{\partial U}{\partial r}\right|_{r = R} =
  K_u\left(u - U|_{r = R}\right)\,, \qquad D_v \left.
    \frac{\partial V}{\partial r}\right|_{r = R} =
  K_v\left(v - V|_{r = R}\right) \,,
\end{equation}
{where $K_u,\,K_v$ are coupling rate constants, also known as
  Langmuir rate constants. Finally, on the domain boundary the
  dynamics of the two surface-bound species are assumed to be
  governed by a system of reaction-diffusion equations with periodic
  boundary conditions in the azimuthal coordinate:}
\begin{equation}\label{eq:surface_RD}
  \frac{\partial u}{\partial t} = \frac{d_u}{R^2}
  \frac{\partial^2 u}{\partial \theta^2} - K_u\left(u - U|_{r=R}\right) +
  f(u,v)\,, \qquad
  \frac{\partial v}{\partial t} = \frac{d_v}{R^2}\frac{\partial^2 v}
  {\partial \theta^2} - K_v\left(v - V|_{r=R}\right) + g(u,v)\,.
\end{equation}
Here $f(u,v)$ and $g(u,v)$ are the given reaction kinetics, while
$d_u,\,d_v$ are surface diffusion coefficients. In the absence of
surface diffusion, this model reduces to the coupled PDE-ODE system
studied through linear stability analysis in \cite{levine2005} for a
slightly different boundary condition and with Gierer-Meinhardt
kinetics on the circular membrane. As a remark, since a
biologically-realistic membrane possesses some width, the coupled
bulk-surface model, defined by
\eqref{eq:bulk_RD}--\eqref{eq:surface_RD}, provides only an
approximation to this more complicated setting in the case where the
width of the membrane is small in comparison with the characteristic
length scale of the bulk domain. {Our coupled bulk-surface model, as introduced above, is given in dimensionless form.  We refer the reader to Appendix \S \ref{sec:dimensionless} for a derivation of the system \eqref{eq:bulk_RD}--\eqref{eq:surface_RD} from a model with physical units.}

{Our primary goal herein is to characterize the dynamics of the coupled system in the weakly nonlinear regime near one of three distinct bifurcations; a Hopf
  bifurcation, for which the spatial mode is trivial with a nonzero
  temporal frequency, a pitchfork bifurcation, for which the spatial
  mode is nontrivial with a zero temporal frequency, and finally a
  codimension-two pitchfork-Hopf bifurcation, which occurs when the
  previous two bifurcations coincide. By deriving amplitude, or
  rather, normal form equations, we will characterize the branching
  behavior in the vicinity of these three bifurcations. Although the
  use of weakly nonlinear analysis to study pattern formation in
  reaction-diffusion systems, convection processes, and fluid flows is
  well-established and ubiquitous in the literature
  (cf.~\cite{cross1993}, \cite{walgraef1997}), the development of a
  weakly nonlinear theory to characterize pattern formation near
  bifurcation points of coupled bulk-surface models requires a careful analysis of the spectral problem for the linearization, where both
  the differential operator and the boundary conditions involve the eigenvalue parameter. This analysis is at the core of performing a weakly nonlinear analysis using a multiple time-scale expansion method. We believe such a spectral problem has not
  been considered previously in the context of two-dimensional pattern formation
  problems. We also give a systematic treatment of the three distinct bifurcations for arbitrary reaction kinetics on the surface.}
  
  In our formulation, we will suppose that when uncoupled from
  the bulk domain, the reaction-diffusion system on the surface
  \eqref{eq:surface_RD} possesses a unique spatially uniform steady
  state, which is linearly stable with respect to any spatial
  perturbation. Consequently, we will restrict the parameter space to
  cases where $d_u = d_v$ in order to avoid the short-range activation
  combined with long-range inhibition paradigm, typical of Turing
  instabilities. {Rather than using the surface diffusion coefficients
  as bifurcation parameters, we will vary the bulk diffusion
  coefficients and the coupling rates, so that the loss of stability
  of the base state results from the diffusive coupling with the bulk
  domain. In terms of these bifurcation parameters that are associated
  with the boundary conditions \eqref{eq:BC}, in \S
  \ref{sec:weakly_nonlinear_theory} we find that multi-scale expansion methods are particularly convenient for deriving amplitude equations characterizing the local branching behavior.}

{Although arbitrary reaction kinetics are employed in our
  analysis, we will apply our weakly nonlinear theory to either the
  well-known Schnakenberg or Brusselator kinetics. In nondimensional
  forms, the Schnakenberg kinetics are}
\begin{equation}\label{eq:schnakenberg}
  f(u,v) = a - u + u^2v \,, \qquad g(u,v) = b - u^2v \,; \qquad
  a,b > 0\,, \quad b - a < (b + a)^3 \,,
\end{equation}
{while the Brusselator kinetics are given by}
\begin{equation}\label{eq:brusselator}
 f(u,v) = a - (b+1)u + u^2v\,, \qquad
 g(u,v) = bu - u^2v \,; \qquad  a > 0\,, \quad 0 < b < a^2 + 1 \,.
\end{equation}
{To validate our weakly nonlinear theory, a combination of
  numerical bifurcation analysis, for the computation of global
  bifurcation branches, and full time-dependent PDE numerical
  simulations are employed. Classical values for the parameters $a$
  and $b$ are used. For the Schnakenberg kinetics, these are $a=0.1$
  and $b=0.9$ (cf.~\cite{madzvamuse2015}). For the Brusselator
  kinetics, $a=3$ will be taken, while different values of $b$, all
  with $b < a^2+1$, will be considered. For these parameters values,
  the uncoupled bulk-surface system without surface diffusion admits a
  unique stable steady state and no patterns appear. In this way, the
  patterns observed arise from the coupling between the bulk and
  surface.  We remark that a symmetry-breaking bifurcation mechanism
  for particular forms of the nonlinearities has also been explored in
  \cite{ratz2012} and \cite{madzvamuse2015} through full PDE
  simulations.}

{The outline of this paper is as follows. In \S
  \ref{sec:weakly_nonlinear_theory}, for arbitrary reaction kinetics,
  we derive amplitude equations (normal forms) near either a Hopf, a
  pitchfork, or a pitchfork-Hopf, bifurcation point of the
  linearization of the base state. In \S
  \ref{sec:bifurcation_analysis}, we analyze these normal forms and
  interpret their equilibria in terms of limit cycles or Turing-type
  patterns of the coupled original system. Subsections
  \ref{subsec:theory_codim_1} and \ref{subsec:theory_codim_2},
  respectively, treat separately the codimension-one and the
  codimension-two cases. Numerical validation of the weakly nonlinear
  theory with the classical Schnakenberg and Brusselator reaction
  kinetics is provided in \S \ref{sec:validation_weakly_nonlinear}. \S
  \ref{sec:global_dynamics} is distinct from the previous sections in
  that, through PDE simulations, it gives a glimpse into novel
  nonlinear patterns that can occur for the coupled system {\em away from}
  bifurcation points. In particular, the dynamics and formation of
  rotating waves is explored for a coupled bulk-surface
  reaction-diffusion system with a slightly more general boundary
  condition than \eqref{eq:BC}. Finally, in \S \ref{sec:discussion},
  we briefly summarize the paper and discuss a few open problems that
  warrant further study.}

\section{Weakly nonlinear theory}\label{sec:weakly_nonlinear_theory}

In this section, the method of multiple time-scales is used to derive
amplitude equations describing the branching behavior near three
distinct bifurcations:
\begin{itemize}
\item Trivial mode $n=0$ undergoes a Hopf bifurcation, at which the
  bifurcating solution is invariant under rotation and reflection
  symmetries.
\item Nontrivial mode $\{n,-n\} \in \Z \backslash \{0\}$ loses
  stability through a pitchfork bifurcation, at which the bifurcating
  solutions are equivariant under rotation and reflection symmetries.
\item Pitchfork-Hopf (Turing-Hopf), when the previous two bifurcations
  occur simultaneously.
\end{itemize}
From the linear stability analysis, curves of codimension-one
bifurcation points and their codimension-two intersection have been
computed in the plane of parameters $(K_v,D_v)$ (see
Fig.~\ref{fig:stability_diagram_schnakenberg} and
Fig.~\ref{fig:stability_diagram_brusselator}). This motivates
introducing a two-parameter bifurcation analysis.

\subsection{Preliminaries}

{Before formulating the multiple time-scale asymptotic
  expansion, the coupled bulk-surface system
  \eqref{eq:bulk_domain}--\eqref{eq:surface_RD} is rewritten as an
  evolution equation, which then facilitates below the introduction of
  an extended linear operator and its adjoint:}
\begin{equation}\label{eq:nonlinear_functional}
 \dot{W} = \bm{F}(W) =
 \begin{pmatrix}
  D_u \Delta U - \sigma_u U \\
  D_v \Delta V - \sigma_v V \\
  \frac{d_u}{R^2}u_{\theta\theta} - K_u\left(u - U\right) + f(u,v) \\
  \frac{d_v}{R^2}v_{\theta\theta} - K_v\left(v - V\right) + g(u,v)
 \end{pmatrix}\,.
\end{equation}
Here $\bm{F}$ is a nonlinear functional acting on $\bm{\WW}$, defined
as a space of vector functions whose components satisfy the
appropriate Langmuir boundary conditions:
\begin{equation}
\bm{\WW} = \left. \left\{ 
W = \begin{pmatrix}
U(r,\theta) \\
V(r,\theta) \\
u(\theta) \\
v(\theta)
\end{pmatrix} \right|
\begin{matrix}
D_u \partial_r U|_{r=R} = K_u \left(u - U|_{r=R}\right) \\
D_v \partial_r V|_{r=R} = K_v \left(v - V|_{r=R}\right)
\end{matrix}
\right\}\,.
\end{equation}
{The radially symmetric steady state (i.e.~the base state) for
\eqref{eq:nonlinear_functional} is given by}
\begin{equation}\label{eq:patternless_solution}
W_e = \begin{pmatrix}
       A_0(0) \frac{I_0(\omega_u r)}{I_0(\omega_u R)} e_1^T\U_e\\
       B_0(0) \frac{I_0(\omega_v r)}{I_0(\omega_v R)} e_2^T\U_e\\
       \U_e
      \end{pmatrix}\,, \qquad 
\begin{cases}      
     \omega_u = \sqrt{\frac{\sigma_u}{D_u}}\,, \quad
     A_0(0) = \frac{K_u I_0(\omega_u R)}{D_u\omega_u I_0^{\prime}(\omega_u R) +
       K_u I_0(\omega_u R)} \\
     \omega_v = \sqrt{\frac{\sigma_v}{D_v}}\,, \quad
     B_0(0) = \frac{K_v I_0(\omega_v R)}{D_v\omega_v I_0^{\prime}(\omega_v R) +
       K_v I_0(\omega_v R)}
      \end{cases}\,,
\end{equation}
where $\U_e = (u_e,v_e)^T$ is the surface steady state vector
satisfying the nonlinear algebraic equation
\begin{equation}
\begin{cases}
 K_u p_0(0)u_e - f(u_e,v_e) = 0 \\
 K_v q_0(0)v_e - g(u_e,v_e) = 0
\end{cases}\,, \qquad
\begin{cases}
  p_0(0) = \frac{D_u \omega_u I_0^{\prime}(\omega_u R)}
  {D_u\omega_u I_0^{\prime}(\omega_u R) + K_u I_0(\omega_u R)} \\
  q_0(0) = \frac{D_v \omega_v I_0^{\prime}(\omega_v R)}
  {D_v\omega_v I_0^{\prime}(\omega_v R) + K_v I_0(\omega_v R)}
\end{cases}\,.
\end{equation}
Here $I_n(z)$ for $n \in \Z$ are the usual modified Bessel
functions. Next, by expanding the nonlinear functional about the
{base state}, we get
\begin{equation}\label{eq:expansion}
  \dot{W} = \underbrace{\bm{F}(W_e)}_{= 0} + \LL(W-W_e) +
  \BB(W - W_e, W - W_e) + \CC(W - W_e, W - W_e, W - W_e) + \ldots\,,
\end{equation}
where $\LL$ is the linearized operator defined by
\begin{equation}\label{eq:linear_operator}
 \LL(W) = 
 \begin{pmatrix}
  D_u \Delta U - \sigma_u U \\
  D_v \Delta V - \sigma_v V \\
  \frac{d_u}{R^2}u_{\theta\theta} - K_u\left(u - U\right) + f_u^e u + f_v^e v \\
  \frac{d_v}{R^2}v_{\theta\theta} - K_v\left(v - V\right) + g_u^e u + g_v^e v
 \end{pmatrix}\,,
 \end{equation}
 while $\BB$ and $\CC$ are, respectively, bilinear and trilinear forms.
 For each $n\in \Z$, the eigenfunctions for the linearized operator are
 given by
\begin{equation}
\WW_n = \begin{pmatrix}
       A_n(\lambda) \frac{I_n(\Omega_u r)}{I_n(\Omega_u R)} e_1^T \bphi_n \\
       B_n(\lambda) \frac{I_n(\Omega_v r)}{I_n(\Omega_v R)} e_2^T \bphi_n \\
       \bphi_n
      \end{pmatrix}e^{in\theta}\,, \qquad \begin{cases}       
        \Omega_u = \sqrt{\frac{\lambda + \sigma_u}{D_u}}\,, \quad
        A_n(\lambda) = \frac{K_u I_n(\Omega_u R)}
        {D_u\Omega_u I_n^{\prime}(\Omega_u R) + K_u I_n(\Omega_u R)} \\
        \Omega_v = \sqrt{\frac{\lambda + \sigma_v}{D_v}}\,,
        \quad B_n(\lambda) = \frac{K_v I_n(\Omega_v R)}
        {D_v\Omega_v I_n^{\prime}(\Omega_v R) + K_v I_n(\Omega_v R)}
      \end{cases}\,.
\end{equation}
The eigenvector $\bphi_n = (\phi_n,\psi_n)^T$ satisfies a homogeneous
linear system 
\begin{equation}\label{eq:lin_hom}
  \left[\bPhi_n(\lambda)\right] \bphi_n =
  \left[ J_e - \lambda I - \left( K_u p_n(\lambda) + \frac{n^2d_u}{R^2}
    \right)E_1 - \left( K_v q_n(\lambda) + \frac{n^2d_v}{R^2} \right)E_2
  \right] \bphi_n = \bzero\,,
\end{equation}
with $J_e, E_1, E_2, p_n(\lambda)$, and $q_n(\lambda)$ defined as
\begin{equation}
J_e = \begin{pmatrix}
       f_u^e & f_v^e \\
       g_u^e & g_v^e
      \end{pmatrix}\,, \quad E_i = e_i e_i^T \in \R^{2\times2}\,, \quad
\begin{cases}
  p_n(\lambda) = \frac{D_u\Omega_u I_n^{\prime}(\Omega_u R)}
  {D_u\Omega_u I_n^{\prime}(\Omega_u R) + K_u I_n(\Omega_u R)} =
  1 - A_n(\lambda) \\
  q_n(\lambda) = \frac{D_v\Omega_v I_n^{\prime}(\Omega_v R)}
  {D_v\Omega_v I_n^{\prime}(\Omega_v R) + K_v I_n(\Omega_v R)} = 1 - B_n(\lambda)
\end{cases}\,.
\end{equation}
Here, the vectors $e_1$ and $e_2$ form the standard orthonormal basis
in the phase space defined by the species $u$ and $v$. A nontrivial
solution to system \eqref{eq:lin_hom} will exist when the following
transcendental equation is satisfied:
\begin{equation}\label{eq:transcendental}
F_n(\lambda) = \det\left[ \bPhi_n(\lambda) \right] = 0\,.
\end{equation}

In the multi-scale analysis below, an application of the solvability
condition requires the formulation of an adjoint linear operator
$\LL^{\star}$, defined by
\begin{equation}
 \LL^{\star}(W^{\star}) = 
 \begin{pmatrix}
  D_u \Delta U^{\star} - \sigma_u U^{\star} \\
  D_v \Delta V^{\star} - \sigma_v V^{\star} \\
  \frac{d_u}{R^2}u_{\theta\theta}^{\star} - K_u\left(u^{\star} - U^{\star}\right) +
  f_u^e u^{\star} + g_u^e v^{\star} \\
  \frac{d_v}{R^2}v_{\theta\theta}^{\star} - K_v\left(v^{\star} - V^{\star}\right) +
  f_v^e u^{\star} + g_v^e v^{\star}
 \end{pmatrix}\,.
\end{equation}
For the special case of Langmuir boundary conditions, the dual space
satisfies $\bm{\WW}^{\star} = \bm{\WW}$, which means that both the boundary
conditions and their adjoint are identical. Furthermore, the adjoint
eigenfunctions yield
\begin{equation}
\WW_n^{\star} = \begin{pmatrix}
  \overline{A_n(\lambda)} \frac{I_n(\overline{\Omega_u} r)}
  {I_n(\overline{\Omega_u} R)} e_1^T \bphi_n^{\star} \\
  \overline{B_n(\lambda)} \frac{I_n(\overline{\Omega_v} r)}
  {I_n(\overline{\Omega_v} R)} e_2^T \bphi_n^{\star} \\
       \bphi_n^{\star}
      \end{pmatrix}e^{in\theta}\,.
\end{equation}
{It is then readily verified that $\WW_n$ and $\WW_n^{\star}$ form an
  orthogonal set of eigenfunctions, satisfying}
\begin{equation}\label{eq:orthogonal}
 \langle \WW_m^{\star}, \WW_n \rangle = 0\,, \quad \text{ if } m \neq n\,,
\end{equation}
where the inner-product in \eqref{eq:orthogonal} is defined by
\begin{equation}\label{eq:inner_product}
  \langle W^{\star}, W \rangle = \int_0^{2\pi}\int_0^R
  \left[ \overline{U^{\star}}U + \overline{V^{\star}}V \right]rdrd\theta +
  \int_0^{2\pi} \left[ \overline{u^{\star}}u + \overline{v^{\star}}v \right]
  R\, d\theta \,. 
\end{equation}
The set of eigenfunctions can be normalized so that
$\langle \WW_n^{\star}, \WW_n \rangle = 1$ for all $n$. Finally, using the
previous definitions of linear operators, eigenfunctions and
inner-product, one may easily verify that the following properties
hold:
\begin{equation}
  \LL \WW_n = \lambda \WW_n\,, \qquad \LL^{\star} \WW_n^{\star} =
  \clambda \WW_n^{\star}\,,
  \qquad \langle W^{\star}, \LL W \rangle = \langle \LL^{\star} W^{\star},
  W \rangle \,.
\end{equation}

{Lastly, we define more precisely the bilinear and trilinear
  forms that arise in the expansion \eqref{eq:expansion}. They are
  defined by}
\begin{equation}
 \BB(W_j,W_k) = 
 \begin{pmatrix}
  0 \\
  0 \\
  B \left(\U_j, \U_k\right)
 \end{pmatrix}\,, \qquad \CC(W_j,W_k,W_l) =
 \begin{pmatrix}
  0 \\
  0 \\
  C \left(\U_j, \U_k, \U_l \right)
 \end{pmatrix}\,.
\end{equation}
The first two components of $\BB$ and $\CC$ vanish since the diffusion
process occurring in the bulk is linear. The reduced bilinear and
trilinear forms $B$ and $C$ are defined by
\begin{equation*}
  B(\U_j,\U_k) = \frac{1}{2}(I \otimes \U_k^T) H_e \U_j\,, \qquad
  C(\U_j,\U_k,\U_l) = \frac{1}{6}(I \otimes \U_l^T) T_e (\U_j \otimes \U_k)\,,
\end{equation*}
where $\otimes$ is the Kronecker product and $H_e,T_e$ are matrices
involving the second and third order partial derivatives
\begin{equation*}
 H_e = \begin{pmatrix}
        f_{uu}^e & f_{uv}^e \\
        f_{uv}^e & f_{vv}^e \\
        g_{uu}^e & g_{uv}^e \\
        g_{uv}^e & g_{vv}^e
       \end{pmatrix}\,, \quad
 T_e = \begin{pmatrix}
        f_{uuu}^e & f_{uuv}^e & f_{uuv}^e & f_{uvv}^e \\
        f_{uuv}^e & f_{uvv}^e & f_{uvv}^e & f_{vvv}^e \\
        g_{uuu}^e & g_{uuv}^e & g_{uuv}^e & g_{uvv}^e \\
        g_{uuv}^e & g_{uvv}^e & g_{uvv}^e & g_{vvv}^e      
       \end{pmatrix}\,.
\end{equation*}
 
\subsection{Multi-scale expansion}

{Let $\mu = (K_v,D_v)^T$ denote the vector of bifurcation
  parameters. As usual, a slow time-scale $\tau = \eps^2t$, with
  $\eps \ll 1$, is introduced. Using the same scaling, the parameters
  are slightly perturbed,}
\begin{equation}\label{eq:expansion_parameter}
 \mu = \mu_0 + \eps^2 \mu_1, \qquad \|\mu_1\| = 1.
\end{equation}
Here $\mu_0$ is the bifurcation point and $\mu_1$ is a unit vector
indicating the direction of the bifurcation. The full system is then
expanded in a regular asymptotic power series around the
{base state} as
\begin{equation}\label{eq:asymptotic}
 W = W_e + \eps W_1 + \eps^2 W_2 + \eps^3 W_3 + \orderFour\,,
\end{equation}
where the subscript here refers to the expansion order
{rather than the mode of the eigenfunction}. Next, by
  inserting \eqref{eq:expansion_parameter} and \eqref{eq:asymptotic}
  into \eqref{eq:expansion}, and collecting powers of $\eps$ we obtain
  that
\begin{equation}\label{eq:collect_powers}
\begin{split}
  & \eps\partial_t W_1 + \eps^2\partial_tW_2 + \eps^3 (\partial_t W_3 +
  \partial_\tau W_1) = \eps \LL W_1 + \eps^2\left(\LL W_2 + \BB(W_1,W_1) +
    \begin{pmatrix} 0 \\ \omega_v^2 V_e(r) e_2^T \mu_1 \\ 0 \\
      -q_0(0)v_e e_1^T \mu_1 \end{pmatrix} \right) \\ 
  & + \eps^3 \left( \LL W_3 + 2\BB(W_1,W_2) + \CC(W_1,W_1,W_1) +
    \begin{pmatrix} 0 \\ \frac{\partial_t + \sigma_v}{D_{v0}}V_1e_2^T\mu_1
 \\ 0 \\ - (v_1 - V_1|_{r=R})e_1^T\mu_1 \end{pmatrix} \right) + \orderFour\,,
\end{split}
\end{equation}
where the vectors $e_1$ and $e_2$ are now the standard orthonormal
basis in the parameter space defined by $K_v$ and $D_v$. The perturbed
boundary conditions satisfy
\begin{equation}\label{eq:collect_powers_BC}
\begin{split}
  \sum_{j=1}^3 \eps^j \left(D_u \partial_r U_j + K_u U_j - K_u u_j \right) +
  \orderFour &= 0\,, \quad r = R\,. \\
\sum_{j=1}^3 \eps^j \left(D_{v0} \partial_r V_j + K_{v0} V_j - K_{v0} v_j \right)
- \left(\eps^2 q_0(0)v_e + \eps^3 (v_1 - V_1) \right)\beta^T\mu_1 +
\orderFour &= 0\,, \quad r = R\,,
\end{split}
\end{equation}
where the vector $\beta$ is defined by
\begin{equation}
 \beta = \begin{pmatrix} 1 \\ -{K_{v0}/D_{v0}} \end{pmatrix}\,.
\end{equation}

\subsection{Weakly nonlinear analysis of patterns}\label{subsec:weakly_nonlinear_analysis_patterns}

The leading order solution corresponds to the {base state}
defined by \eqref{eq:patternless_solution} evaluated at the
bifurcation point $\mu = \mu_0$. Next, by collecting terms at
$\orderOne$ we get the linearized problem
\begin{equation}
 \partial_t W_1 = \LL(\mu_0; W_1)\,, \qquad 
 \begin{cases}
  D_u \partial_r U_1 = K_u(u_1 - U_1) \\
  D_{v0} \partial_r V_1 = K_{v0}(v_1 - V_1)
 \end{cases}, \quad r = R\,.
\end{equation}
Here the notation $\LL(\mu_0;\cdot)$ indicates that the linear
operator is evaluated at the bifurcation point. The solution of the
linearized {system} depends on the type of bifurcation and the
spatial mode considered. {We will consider the following
three cases:}
\begin{itemize}
\item \textbf{Hopf bifurcation}. The critical eigenvalues and spatial mode
  are respectively $\lambda = \pm i\lambda_I$ and $n=0$, which yields
\begin{equation}\label{eq:linearized_hopf}
   W_1 = \WW_0A_0(\tau)e^{i\lambda_I t} +
   \overline{\WW_0}\overline{A_0(\tau)}e^{-i\lambda_I t} \,,
\end{equation}
where the eigenfunction $\WW_0$ is evaluated at $\mu_0$ and $\lambda = i\lambda_I$.
\item \textbf{Pitchfork bifurcation}. The critical eigenvalue is
  $\lambda = 0$ and because of the reflection symmetry, if $n \neq 0$
  is a critical spatial mode, then {so is $-n$}. Moreover,
  it is known that the center manifold preserves the symmetries of the
  system. Therefore, the center eigenspace and manifold are also
  two-dimensional, and the solution in the linear regime has the form
 \begin{equation}\label{eq:linearized_turing}
  W_1 = \WW_nA_n(\tau) + \WW_{-n}\overline{A_n(\tau)}\,,
 \end{equation}
 where $\WW_{\pm n}$ are evaluated at $\mu_0$ and $\lambda = 0$.
\item \textbf{Pitchfork-Hopf bifurcation}. The critical spatial modes
  are $\left\{0,n,-n\right\}$, with $n\neq0$, and the bifurcating
  eigenvalues of the linearized problem are $\left\{\pm i\lambda_I,0\right\}$. This
  yields a four-dimensional center eigenspace of the form
 \begin{equation}\label{eq:linearized_turing_hopf}
   W_1 = \WW_0A_0(\tau)e^{i\lambda_I t} + \overline{\WW_0}
   \overline{A_0(\tau)}e^{-i\lambda_I t} + \WW_nA_n(\tau) +
   \WW_{-n}\overline{A_n(\tau)}\,.
 \end{equation}
 Again, all the eigenfunctions above are evaluated at the bifurcation
 point and the critical set of eigenvalues.
\end{itemize}
The goal of our analysis is to derive evolution equations for the
complex amplitudes $A_0$ and $A_n$, {where} the subscript
here indicates which spatial mode has exchanged stability through the
bifurcation.

{By collecting terms of order ${\mathcal O}(\epsilon^2)$ in
  \eqref{eq:collect_powers}, we obtain}
\begin{equation}\label{eq:order_2}
 \partial_t W_2 = \LL(\mu_0;W_2) + \BB(W_1,W_1) +
 \begin{pmatrix}
 0 \\
 \omega_v^2V_e(r)e_2^T\mu_1 \\
 0 \\ 
 -q_0(0)v_e e_1^T\mu_1
 \end{pmatrix}\,,
\end{equation}
{together with the appropriate boundary conditions, as
  obtained from \eqref{eq:collect_powers_BC}:}
\begin{equation}
 \begin{cases}
  D_u \partial_r U_2 = K_u(u_2 - U_2) \\
  D_{v0} \partial_r V_2 = K_{v0} (v_2 - V_2) + q_0(0)v_e\beta^T\mu_1
 \end{cases}\,, \quad r = R\,.
\end{equation}
The evaluation of the quadratic terms $\BB(W_1,W_1)$ will depend on
whether we consider the bifurcations \eqref{eq:linearized_hopf},
\eqref{eq:linearized_turing} or
\eqref{eq:linearized_turing_hopf}. Below, the nontrivial part of the
bilinear form is stated explicitly for each case.
\begin{itemize} 
 \item Hopf bifurcation:
 \begin{equation}
   B(\U_1,\U_1) = A_0^2 B(\bphi_0,\bphi_0) e^{2i\lambda_I t} +
   2|A_0|^2B(\bphi_0, \overline{\bphi_0}) +
   \overline{A_0}^2 B(\overline{\bphi_0},\overline{\bphi_0}) e^{-2i\lambda_I t}\,.
 \end{equation}
 \item Pitchfork bifurcation:
 \begin{equation}
   B(\U_1,\U_1) = A_n^2 B(\bphi_n,\bphi_n) e^{2in\theta} +
   2|A_n|^2B(\bphi_n, \bphi_{-n}) + \overline{A_n}^2 B(\bphi_{-n},\bphi_{-n})
   e^{-2in\theta}\,.
 \end{equation}
 \item Pitchfork-Hopf bifurcation:
 \begin{equation}
 \begin{split}
   B(\U_1,\U_1) &= A_0^2 B(\bphi_0,\bphi_0) e^{2i\lambda_I t} +
   2|A_0|^2B(\bphi_0, \overline{\bphi_0}) + \overline{A_0}^2
   B(\overline{\bphi_0},\overline{\bphi_0}) e^{-2i\lambda_I t} \\
   &+ A_n^2 B(\bphi_n,\bphi_n) e^{2in\theta} + 2|A_n|^2B(\bphi_n, \bphi_{-n}) +
   \overline{A_n}^2 B(\bphi_{-n},\bphi_{-n}) e^{-2in\theta} \\
   &+ 2A_0A_n B(\bphi_0,\bphi_n)e^{i(n\theta + \lambda_I t)} + 2A_0 \overline{A_n}
   B(\bphi_0,\bphi_{-n})e^{i(-n\theta + \lambda_I t)} \\
   &+ 2\overline{A_0}A_n B(\overline{\bphi_0},\bphi_n)e^{i(n\theta - \lambda_I t)}
   + 2\overline{A_0}\overline{A_n}
   B(\overline{\bphi_0},\bphi_{-n})e^{-i(n\theta + \lambda_I t)}\,.
 \end{split}
 \end{equation}
\end{itemize}
Once again, because of the reflection symmetry, we have that
$\bphi_n = \bphi_{-n}$. By examining these bilinear forms, the
following ansatz can be formulated for the solution of the system
\eqref{eq:order_2}:
\begin{itemize}
 \item Hopf bifurcation:
 \begin{equation}
   W_2 = W_{0000} + A_0^2 W_{2000}e^{2i\lambda_I t} + |A_0|^2W_{1100} +
   \overline{A_0}^2W_{0200}e^{-2i\lambda_I t}\,.
 \end{equation}
 \item Pitchfork bifurcation:
 \begin{equation}
  W_2 = W_{0000} + A_n^2 W_{0020} + |A_n|^2W_{0011} + \overline{A_n}^2W_{0002}\,.
 \end{equation}
 \item Pitchfork-Hopf bifurcation:
 \begin{equation}
  \begin{split}
    W_2 &= W_{0000} + A_0^2 W_{2000}e^{2i\lambda_I t} + |A_0|^2W_{1100} +
    \overline{A_0}^2W_{0200}e^{-2i\lambda_I t} \\
  &+ A_n^2 W_{0020} + |A_n|^2W_{0011} + \overline{A_n}^2W_{0002} \\
  &+ A_0A_nW_{1010}e^{i\lambda_I t} + A_0\overline{A_n}W_{1001} e^{i\lambda_I t} +
  \overline{A_0}A_nW_{0110}e^{-i\lambda_I t} +
  \overline{A_0}\overline{A_n}W_{0101} e^{-i\lambda_I t }\,.
 \end{split}
 \end{equation}
\end{itemize}

Next, we briefly outline the computation of the term $W_{0000}$. This
term arises from the perturbation of the bifurcation parameters within
the {base state}, and satisfies
\begin{equation}\label{eq:W_0000}
 \LL(\mu_0;W_{0000}) +  \begin{pmatrix}
 0 \\
 \omega_v^2V_e(r)e_2^T\mu_1 \\
 0 \\ 
 -q_0(0)v_e e_1^T\mu_1
 \end{pmatrix} = 0\,, \quad
 \begin{cases}
  D_u \partial_r U_{0000} = K_u(u_{0000} - U_{0000}) \\
  D_{v0} \partial_r V_{0000} = K_{v0} (v_{0000} - V_{0000}) + q_0(0)v_e\beta^T\mu_1
 \end{cases}, \quad r = R\,.
\end{equation}
Solving for $U_{0000}$, one obtains the same expression as for the
steady state profile, given by
\begin{equation}
 U_{0000} = A_0(0) \frac{I_0(\omega_u r)}{I_0(\omega_u R)} e_1^T \U_{0000}\,.
\end{equation}
Since the equation for $V_{0000}$ is forced by a multiple of the
steady state solution, the reduction of order method is used to yield
the following ansatz:
\begin{equation}
  V_{0000} = (\gamma_0 + \gamma_1(r))\frac{I_0(\omega_v r)}{I_0(\omega_v R)}\,,
  \quad \text{with}\quad  \gamma_1(0) = 0\,,
\end{equation}
where $\gamma_1(r)$ is found to satisfy the second-order differential equation
\begin{equation}\label{eq:gamma_1}
  \gamma_1^{\prime\prime}(r)I_0(\omega_vr) +
  \gamma_1^{\prime}(r)\left(2\omega_vI_1(\omega_vr) +
    \frac{1}{r}I_0(\omega_vr)\right) +
  \frac{\omega_v}{D_{v0}}e_2^T\mu_1B_0(0)I_0(\omega_vr)v_e = 0\,.
\end{equation}
{The ODE \eqref{eq:gamma_1} is readily solved using
  $rI_0(\omega_vr)$ as an integrating factor. By integrating twice, we obtain
  that}
\begin{equation}
  \gamma_1(r) = - \frac{e_2^T\mu_1B_0(0)v_e}{D_{v0}}\int_0^{\omega_v r}\rho
  \left[ 1 - \left(\frac{I_1(\rho)}{I_0(\rho)}\right)^2 \right]\,d\rho\,.
\end{equation}
Then, upon application of the perturbed boundary condition, we
determine the constant $\gamma_0$ as
\begin{equation}
  \gamma_0 = B_0(0)v_{0000} + \beta^T\mu_1 \frac{q_0(0)B_0(0)}{K_{v0}}v_e -
  \gamma_1(R) - \frac{D_{v0}B_0(0)}{K_{v0}}\gamma_1^{\prime}(R)\,.
\end{equation}
Next, the evaluation of $U_{0000}$ and $V_{0000}$ on the boundary leads to
\begin{equation}\label{eq:U_V_0000}
 \left. \begin{pmatrix}
  U_{0000} \\
  V_{0000}
 \end{pmatrix}\right|_{r=R} =
 \begin{pmatrix}
  A_0(0) u_{0000} \\
  B_0(0) v_{0000} + \Delta^T \mu_1
 \end{pmatrix}\,,
\end{equation}
where the coefficient of the detuning vector is given explicitly by
\begin{equation}
  \Delta = \frac{q_0(0)B_0(0)}{K_{v0}} e_1 + \frac{\omega_v^2 K_{v0}R
    (I_0^2(\omega_vR)-I_1^2(\omega_v R))/2 - \omega_v K_{v0}I_0(\omega_v R)
    I_1(\omega_v R)}{\left(D_{v0}\omega_vI_1(\omega_v R) +
      K_{v0}I_0(\omega_v R)\right)^2}e_2\,.
\end{equation}
{Finally, the substitution of \eqref{eq:U_V_0000} into the constraint
\eqref{eq:W_0000} for the membrane components determines
$\U_{0000}$ as}
\begin{equation}\label{eq:U_0000}
  \bPhi_0(\mu_0;0)\U_{0000} = \alpha^T\mu_1 E_2\U_e \qquad
  \Rightarrow \qquad \U_{0000} = \alpha^T\mu_1 [\bPhi_0(\mu_0;0)]^{-1}E_2\U_e\,.
\end{equation}
Here the vector coefficient $\alpha$ is defined by
\begin{align}
 \alpha &= q_0(0)e_1 - K_{v0}\Delta \nonumber \\
        &= (q_0(0))^2e_1 + \frac{\omega_v K_{v0}^2}
          {\left(D_{v0}\omega_v I_1(\omega_v R)
          + K_{v0}I_0(\omega_v R)\right)^2}
          \left( I_0(\omega_v R)I_1(\omega_v R) + \frac{\omega_v R}{2}
          (I_1^2(\omega_v R) - I_0^2(\omega_v R)) \right) e_2\,.
\end{align}
Many of the nontrivial $W_{jklm}$ can be found using the spatial and
temporal reflection symmetries of the reduced system. Here, the linear
{inhomogeneous problems} to be solved are listed below.

Starting with the Hopf bifurcation, where $W_{0200} = \overline{W_{2000}}$,
it is readily found that $W_{2000}$ satisfies
\begin{equation}
  \LL\left(\mu_0; W_{2000}\right) - 2i\lambda_I W_{2000} = -
  \BB(\WW_0,\WW_0) \qquad \Rightarrow \qquad W_{2000} =   
  \begin{pmatrix} 
  A_0(2i\lambda_I) \frac{I_0(\Omega_{2u}r)}{I_0(\Omega_{2u}R)} e_1^T \U_{2000} \\
  B_0(2i\lambda_I) \frac{I_0(\Omega_{2v}r)}{I_0(\Omega_{2v}R)} e_2^T \U_{2000} \\
  \U_{2000}
 \end{pmatrix}
\end{equation}
where $\Omega_{2u}$ and $\Omega_{2v}$ are defined by 
\begin{equation*}
  \Omega_{2u} = \sqrt{\frac{\sigma_u + 2i\lambda_I}{D_u}}\,, \qquad
  \Omega_{2v} = \sqrt{\frac{\sigma_v + 2i\lambda_I}{D_v}}\,.
\end{equation*}
Next, solving for $W_{1100}$ leads to 
\begin{equation}
  \LL\left(\mu_0; W_{1100}\right) = - 2\BB(\WW_0,\overline{\WW_0}) \qquad
  \Rightarrow \qquad W_{1100} =   
  \begin{pmatrix} 
  A_0(0) \frac{I_0(\omega_u r)}{I_0(\omega_u R)} e_1^T \U_{1100} \\
  B_0(0) \frac{I_0(\omega_v r)}{I_0(\omega_v R)} e_2^T \U_{1100} \\
  \U_{1100}
 \end{pmatrix}\,.
\end{equation}
Finally, $\U_{2000}$ and $\U_{1100}$ each satisfy the following
two-dimensional linear systems:
\begin{equation}
  \left[\bPhi_0(\mu_0;2i\lambda_I)\right]\U_{2000} = -B(\bphi_0,\bphi_0)\,,
\qquad \left[\bPhi_0(\mu_0;0)\right]\U_{1100} =-2B(\bphi_0,\overline{\bphi_0})\,.
\end{equation}

With regards to the pitchfork bifurcation, reflection symmetry
guarantees that $W_{0002} = \overline{W_{0020}}$. Hence, the systems
to be solved and their solutions are given by
\begin{equation}
  \LL\left(\mu_0; W_{0020}\right) = - \BB(\WW_n,\WW_n)\qquad \Rightarrow
  \qquad W_{0020} = 
 \begin{pmatrix}
  A_{2n}(0) \frac{I_{2n}(\omega_u r)}{I_{2n}(\omega_u R)} e_1^T \U_{0020} \\
  B_{2n}(0) \frac{I_{2n}(\omega_v r)}{I_{2n}(\omega_v R)} e_2^T \U_{0020} \\
  \U_{0020}
 \end{pmatrix}e^{2in\theta}\,,
\end{equation}
and 
\begin{equation}
  \quad \LL\left(\mu_0; W_{0011}\right) = - 2\BB(\WW_n,\WW_{-n})
 \qquad  \Rightarrow \qquad W_{0011} = 
 \begin{pmatrix}
  A_{0}(0) \frac{I_{0}(\omega_u r)}{I_{0}(\omega_u R)} e_1^T \U_{0011} \\
  B_{0}(0) \frac{I_{0}(\omega_v r)}{I_{0}(\omega_v R)} e_2^T \U_{0011} \\
  \U_{0011}
 \end{pmatrix}\,.
\end{equation}
{Here $\U_{0020}$ and $\U_{0011}$ satisfy the reduced linear
systems}
\begin{equation}
 \left[\bPhi_{2n}(\mu_0;0)\right]\U_{0020} = -B(\bphi_n,\bphi_n)\,, \qquad
 \left[\bPhi_0(\mu_0;0)\right]\U_{0011} = -2B(\bphi_n,\bphi_{-n})\,.
\end{equation}

The $W_{jklm}$ common to both the codimension-one and codimension-two
bifurcations remain the same. In addition to those terms, one needs to
solve for the mixed coefficients $W_{1010}$ and $W_{1001}$ as follows:
\begin{equation}
  \LL\left(\mu_0; W_{1010}\right) - i\lambda_I W_{1010} = - 2\BB(\WW_0,\WW_n)
  \quad \Rightarrow \quad W_{1010} = 
 \begin{pmatrix}
  A_n(i\lambda_I) \frac{I_{n}(\Omega_u r)}{I_{n}(\Omega_u R)} e_1^T \U_{1010} \\
  B_n(i\lambda_I) \frac{I_{n}(\Omega_v r)}{I_{n}(\Omega_v R)} e_2^T \U_{1010} \\
  \U_{1010}
 \end{pmatrix}e^{in\theta}\,,
\end{equation}
and 
\begin{equation}
  \LL\left(\mu_0; W_{1001}\right) - i\lambda_I W_{1001} = - 2\BB(\WW_0,\WW_{-n})
  \quad \Rightarrow \quad W_{1001} = 
 \begin{pmatrix}
  A_n(i\lambda_I) \frac{I_{n}(\Omega_u r)}{I_{n}(\Omega_u R)} e_1^T \U_{1001} \\
  B_n(i\lambda_I) \frac{I_{n}(\Omega_v r)}{I_{n}(\Omega_v R)} e_2^T \U_{1001} \\
  \U_{1001}
 \end{pmatrix}e^{-in\theta}\,,
\end{equation}
where $\U_{1010} = \U_{1001}$ (because $\bphi_{n} = \bphi_{-n}$), which
satisfies the further two-dimensional linear system
\begin{equation}
 \left[\bPhi_n(\mu_0;i\lambda_I)\right]\U_{1010} = -2B(\bphi_0,\bphi_n)\,.
\end{equation}
The remaining coefficients are found {trivially} using
symmetries, and are
\begin{equation}
 W_{0101} = \overline{W_{1010}}\,, \qquad W_{0110} = \overline{W_{1001}}\,.
\end{equation}

\subsection{Solvability condition and amplitude equations}

Upon collecting terms of $\orderThree$ in \eqref{eq:collect_powers}, we
obtain that
\begin{equation}\label{eq:order_3}
  \partial_t W_3 - \LL(\mu_0; W_3) = -\partial_\tau W_1 + 2\BB(W_1,W_2) +
  \CC(W_1,W_1,W_1) + \begin{pmatrix}
    0\\  e_2^T\mu_1 \frac{\partial_t + \sigma_v}{D_{v0}}V_1 \\ 0 \\
    -e_1^T\mu_1(v_1 - V_1|_{r=R}) \end{pmatrix}\,,
\end{equation}
while the $\orderThree$ terms in the boundary conditions
\eqref{eq:collect_powers_BC} yield
\begin{equation}\label{eq:BC_order_3}
 \left. \partial_r \begin{pmatrix}
             D_u U_3 \\
             D_{v0} V_3 
            \end{pmatrix}\right|_{r=R} =
            \begin{pmatrix}
             K_u(u_3-U_3|_{r=R}) \\
             K_{v0}(v_3-V_3|_{r=R})
            \end{pmatrix} +
            \begin{pmatrix}
             0 \\
             \beta^T\mu_1 (v_1 - V_1|_{r=R})
            \end{pmatrix}.
\end{equation}
Next, suitable ansatzes are formulated based on the method of
undetermined coefficients. Nonlinear evolution equations for the
amplitudes $A_0(\tau)$ and $A_n(\tau)$ will arise from the application
of a solvability condition on \eqref{eq:order_3} and
\eqref{eq:BC_order_3}.
\begin{itemize}
\item \textbf{Hopf bifurcation}. Since $n=0$ is the unstable mode, the
  solution is radially symmetric of the form
 \begin{equation}\label{eq:ansatz_hopf}
   W_3 = Xe^{i\lambda_I t} + \overline{X}e^{-i\lambda_I t}\,, \quad
   X = \begin{pmatrix} x_1(r) \\ x_2(r) \\ x_3 \\ x_4 \end{pmatrix}.
 \end{equation}
\item \textbf{Pitchfork bifurcation}. The bifurcating branch is
  stationary and spatially inhomogeneous (i.e.~angularly dependent)
 \begin{equation}\label{eq:ansatz_pitchfork}
   W_3 = Y, \quad Y = \begin{pmatrix} y_1(r,\theta) \\ y_2(r,\theta) \\
     y_3(\theta) \\ y_4(\theta) \end{pmatrix}.
 \end{equation}
\item \textbf{Pitchfork-Hopf bifurcation}. With $X$ and $Y$ defined as
  in the other cases, we have
 \begin{equation}
  W_3 = Xe^{i\lambda_I t} + \overline{X}e^{-i\lambda_I t} + Y\,.
 \end{equation}
\end{itemize}
When treating the Hopf bifurcation, equations \eqref{eq:order_3} and
\eqref{eq:BC_order_3} represent a forced oscillatory
system. {Typically, the presence of forcing with resonant
  terms generates secular growth of the solution. However, since
  boundedness of the solution is required on the fast time-scale,
  these secular terms must be eliminated for self-consistency of the
  multiple time-scale asymptotic expansion. This elimination is done
  using a solvability condition.}

Upon substituting \eqref{eq:ansatz_hopf} into \eqref{eq:order_3},
and equating coefficients of $e^{i\lambda_I t}$, we get
\begin{equation}
\begin{split}\label{eq:X_hopf}
  &i\lambda_I X - \LL(\mu_0;X) = - \WW_0\frac{dA_0}{d\tau} +
  \left(2\BB(\WW_0,W_{0000}) +
    \begin{pmatrix} 0 \\ e_2^T\mu_1 \Omega_v^2B_0(i\lambda_I)
      \frac{I_0(\Omega_v r)}{I_0(\Omega_v R)} \\ 0 \\
      -e_1^T\mu_1 q_0(i\lambda_I) \end{pmatrix}
    e_2^T\bphi_0 \right)A_0 \\
  &+ \left(2\BB(\WW_0,W_{1100}) + 2\BB\left(\overline{\WW_0},W_{2000}\right) +
    3\CC\left(\WW_0,\WW_0, \overline{\WW_0}\right)\right)|A_0|^2A_0\,,
\end{split}
\end{equation}
where $X$ satisfies the perturbed boundary condition given by
\begin{equation}\label{eq:BC_hopf}
 \left[\left. \partial_r \begin{pmatrix}
             D_u x_1 \\
             D_{v0} x_2 
            \end{pmatrix}\right|_{r=R} -
            \begin{pmatrix}
             K_u(x_3-x_1|_{r=R}) \\
             K_{v0}(x_4-x_2|_{r=R})
            \end{pmatrix}\right] = 
            \begin{pmatrix}
             0 \\
             \beta^T\mu_1 q_0(i\lambda_I) e_2^T\bphi_0
            \end{pmatrix}A_0\,.
\end{equation}
 
Since its bifurcating solution is stationary, a different argument
must be invoked when treating the pitchfork bifurcation. Again, the
substitution of \eqref{eq:ansatz_pitchfork} into \eqref{eq:order_3}
leads to
\begin{equation}
\begin{split}\label{eq:Y_turing}
  & -\LL(\mu_0;Y) = - \WW_n\frac{dA_n}{d\tau} + \left(2\BB(\WW_n,W_{0000}) +
    \begin{pmatrix} 0 \\ e_2^T\mu_1 \omega_v^2B_n(0)
      \frac{I_n(\omega_v r)}{I_n(\omega_v R)} \\ 0 \\ -e_1^T\mu_1 q_n(0)
    \end{pmatrix} e_2^T\bphi_n e^{in\theta} \right)A_n \\
  & + \left(2\BB(\WW_n,W_{0011}) + 2\BB\left(\WW_{-n},W_{0020}\right) +
    3\CC\left(\WW_n,\WW_n,\WW_{-n}\right)\right)|A_n|^2A_n +
  \mathcal{O}\left(\overline{A_n},\overline{A_n}^3,A_n^3\right)\,,
\end{split}
\end{equation}
where $Y$ also satisfies the nontrivial boundary condition
\begin{equation}\label{eq:BC_turing}
 \left[\left. \partial_r \begin{pmatrix}
             D_u y_1 \\
             D_{v0} y_2 
            \end{pmatrix}\right|_{r=R} -
            \begin{pmatrix}
             K_u(y_3-y_1|_{r=R}) \\
             K_{v0}(y_4-y_2|_{r=R})
            \end{pmatrix}\right] = 
            \begin{pmatrix}
             0 \\
             \beta^T\mu_1 q_n(0) e_2^T\bphi_n
            \end{pmatrix}A_n e^{in\theta} + \mathcal{O}(\overline{A_n})\,.
\end{equation}
Because of their orthogonality property, the contribution from other
circular modes will vanish when taking the inner-product with the
adjoint eigenfunction $\WW_{n}^{\star}$. Hence, these terms have not
been explicitly stated.

{The same procedure} is applied to the {codimension-two} case
where a pitchfork bifurcation interacts with a Hopf bifurcation. Equating
coefficients of $e^{i\lambda_It}$ leads to a linear inhomogeneous
equation for the coefficient $X$ given by
\begin{equation}
\begin{split}\label{eq:X_turing_hopf}
  &i\lambda_I X - \LL(\mu_0;X) = - \WW_0\frac{dA_0}{d\tau} +
  \left(2\BB(\WW_0,W_{0000}) + \begin{pmatrix} 0 \\
      e_2^T\mu_1 \Omega_v^2B_0(i\lambda_I)
      \frac{I_0(\Omega_v r)}{I_0(\Omega_v R)} \\ 0 \\
      -e_1^T\mu_1 q_0(i\lambda_I) \end{pmatrix} e_2^T\bphi_0 \right)A_0 \\
  &+ \left(2\BB(\WW_0,W_{1100}) + 2\BB\left(\overline{\WW_0},W_{2000}\right)
    + 3\CC\left(\WW_0,\WW_0, \overline{\WW_0}\right)\right)|A_0|^2A_0 \\
  &+ \left(2\BB(\WW_0,W_{0011}) + 2\BB(\WW_n,W_{1001}) + 2\BB(\WW_{-n},W_{1010})
    + 6\CC(\WW_0,\WW_n,\WW_{-n}) \right) |A_n|^2A_0 \\
 &+ \mathcal{O}\left(A_0A_n^2,A_0\overline{A_n}^2\right)\,,
\end{split}
\end{equation}
which is subject to the same boundary condition as for the Hopf
bifurcation (see equation \eqref{eq:BC_hopf}).

The term $Y$, which is constant on the fast time-scale, satisfies 
\begin{equation}\label{eq:Y_turing_hopf}
 \begin{split}
   & -\LL(\mu_0;Y) = - \WW_n\frac{dA_n}{d\tau} + \left(2\BB(\WW_n,W_{0000}) +
     \begin{pmatrix} 0 \\ e_2^T\mu_1 \omega_v^2B_n(0)
       \frac{I_n(\omega_v r)}{I_n(\omega_v R)} \\ 0 \\ -e_1^T\mu_1 q_n(0)
     \end{pmatrix} e_2^T\bphi_n e^{in\theta} \right)A_n \\
   & + \left(2\BB\left(\WW_n,W_{0011}\right) + 2\BB\left(\WW_{-n},W_{0020}\right)
     + 3\CC\left(\WW_n,\WW_n,\WW_{-n}\right)\right)|A_n|^2A_n \\
   & + \left(2\BB(\WW_n,W_{1100}) + 2\BB(\WW_0,W_{0110}) +
     2\BB\left(\overline{\WW_0},W_{1010}\right) +
     6\CC\left(\WW_0,\overline{\WW_0},\WW_n\right)\right) |A_0|^2A_n \\
   &+ \mathcal{O}\left(\overline{A_n},\overline{A_n}^3,A_n^3,|A_0|^2
     \overline{A_n}\right)\,,
 \end{split}
\end{equation}
and is subject to the same boundary condition as for the pitchfork bifurcation
(see equation \eqref{eq:BC_turing}).

The next two lemmas, involving our solvability conditions,
{provide necessary conditions} under which systems
\eqref{eq:X_hopf}, \eqref{eq:Y_turing}, \eqref{eq:X_turing_hopf} and
\eqref{eq:Y_turing_hopf}, together with the appropriate boundary conditions
\eqref{eq:BC_hopf} and \eqref{eq:BC_turing}, admit a solution. Lemma
\ref{lemma:general_solvability_condition} is general, while Lemma
\ref{lemma:solvability_condition} is specific to each bifurcation.

\begin{lemma}\label{lemma:general_solvability_condition}
{Let $\{\WW_n,\WW_n^{\star}\}$ be an orthogonal pair composed of an
  eigenfunction and its adjoint, and let $\lambda_c$ denote the
  critical eigenvalue at a given bifurcation point
  $\mu = \mu_0 \in \R^p$ ($p$ independent bifurcation parameters). Then,}
 \begin{equation}\label{eq:eigenfunctions_properties}
   \LL(\mu_0;\WW_n) = \lambda_c \WW_n\,, \qquad \LL^{\star}(\mu_0;\WW_n^{\star})
   = \overline{\lambda_c} \WW_n^{\star}\,, \qquad \lambda_c = \begin{cases}
    i\lambda_I & n = 0 \\
    0 & n \neq 0
   \end{cases}\,.
 \end{equation}
Consider the following linear inhomogeneous system,
 \begin{equation}\label{eq:linear_non_homogeneous}
  \lambda_c X - \LL(\mu_0;X) = \bm{\FF}
 \end{equation}
 where
 $X = \left(x_1(r,\theta),x_2(r,\theta),x_3(\theta),x_4(\theta)\right)^T$
 is subject to the inhomogeneous boundary condition
  \begin{equation}\label{eq:non_homogeneous_BC}
   \left[\left. \partial_r \begin{pmatrix}
             D_u x_1 \\
             D_v x_2 
            \end{pmatrix}\right|_{r=R} -
            \begin{pmatrix}
             K_u(x_3-x_1|_{r=R}) \\
             K_v(x_4-x_2|_{r=R})
            \end{pmatrix}\right] = 
            \begin{pmatrix}
             \xi(\theta) \\
             \eta(\theta)
            \end{pmatrix}.
  \end{equation}
  Then, a necessary condition for \eqref{eq:linear_non_homogeneous} to
  have a solution $X$ subject to  \eqref{eq:non_homogeneous_BC} is that
 \begin{equation}\label{eq:result}
   \left\langle \WW_n^{\star}, \bm{\FF} \right\rangle +
   \int_{\partial\Omega} \overline{U_n^{\star}}\xi \, d\sigma +
   \int_{\partial\Omega} \overline{V_n^{\star}}\eta \,d\sigma = 0\,,
 \end{equation}
 where $U_n^{\star}$ and $V_n^{\star}$ are the bulk components of the
 adjoint eigenfunction $\WW_n^{\star}$.
 \begin{proof}
   The result follows from a careful application of the Fredholm
   alternative, where the inhomogeneous boundary condition is taken
   into account. We take the inner-product with the adjoint
   eigenfunction on each side of \eqref{eq:linear_non_homogeneous} to get
 \begin{equation}\label{eq:step_1}
   \left\langle \WW_n^{\star}, \lambda_c X - \LL(\mu_0;X) \right\rangle =
   \lambda_c \left\langle \WW_n^{\star}, X \right\rangle -
   \left\langle \WW_n^{\star}, \LL(\mu_0;X) \right\rangle =
   \left\langle \WW_n^{\star}, \bm{\FF} \right\rangle \,.
 \end{equation}
 Next, by using the definition of the inner-product given in
 \eqref{eq:inner_product}, one can write
 \begin{align}
   & \langle \WW_n^{\star}, \LL(\mu_0; X)\rangle =
   \\ &\underbrace{\int_\Omega \overline
        {\begin{pmatrix} U_n^{\star} \\ V_n^{\star} \end{pmatrix}}^T
   \begin{pmatrix} D_u \Delta x_1 - \sigma_u x_1 \\
     D_v \Delta x_2 - \sigma_v x_2 \end{pmatrix} dA}_{{\star}}
   + \underbrace{\int_{\partial\Omega} \overline{\begin{pmatrix} u_n^{\star}
         \\ v_n^{\star} \end{pmatrix}}^T
   \begin{pmatrix} d_u \Delta_s x_3 - K_u \left(x_3 - x_1 \right) +
     f_u^e x_3 + f_v^e x_4 \\ d_v\Delta_s x_4 - K_v \left(x_4 - x_2 \right) +
     g_u^e x_3 + g_v^e x_4  \end{pmatrix} \, d\sigma}_{{\star}{\star}},
 \end{align}
 where $\Omega$ is a disk of radius $R$ and
 $\Delta_s = \partial_{\sigma\sigma}$ is the 1-D Laplace-Beltrami
 operator, with $\sigma = R \theta$ being the arc-length
 parameterizing the boundary. Next, using Green's second identity, the
 first integral $({\star})$ can be rewritten as follows:
 \begin{equation}\label{eq:part_1}
   \int_{\Omega}\overline{\begin{pmatrix} D_u \Delta U_n^{\star} -
       \sigma_u U_n^{\star} \\ D_v \Delta V_n^{\star} - \sigma_v V_n^{\star}
     \end{pmatrix}}^T \begin{pmatrix} x_1 \\ x_2 \end{pmatrix} dA
   + \int_{\partial\Omega} \left( \overline{\begin{pmatrix} U_n^{\star} \\
         V_n^{\star} \end{pmatrix}}^T \begin{pmatrix} K_u(x_3 - x_1) + \xi
       \\ K_v(x_4 - x_2) + \eta \end{pmatrix} - \begin{pmatrix} x_1
       \\ x_2 \end{pmatrix}^T \overline{\begin{pmatrix} K_u(u_n^{\star} -
   U_n^{\star}) \\ K_v(v_n^{\star} - V_n^{\star}) \end{pmatrix}}\right)d\sigma\,.
 \end{equation}
 In addition, the second integral can be treated using Lagrange's
 identity and the angular periodicity of surface bound components, as
 \begin{equation}\label{eq:part_2}
   {\star\star} = \int_{\partial\Omega} \overline{\begin{pmatrix} d_u\Delta_s
       u_n^* - K_u u_n^* + f_u^e u_n^{\star} + g_u^e v_n^{\star} \\
       d_u\Delta_s u_n^{\star} - K_v v_n^{\star} + f_v^e u_n^{\star} +
       g_v^e v_n^{\star} \end{pmatrix}}^T \begin{pmatrix} x_3 \\ x_4
   \end{pmatrix} d\sigma + \int_{\partial\Omega} \overline{\begin{pmatrix}
       K_u u_n^{\star} \\ K_v v_n^{\star} \end{pmatrix}}^T \begin{pmatrix}
     x_1 \\ x_2 \end{pmatrix} d\sigma \,.
 \end{equation}
 {Next, upon adding \eqref{eq:part_1} and \eqref{eq:part_2},
   we obtain after some algebra the following expression involving the
   adjoint linear operator:}
 \begin{align}\label{eq:step_2}
   \langle \WW_n^{\star}, \LL(\mu_0; X)\rangle &=
 \int_{\Omega}\overline{\begin{pmatrix} D_u \Delta U_n^{\star} -
     \sigma_u U_n^{\star} \\ D_v \Delta V_n^{\star} - \sigma_v V_n^{\star}
   \end{pmatrix}}^T \begin{pmatrix} x_1 \\ x_2 \end{pmatrix} dA \nonumber \\
 & + \int_{\partial\Omega} \overline{\begin{pmatrix} d_u\Delta_s u_n^{\star} -
     K_u \left(u_n^{\star} - U_n^{\star}\right) + f_u^e u_n^{\star} + g_u^e v_n^{\star}
     \\ d_u\Delta_s u_n^{\star} - K_v \left(v_n^{\star} - V_n^{\star}\right) +
     f_v^e u_n^{\star} + g_v^e v_n^{\star} \end{pmatrix}}^T
   \begin{pmatrix} x_3 \\ x_4 \end{pmatrix} d\sigma + \int_{\partial\Omega}
   \left(\overline{U_n^{\star}}\xi + \overline{V_n^{\star}}\eta \right)d\sigma
   \nonumber \\ &= \langle \LL^{\star}\left(\mu_0;\WW_n^{\star}\right) ,
  X \rangle + \int_{\partial\Omega}\left( \overline{U_n^{\star}}\xi +
      \overline{V_n^{\star}}\eta \right)d\sigma \nonumber \\
  &= \lambda_c \langle \WW_n^{\star}, X \rangle +
    \int_{\partial\Omega}\left( \overline{U_n^{\star}}\xi +
    \overline{V_n^{\star}}\eta \right) d\sigma\,.
 \end{align}
 The result \eqref{eq:result} is readily obtained after the
 substitution of \eqref{eq:step_2} back into \eqref{eq:step_1}.
 \end{proof}
\end{lemma}

\begin{lemma}\label{lemma:solvability_condition}
  (\textbf{Solvability condition}). {The imposition of the
    solvability condition for each of the three bifurcations leads to
    the following amplitude equations}:
 \begin{itemize}
 \item \textbf{Hopf bifurcation}. {A necessary condition for the
   inhomogeneous system \eqref{eq:X_hopf} and \eqref{eq:BC_hopf} to
   have a solution $X$ is that the amplitude $A_0(\tau)$ satisfies the
   following ODE:}
  \begin{align}
    & -2\pi R B_0(i\lambda_I)q_0(i\lambda_I)e_2^T\overline{\bphi_0^{\star}}
      e_2^T\bphi_0 \beta^T\mu_1 A_0 = -\frac{dA_0}{d\tau} +
      2\left\langle \WW_0^{\star}, \BB(\WW_0,W_{0000}) \right\rangle A_0
      \nonumber \\
    &+ 2\pi R e_2^T\overline{\bphi_0^{\star}}e_2^T\bphi_0
      \left( (B_0(i\lambda_I))^2 \Omega_v^2 \frac{R}{2}
      \left( 1 - \left(\frac{I_1(\Omega_v R)}{I_0(\Omega_v R)}\right)^2 \right)
      e_2^T\mu_1 - q_0(i\lambda_I) e_1^T\mu_1 \right)A_0 \nonumber \\
    &+ \left\langle \WW_0^{\star}, 2\BB\left(\WW_0,W_{1100}\right) +
      2\BB\left(\overline{\WW_0},W_{2000}\right) + 3\CC\left(\WW_0,\WW_0,
      \overline{\WW_0}\right) \right\rangle |A_0|^2A_0\,.
  \end{align}
\item \textbf{Pitchfork bifurcation}. {A necessary condition for the
  inhomogeneous system \eqref{eq:Y_turing} and \eqref{eq:BC_turing} to
  have a solution $Y$ is that the amplitude $A_n(\tau)$ satisfies the
  following ODE:}
  \begin{align}
    & -2\pi R B_n(0)q_n(0)e_2^T\bphi_n^{\star}e_2^T\bphi_n \beta^T\mu_1 A_n =
      -\frac{dA_n}{d\tau} + 2\left\langle \WW_n^{\star}, \BB(\WW_n,W_{0000})
      \right\rangle A_n \nonumber \\ 
    &+ 2\pi R e_2^T\bphi_n^{\star}e_2^T\bphi_n \left( (B_n(0))^2 \omega_v^2
      \frac{R}{2}\left(1 - \frac{I_{n-1}(\omega_v R)I_{n+1}(\omega_v R)}
      {(I_n(\omega_v R))^2} \right) e_2^T\mu_1 - q_n(0) e_1^T\mu_1 \right)
      A_n \nonumber \\
    &+ \left\langle \WW_n^{\star}, 2\BB\left(\WW_n,W_{0011}\right) +
      2\BB\left(\WW_{-n},W_{0020}\right) +
      3\CC\left(\WW_n,\WW_n,\WW_{-n}\right) \right\rangle |A_n|^2A_n\,.
  \end{align}
\item \textbf{Pitchfork-Hopf bifurcation}. {A necessary condition for
  the inhomogeneous system \eqref{eq:X_hopf}, \eqref{eq:Y_turing},
  \eqref{eq:BC_hopf}, and \eqref{eq:BC_turing} to have solutions $X$
  and $Y$ is that the amplitudes $A_0(\tau)$ and $A_n(\tau)$ satisfy
  the following system of ODEs:}
  \begin{align}
    &-2\pi R B_0(i\lambda_I)q_0(i\lambda_I)e_2^T\overline{\bphi_0^{\star}}
      e_2^T\bphi_0 \beta^T\mu_1 A_0 = -\frac{dA_0}{d\tau} + 2\left\langle
      \WW_0^{\star}, \BB(\WW_0,W_{0000}) \right\rangle A_0 \nonumber \\
    &+ 2\pi R e_2^T\overline{\bphi_0^{\star}}e_2^T\bphi_0
      \left( (B_0(i\lambda_I))^2 \Omega_v^2 \frac{R}{2}
      \left( 1 - \left(\frac{I_1(\Omega_v R)}{I_0(\Omega_v R)}\right)^2
      \right) e_2^T\mu_1 - q_0(i\lambda_I) e_1^T\mu_1 \right)A_0 \nonumber \\
    &+ \left\langle \WW_0^{\star}, 2\BB\left(\WW_0,W_{1100}\right) +
      2\BB\left(\overline{\WW_0},W_{2000}\right) + 3\CC\left(\WW_0,\WW_0,
      \overline{\WW_0}\right) \right\rangle |A_0|^2A_0 \nonumber \\
    &+ \left\langle \WW_0^{\star}, 2\BB(\WW_0,W_{0011}) + 2\BB(\WW_n,W_{1001}) +
  2\BB(\WW_{-n},\WW_{1010}) + 6\CC(\WW_0,\WW_n,\WW_{-n}) \right\rangle|A_n|^2A_0
  \end{align}
  and 
  \begin{align}
    & -2\pi R B_n(0)q_n(0)e_2^T\bphi_n^{\star}e_2^T\bphi_n \beta^T\mu_1 A_n
      = -\frac{dA_n}{d\tau} + 2\left\langle \WW_n^{\star},
      \BB(\WW_n,W_{0000})\right\rangle A_n \nonumber \\ 
    &+ 2\pi R e_2^T\bphi_n^{\star}e_2^T\bphi_n
      \left( (B_n(0))^2 \omega_v^2 \frac{R}{2}
      \left(1 - \frac{I_{n-1}(\omega_v R)I_{n+1}(\omega_v R)}
      {(I_n(\omega_v R))^2} \right) e_2^T\mu_1 - q_n(0) e_1^T\mu_1 \right) A_n
      \nonumber \\
    &+ \left\langle \WW_n^{\star}, 2\BB\left(\WW_n,W_{0011}\right) +
      2\BB\left(\WW_{-n},W_{0020}\right) + 3\CC\left(\WW_n,\WW_n,\WW_{-n}\right)
      \right\rangle |A_n|^2A_n \nonumber \\
    &+ \left\langle \WW_n^{\star}, 2\BB(\WW_n,W_{1100}) +
      2\BB\left(\WW_0,W_{0110}\right) +
      2\BB\left(\overline{\WW_0},W_{1010}\right) +
      6\CC\left(\WW_0,\overline{\WW_0},\WW_n\right) \right\rangle|A_0|^2A_n\,.
  \end{align}
 \end{itemize}
 In each of these three cases, $\WW_0^{\star}$ and $\WW_n^{\star}$ are
 adjoint eigenfunctions associated with the adjoint linear
 operator. Hence, they satisfy the following relations:
\begin{equation}
  \LL^{\star}(\mu_0,\WW_0^{\star}) = -i\lambda_I\WW_0^{\star}\,, \qquad
  \LL^{\star} (\mu_0,\WW_n^{\star}) = 0\,.
\end{equation}
\begin{proof}
  The results stated above are obtained from a direct application of
  Lemma \ref{lemma:general_solvability_condition}.
\end{proof}
\end{lemma}

{The rearrangement of each of the differential equations
  given in Lemma \ref{lemma:solvability_condition} provides a system
  of amplitude equations describing the branching behavior in the
  vicinity of the bifurcation point, in the limit $\eps\to 0$. For each
  of the three bifurcations the amplitude equations have the following
  form:}
\begin{itemize}
 \item \textbf{Hopf bifurcation}.
 \begin{equation}\label{eq:amplitude_hopf}
  \frac{dA_0}{d\tau} = g_{1000}^T\mu_1 A_0 + g_{2100} |A_0|^2A_0\,.
 \end{equation}
 \item \textbf{Pitchfork bifurcation}.
 \begin{equation}\label{eq:amplitude_pitchfork}
  \frac{dA_n}{d\tau} = g_{0010}^T\mu_1 A_n + g_{0021} |A_n|^2A_n\,.
 \end{equation}
\item \textbf{Pitchfork-Hopf bifurcation}. In this case, one shall
  analyze a system of two amplitude equations similar to their
  codimension-one analog. In each equation, the additional term
  corresponds to the mixed-mode term (all the other terms remain the
  same as for their codimension-one counterpart).
 \begin{subequations}\label{eq:amplitude_pitchfork_hopf}
  \begin{align}
    \frac{dA_0}{d\tau} &= g_{1000}^T\mu_1 A_0 + g_{2100} |A_0|^2A_0 +
                         g_{1011}|A_n|^2A_0 \,, \\
    \frac{dA_n}{d\tau} &= g_{0010}^T\mu_1 A_n + g_{0021} |A_n|^2A_n +
                         g_{1110}|A_0|^2A_n\,.
  \end{align}
 \end{subequations}
\end{itemize}

{Explicit expressions for the coefficients of the nonlinear terms in
the amplitude equations are}
\begin{subequations}\label{eq:nonlinear_coefficients}
 \begin{align}
   g_{2100} &= \left\langle \WW_0^{\star}, 2\BB\left(\WW_0,W_{1100}\right) +
              2\BB\left(\overline{\WW_0},W_{2000}\right) +
              3\CC\left(\WW_0,\WW_0, \overline{\WW_0}\right) \right\rangle, \\
   g_{1011} &= \left\langle \WW_0^{\star}, 2\BB(\WW_0,W_{0011}) +
              2\BB(\WW_n,W_{1001}) + 2\BB(\WW_{-n},\WW_{1010}) +
              6\CC(\WW_0,\WW_n,\WW_{-n}) \right\rangle, \\
   g_{0021} &= \left\langle \WW_n^{\star}, 2\BB\left(\WW_n,W_{0011}\right) +
              2\BB\left(\WW_{-n},W_{0020}\right) +
              3\CC\left(\WW_n,\WW_n,\WW_{-n}\right) \right\rangle, \\
   g_{1110} &= \left\langle \WW_n^{\star}, 2\BB(\WW_n,W_{1100}) +
              2\BB\left(\WW_0,W_{0110}\right) +
              2\BB\left(\overline{\WW_0},W_{1010}\right) +
              6\CC\left(\WW_0,\overline{\WW_0},\WW_n\right) \right\rangle\,,
 \end{align}
\end{subequations}
where $g_{2100},g_{1011} \in \C$ and $g_{0021},g_{1110} \in \R$. The
coefficients of the linear terms consist of the projection of the
vectors $g_{1000} \in \C^2$ and $g_{0010} \in \R^2$ onto the detuning
unit vector $\mu_1$. They are given by
\begin{subequations}\label{eq:linear_coefficients}
\begin{align}
  g_{1000} &= 2\pi R e_2^T\overline{\bphi_0^{\star}}e_2^T\bphi_0
             \left( B_0(i\lambda_I)q_0(i\lambda_I)\beta +
             (B_0(i\lambda_I))^2 \Omega_v^2 \frac{R}{2}
             \left( 1 - \left(\frac{I_1(\Omega_v R)}{I_0(\Omega_v R)}\right)^2
             \right) e_2 - q_0(i\lambda_I) e_1\right) \nonumber \\
           & + 4\pi R \overline{\bphi_0^{\star}}^T
             B\left(\bphi_0,\tilde{\U}_{0000}\right) \alpha\,, \\
  g_{0010} &= 2\pi R e_2^T\bphi_n^{\star}e_2^T\bphi_n \left(B_n(0)q_n(0)\beta +
             (B_n(0))^2 \omega_v^2 \frac{R}{2}\left(1 - \frac{I_{n-1}(\omega_v R)
             I_{n+1}(\omega_v R)}{(I_n(\omega_v R))^2} \right) e_2 -
             q_n(0) e_1 \right) \nonumber \\
& + 4\pi R {\bphi_n^{\star}}^TB\left(\bphi_n,\tilde{\U}_{0000}\right) \alpha\,,
\end{align}
\end{subequations}
where $\tilde{\U}_{0000} = [\bPhi_0(\mu_0;0)]^{-1}E_2\U_e$ appears in
the expression for $\U_{0000}$ (see equation \eqref{eq:U_0000}). The
two quantities are related to each other by
\begin{equation}
 \U_{0000} = \tilde{\U}_{0000}\alpha^T\mu_1\,.
\end{equation}
{In \S \ref{sec:validation_weakly_nonlinear} below, the
  solution behaviors in the weakly nonlinear regime, as predicted
  after the numerical evaluation of the coefficients in the amplitude
  equations, will be compared with numerical PDE simulations of the
  full coupled bulk-surface reaction-diffusion model.}

{We remark that the generic form for the coefficients of the
  cubic terms of \eqref{eq:amplitude_hopf} -
  \eqref{eq:amplitude_pitchfork_hopf} is well-known in the
  literature. Since the pitchfork bifurcation is simply a Hopf
  bifurcation with a zero-crossing eigenvalue, these coefficients also
  arise from the weakly nonlinear analysis of a codimension-two
  Hopf-Hopf bifurcation, in which the pair of critical eigenfunctions
  is fully complex. We refer the reader to the appendix of
    \cite{gou2016DH}, where a center manifold reduction was used to
    derive a two-dimensional system of amplitude equations at a
    double-Hopf bifurcation.}
    
{We have performed a
    systematic derivation yielding explicit formulae for the
    normal form coefficients for the three distinct bifurcations in
    our coupled bulk-surface model. This involved a
    careful analysis of the underlying spectral
    problem. These normal form results can be used for any nonlinear
    surface reaction kinetics specified on the circular boundary. 
    The bifurcation parameters employed here, being the
    bulk diffusion coefficient $D_v$ and the coupling rate constant
    $K_v$, are used below to understand the specific role of
    bulk-surface coupling. In our derivation above we have allowed for
    an arbitrary sweep of the two-parameter vector $(K_v,D_v)$ across
    linear stability boundaries. Our choice of performing an
    asymptotic multiple time-scale analysis, as opposed to a center
    manifold reduction, was partially motivated by computational convenience and flexibility. The main advantage of multi-scale theory concerns the ease with 
    which bifurcation parameters can be treated if they appear in the
    boundary conditions \eqref{eq:BC}.
    By simply keeping track of the terms at each order
    within the boundary conditions, the method of multiple time-scales
    provides a direct approach to compute the normal form.}

\section{Bifurcation analysis of amplitude equations}\label{sec:bifurcation_analysis}

{In this section, the equilibria of the amplitude equations
  derived in the preceding section are analyzed. Under nondegeneracy
  conditions, the stability properties of the steady states in the
  normal form are {preserved} in the full model. Bifurcations
  of codimension-one and two are treated separately below.}

\subsection{Codimension-one: Hopf and pitchfork bifurcations}\label{subsec:theory_codim_1}

For a Hopf bifurcation, the substitution of
$A_0(\tau) = \rho_0(\tau)e^{i\theta_0(\tau)}$ within
\eqref{eq:amplitude_hopf} yields a coupled system of ODEs for the
{magnitude $\rho_0$ and the phase $\theta_0$, given by
\begin{equation}\label{eq:rho}
  \frac{d\rho_0}{d\tau} = \left[\Re\left(g_{1000}\right)\right]^T\mu_1 \rho_0
  + \Re\left(g_{2100}\right)\rho_0^3 \,, \qquad
  \frac{d\theta_0}{d\tau} = \left[\Im\left(g_{1000}\right)\right]^T\mu_1 +
  \Im\left(g_{2100}\right)\rho_0^2\,.
\end{equation}
When it exists, a steady state $\rho_{0e}$ of \eqref{eq:rho} is given by}
\begin{equation}\label{eq:rho0}
  \rho_{0e} = \sqrt{-\frac{\left[\Re\left(g_{1000}\right)\right]^T\mu_1}
    {\Re\left(g_{2100}\right)}}\,,
\end{equation}
where $\mu_1$ is a free two-dimensional unit vector. The
transversality condition is violated when $\mu_1$ is tangent to the
Hopf stability boundary at $\mu_0 \in \R^2$, for which $\rho_{0e}$
from \eqref{eq:rho0} vanishes. Consequently, it can easily be argued
that the vector $\Re(g_{1000})$ is normal to the stability curve and
that a natural choice for the orientation of $\mu_1$ is
\begin{equation}
  \mu_1 = - \frac{\Re(g_{2100})}{|\Re(g_{2100})|}\frac{\Re(g_{1000})}
  {\|\Re(g_{1000})\|} \,.
\end{equation}
{Such a choice provides the maximal magnitude $\rho_{0e}$ given by}
\begin{equation}\label{eq:rho_0_e}
 \rho_{0e} = \sqrt{\frac{\|\Re\left(g_{1000}\right)\|}{|\Re\left(g_{2100}\right)|}}.
\end{equation}
The nontrivial fixed points $\rho_{0e}$ in \eqref{eq:rho0} correspond
to limit cycle of the complex amplitude $A_0(\tau)$. The stability of
the limit cycles is given by the sign of the real part of the cubic
term coefficient in \eqref{eq:rho}. Hence, one can distinguish between
supercritical and subcritical bifurcations {in the usual way:
\begin{equation}
  \Re(g_{2100}) < 0 \,\,\, \mbox{(Supercritical Hopf)}\,, \qquad
  \Re(g_{2100}) > 0 \,\,\, \mbox{(Subcritical Hopf)}\,.
\end{equation}
}Next, from substituting $\rho_{0e}$ into the
equation for the phase in (\ref{eq:rho}), the steady state phase
$\theta_0$ is
\begin{equation}
  \theta_0(\tau) = \theta_0(0) + \tilde{\theta}_0 \tau\,, \qquad
  \text{with} \qquad \tilde{\theta}_0 =
  \left[\Im\left(g_{1000}\right)\right]^T\mu_1 +
  \Im\left(g_{2100}\right)\rho_{0e}^2\,,
\end{equation}
where $\theta_0(0) \in \R$ denotes the phase shift symmetry of
periodic solutions.

The following lemma uses the nontrivial steady state from the
amplitude equation associated with the Hopf bifurcation to give
approximate periodic solutions of the full coupled bulk-surface PDE
model in the weakly nonlinear regime:

\begin{lemma}\label{lemma:family_periodic_weakly}
  \textbf{(Periodic solutions in the weakly nonlinear regime)}. Let
  $g_{2100} \in \C$ be the cubic term coefficient in 
  \eqref{eq:amplitude_hopf}, and assume that its real part is nonzero,
  hence excluding degenerate cases. Then, in the limit 
  $\eps \to 0$ with $\eps = \sqrt{\|\mu - \mu_0\|}$ denoting the
  square-root of the distance with the Hopf bifurcation
  point, a leading-order approximate family of periodic solutions
  is given by
 \begin{equation}\label{eq:periodic_solution}
   W(t) = W_e + \eps \rho_{0e} \left[ \WW_0 e^{i\left(\lambda_I t + \theta_0(0)\right)}
     + \overline{\WW_0} e^{-i\left(\lambda_It + \theta_0(0)\right)} \right] + \orderTwo,
 \end{equation}
 for any $\theta_0(0) \in \R$ and with $\rho_{0e}$ defined by \eqref{eq:rho_0_e}.
 
{For the surface-bound activator species, let
 $u_\text{amp}$ denote the amplitude of the bifurcating limit cycle
 near the Hopf bifurcation point. A leading-order
  approximation for $u_\text{amp}$ is}
 \begin{equation}\label{eq:u_amp}
   u_{\text{amp}} = \max_{0 \leq t < T_p} \left\{ |u(t) - u_e| \right\}=
   2\eps \rho_{0e}|e_1^T\bphi_0| + \orderTwo,
 \end{equation}
{where the period $T_p$ satisfies
 \begin{equation}\label{eq:period}
   T_p = \frac{2\pi}{\lambda_I} + \orderTwo.
 \end{equation}
 Finally, the periodic solution in \eqref{eq:periodic_solution}
is asymptotically stable when $\Re(g_{2100}) < 0$ (supercritical Hopf)
and it is unstable for $\Re(g_{2100}) > 0$ (subcritical
Hopf).}
\end{lemma}

{Next, we consider the pitchfork bifurcation. Since the coefficients
$g_{0010}$ and $g_{0021}$ in \eqref{eq:amplitude_pitchfork} are real,
we set $A_n(\tau) = \rho_n(\tau)e^{i\theta_n(\tau)}$ into
\eqref{eq:amplitude_pitchfork} to obtain a decoupled system of ODEs
for $\rho_n$ and $\theta_n$ given by
\begin{equation}\label{eq:rho_theta_n}
  \frac{d\rho_n}{d\tau} = g_{0010}^T\mu_1 \rho_n + g_{0021}\rho_n^3\,,
  \qquad \frac{d\theta_n}{d\tau} = 0 \,.
\end{equation}
When they exist, fixed points of \eqref{eq:rho_theta_n} are}
\begin{equation}
  \rho_{ne} = \sqrt{-\frac{g_{0010}^T \mu_1}{g_{0021}}}\,, \qquad \theta_{n}(\tau)
  \equiv \theta_n \in \R\,,
\end{equation}
where the constant $\theta_n$ accounts for the rotational equivariance
of the pattern. The orientation of the unit vector {$\mu_1$ is
chosen in a similar way as for the Hopf bifurcation, with the
"natural choice" being}
\begin{equation}
 \mu_1 = - \frac{g_{0021}}{|g_{0021}|}\frac{g_{0010}}{\|g_{0010}\|} \,.
\end{equation}
This yields the following expression for $\rho_{ne}$:
\begin{equation}\label{eq:rho_n_e}
 \rho_{ne} = \sqrt{\frac{\|g_{0010}\|}{|g_{0021}|}}\,.
\end{equation}

Again, the pitchfork bifurcation is classified according to the
stability of the nontrivial steady state of the amplitude equation. It
{readily follows that
\begin{equation}
  g_{0021} < 0  \,\,\, \mbox{(Supercritical pitchfork)} \,, \qquad
  g_{0021} > 0 \,\,\, \mbox{(Subcritical pitchfork)}\,.
\end{equation}
}
Finally, this section is concluded with the analog of Lemma
\ref{lemma:family_periodic_weakly} for Turing-type patterns arising
from pitchfork bifurcations in the full coupled bulk-surface PDE model.

\begin{lemma}
  \textbf{(Spatially inhomogeneous equilibria in the weakly nonlinear
    regime)}. Let $g_{0021} \in \R$ be the cubic term coefficient in
  \eqref{eq:amplitude_pitchfork}, and assume that it is
  nonzero, hence excluding degenerate cases. {Then, in the limit 
  $\eps \to 0$ with $\eps = \sqrt{\|\mu - \mu_0\|}$ denoting the
  square-root of the distance from the \textbf{pitchfork} bifurcation
  point, a leading-order approximate family of spatially inhomogeneous
  equilibria is given by}
 \begin{equation}\label{eq:patterned_solution}
   W = W_e + \eps \rho_{ne} \left[ \WW_n e^{i\theta_n} +
     \WW_{-n} e^{-i\theta_n} \right] + \orderTwo,
 \end{equation}
 for any $\theta_n \in \R$ and with $\rho_{ne}$ defined by \eqref{eq:rho_n_e}. 
 
{For the surface-bound activator species, let $u_\text{amp}$ be the
  amplitude of the bifurcating Turing-type pattern near the pitchfork
  bifurcation point. A leading-order approximation for $u_\text{amp}$
 is}
 \begin{equation}
   u_{\text{amp}} = \max_{0 \leq \theta < 2\pi} \left\{ |u(\theta) - u_e| \right\}
   = 2\eps \rho_{ne}|e_1^T\bphi_n| + \orderTwo.
 \end{equation}
 
 {Finally, the patterned solution given by
   \eqref{eq:patterned_solution} is asymptotically stable when
   $g_{0021} < 0$ (supercritical pitchfork), and it is unstable for
   $g_{0021} > 0$ (subcritical pitchfork).}
\end{lemma}

\subsection{Codimension-two: Pitchfork-Hopf bifurcation}\label{subsec:theory_codim_2}

In this subsection, the equilibria of the coupled amplitude equations
\eqref{eq:amplitude_pitchfork_hopf} are analyzed for general values of
its coefficients. For this purpose, it is appropriate to rescale the
time variable into the original fast time-scale $t$. Hence, letting
$z_0(t) = \eps A_0(\eps^2 t)e^{i\lambda_It}$ and
$z_n(t) = \eps A_n(\eps^2t)$, and recalling that
$\mu - \mu_0 = \eps^2 \mu_1$, the following Poincaré normal form can
readily be obtained:
\begin{subequations}
 \begin{align}
   \dot{z}_0 &= \left(i\lambda_I + g_{1000}^T(\mu - \mu_0) \right)z_0
               + g_{2100}|z_0|^2z_0 + g_{1011}|z_n|^2z_0\,, \\
 \dot{z}_n &= g_{0010}^T(\mu - \mu_0)z_n + g_{1110}|z_0|^2z_n +g_{0021}|z_n|^2z_n\,.
 \end{align}
\end{subequations}
Since $g_{2100},g_{1011}$ are complex-valued, while
$g_{0021},g_{1110}$ are real-valued, cylindrical polar coordinates are
appropriate for this normal form. Letting $z_0 = r e^{i\phi}$,
$z_n = w \geq 0$, and $\mu = \mu_0$ we obtain a system of ODEs for the
{magnitudes $w,r$ and the phase $\phi$, given by
\begin{equation}\label{eq:normal_form_phase}
  \dot{w} = g_{0021}w^3 + g_{1110}wr^2\,, \qquad
  \dot{r} = \Re(g_{1011})w^2r + \Re(g_{0021})r^3 \,, \qquad
   \dot{\phi} = \lambda_I + \Im(g_{2100})r^2 + \Im(g_{1011})w^2\,.
\end{equation}
Since the third equation above is decoupled from the others,
  it need not be considered. The first two ODEs in
 (\ref{eq:normal_form_phase}) can then be conveniently written as
\begin{equation}\label{eq:normal_form}
  \dot{w} = p_{11}w^3 + p_{12}wr^2\,, \qquad
  \dot{r} = p_{21}w^2r + p_{22}r^3\,,
\end{equation}
}where $p_{11} = g_{0021}, p_{12} = g_{1110}, p_{21} = \Re(g_{1011})$
and $p_{22} = \Re(g_{2100})$. It is assumed that these coefficients
are nonzero and that $p_{11}p_{22} - p_{12}p_{21} \neq 0$, which is a
necessary condition for the existence of the mixed-mode steady state.

To relate our results with those in \cite{guckenheimer1983}, we
{augment \eqref{eq:normal_form} with linear terms to obtain
\begin{equation}\label{eq:perturbed_normal_form}
  \dot{w} = w(\delta_1 + p_{11}w^2 + p_{12}r^2)\,, \qquad
  \dot{r} = r(\delta_2 + p_{21}w^2 + p_{22}r^2) \,,
\end{equation}
}where $\delta_1$ and $\delta_2$ are generic unfolding
parameters. This recovers the well-known canonical truncated system of
amplitude equations at a codimension-two pitchfork-Hopf and
double-Hopf bifurcations. We refer the reader to
\cite{guckenheimer1983} and \cite{kuznetsov2004}, where the phase
portraits of \eqref{eq:perturbed_normal_form} is classified. In the
discussion below, the approach and classification from
\cite{guckenheimer1983} is followed.

We first reduce the number of parameters in
\eqref{eq:perturbed_normal_form} by setting
$\bar{w} = \sqrt{|p_{11}|}w$ and $\bar{r} = \sqrt{|p_{22}|}r$. After a
possible time rescaling if $p_{11} < 0$, and further dropping the bars
to simplify the notation, {this change of variable yields
\begin{equation}\label{eq:rescaled_perturbed_normal_form}
  \dot{w} = w(\delta_1 + w^2 + \gamma r^2)\,, \qquad
  \dot{r} = r(\delta_2 + \eta w^2 + d r^2)\,,
\end{equation}
}where $\gamma, \eta$ and $d$ are given by
\begin{equation}\label{eq:scaling}
  \gamma = \frac{p_{12}}{|p_{22}|}\,, \qquad \eta = \frac{p_{21}}{|p_{11}|}\,,
  \qquad d = \frac{p_{22}}{|p_{22}|} = \pm 1\,.
\end{equation}
Depending on the signs of the four quantities
$d,\gamma,\eta$ and
$d-\gamma\eta$, it is possible to distinguish between 12 topologically
different stability diagrams in the plane of generic parameters
$(\delta_1,\delta_2)$ (see \cite{guckenheimer1983}). In addition to
the trivial equilibrium $E_0 =
(0,0)$, \eqref{eq:rescaled_perturbed_normal_form} possesses up to
three additional steady states. Two of those equilibria are located on
the coordinate axes and are given by
\begin{equation}
  E_1 = \left(\sqrt{-\delta_1},0\right) \quad \text{for } \delta_1 < 0\,,
  \qquad \text{and} \quad E_2 = \left(0,\sqrt{-\frac{\delta_2}{d}}\right)
  \quad \text{for } \frac{\delta_2}{d} < 0\,,
\end{equation}
while the last one corresponds to the mixed-mode equilibrium defined by
\begin{equation}
  E_3 = \left(\sqrt{\frac{\gamma\delta_2 - d \delta_1}{d - \gamma \eta}},
    \sqrt{\frac{\eta\delta_1 - \delta_2}{d - \gamma \eta}} \right) \qquad
  \text{for } \quad \frac{\gamma\delta_2 - d \delta_1}{d - \gamma \eta},
  \frac{\eta\delta_1 - \delta_2}{d - \gamma \eta} > 0 \,.
\end{equation}
{With regards to the coupled bulk-surface model, these equilibria
have a precise meaning in terms} of oscillatory and patterned
solutions. They are classified below in Table
\ref{table:steady_states}.

\begin{table}[htbp]
\centering
\caption{\label{table:steady_states} Classification of steady states
  of \eqref{eq:rescaled_perturbed_normal_form} and solution
  correspondence.}
\begin{tabular}{|c|c|}
\hline
\textbf{Amplitude solution} & \textbf{Full model solution} \\ \hline
Trivial steady state $E_0$ & Base-state solution $W_e$ \\ \hline
First mode $E_1$ & Spatially inhomogeneous steady state \\ \hline
Second mode $E_2$ & Periodic solution around the base state solution \\ \hline
Mixed-mode $E_3$ & Oscillations around a patterned solution (breather solution) \\ \hline
\end{tabular}
\end{table}

{We now summarize the bifurcations arising in the
truncated system of amplitude equations
\eqref{eq:rescaled_perturbed_normal_form}. Firstly,} the single
mode equilibria $E_1$ and $E_2$ bifurcate from the origin on the lines
\begin{equation}\label{eq:H_1_2}
  H_1 = \left\{(\delta_1,\delta_2) | \delta_1 = 0\right\}\,,
  \qquad H_2 = \left\{(\delta_1,\delta_2) | \delta_2 = 0\right\},
\end{equation}
and, respectively, exist for $\delta_1<0$ and for
${\delta_2/d} < 0$. Next, the mixed-mode equilibrium $E_3$
bifurcates from each of the single mode equilibria on the
semi-infinite pitchfork lines, given by
\begin{subequations}\label{eq:pitchfork_lines}
\begin{align}
  T_1 &= \left\{(\delta_1,\delta_2) \left| \,\, \gamma \delta_2 = d\delta_1\,,
        \,\,\, \delta_1 > 0 \text{ or } \delta_1 < 0 \right. \right\}, \\
  T_2 &= \left\{(\delta_1,\delta_2) \left| \,\, \delta_2 = \eta \delta_1\,,
        \,\,\, \delta_1 > 0 \text{ or } \delta_1 < 0 \right. \right\}\,,
\end{align}
\end{subequations}
whose orientations (whether for each case $\delta_1$ is positive or
negative) are chosen such that $T_1$ and $T_2$ form the boundaries of
the existence region of $E_3$.

Cases for which $d = 1$ are known to be simple since the truncated
system of amplitude equations
\eqref{eq:rescaled_perturbed_normal_form} does not possess any limit
cycles. Such a situation corresponds to both pitchfork and Hopf
bifurcations being either subcritical or supercritical, with no
fifth-order terms being needed in the normal form. More intricate
cases {arise} when $d = -1$ and $d-\gamma\eta > 0$, at which
$E_3$ bifurcates through a degenerate Hopf bifurcation on the
semi-infinite line given by
\begin{equation}\label{eq:hopf_line}
 C = \left\{ (\delta_1,\delta_2) \left| \delta_2 = \frac{d(1-\eta)}{\gamma - d} \delta_1, \quad \delta_1 > 0 \text{ or } \delta_1 < 0 \right. \right\}.
\end{equation}
The stability of the bifurcating limit cycle is typically determined
{by including} one fifth-order term to the normal form. More
details on this challenging computation are found in the classic
reference \cite{guckenheimer1983}.

{In the next lemma, an affine transformation mapping parameter spaces
is defined.}

\begin{lemma}\label{lemma:map}
  \textbf{(Mapping parameter spaces)}. Let $\TT$ be an affine
  transformation in $\R^2$ that maps the generic parameter space
  defined by $\delta_1$ and $\delta_2$ to the original bifurcation
  parameter space defined by $K_v$ and $D_v$. Then, it must satisfy
 \begin{align}\label{eq:affine_map}
   \TT(\delta_1,\delta_2) &= \mu_0 + \RR\left(\pm\frac{\pi}{2}\right)
 \begin{bmatrix} g_{0010} & \Re(g_{1000}) \end{bmatrix}
 \RR\left(\pm\frac{\pi}{2}\right) \begin{pmatrix} \delta_1 \\ \delta_2
 \end{pmatrix} \nonumber \\
  &= \mu_0 + \begin{bmatrix} \RR\left(\dfrac{\pi}{2}\right)\Re(g_{1000})
    & \RR\left(-\dfrac{\pi}{2}\right)g_{0010} \end{bmatrix}
      \begin{pmatrix} \delta_1 \\ \delta_2 \end{pmatrix},
 \end{align}
 \noindent where $\mu_0 = (K_{v0},D_{v0})^T$ is the codimension-two
 bifurcation point and $\RR(\varphi)$ is the anti-clockwise rotation
 matrix in the Euclidean plane defined by
\begin{equation}\label{eq:rotation_matrix}
 \RR (\varphi) = \begin{pmatrix}
                          \cos(\varphi) & - \sin(\varphi) \\
                          \sin(\varphi) & \cos(\varphi)
                         \end{pmatrix}.
\end{equation}
\begin{proof}
  The proof follows from the fact that $g_{0010}$ and $\Re(g_{1000})$
  are, respectively, normal to the pitchfork $n=1$ and Hopf $n=0$
  stability boundaries.
\end{proof}
\end{lemma}

{This subsection concludes with Lemma
  \ref{lemma:mixed_mode_solution}, where the mixed-mode solution in
  the weakly nonlinear regime is defined. Its stability will be
  discussed in \S \ref{subsec:codimension_two_bifurcation} for two
  pitchfork-Hopf bifurcations involving distinct reaction kinetics.}

\begin{lemma}\label{lemma:mixed_mode_solution}
  \textbf{(Mixed-mode solution in the weakly nonlinear regime)}.  Let
  $p_{11} = g_{0021}$, $p_{12} = g_{1110}$, $p_{21} = \Re(g_{1011})$
  and $p_{22} = \Re(g_{2100})$ be the four cubic term coefficients of
  system \eqref{eq:amplitude_pitchfork_hopf}. {If none of these
  coefficients vanish, and if $p_{22}p_{11} - p_{12}p_{21}$
  is nonzero, then in the limit $\eps \to 0$, with
  $\eps = \sqrt{\|\mu - \mu_0\|}$ denoting the square-root of the
  distance with the \textbf{pitchfork-Hopf} bifurcation point}, a
  leading-order approximate family of spatio-temporal oscillatory
  solutions is given by
 \begin{equation}\label{eq:spatio_temporal_oscillations}
   W(t) = W_e + \eps \left( \rho_{ne} \left[ \WW_n e^{i\theta_n} +
  \text{c.c.} \right] + \rho_{0e}
\left[ \WW_0 e^{i\left(\lambda_I t + \theta_0(0)\right)} + \text{c.c.} \right]
\right)  + \orderTwo,
 \end{equation}
 for any $\theta_n, \theta_0(0)\in \R$, {as a result of the
   azimuthal and temporal phase shift symmetries}. The pair
   $(\rho_{ne}, \rho_{0e})$ corresponds to the mixed-mode equilibrium
   and is given by
 \begin{equation}
   (\rho_{ne},\rho_{0e}) = \left(\sqrt{\frac{p_{12}\left[\Re(g_{1000})\right]^T
    \mu_1 - p_{22}\left[g_{0010}\right]^T\mu_1}{p_{22}p_{11} - p_{12}p_{21}}},
     \sqrt{\frac{p_{21}\left[g_{0010}\right]^T\mu_1 -
 p_{11}\left[\Re(g_{1000})\right]^T\mu_1}{p_{22}p_{11} - p_{12}p_{21}}} \right)\,,
 \end{equation}
 with the detuning vector $\mu_1$ chosen such that
 $\mu = \mu_0 + \eps^2\mu_1$ is within the mixed-mode region of
 existence. In particular, it can be taken to be parallel to the
 bisector of this region.
 
{For the surface-bound activator species, let
 $u_\text{amp}$ be the amplitude of the bifurcating spatio-temporal
 pattern near the pitchfork-Hopf bifurcation point. A
 leading-order approximation for it is given by}
 \begin{equation}
   u_{\text{amp}} = \max_{\substack{0 \leq \theta < 2\pi\\ 0 \leq t < T_p}}
   \left\{ |u(\theta,t) - u_e| \right\} = 2\eps \left( \rho_{ne}|e_1^T\bphi_n|
     + \rho_{0e}|e_1^T\bphi_0|\right) + \orderTwo\,.
 \end{equation}
\end{lemma}

\section{Validation of weakly nonlinear theory}\label{sec:validation_weakly_nonlinear}

{In this section we validate the weakly nonlinear theory
  developed in \S \ref{sec:weakly_nonlinear_theory} and \S
  \ref{sec:bifurcation_analysis} by comparing predictions of this
  theory with either numerical bifurcation results or full numerical
  time-dependent PDE solutions. In the comparisons we will consider
  both the Schnakenberg \eqref{eq:schnakenberg} and Brusselator
  \eqref{eq:brusselator} boundary reaction kinetics.  Since the
  parameter space is large, the bulk domain is restricted to the unit
  disk. For the uncoupled case (with $K_u = K_v = 0$), the surface
  diffusion coefficients and reaction kinetic parameters are chosen to
  ensure that there is a unique stable patternless solution for the
  reaction-diffusion system on the domain boundary. In addition, since
  typical surface diffusion coefficients are smaller than their bulk
  diffusion counterparts in applications, the condition $d_u \leq D_u$
  and $d_v \leq D_v$ shall be imposed. Ultimately, the bifurcation
  analysis will illustrate how varying the ratio of bulk diffusivity
  and coupling coefficients can destabilize the system and lead to
  novel spatio-temporal dynamics.}

\subsection{Codimension-one bifurcation}\label{subsec:codim_1_bif}

\begin{figure}[htbp]
\centering
\begin{subfigure}{0.47\linewidth}
\includegraphics[width=\linewidth,height=5.0cm]{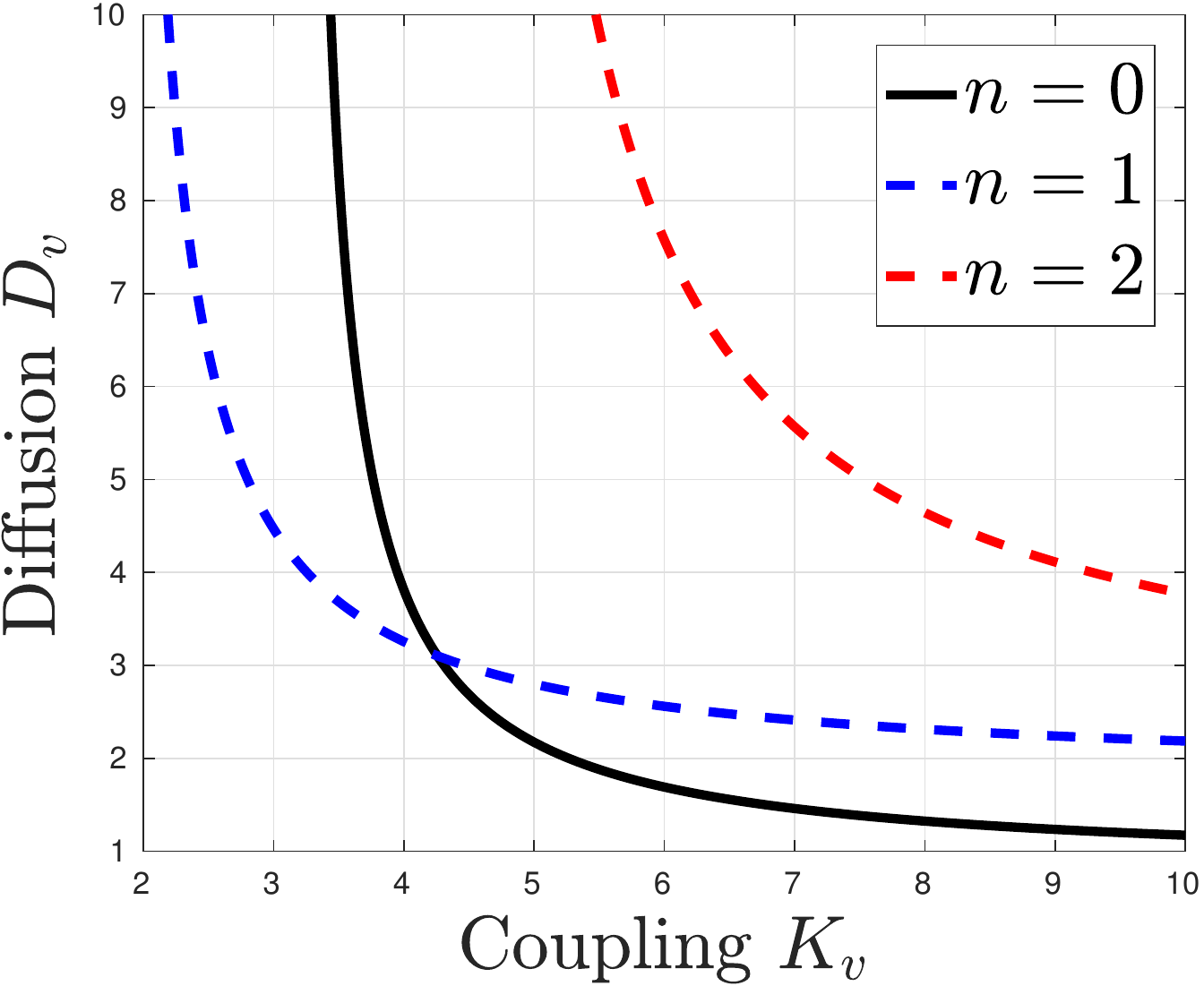}
\end{subfigure}
\begin{subfigure}{0.47\linewidth}
\includegraphics[width=\linewidth,height=5.0cm]{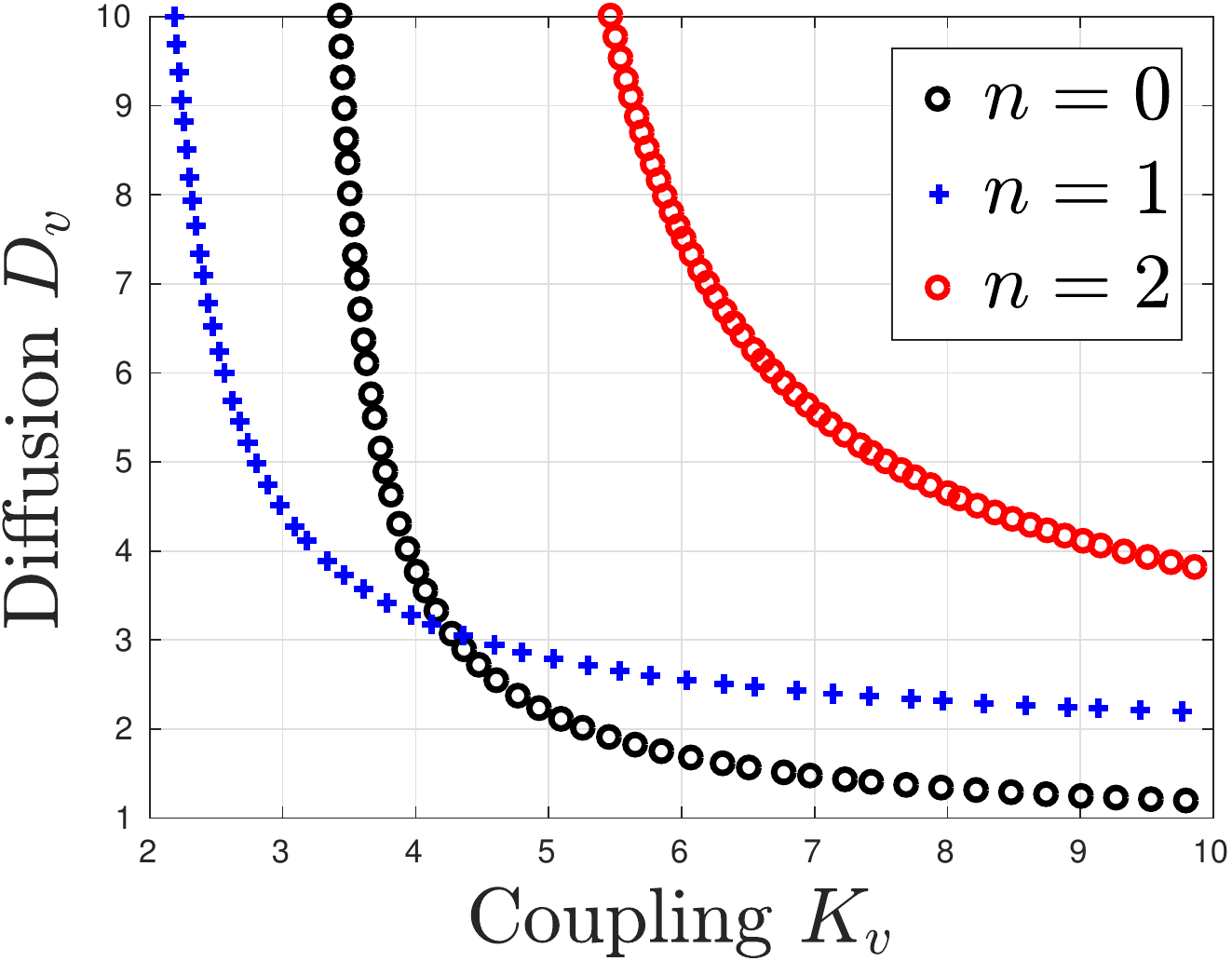}
\end{subfigure}
\caption{\label{fig:stability_diagram_schnakenberg} Linear stability diagram
  in the plane of parameters $(K_v,D_v)$ with Schnakenberg reaction
  kinetics. Other parameters are
  $R = 1,\,D_u = 1,\,\sigma_u = \sigma_v = 0.01,\, K_u = 0.1,\, d_u =
  d_v = 0.1,\, a = 0.1,\, b = 0.9\,$. In the right panel, the symbol
  "o" indicates supercritical while "$+$" indicates subcritical.}
\end{figure}

\begin{figure}[htbp]
\centering
\begin{subfigure}{0.47\linewidth}
\includegraphics[width=\linewidth,height=5.0cm]{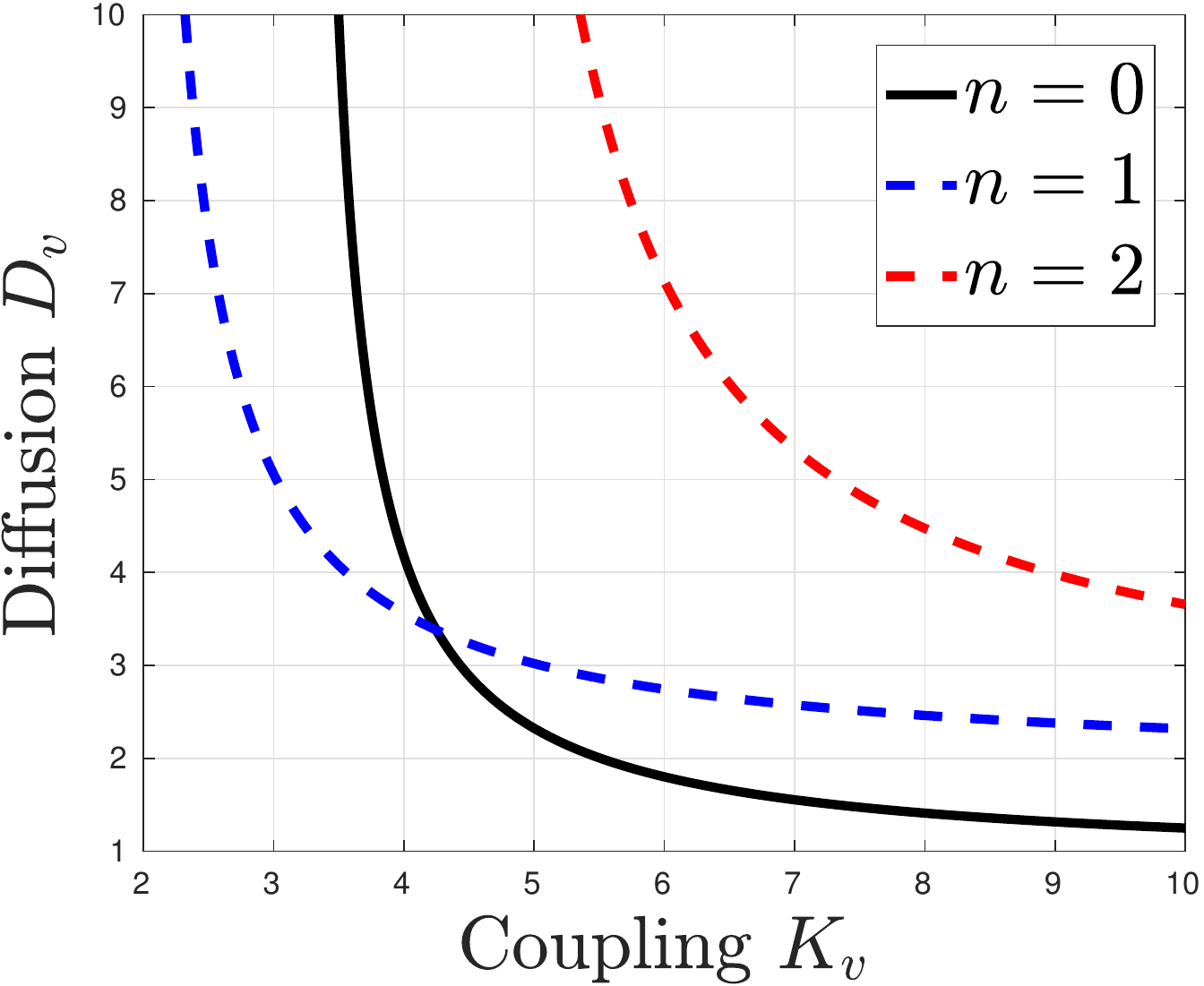}
\end{subfigure}
\begin{subfigure}{0.47\linewidth}
\includegraphics[width=\linewidth,height=5.0cm]{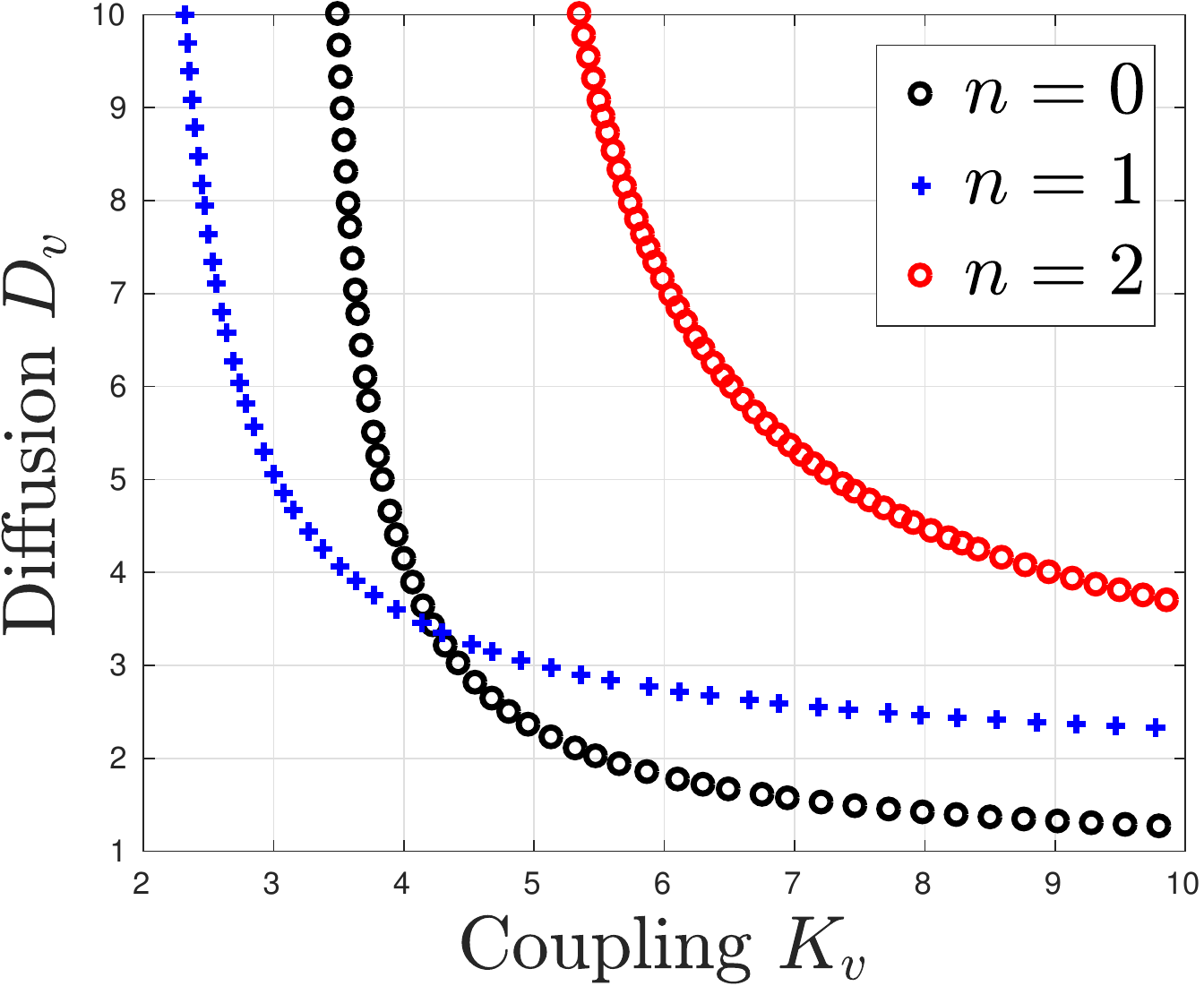}
\end{subfigure}
\caption{\label{fig:stability_diagram_brusselator} Stability diagram
  in the plane of parameters $(K_v,D_v)$ with Brusselator reaction
  kinetics. Other parameters are
  $R = 1,\, D_u = 1,\, \sigma_u = \sigma_v = 0.01,\, K_u = 0.1,\, d_u
  = d_v = 0.5,\, a = 3,\, b = 7.5\,$. In the right panel, the symbol
  "o" indicates supercritical while "$+$" indicates subcritical.}
\end{figure}

{By numerically computing the roots of the eigenvalue
  relation \eqref{eq:transcendental}, we can readily determine a
  linear stability diagram in the $D_v$ versus $K_v$ parameter
  space for the Schnakenberg and Brusselator boundary kinetics.  For a
  circular harmonic mode with either $n=0,1,2$, the marginal stability
  curves in $D_v$ versus $K_v$ parameter space, for a fixed set of
  parameters, are shown in
  Fig.~\ref{fig:stability_diagram_schnakenberg} and
  Fig.~\ref{fig:stability_diagram_brusselator} for the Schnakenberg
  and Brusselator kinetics, respectively. In each phase diagram, the
  region of linear stability is located to the left of all the
  curves. Our computations show that the trivial mode $n=0$ loses
  stability through a Hopf bifurcation, while for the nontrivial modes
  $n=1,2$, stability is lost via a Turing bifurcation (zero-crossing
  eigenvalue).}

{Some qualitative trends are suggested from these linear
  stability phase diagrams. Firstly, when both bulk diffusion
  coefficients have the same order of magnitude, the Hopf bifurcation
  of the trivial mode is the dominant instability. It occurs when the
  coupling of the inhibitor is much larger than that of the activator,
  i.e.~${K_v/K_u} \approx 10^2$. Secondly, when this ratio decreases
  while the ratio ${D_v/D_u}$ increases, the primary instability
  switches to the Turing bifurcation of the first nontrivial mode
  $(n=1)$.} This is reminiscent of the classical Turing paradigm for
pattern formation, whereby the inhibitor is required to diffuse faster
than the activator in order for a pattern to form.

{The redrawn phase diagrams in the right panels of
  Fig.~\ref{fig:stability_diagram_schnakenberg} and
  Fig.~\ref{fig:stability_diagram_brusselator}, indicating the local
  branching behavior of the bifurcation, was obtained after
  numerically evaluating the cubic term coefficients in the normal
  forms \eqref{eq:amplitude_hopf} and \eqref{eq:amplitude_pitchfork}.}
For both reaction kinetics, supercritical Hopf and subcritical Turing
bifurcations are predicted to occur as the stability curves associated
with the modes $n=0,1$ are crossed. Although the transversal crossing
of the rightmost curve associated with the mode $n=2$ is predicted to
correspond to a supercritical Turing bifurcation, little attention is
given to it in the subsequent discussion since it corresponds to a
secondary instability.

{We emphasize that our conclusions only hold for the current
  set of fixed parameters given in the captions of
  Fig.~\ref{fig:stability_diagram_schnakenberg} and
  Fig.~\ref{fig:stability_diagram_brusselator}. In \S
  \ref{subsubsec:subsub_turing} we show that a conclusion of super- or
  subcriticality can depend on the specific choices of reaction
  kinetic parameters and surface diffusion coefficients.}

\subsubsection{Periodic solutions arising from Hopf bifurcations}\label{subsubsec:subsub_hopf}

{In this subsection the loss of stability through the
  supercritical Hopf bifurcation, as predicted by our weakly nonlinear
  analysis in Fig.~\ref{fig:stability_diagram_schnakenberg} and
  Fig.~\ref{fig:stability_diagram_brusselator}, is
  investigated. Numerical continuation methods combined with PDE
  simulations are used to study the dynamics in the weakly nonlinear
  regime. As a result of azimuthal invariance, a simple 1-D finite
  difference method-of-lines approach is used to spatially
  discretize the coupled bulk-surface system, with the mean value
  theorem applied to derive an ODE for the bulk species at the origin
  that avoids the singularity inherent to the Laplacian in polar
  coordinates. Details of the discretization process are given in
  Appendix \ref{sec:numerical_methods}.}

\begin{figure}[htbp]
\begin{subfigure}{0.47\linewidth}
\includegraphics[width=\linewidth,height=5.0cm]{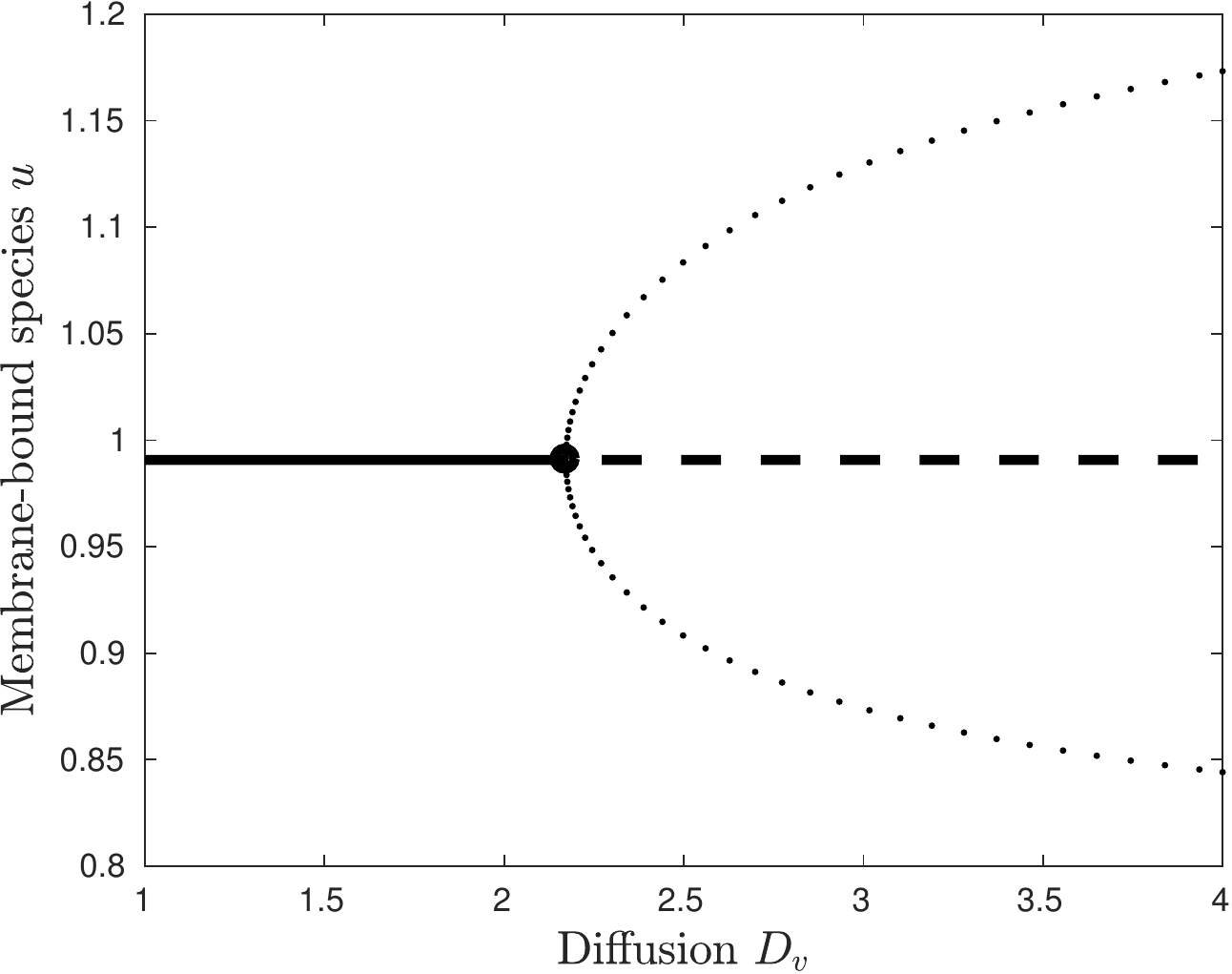}
\end{subfigure}
\begin{subfigure}{0.47\linewidth}
\includegraphics[width=\linewidth,height=5.0cm]{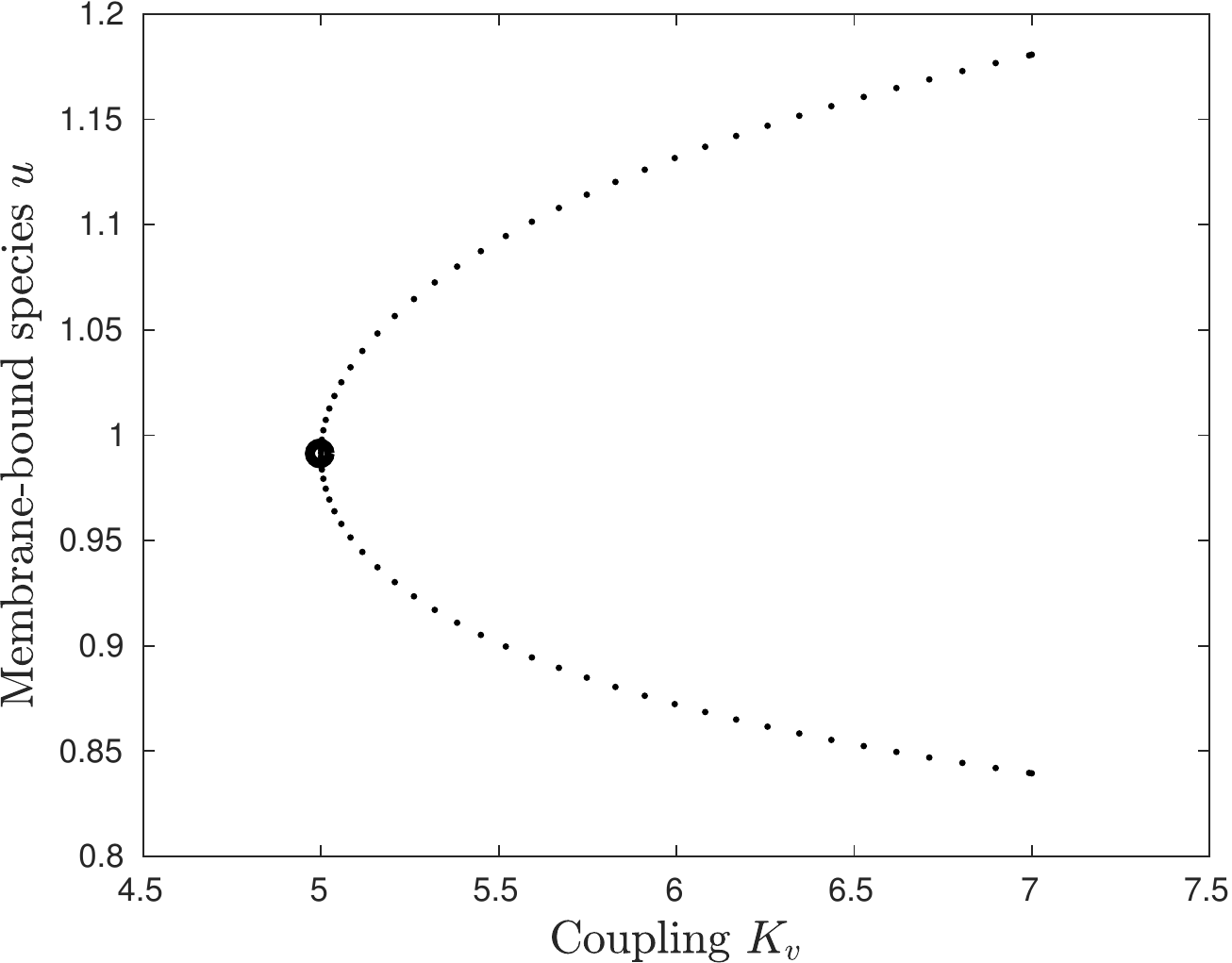}
\end{subfigure}
\caption{\label{fig:global_branches} Global periodic solution branches
  computed with AUTO \cite{doedel2007} past a supercritical Hopf
  bifurcation for the Schnakenberg kinetics. {The continuation
  parameter in the left panel is $D_v$ with $K_v=5$, while in the
  right panel $K_v$ is used with $D_v\approx 2.17$. $N=200$ points 
  discretize the radial direction.}}
\end{figure}

{We first study the loss of stability on the vertical line
  $K_v = 5$ from Fig.~\ref{fig:stability_diagram_schnakenberg} that
  intersects the $n=0$ Hopf curve at $D_v\approx 2.17$}. Plots of
global periodic solution branches using either $K_v$ or $D_v$ as the
bifurcation parameter are displayed in
Fig.~\ref{fig:global_branches}. {These numerical bifurcation
  diagrams confirm the prediction of the weakly nonlinear theory of a
  loss of stability through a supercritical Hopf bifurcation.}

\begin{figure}[htbp]
\centering
\begin{subfigure}{0.47\linewidth}
\includegraphics[width=\linewidth,height=5.0cm]{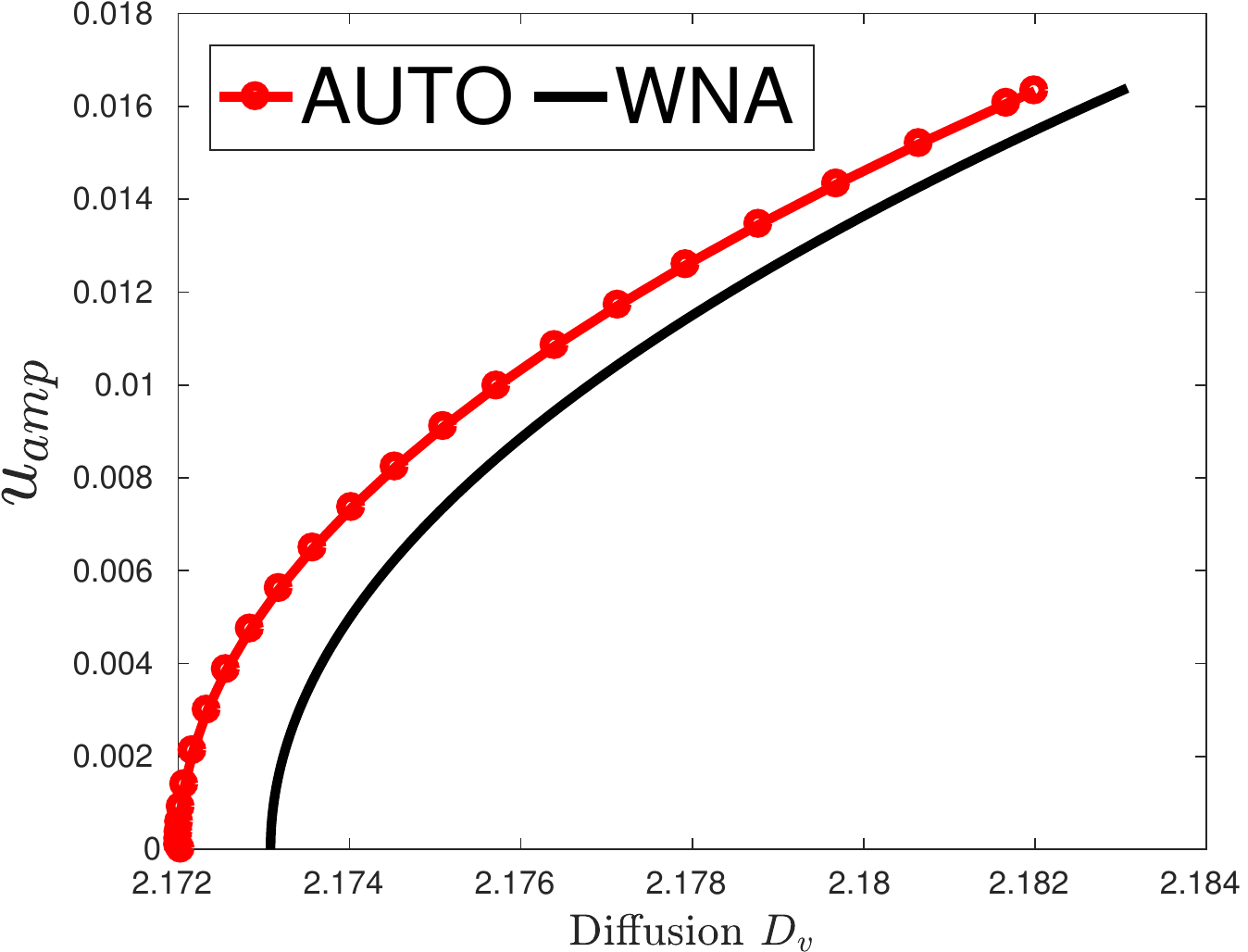}
\end{subfigure}
\begin{subfigure}{0.47\linewidth}
\includegraphics[width=\linewidth,height=5.0cm]{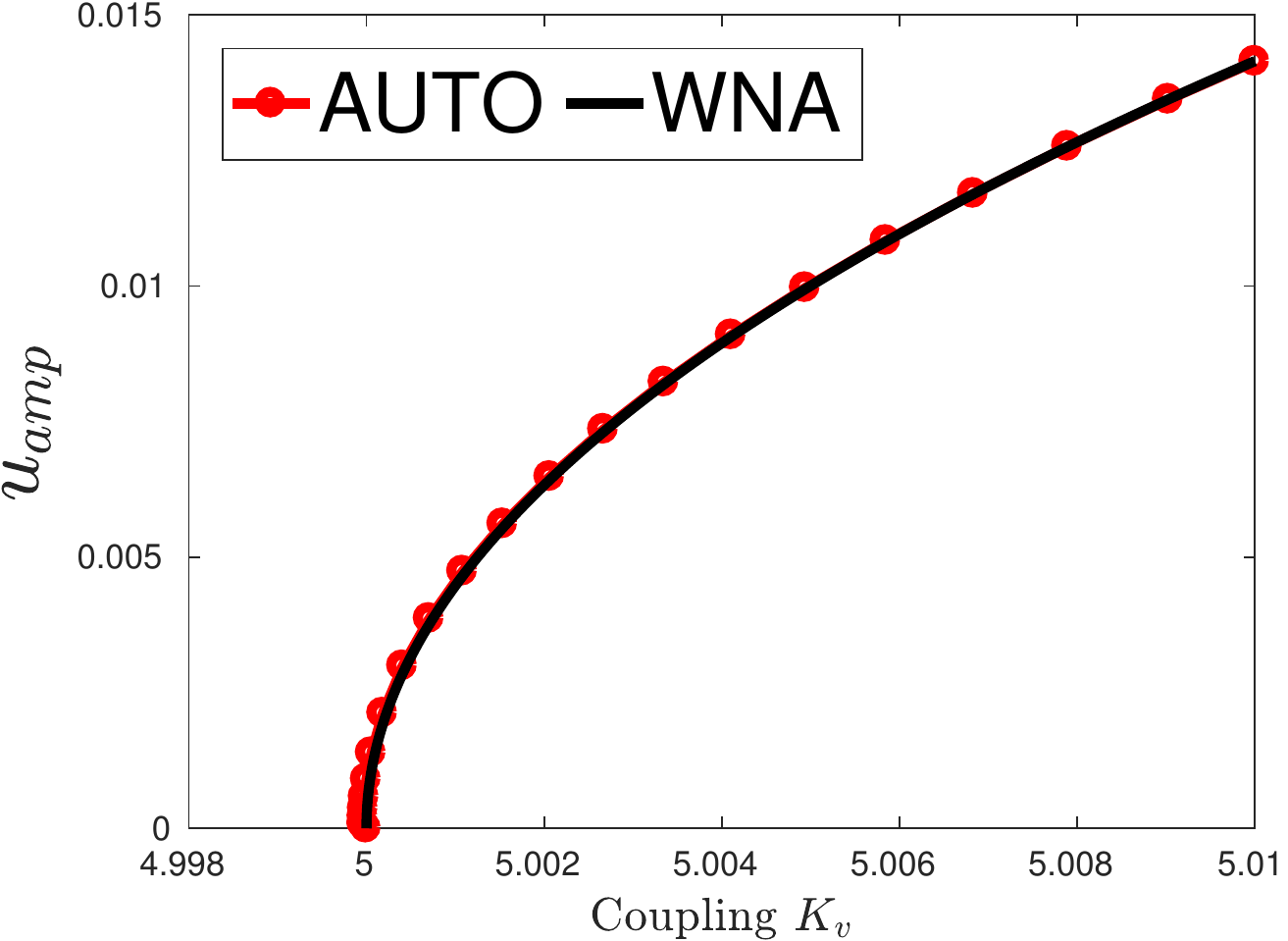}
\end{subfigure}
\caption{\label{fig:amplitude} Amplitude of periodic solutions in the
  weakly nonlinear regime for the Schnakenberg reaction kinetics. The
  red curve is computed with AUTO using $N=200$ equidistant mesh
  points in the radial direction, while the black curve is obtained
  directly from the normal form \eqref{eq:u_amp} for $0\leq\eps\leq 0.1$.}
\end{figure}

\begin{figure}[htbp]
\begin{subfigure}{0.47\linewidth}
\includegraphics[width=\linewidth,height=5.0cm]{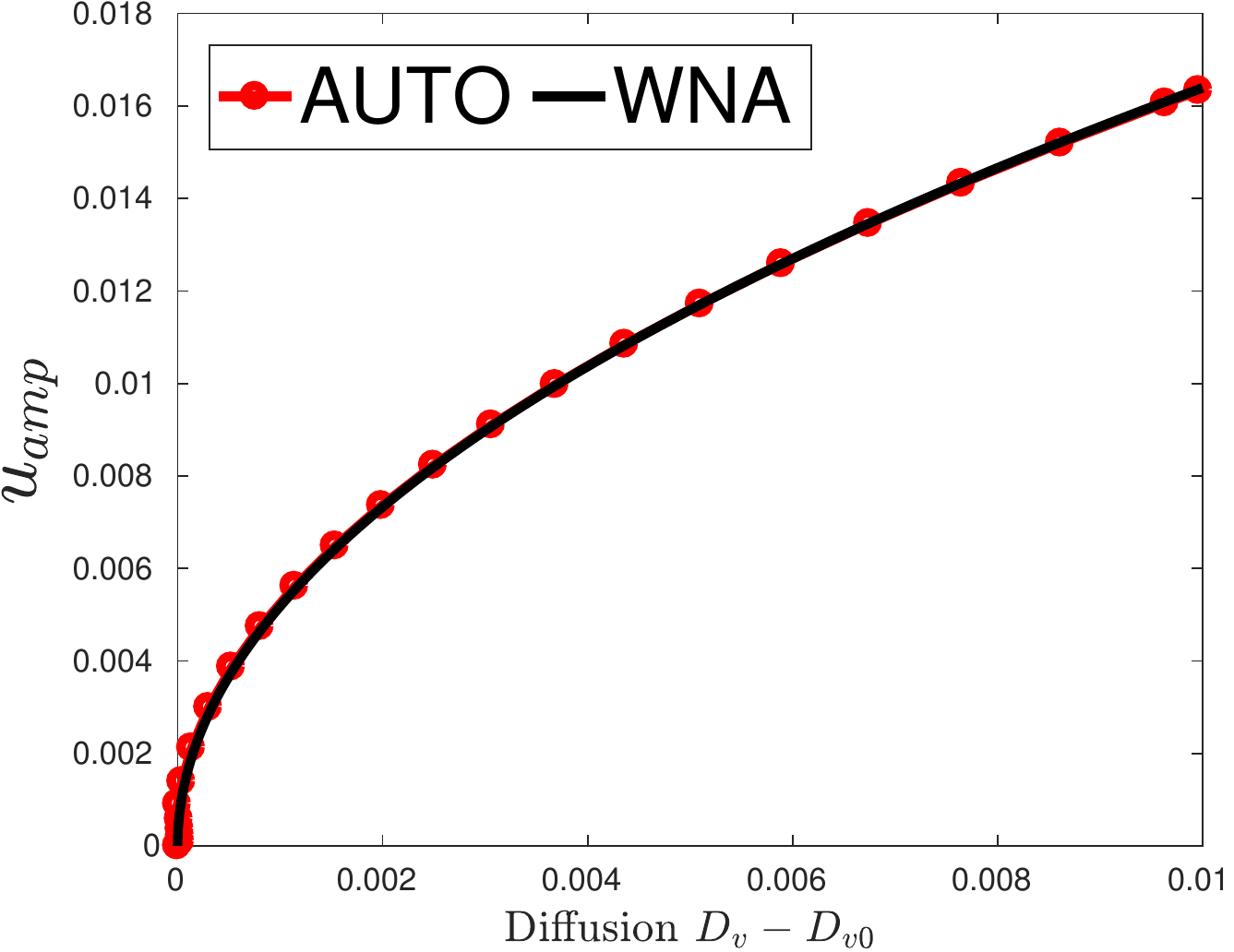}
\end{subfigure}
\begin{subfigure}{0.47\linewidth}
\includegraphics[width=\linewidth,height=5.0cm]{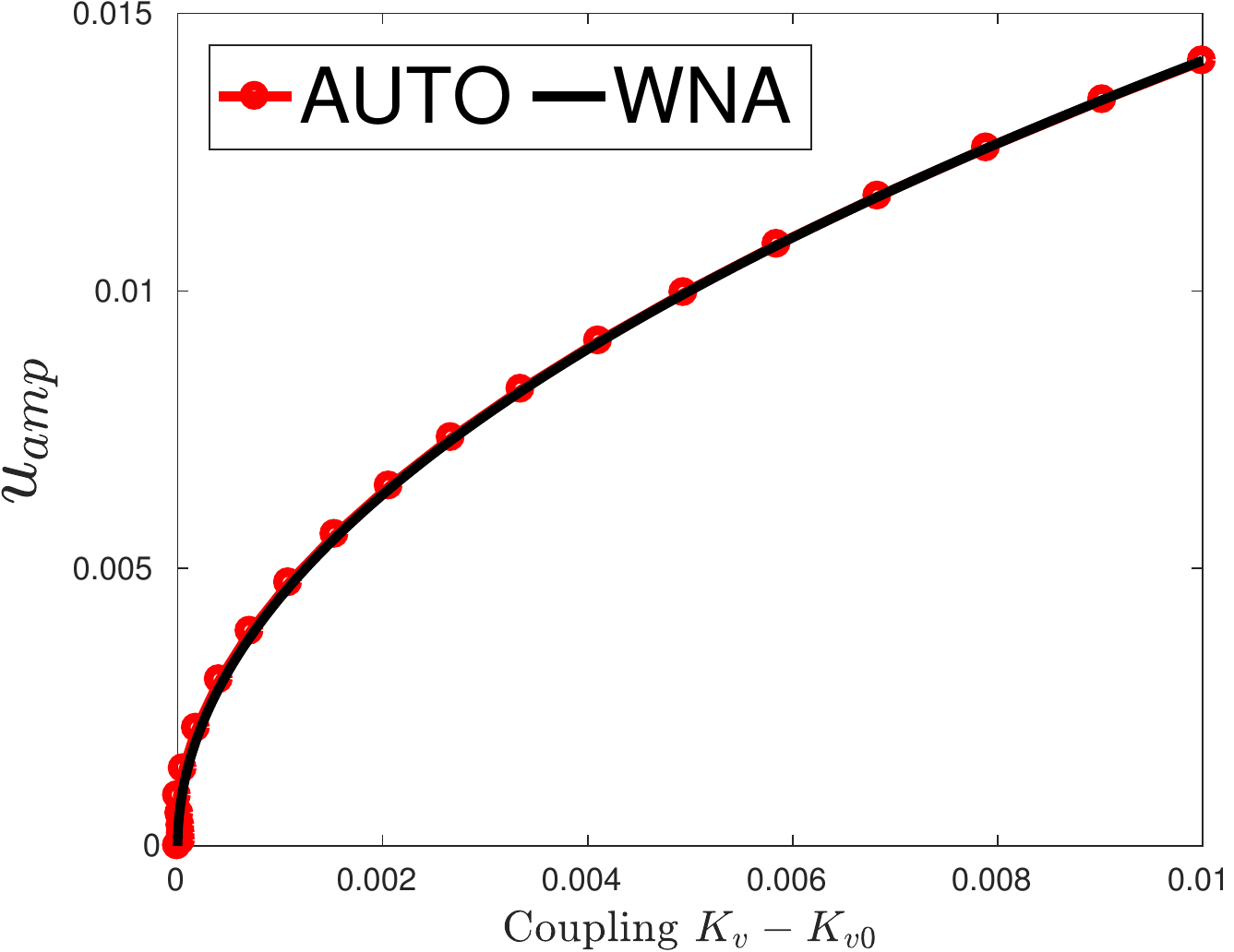}
\end{subfigure}
\caption{\label{fig:amplitude_shifted}Translation to the origin of the
  branches in Fig.~\ref{fig:amplitude}. The curvature of the bifurcating
  branches is correctly approximated by the weakly nonlinear theory.}
\end{figure}

{Next, in the vicinity of the bifurcation point,
  \eqref{eq:u_amp} of Lemma \ref{lemma:family_periodic_weakly} is used
  to predict the amplitude of the limit cycle. A graphical comparison
  between this predicted amplitude and corresponding numerical results
  computed with AUTO is shown in Fig.~\ref{fig:amplitude}. In the left
  panel of Fig.~\ref{fig:amplitude} a slight shift in the bifurcation
  point caused by spatial discretization errors is observed.  As the
  mesh is refined, the gap between the bifurcation points computed
  from the spatially discretized system and directly from the
  transcendental equation \eqref{eq:transcendental} by solving for
  pure imaginary roots is expected to shrink. This shift is not as
  apparent} in the right panel of Fig.~\ref{fig:amplitude} since the
bulk diffusivity $D_v$ was used to locate the Hopf bifurcation
point. From Fig.~\ref{fig:amplitude_shifted}, the two branches
essentially coincide after translation to the origin.

\begin{figure}[htbp]
\centering
\begin{subfigure}{0.47\linewidth}
\includegraphics[width=\linewidth,height=5.0cm]{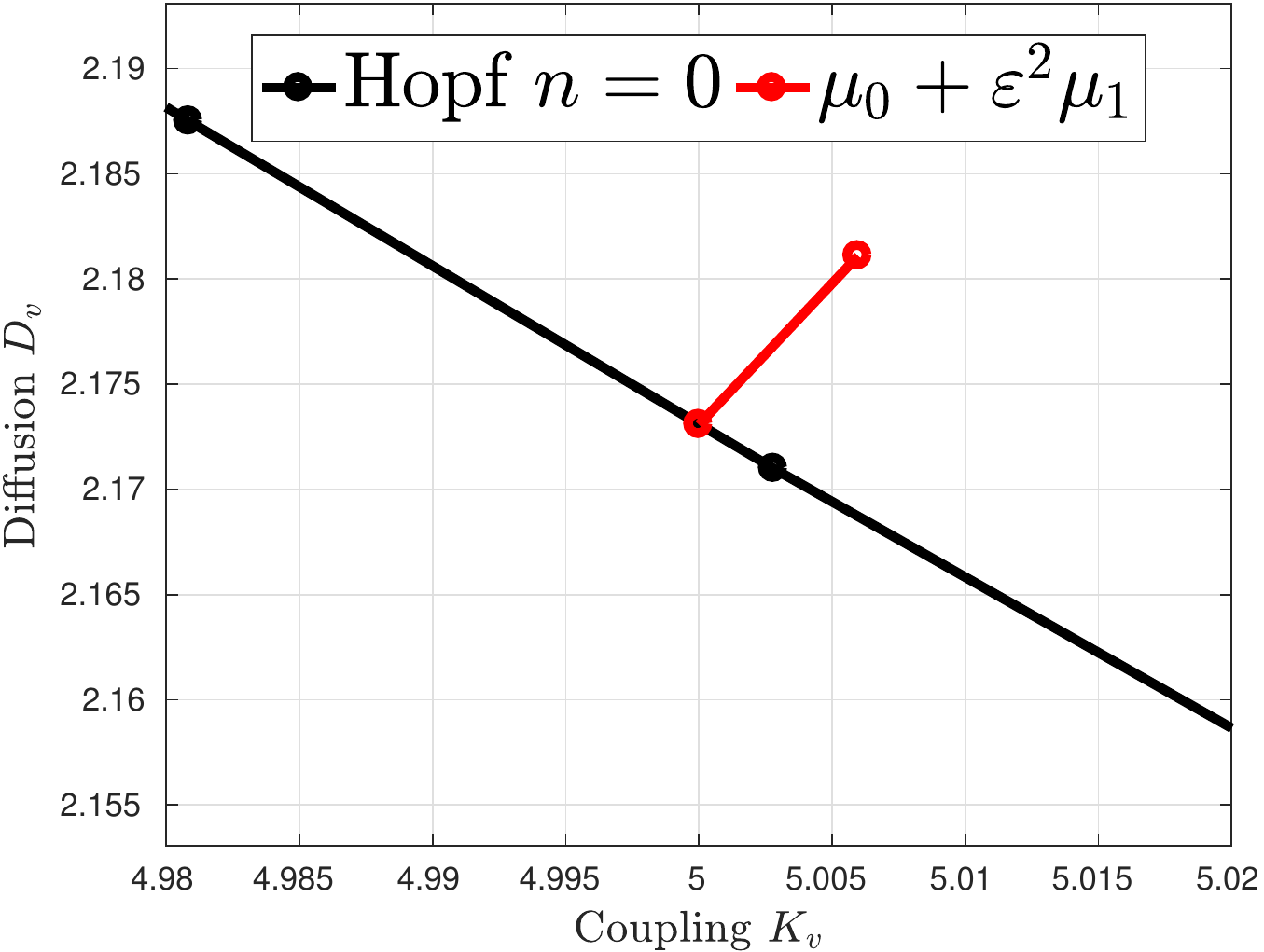}
\end{subfigure}
\begin{subfigure}{0.47\linewidth}
\includegraphics[width=\linewidth,height=5.0cm]{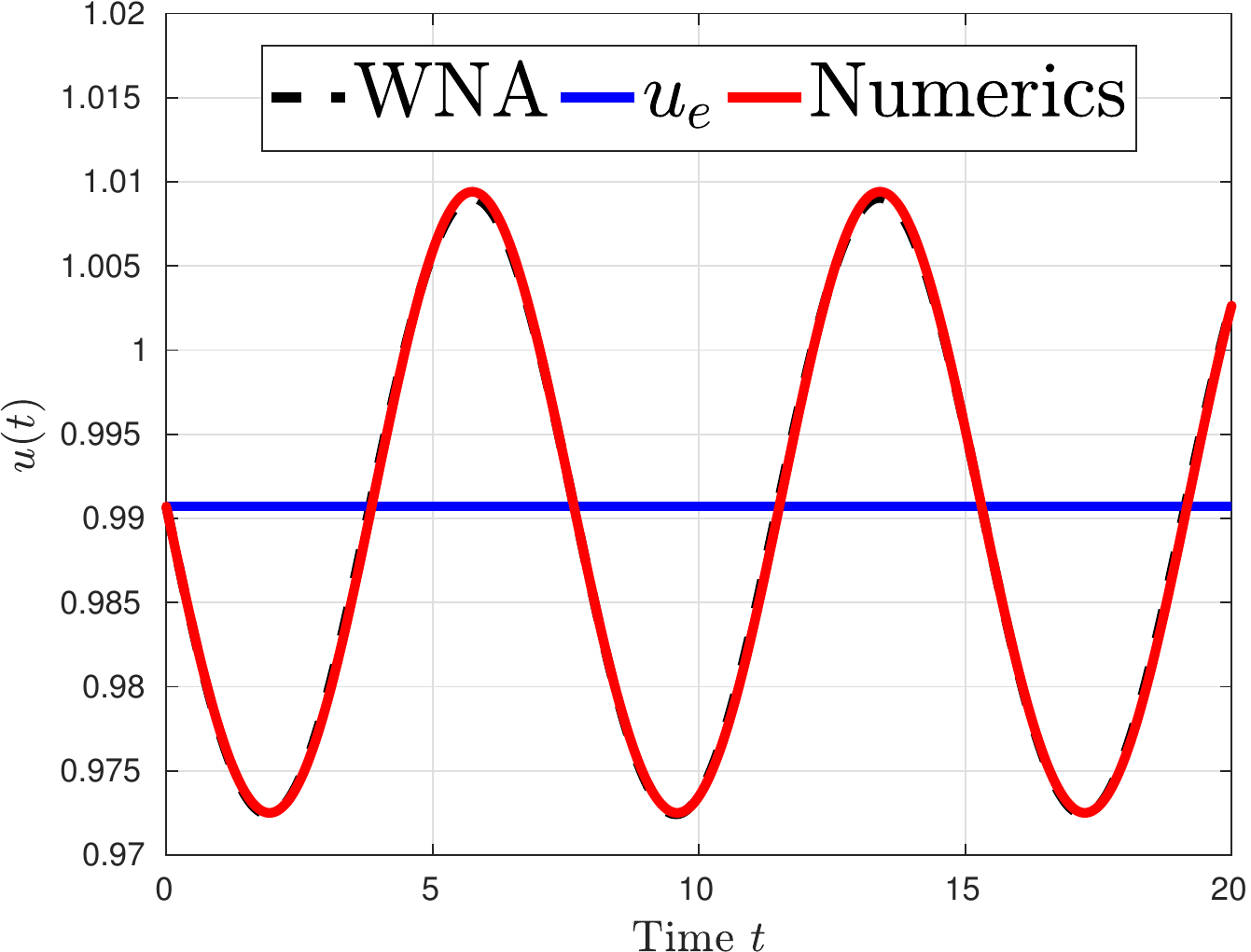}
\end{subfigure}
\caption{\label{fig:schnakenberg_K_v_D_v} Full numerical simulation of
  the reduced PDE-ODE model with Schnakenberg kinetics. The right
  panel shows the membrane-bound activator $u(t)$ (red curve) which
  oscillates around the equilibrium solution (blue line). Notice here
  the good agreement with the solution in the weakly nonlinear regime
  (black dashed coinciding curve). Implicit-explicit time-stepping (SBDF2)
  \cite{ruuth1995} is employed from an initial condition given by
  \eqref{eq:periodic_solution} with $t=0$,
  $\theta_0(0) = \frac{\pi}{2}$ and $\eps = 0.1$. The bifurcation
  point and detuning vector respectively satisfy $\mu_0=(5,2.17)^T$
  and $\mu_1=(0.6,0.8)^T$. The parameter values for the simulation are
  indicated in the left panel by a red dot.}
\end{figure}

{As a further validation of the weakly nonlinear theory, we
  now compare the predicted period of oscillations near the Hopf point
  with corresponding numerical results extracted from PDE simulations.
  As shown in the right panel of Fig.~\ref{fig:schnakenberg_K_v_D_v}
  for a particular choice of detuning vector $\mu_1$ normal to the
  stability boundary (see the left panel of
  Fig.~\ref{fig:schnakenberg_K_v_D_v}), the numerically computed
  period of oscillations is $T_p \approx 7.65$. This value agrees well
  with the result \eqref{eq:period} from the weakly nonlinear theory.}

\begin{figure}[htbp]
\begin{subfigure}{0.32\linewidth}
\includegraphics[width=\linewidth,height=4.8cm]{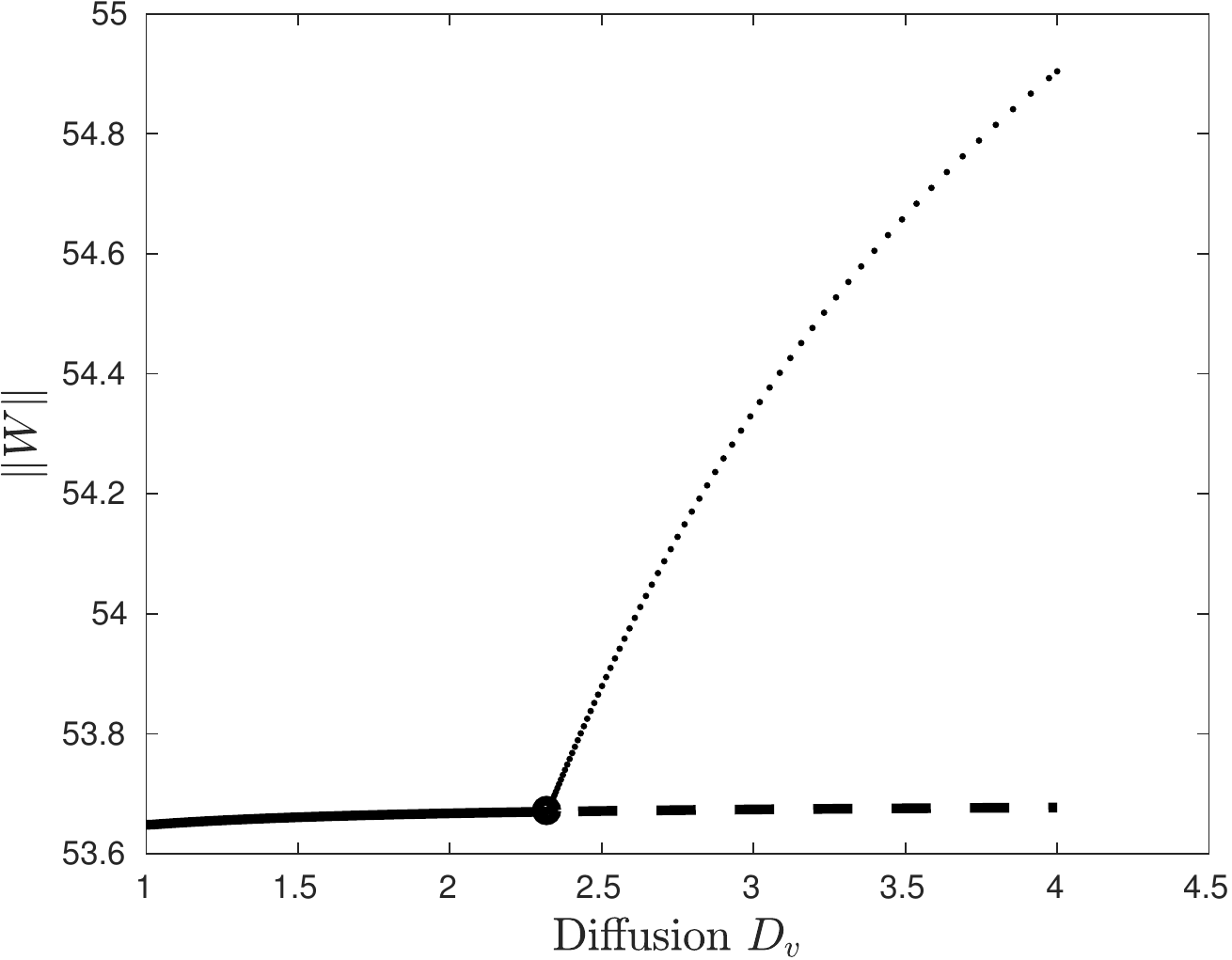}
\end{subfigure}
\begin{subfigure}{0.32\linewidth}
\includegraphics[width=\linewidth,height=4.8cm]{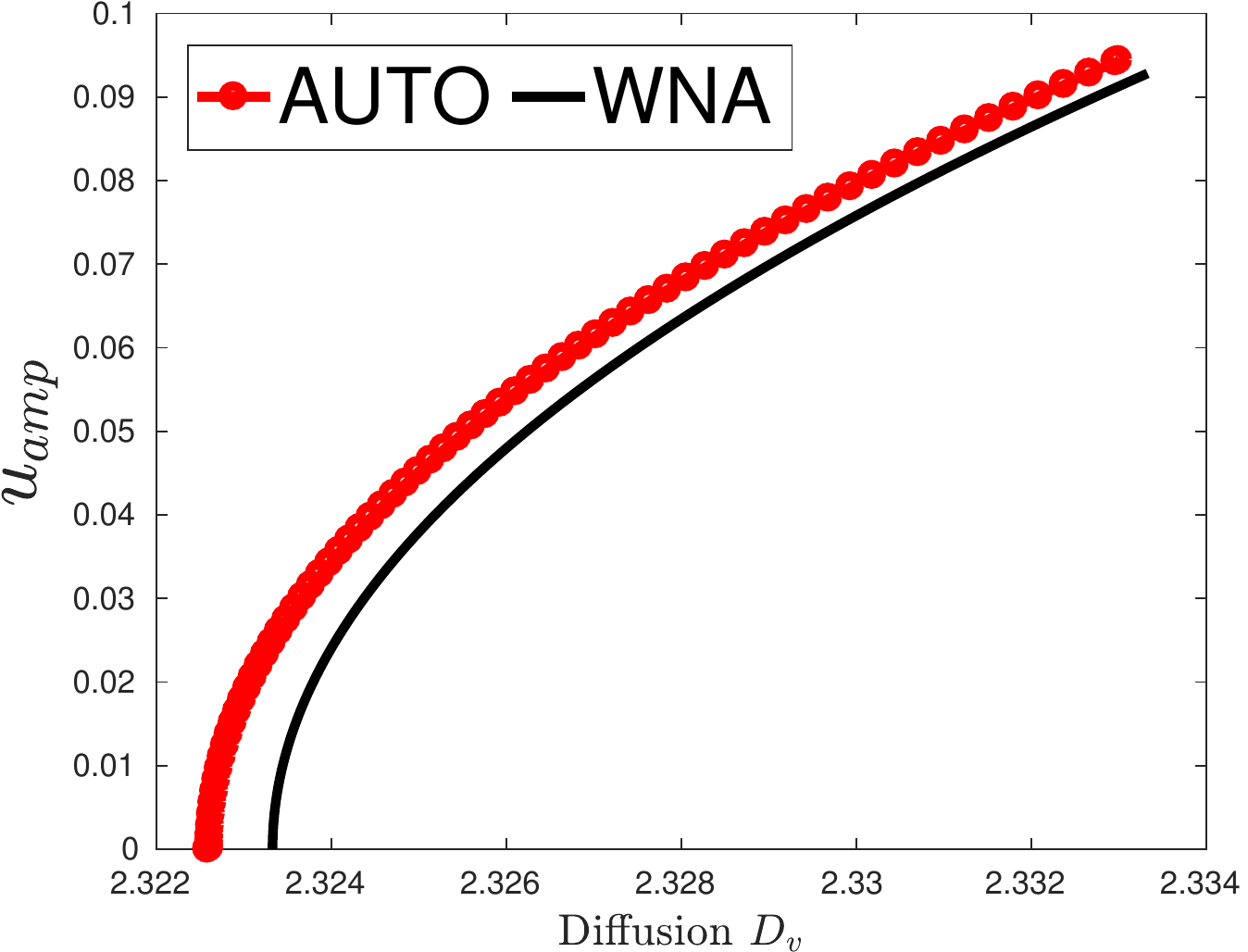}
\end{subfigure}
\begin{subfigure}{0.32\linewidth}
\includegraphics[width=\linewidth,height=4.8cm]{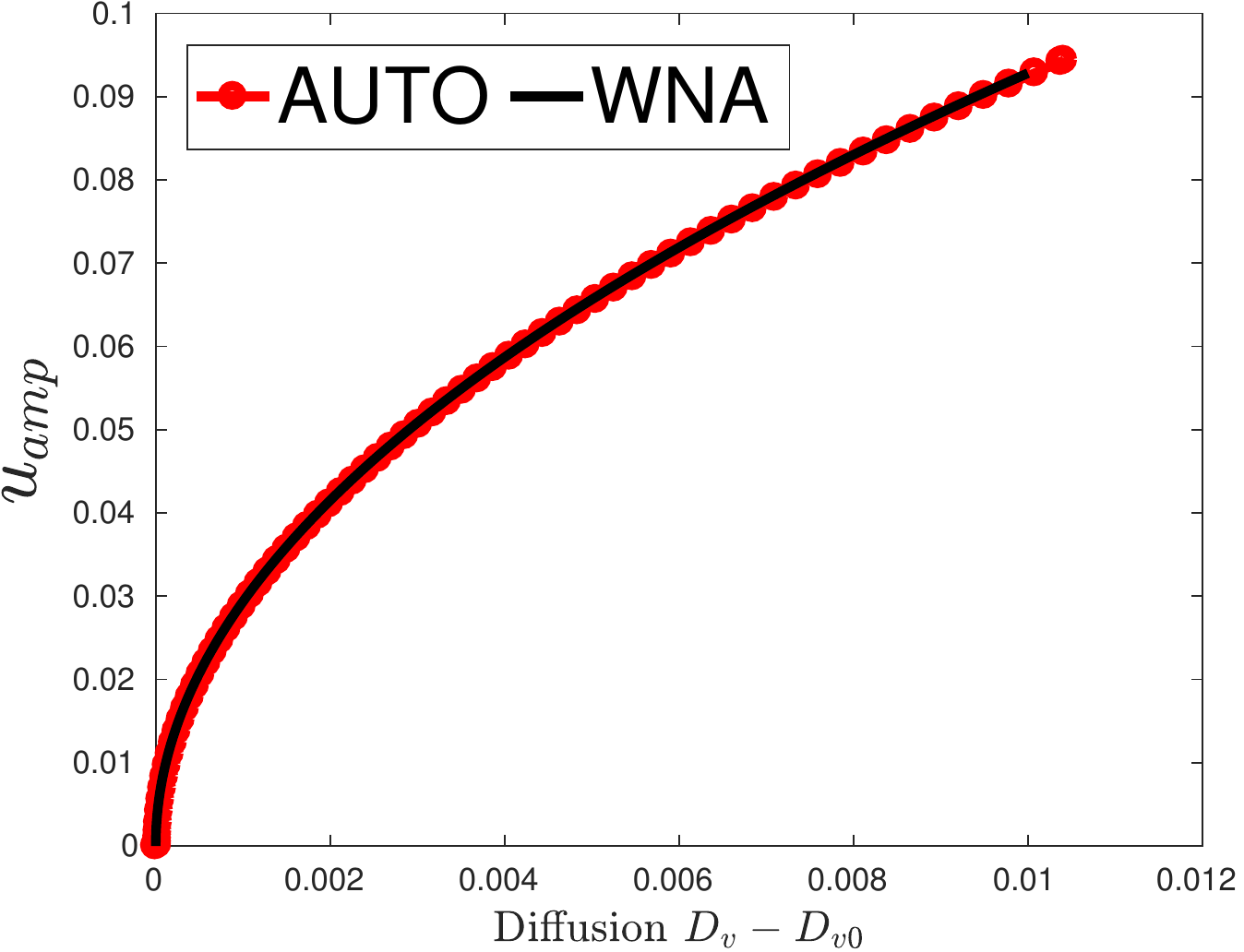}
\end{subfigure}
\caption{\label{fig:branches_brusselator} Periodic solution branches
  past a supercritical Hopf bifurcation with the Brusselator reaction
  kinetics. $N=200$ equidistant meshpoints discretize the radial
  direction for the radially symmetric reduced PDE-ODE model.  Left
  panel: plot of the norm of the global solution branches computed with AUTO. Middle and right panels: local branching
  behavior predicted from the weakly nonlinear theory and from the
  bifurcation software AUTO are favorably compared.}
\end{figure}

\begin{figure}[htbp]
\centering
\begin{subfigure}{0.47\linewidth}
\includegraphics[width=\linewidth,height=5.0cm]{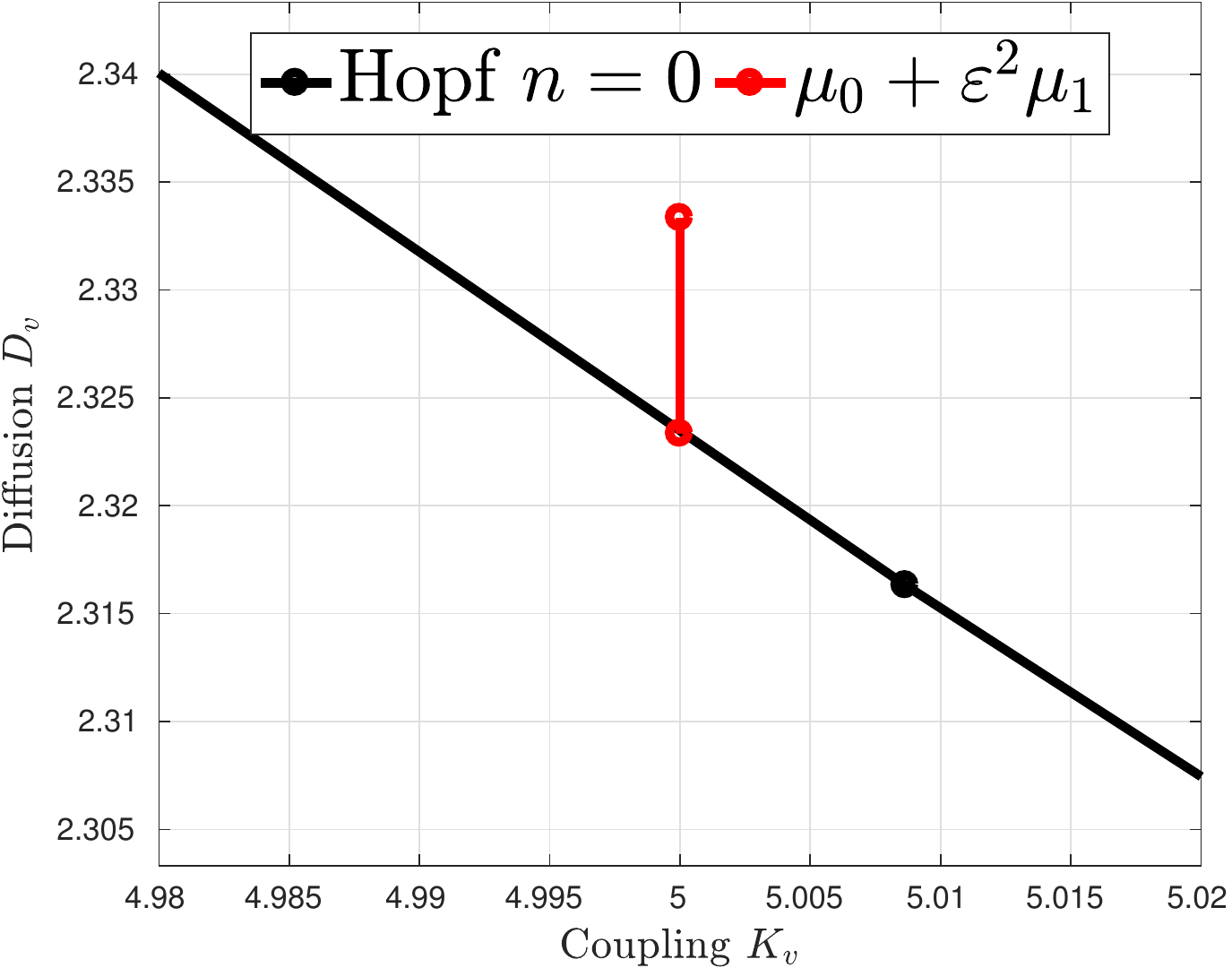}
\end{subfigure}
\begin{subfigure}{0.47\linewidth}
\includegraphics[width=\linewidth,height=5.0cm]{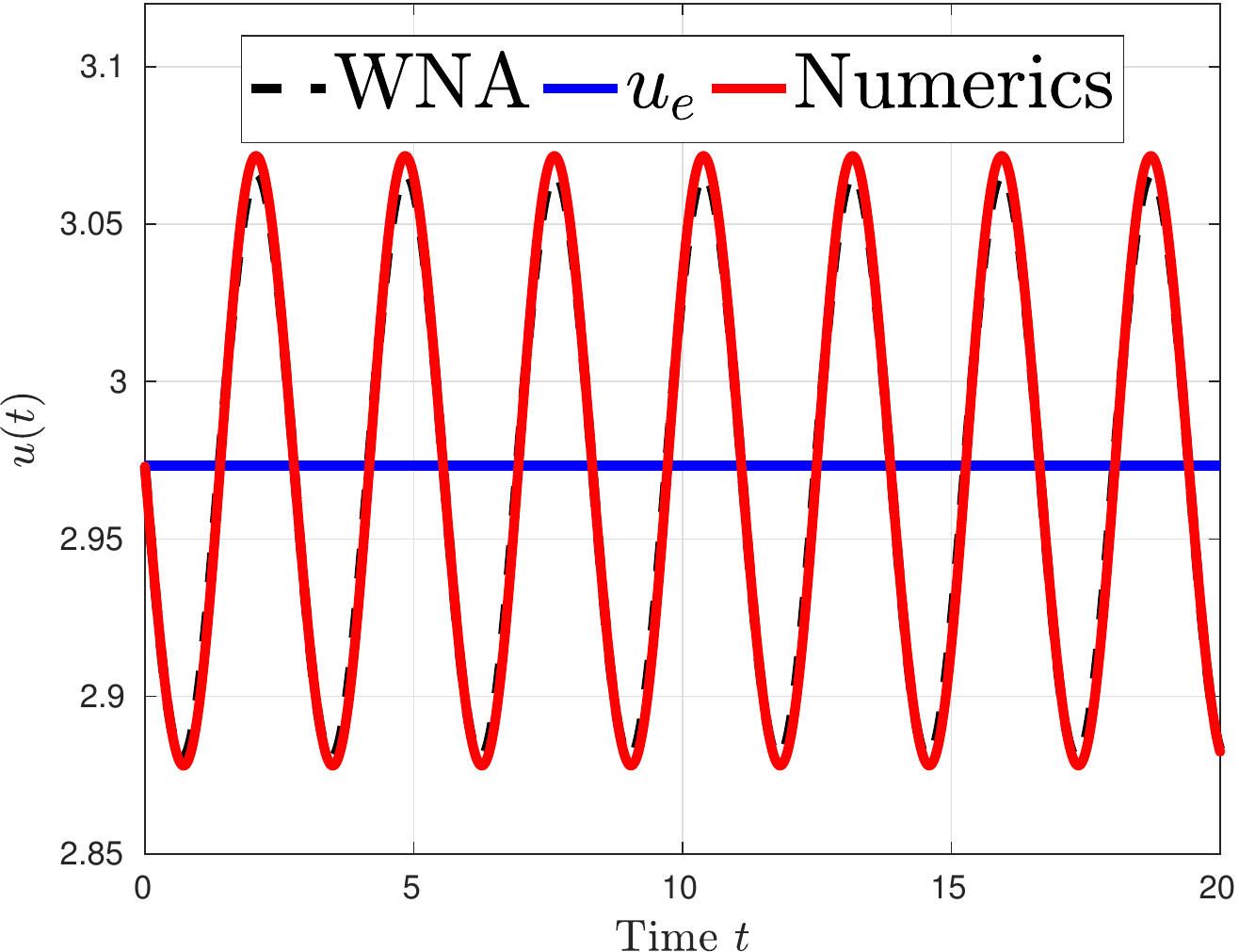}
\end{subfigure}
\caption{\label{fig:brusselator_D_v} Full numerical simulation of the
  reduced PDE-ODE model with Brusselator kinetics. The same numerical
  method and initial condition used in Fig.~\ref{fig:schnakenberg_K_v_D_v}
  are employed here. The parameters are
  $\eps=0.1$, $\mu_0 = (5,2.32)^T$ and $\mu_1 = (0,1)^T$. The red dot
  in the left panel indicates $\mu_0 + \eps^2\mu_1$, where the
  simulation is performed.}
\end{figure}

Similar numerical experiments can be performed with the Brusselator
reaction kinetics for the parameter set in the caption of
Fig.~\ref{fig:stability_diagram_brusselator}. {The results
  are qualitatively similar, with both the numerical results and the
  weakly nonlinear analysis predicting a loss of stability through a
  supercritical Hopf bifurcation.} The periodic solution branches are
displayed in Fig.~\ref{fig:branches_brusselator} for the
membrane-bound activator. Despite the slight shift in the bifurcation
point, we observe a good agreement between the curvature of the
branches computed with AUTO and from the weakly nonlinear theory
\eqref{eq:u_amp}. {Moreover, at the bifurcation point, the
  magnitude of the eigenvalues is larger for the Brusselator than for
  the Schnakenberg model. This leads to a smaller oscillation period
  $T_p \approx 2.7719$ for the Brusselator}. The corresponding full
numerical simulation of the reduced PDE-ODE model is given in
Fig.~\ref{fig:brusselator_D_v} for $\eps = 0.1$ and $\mu_1 = (0,1)^T$.

\begin{figure}[htbp]
\centering
\begin{subfigure}{0.47\linewidth}
\includegraphics[width=\linewidth,height=5.0cm]{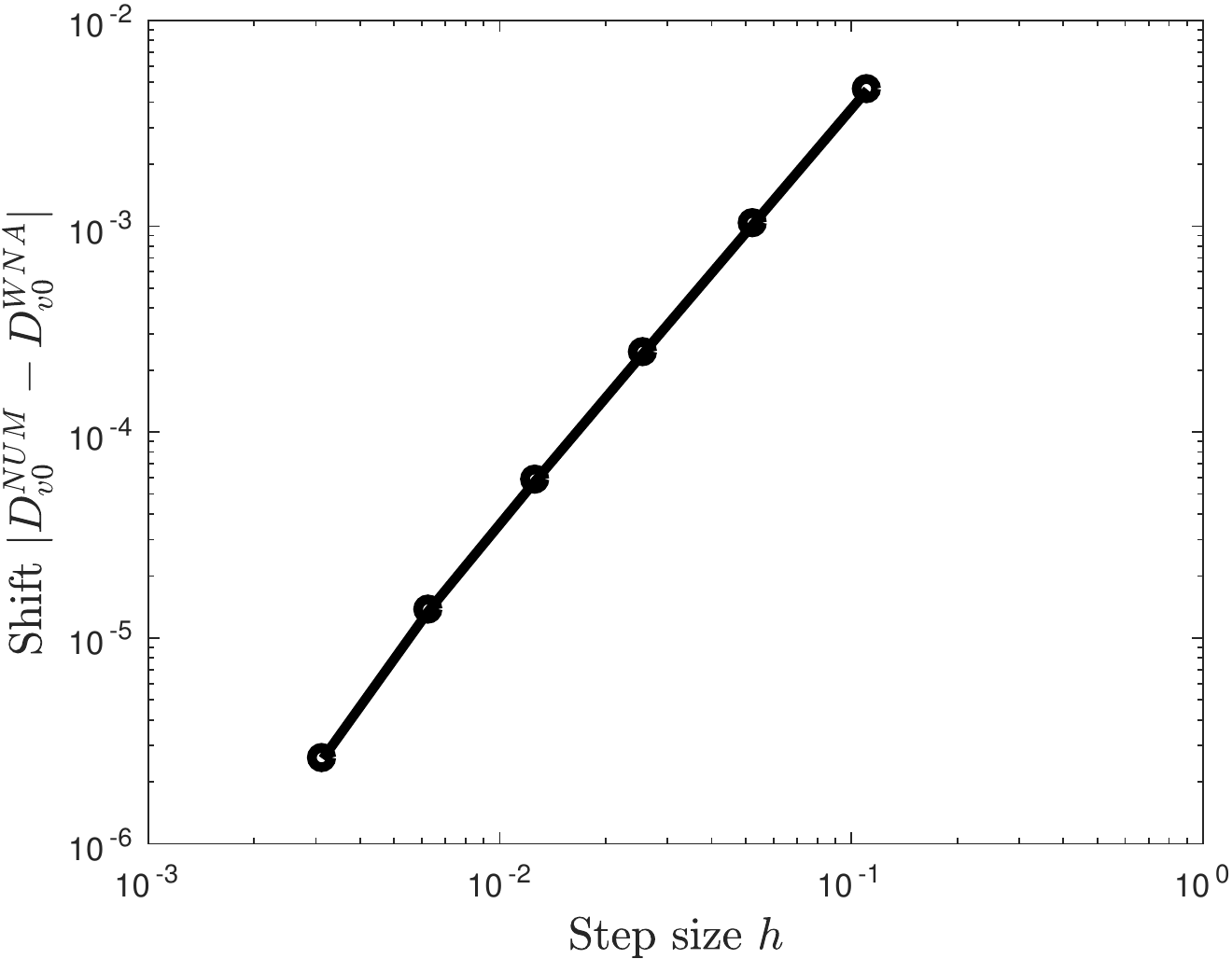}
\end{subfigure}
\begin{subfigure}{0.47\linewidth}
\includegraphics[width=\linewidth,height=5.0cm]{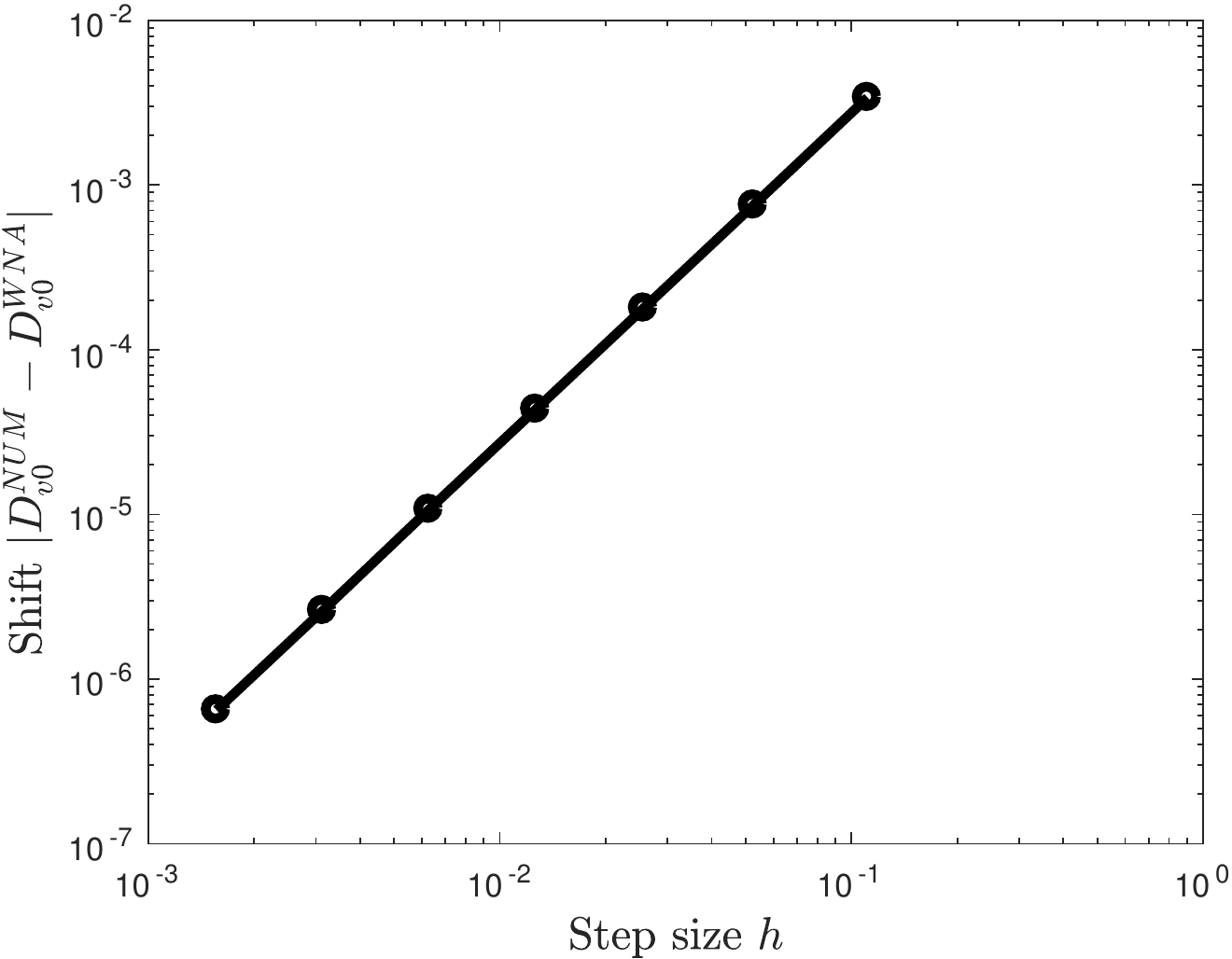}
\end{subfigure}
\caption{\label{fig:convergence_study} Convergence of numerically
  computed Hopf bifurcation points $D_{v0}$ as the step size decreases
  for the Schnakenberg (left panel) and Brusselator (right panel)
  reaction kinetics with $K_v = 5$ and all the other parameters being
  the same as in Fig.~\ref{fig:stability_diagram_schnakenberg} and
  Fig.~\ref{fig:stability_diagram_brusselator}. Horizontal and vertical
  axis are displayed in a log scale. The computation of the
  bifurcation point in the spatially discretized system is performed
  with the software \textsc{coco} \cite{danko2013}. The reference
  bifurcation point is obtained directly by solving the transcendental
  equation \eqref{eq:transcendental} for a pair of purely imaginary
  roots $\pm i\lambda_I$ and, therefore, is more accurate.}
\end{figure}

{As a result of the slight} discrepancy between the
bifurcation points computed in the spatially discretized system as
compared to the weakly nonlinear theory, it is misleading to compare
the result of a simulation with the analytical solution at a fixed
parameter value. Therefore, a numerical convergence study as the mesh
size $h = {R/(N-1)}$ decreases, with $N$ being the number of nodes
used to discretize the radial direction (assuming azimuthal symmetry),
is provided in Fig.~\ref{fig:convergence_study}.  {Since
  second-order centered differences are used, a quadratic rate of
  convergence is expected.} Letting $\mu_0^{num}$ and $\mu_0^{wna}$
denote, respectively, the bifurcation points in the spatially
discretized versus continuous systems, we expect that
\begin{equation}\label{eq:error}
 \|\mu_0^{num} - \mu_0^{wna}\|_2 \leq \mathcal{O}(h^2) \,,
\end{equation}
as $h$ tends to zero. This (roughly) quadratic convergence is
confirmed in Fig.~\ref{fig:convergence_study} where we computed the
slope $\gamma$ of the two curves, characterizing the convergence
rates, as
\begin{equation}
  \text{Schnakenberg (left): } \gamma \approx 2.078\,, \qquad
  \text{Brusselator (right): } \gamma \approx 2.008\,.
\end{equation}

\begin{figure}[htbp]
\centering
\begin{subfigure}{0.47\linewidth}
\includegraphics[width=\linewidth,height=5.0cm]{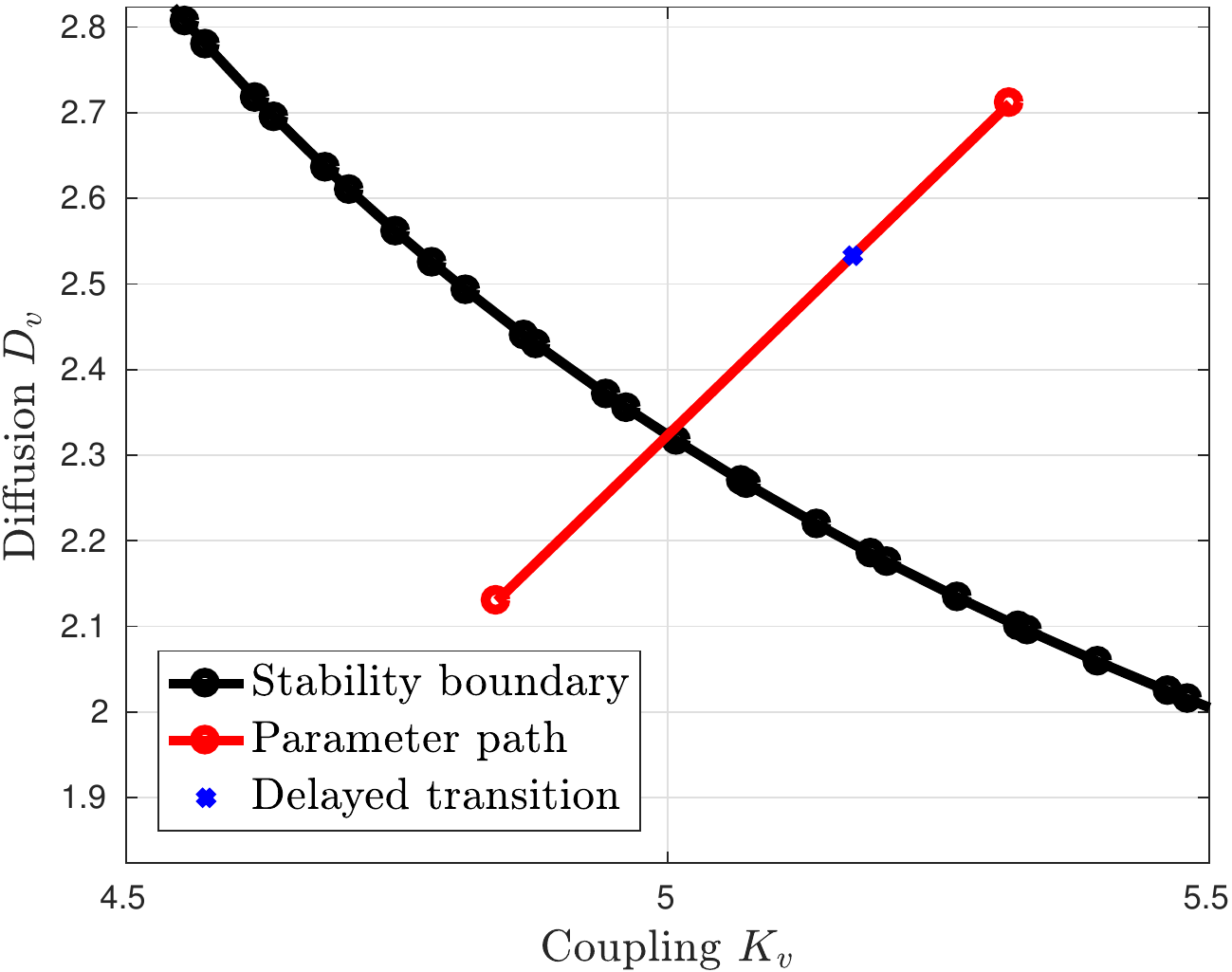}
\end{subfigure}
\begin{subfigure}{0.47\linewidth}
\includegraphics[width=\linewidth,height=5.0cm]{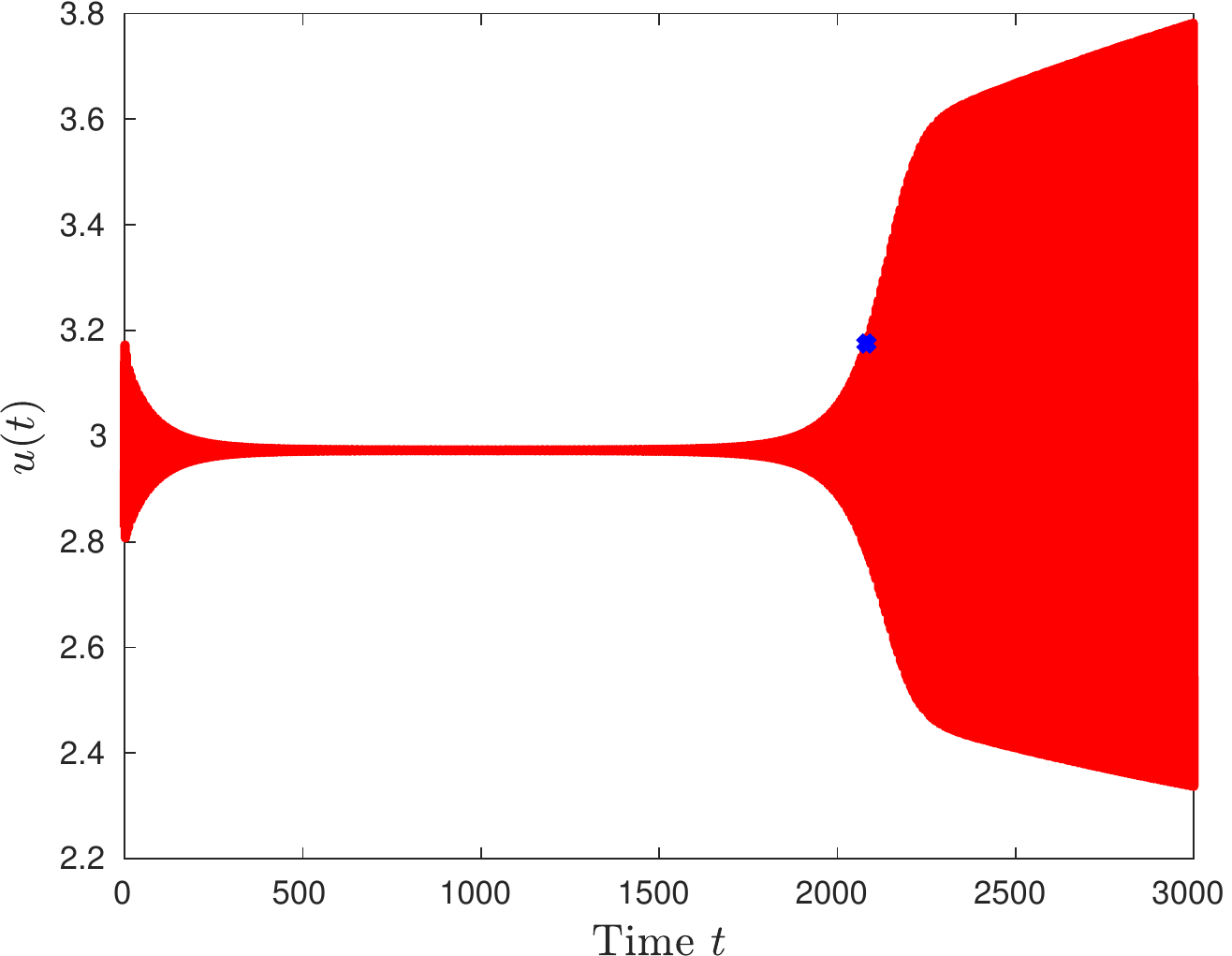}
\end{subfigure}
\caption{\label{fig:ramping_brusselator} Delayed Hopf bifurcation for
  the Brusselator kinetics with $\kappa = 10^{-3}$, $\eps = 0.5$,
  $\mu_0 = (5,2.32)^T$ and $\mu_1 = (0.63,0.78)^T$ (see
  \eqref{eq:time_parameters}). {The left panel shows the path
    of the time-dependent parameter sweep, with the initial and final
    points being in the stability (bottom left) and instability (up
    right) regions, respectively.} Numerical results for the
  membrane-bound activator $u(t)$ are shown in the right panel. The
  ``static'' bifurcation point is reached when $t = 1000$, but the
  transition to a periodic solution is delayed. The reduced PDE-ODE
  model (because of azimuthal invariance) is discretized using the
  method of lines. {The stiff ODE MATLAB solver
    \textit{ode23s} is used for the numerical time integration.}}
\end{figure}

{Next, we illustrate a delayed bifurcation behavior for the
  onset of oscillations.  It is well-known that a slow sweep of a
  parameter through a Hopf bifurcation point can cause delayed
  transitions to oscillatory dynamics in systems of ODEs
  \cite{erneux1989}.} In order to observe such a delayed Hopf
bifurcation effect in our PDE setting, the following numerical
experiment was performed: We let $\kappa \ll 1$ be a second small
parameter and $\mu(t) \in \R^2$ be a time-dependent vector of
bifurcation parameters such that
\begin{equation}\label{eq:time_parameters}
 \mu(t) = \mu_0 + (1-\kappa t)\eps^2\mu_1.
\end{equation}
Here, the parameter $\eps$ does not need to be particularly small
because the weakly nonlinear results from \S
\ref{sec:weakly_nonlinear_theory} and \S
\ref{sec:bifurcation_analysis} do not apply to time-dependent
bifurcation parameters. Full numerical results of the delayed
bifurcation are shown in Fig.~\ref{fig:ramping_brusselator}. The
initial vector of bifurcation parameters is
$\mu(0) = \mu_0 - \eps^2 \mu_1$, for which the solution decays
exponentially. The ``static'' bifurcation point is reached when
$t = \kappa^{-1}$, after which the instability region is entered and
sustained oscillations are expected. {However, because of the
  slow parameter sweep, the transition to oscillations is delayed; an
  effect clearly observed in Fig.~\ref{fig:ramping_brusselator}.}

\begin{figure}[htbp]
\centering
\begin{subfigure}{0.47\linewidth}
\includegraphics[width=\linewidth,height=5.0cm]{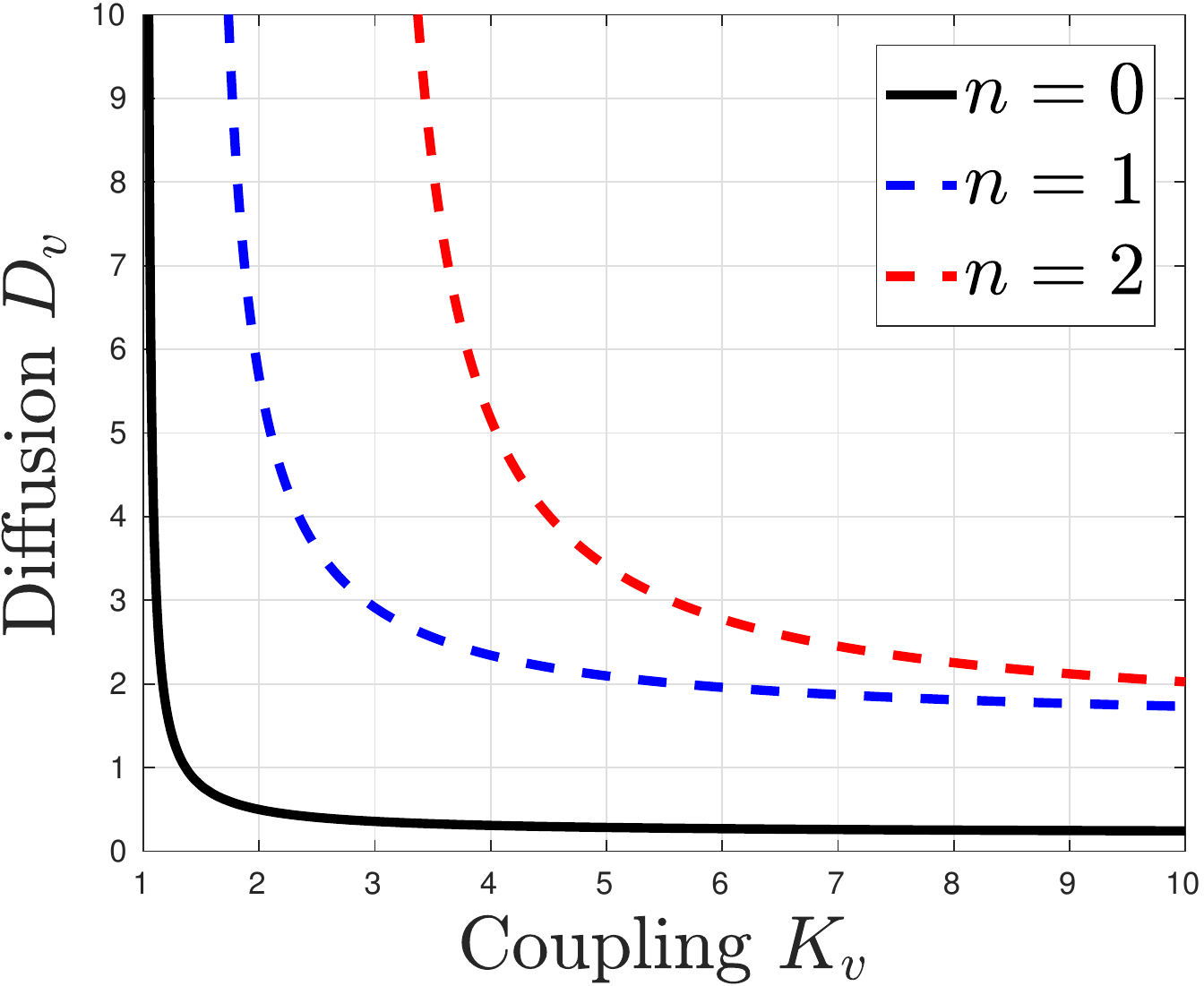}
\end{subfigure}
\begin{subfigure}{0.47\linewidth}
\includegraphics[width=\linewidth,height=5.0cm]{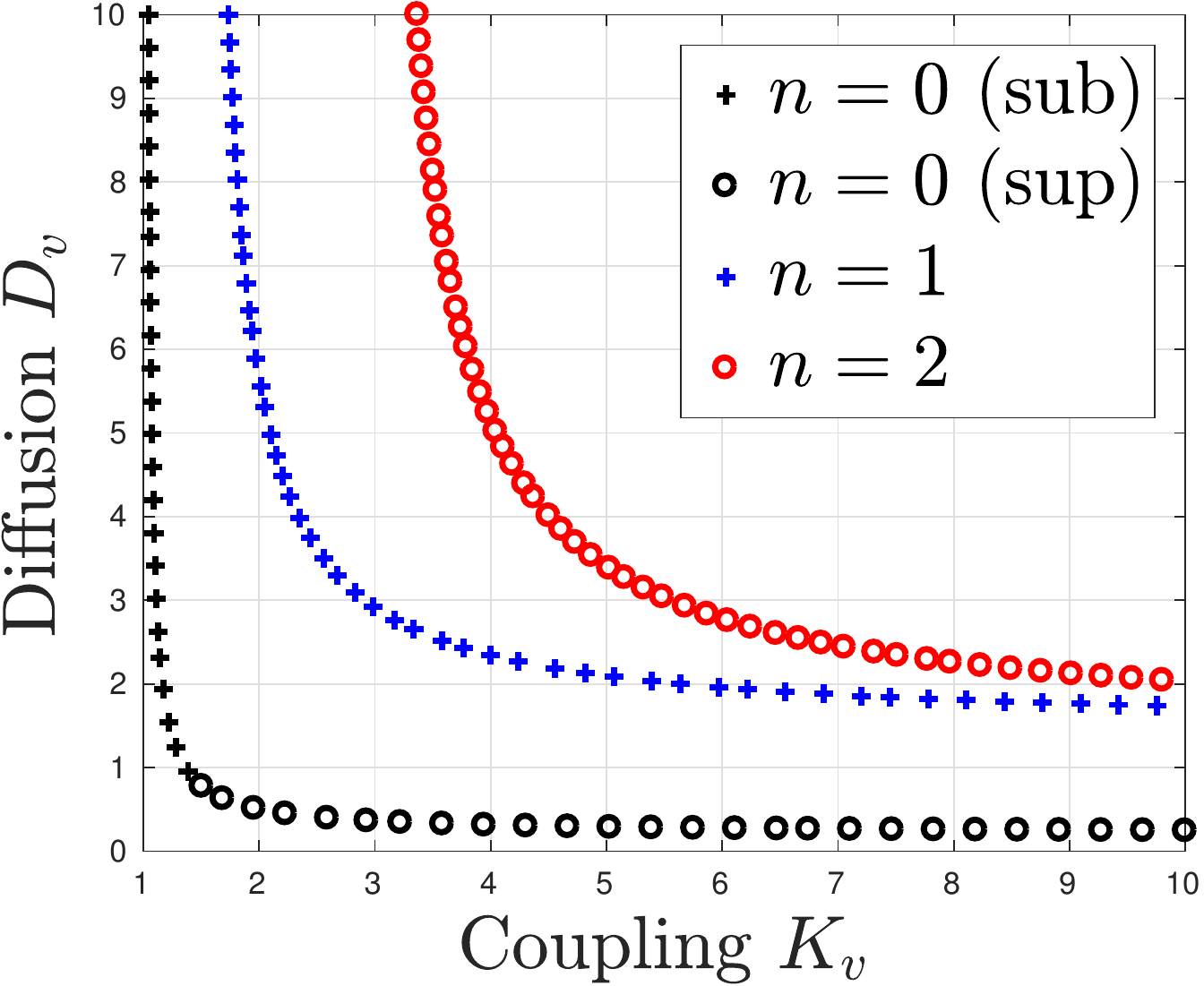}
\end{subfigure}
\caption{\label{fig:brusselator_EAD_NO_sub} Stability curves
  associated with the modes $n=0,\, 1,\, 2$ in the plane of parameters
  $(K_v,D_v)$ for the Brusselator kinetics. Other parameter values are
  $R=1,\, D_u=1,\, \sigma_u=\sigma_v=0.01,\, K_u=0.1,\, d_u=d_v =
  0.5,\, a=3$ and $b=8.7$. The region of linear stability is located
  to the left of the $n=0$ stability boundary, which corresponds to a
  locus of Hopf bifurcations. In the right panel, the symbol 'o'
  indicates a supercritical bifurcation while '+' indicates a
  subcritical bifurcation. {In the right panel, notice the
    transition from sub to super-criticality along the $n=0$ Hopf
    locus (black curve)}.}
\end{figure}

\begin{figure}[htbp]
\centering
\begin{subfigure}{0.47\linewidth}
\includegraphics[width=\linewidth,height=5.0cm]{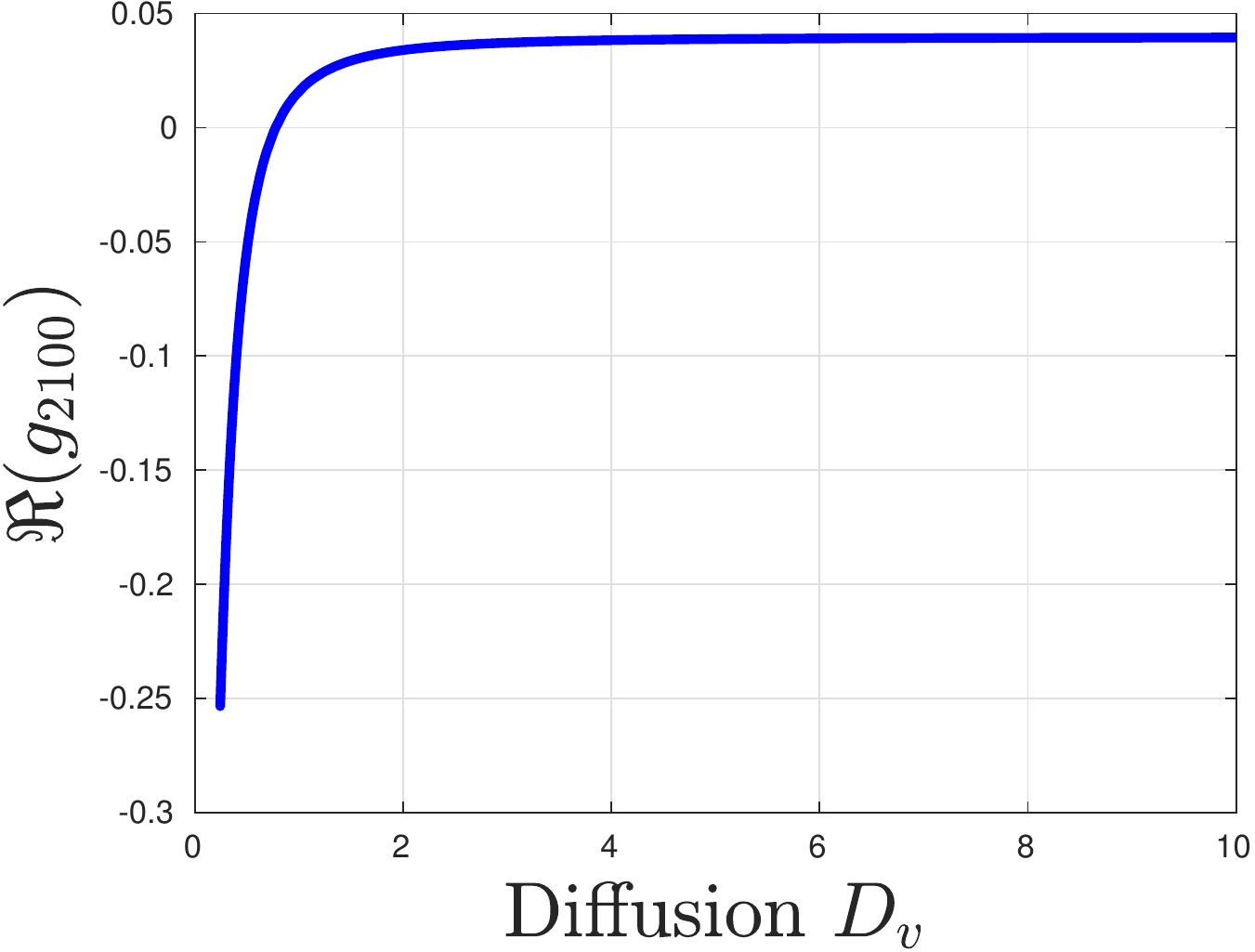}
\end{subfigure}
\caption{\label{fig:Reg2100} Transition from a supercritical to a
  subcritical Hopf bifurcation for the Brusselator
  kinetics. {The plot shows that the real part of the coefficient
  $g_{2100}$ in \eqref{eq:amplitude_hopf}, when numerically evaluated
  along the Hopf stability curve in
  Fig.~\ref{fig:brusselator_EAD_NO_sub}, exhibits a sign-change.}}
\end{figure}

{Next, we show that with Brusselator kinetics, there is a
  parameter regime where the Hopf bifurcation switches from
  supercritical to subcritical. A new linear stability phase diagram
  illustrating this transition is shown in
  Fig.~\ref{fig:brusselator_EAD_NO_sub}. For this parameter set, the
  Hopf locus, provided by the trivial $n=0$ mode (black curve), is the
  primary instability. The same parameter values as given in the
  caption of Fig.~\ref{fig:stability_diagram_brusselator} are used
  here, with the only difference being that the Brusselator kinetic
  parameter $b$ has been increased from $b=7.5$ to $b=8.7$. For this
  parameter set, we conclude that ratios of bulk diffusivity and
  coupling given by ${D_v/D_u} \gg 1$ and ${K_v/K_u} \approx 1$ are
  sufficient for the Hopf bifurcation to be
  subcritical. Alternatively, when ${D_v/D_u} \approx 1$ and
  ${K_v/K_u} \gg 1$, the Hopf bifurcation is supercritical. This
  criticality change is also shown in Fig.~\ref{fig:Reg2100}, where we
  clearly observe that the real part of the cubic term coefficient in
  \eqref{eq:amplitude_hopf} changes sign at some point along the Hopf locus.}

\begin{figure}[htbp]
\centering
\begin{subfigure}{0.47\linewidth}
\includegraphics[width=\linewidth,height=5.0cm]{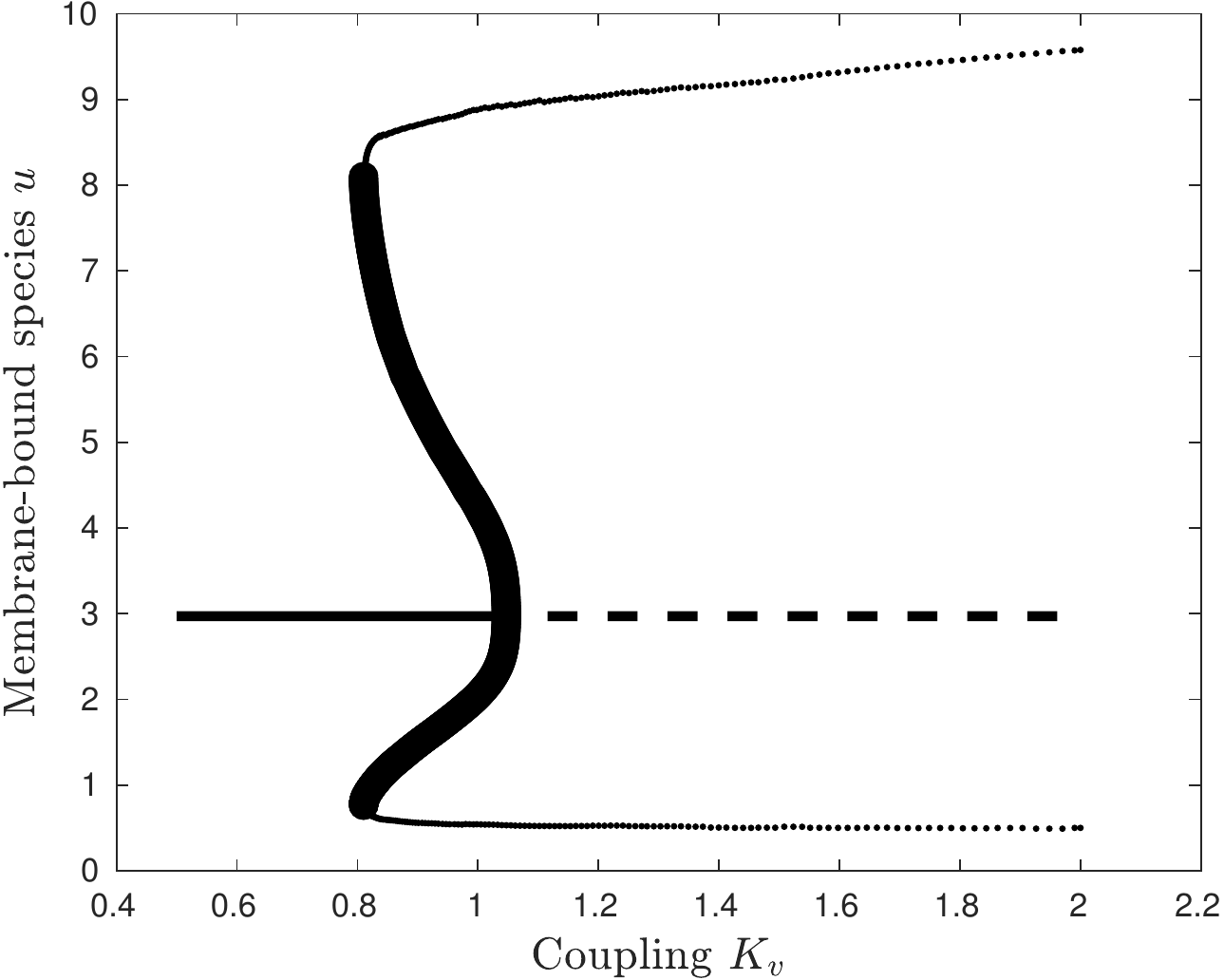}
\end{subfigure}
\begin{subfigure}{0.47\linewidth}
\includegraphics[width=\linewidth,height=5.0cm]{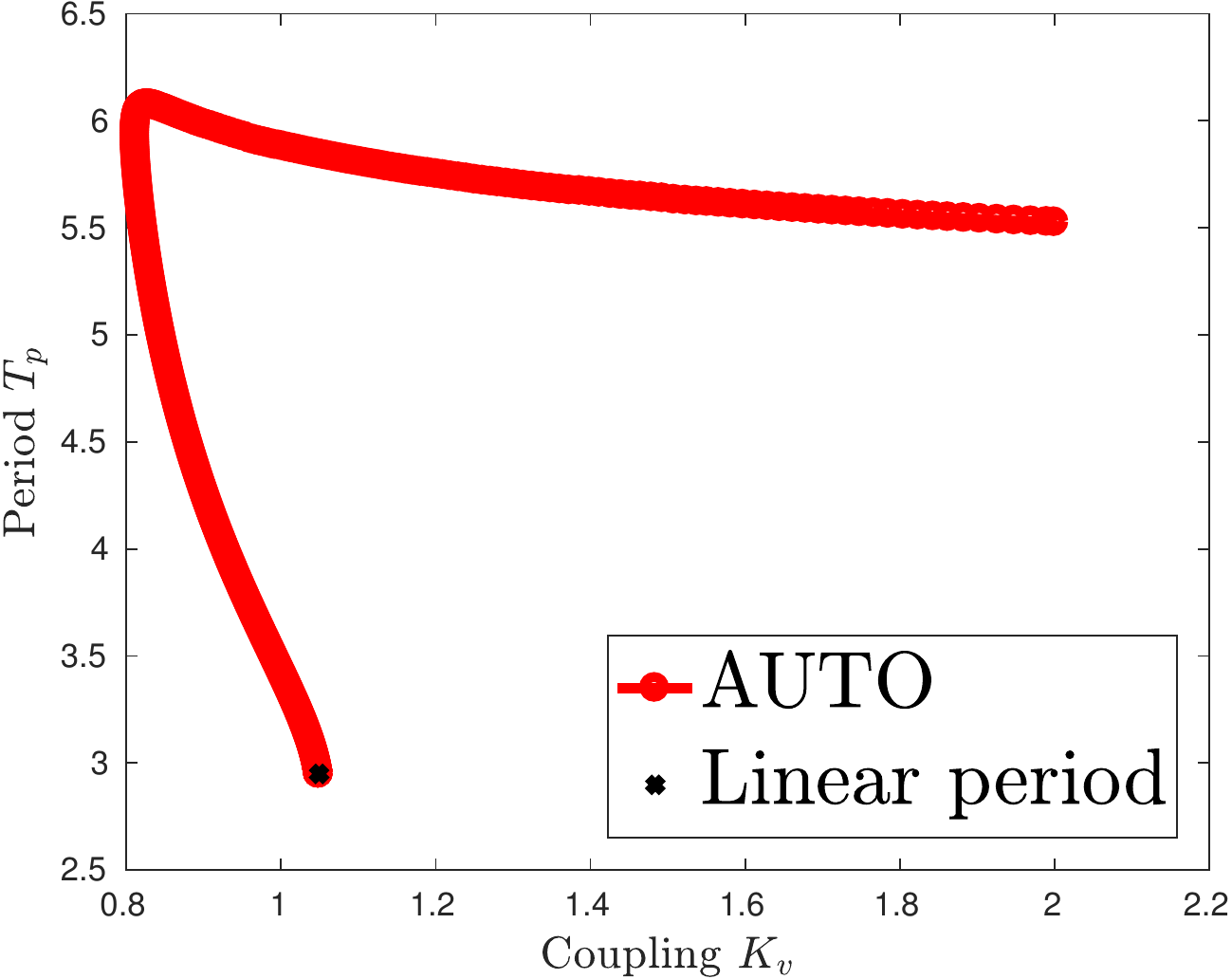}
\end{subfigure}
\caption{\label{fig:global_branch_sub} The left panel shows the global
  bifurcating branch for the membrane-bound activator species. The
  right panel shows the numerically-computed period {as a
    function of the coupling $K_v$ when $D_v=9$ (other parameters as
    in the caption of Fig.~\ref{fig:brusselator_EAD_NO_sub}).} Notice
  that the initial point matches the linear period predicted by the
  asymptotic theory, indicated by a black "x". The computation is
  performed with AUTO using $N=200$ radial grid points in the bulk.}
\end{figure}

{To confirm this transition to a subcritical Hopf bifurcation
  as predicted from our weakly nonlinear theory, we perform numerical
  simulations near the intersection point of the Hopf locus in
  Fig.~\ref{fig:brusselator_EAD_NO_sub} and the horizontal line
  $D_v = 9$.} The global bifurcating branch computed with AUTO is
shown in the left panel of Fig.~\ref{fig:global_branch_sub}. In the
right panel of Fig.~\ref{fig:global_branch_sub} we plot the
corresponding numerically computed period of oscillations. These
numerical results confirm the predicted loss of stability through a
subcritical Hopf bifurcation. The results also suggest bistability
between a large amplitude limit cycle and the steady state solution in
a small parameter window prior to the bifurcation
point. {From the left panel of
  Fig.~\ref{fig:global_branch_sub}, the stable and unstable branches
  of periodic solution merge at a fold point around
  $K_v \approx 0.8$.}

\begin{figure}[htbp]
\centering
\begin{subfigure}{0.47\linewidth}
\includegraphics[width=\linewidth,height=5.0cm]{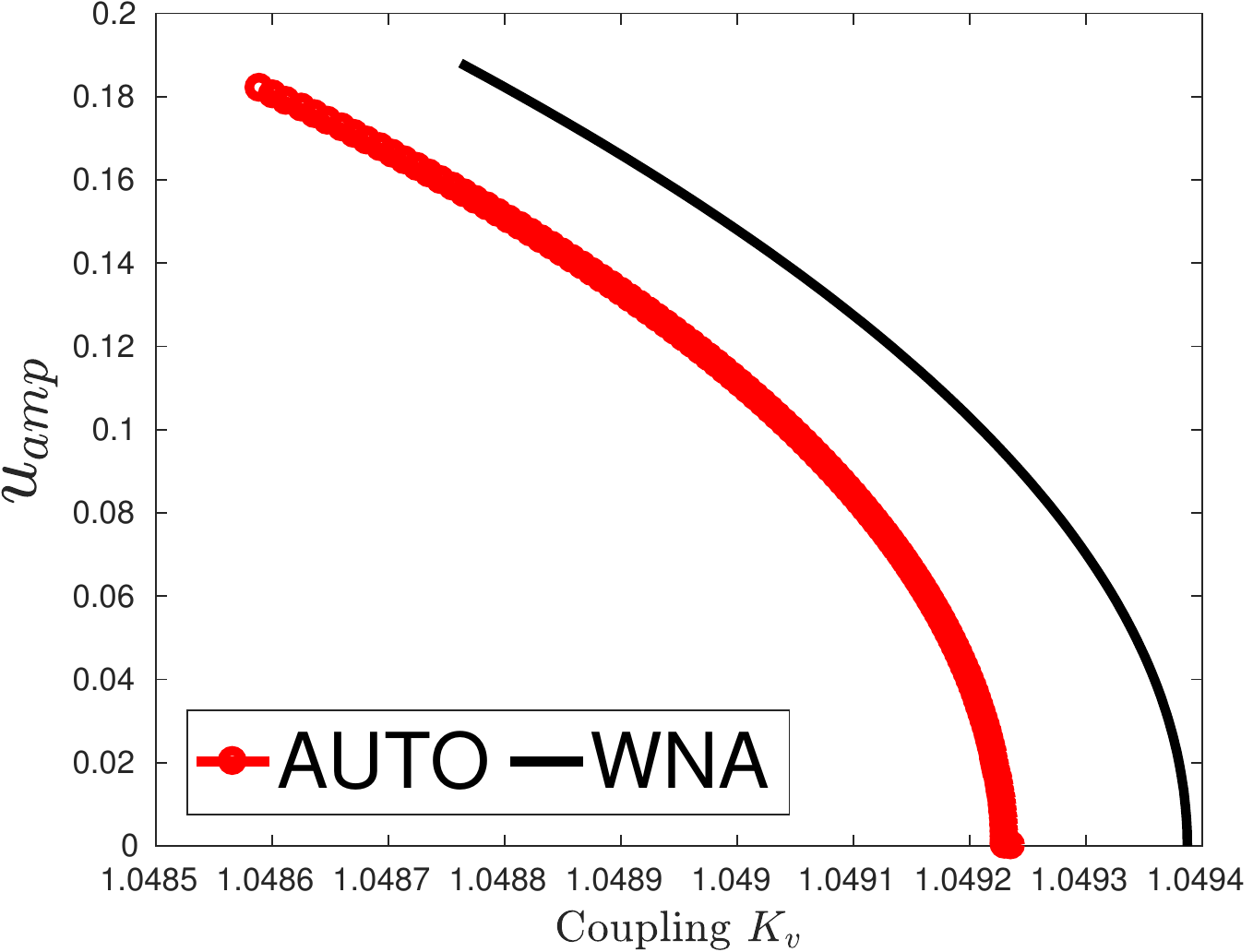}
\end{subfigure}
\caption{\label{fig:amplitude_NO} Local unstable periodic solution
  branch past a subcritical Hopf bifurcation for the Brusselator
  kinetics {with $D_v=9$ (other parameters as in the caption of
  Fig.~\ref{fig:brusselator_EAD_NO_sub}).} The red curve is obtained
  through numerical continuation using AUTO, while the black curve is
  the amplitude $u_{amp}$ as predicted by the weakly nonlinear theory
  in \eqref{eq:u_amp} with $\eps = 2.5 \times 10^{-2}$.}
\end{figure}

In Fig.~\ref{fig:amplitude_NO}, we compare the numerically computed
amplitude of {oscillations against results from our weakly
  nonlinear theory for $\eps = 0.025$.} Despite the slight shift
between the bifurcation points, good agreement is once again
obtained. However, we notice that the range over which the two
branches coincide is much more narrow than for their supercritical
counterparts. {This is likely due to the real part of the
  cubic term coefficient in \eqref{eq:amplitude_hopf} having a rather
  small magnitude when $D_v = 9$ (see Fig.~\ref{fig:Reg2100}),
  which suggests that the unresolved quintic term in the normal form may
  be quantitatively significant.}

\begin{figure}[htbp]
\centering
\begin{subfigure}{0.32\linewidth}
\includegraphics[width=\linewidth,height=5.0cm]{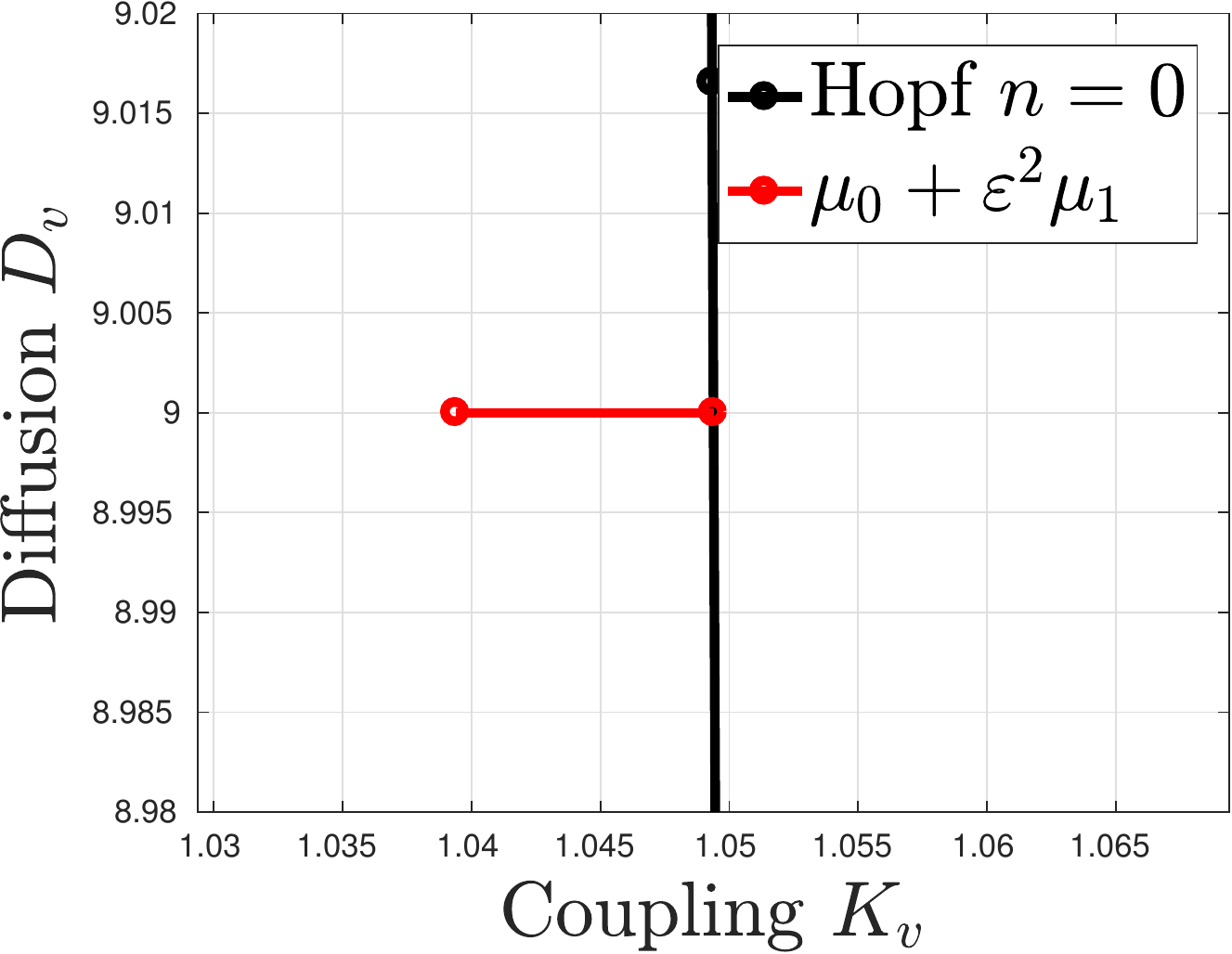}
\end{subfigure}
\begin{subfigure}{0.32\linewidth}
\includegraphics[width=\linewidth,height=5.0cm]{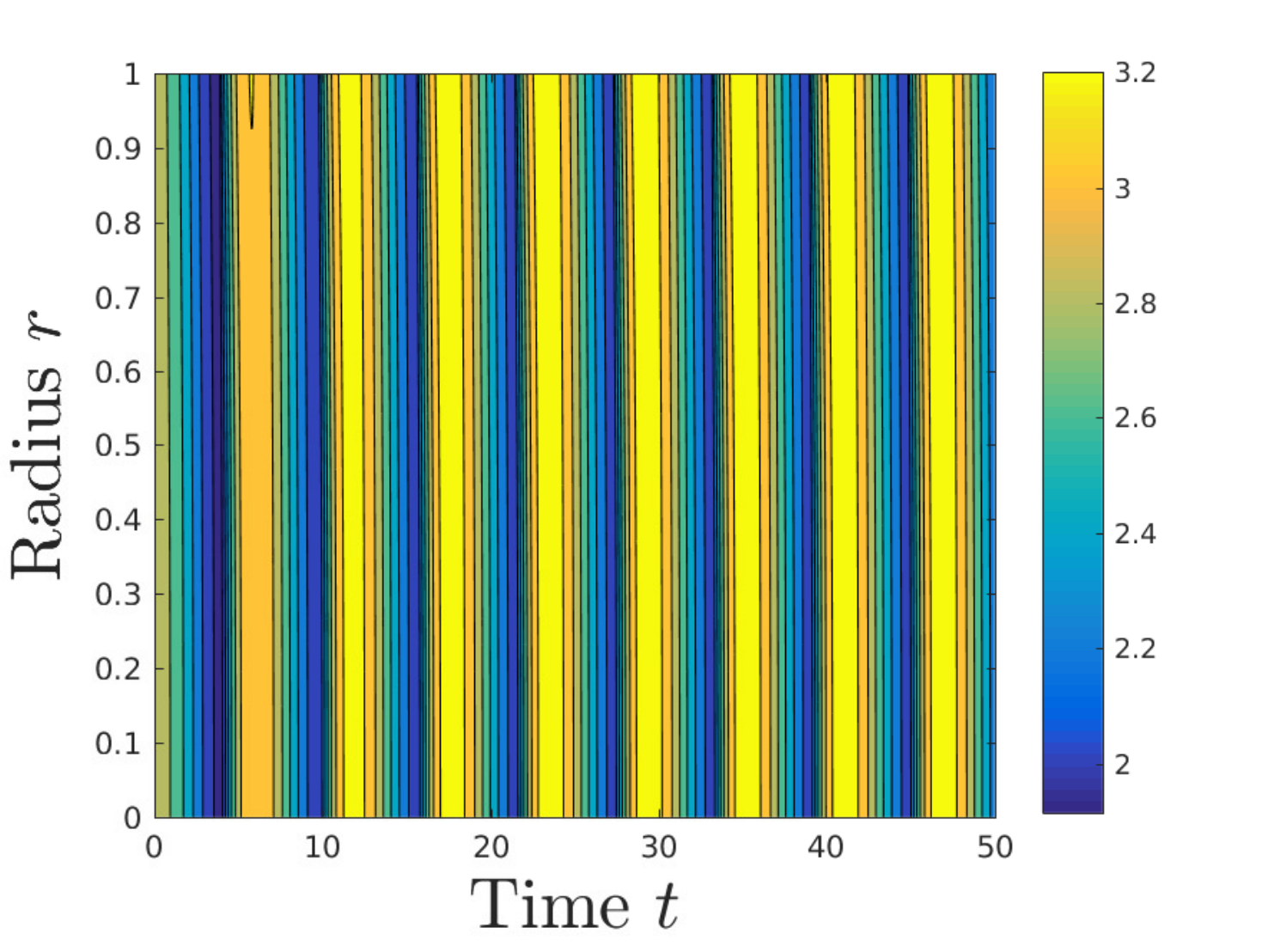}
\end{subfigure}
\begin{subfigure}{0.32\linewidth}
\includegraphics[width=\linewidth,height=5.0cm]{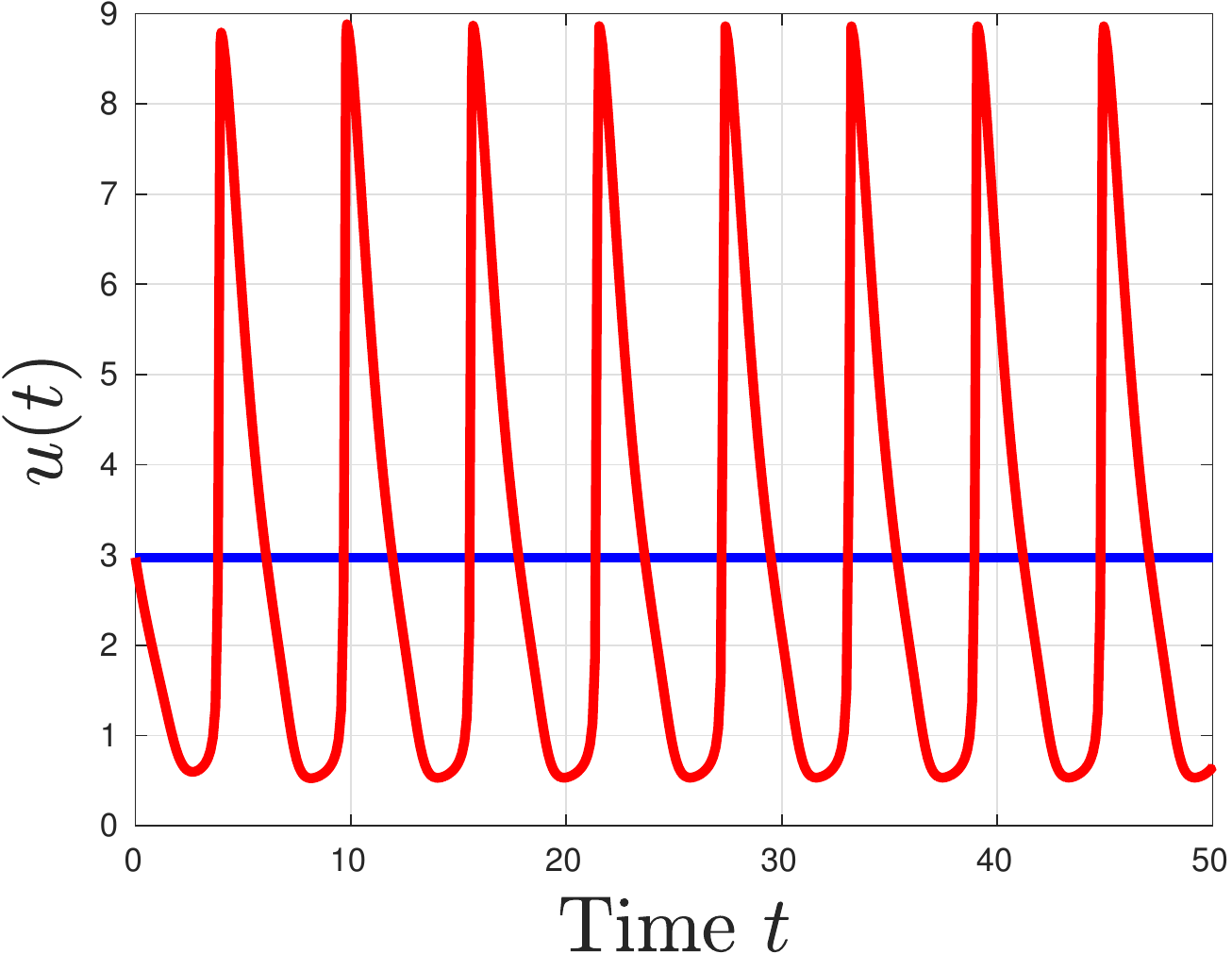}
\end{subfigure}
\caption{\label{fig:U_u_sub} Highly nonlinear
  {relaxation-type} oscillations near a subcritical Hopf
  bifurcation for the Brusselator kinetics. As shown in the left
  panel, the simulation is performed on the horizontal line $D_v = 9$
  with $\eps = 0.1$. In the middle panel, the numerically computed
  spatio-temporal bulk oscillations are plotted for the activator
  species $U(r,t)$, with the radius $r$ on the vertical axis and time
  $t$ on the horizontal axis. The right panel shows the
  corresponding relaxation oscillations for the membrane-bound
  activator species. The initial condition corresponds to the unstable
  periodic solution in the weakly nonlinear regime as given by
  \eqref{eq:periodic_solution} with
  $\theta_0(0) = \frac{\pi}{2},\, t=0,\, \eps=0.1$.}
\end{figure}

{Finally, in Fig.~\ref{fig:U_u_sub} we show numerical PDE
  results of large amplitude relaxation oscillations that can occur on
  the horizontal line $D_v = 9$ in
  Fig.~\ref{fig:brusselator_EAD_NO_sub}.  These oscillations,
  characterized by sharp variations followed by a rest period, are
  often observed in simpler ODE models having a subcritical Hopf
  bifurcation. They are qualitatively distinct from the harmonic-type
  oscillations in Fig.~\ref{fig:schnakenberg_K_v_D_v} and
  Fig.~\ref{fig:brusselator_D_v}.}

\vfill\eject
\subsubsection{Turing patterns arising from pitchfork bifurcations}\label{subsubsec:subsub_turing}

{In this subsection the formation of spatially inhomogeneous
  steady states close to a pitchfork (or Turing) bifurcation is
  investigated numerically. Since there is no azimuthal invariance
near such bifurcations, one cannot use 1-D finite differences to
spatially discretize the system as in \S
\ref{subsubsec:subsub_hopf}.} Hence, the full spatial structure of the
model must be considered, and its discretization is done with the
finite element method as implemented by the PDE Toolbox of MATLAB
\cite{matlab2016}. To be more precise, linear triangular elements are
used to discretize the bulk domain, while the 1-D Laplace-Beltrami
boundary diffusion operators is handled with second-order centered
differences using the nodes attached to the boundary.

{As mentioned in \cite{sand2007}}, the computation of general
equilibria to system of elliptic PDEs posed on arbitrary 2-D or 3-D
domains is a challenging task. For these domains, the spatial
discretization yields large and sparse systems of nonlinear equations
{for which traditional software like AUTO} \cite{doedel2007}
and MATCONT \cite{matcont2003} are of limited use. It is to address
these issues that a number of new MATLAB packages such as
\textit{pde2path} \cite{pde2path2014} and Computational Continuation
Core (\textsc{coco}) \cite{danko2013} have emerged in the research
community. While \textit{pde2path} has been specifically designed for
systems of elliptic PDEs, it cannot handle nonstandard boundary
conditions like those encountered in bulk-surface coupled models. The
base state can be computed using the \textit{Equilibrium Point}
toolbox from \textsc{coco}, but our attempt to compute the global
bifurcating branch at a pitchfork bifurcation point has been
unsuccessful. From a numerical bifurcation analysis perspective, the
situation is rather {degenerate} since rotational symmetry
results in two critical eigenfunctions at the pitchfork bifurcation
point. Consequently, the results exhibited in this section mostly rely
on full numerical {time-dependent} simulations using
Implicit-Explicit time-stepping; an approach that unfortunately only
reveals stable steady states (either patterned or patternless). In
practice, the simulation is stopped when the relative distance between
the current and the previous time step becomes smaller than some given
tolerance. {More details regarding the spatial discretization
  and the numerical methods are given in Appendix
  \ref{sec:numerical_methods}.}

\begin{figure}[htbp]
\centering
\begin{subfigure}{0.47\linewidth}
\includegraphics[width=\linewidth,height=5.0cm]{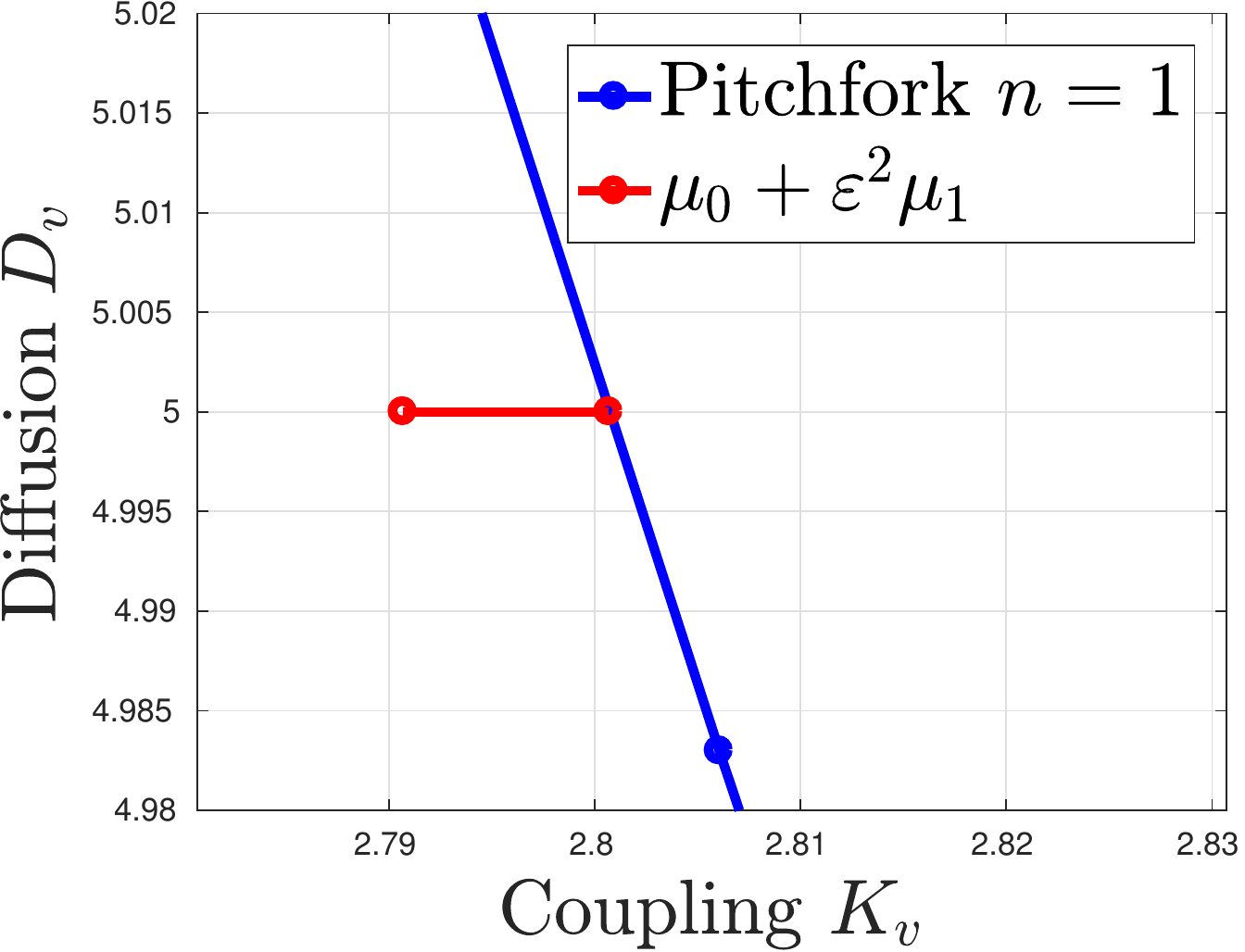}
\end{subfigure}
\begin{subfigure}{0.47\linewidth}
\includegraphics[width=\linewidth,height=5.0cm]{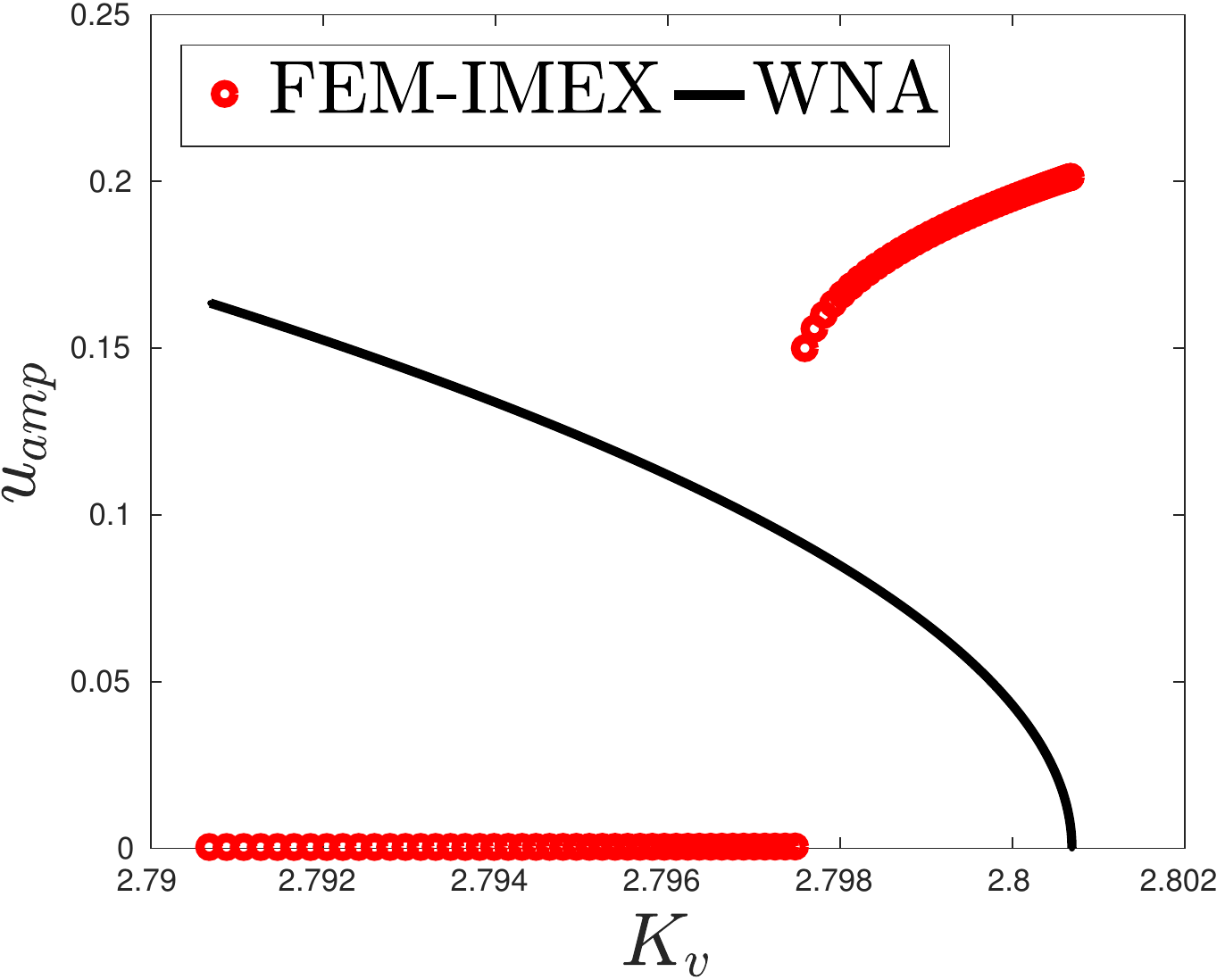}
\end{subfigure}
\caption{\label{fig:subcritical_pitchfork_schnakenberg} Subcritical
  pitchfork bifurcation with the Schnakenberg kinetics and $D_v = 5$
  (the other parameters are given in the caption of
  Fig.~\ref{fig:stability_diagram_schnakenberg}). The left panel is a
  magnified version of the stability diagram near the bifurcation
  point for $\eps = 0.1$. The stability region is located to the right
  of the blue curve, while the red curve indicates the parameter
  path. The right panel displays the maximal amplitude of the
  membrane-bound activator, with the black curve obtained from the
  weakly nonlinear theory \eqref{eq:amplitude_pitchfork} while the red
  curve is computed through successive numerical simulations. The
  boundary of the circular bulk domain is discretized with $N=200$
  nodes.}
\end{figure}

\begin{figure}[htbp]
\centering
\begin{subfigure}{0.47\linewidth}
\includegraphics[width=\linewidth,height=5.0cm]{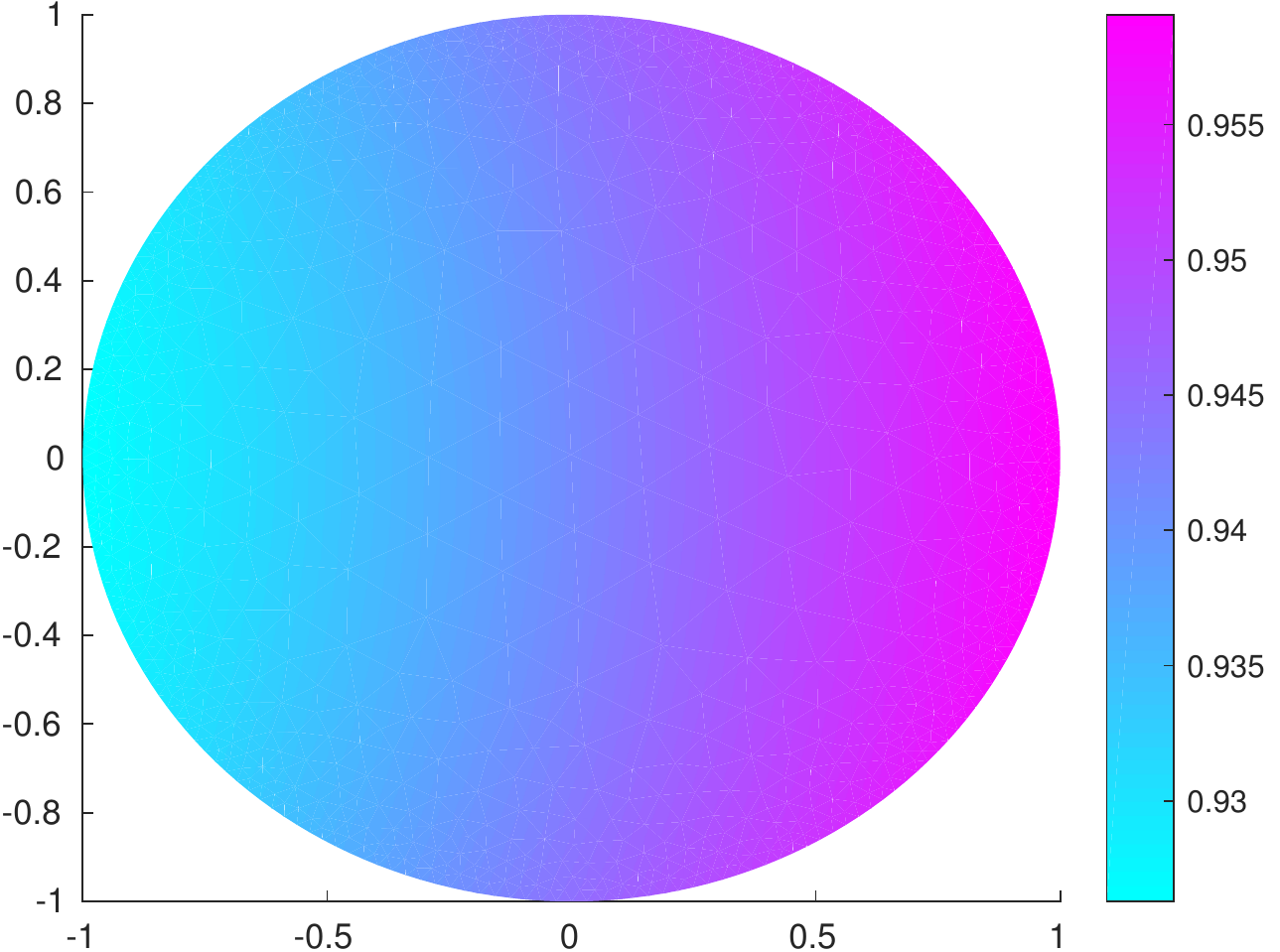}
\end{subfigure}
\begin{subfigure}{0.47\linewidth}
\includegraphics[width=\linewidth,height=5.0cm]{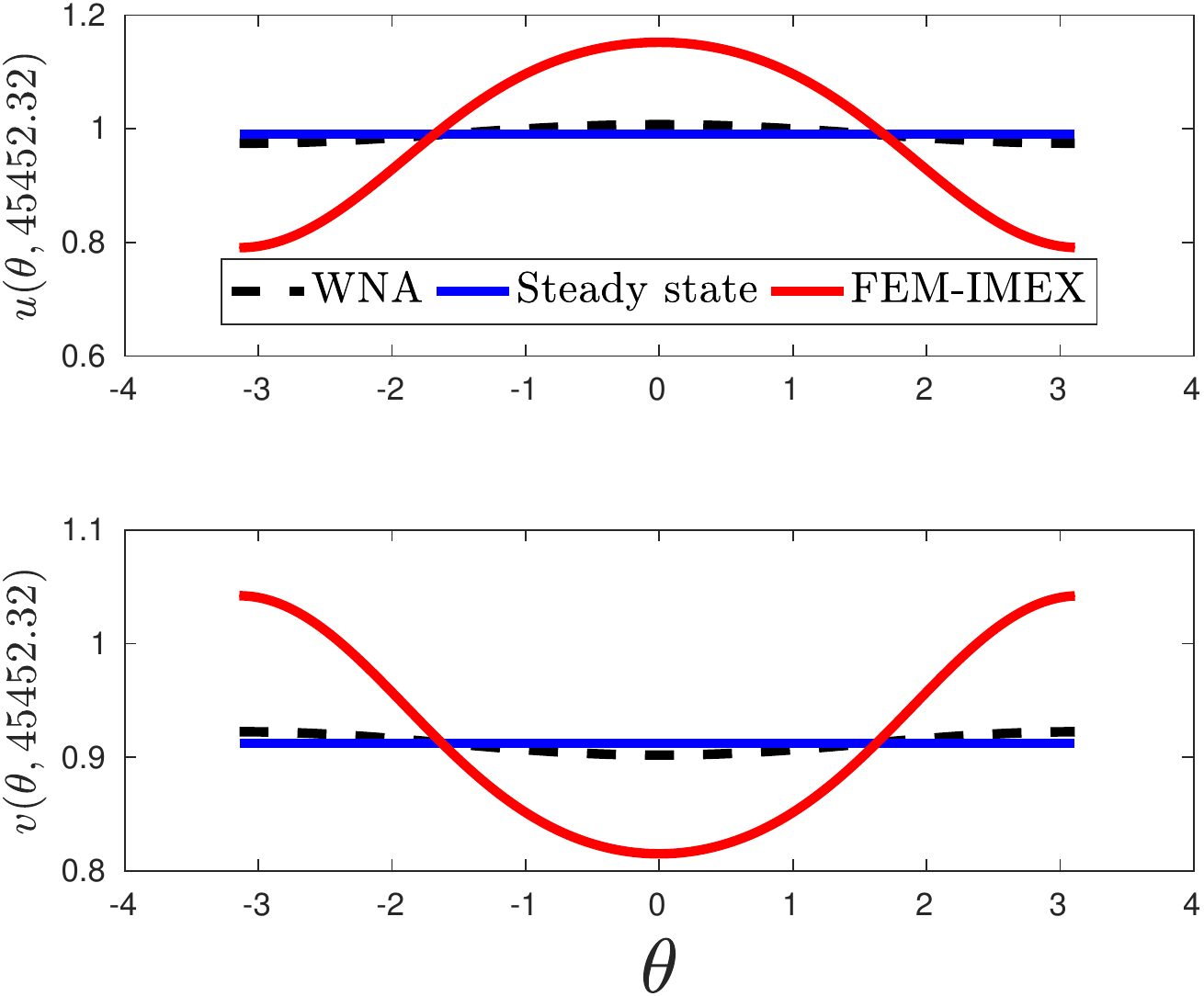}
\end{subfigure}
\caption{\label{fig:pattern_schnakenberg} A numerically computed
  stable pattern for $t \approx 45452$ in the vicinity of a
  subcritical pitchfork bifurcation under Schnakenberg kinetics. The
  black dashed curves in the right panel correspond to the unstable
  membrane-bound patterns, while the red curves correspond to the
  stable patterns. The solution in the weakly nonlinear regime, as
  given by \eqref{eq:patterned_solution} with $\eps=0.01$, $n=1$ and
  $\theta_n = 0$, is used as an initial condition. Because the
  critical eigenvalues are very small near the bifurcation point, the
  numerical solution {only very} slowly reaches the stable patterned
  state.}
\end{figure}

First, the loss of stability of the modes $n = 1$ through subcritical
pitchfork bifurcations as the coupling rate $K_v$ increases is
investigated numerically. The reader is referred to
Fig.~\ref{fig:stability_diagram_schnakenberg} and
Fig.~\ref{fig:stability_diagram_brusselator}, where we consider the
horizontal line $D_v = 5$ {and its intersection with the
  $n=1$ pitchfork curve.} In the right panel of
Fig.~\ref{fig:subcritical_pitchfork_schnakenberg} the amplitude of the
membrane-bound activator species for the Schnakenberg kinetics is
shown. The black curve is the unstable bifurcating branch and is only
valid locally. In theory, the unstable (black) and the stable (red)
branches should merge at a turning (or fold) point near
$K_v \approx 2.7976$. Such a feature cannot be detected with the
weakly nonlinear analysis from \S \ref{sec:weakly_nonlinear_theory} or
with direct time-stepping numerical simulations. Instead, numerical
continuation methods must be employed. Having discussed the challenges
associated with such a task earlier in this subsection, the
computation of the full branch is an open problem.  Nevertheless, the
solution in the weakly nonlinear regime can be used as an initial
condition for a direct numerical simulation, with the anticipation
that it evolves to the stable branch. The result of such an experiment
is shown in Fig.~\ref{fig:pattern_schnakenberg} for $\eps = 0.01$.

\begin{figure}[htbp]
\centering
\begin{subfigure}{0.47\linewidth}
\includegraphics[width=\linewidth,height=5.0cm]{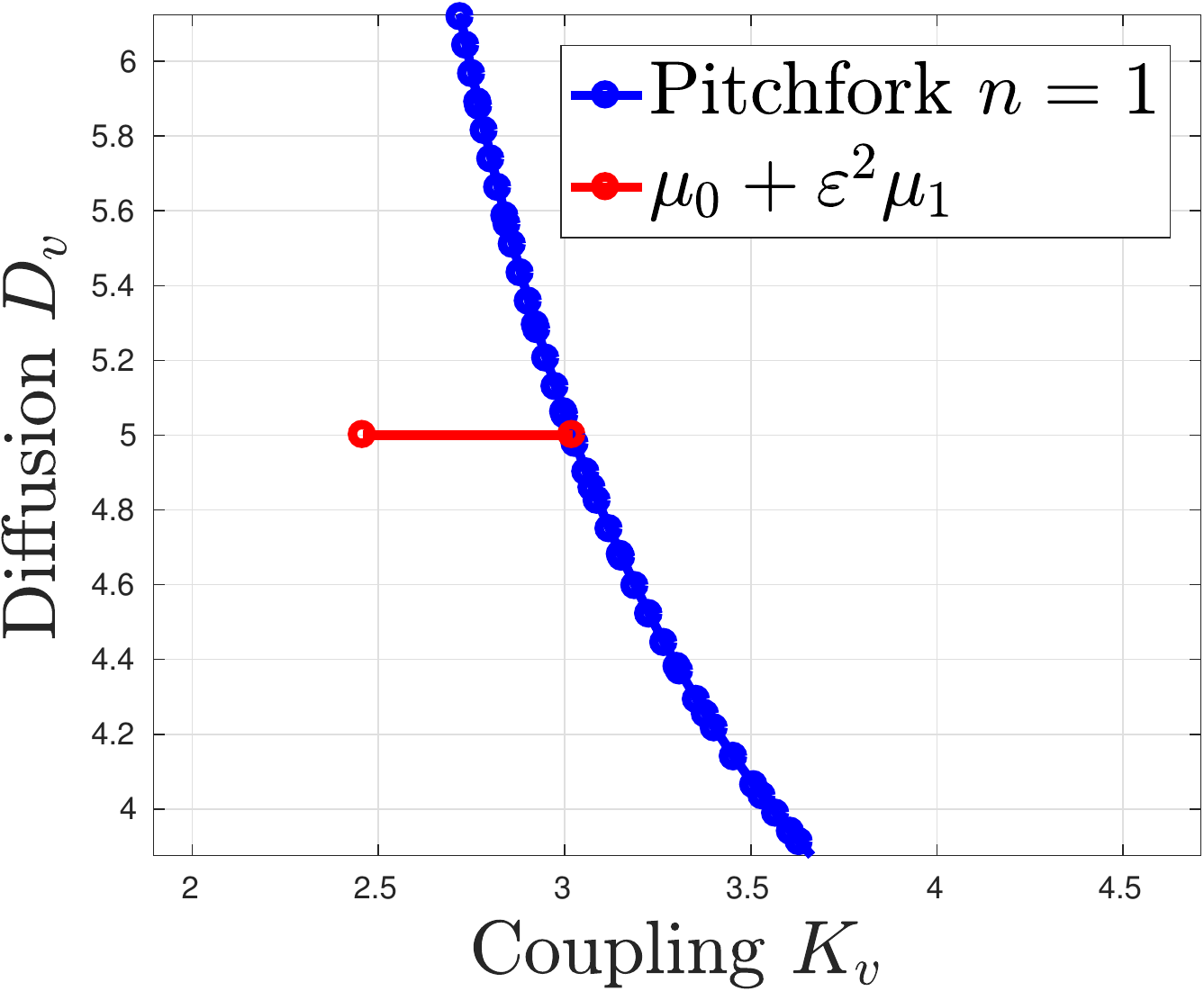}
\end{subfigure}
\begin{subfigure}{0.47\linewidth}
\includegraphics[width=\linewidth,height=5.0cm]{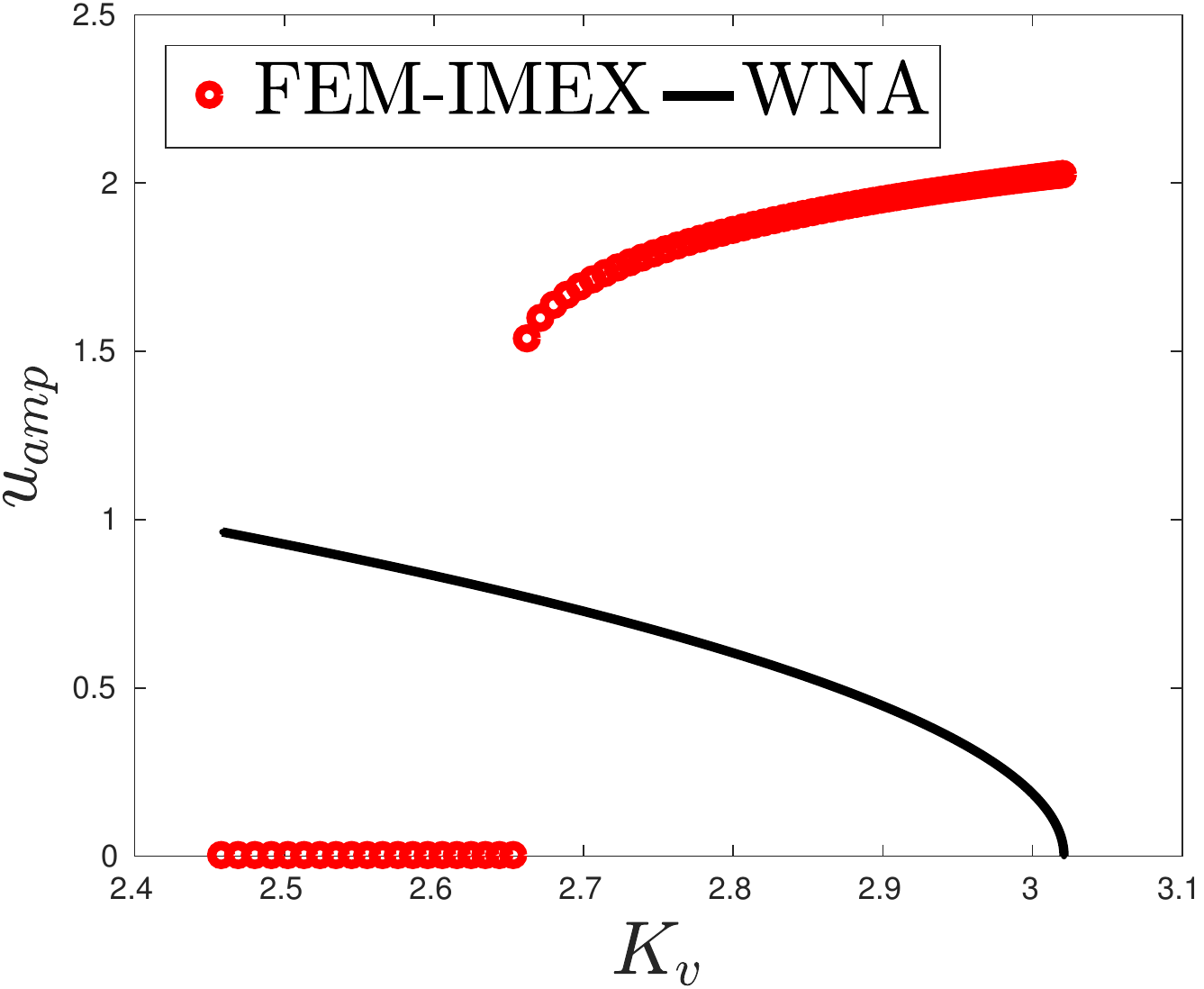}
\end{subfigure}
\caption{\label{fig:subcritical_pitchfork_brusselator} Subcritical
  pitchfork bifurcation with the Brusselator kinetics, $D_v = 5$, and
  $\eps = 0.75$ (the other parameters are given in the caption of
  Fig.~\ref{fig:stability_diagram_brusselator}). Again, the unstable branch
  goes backward, under which the {base state solution} is linearly
  stable. The boundary of the circular bulk domain is discretized with
  $N=200$ nodes.}
\end{figure}

\begin{figure}[htbp]
\centering
\begin{subfigure}{0.47\linewidth}
\includegraphics[width=\linewidth,height=5.0cm]{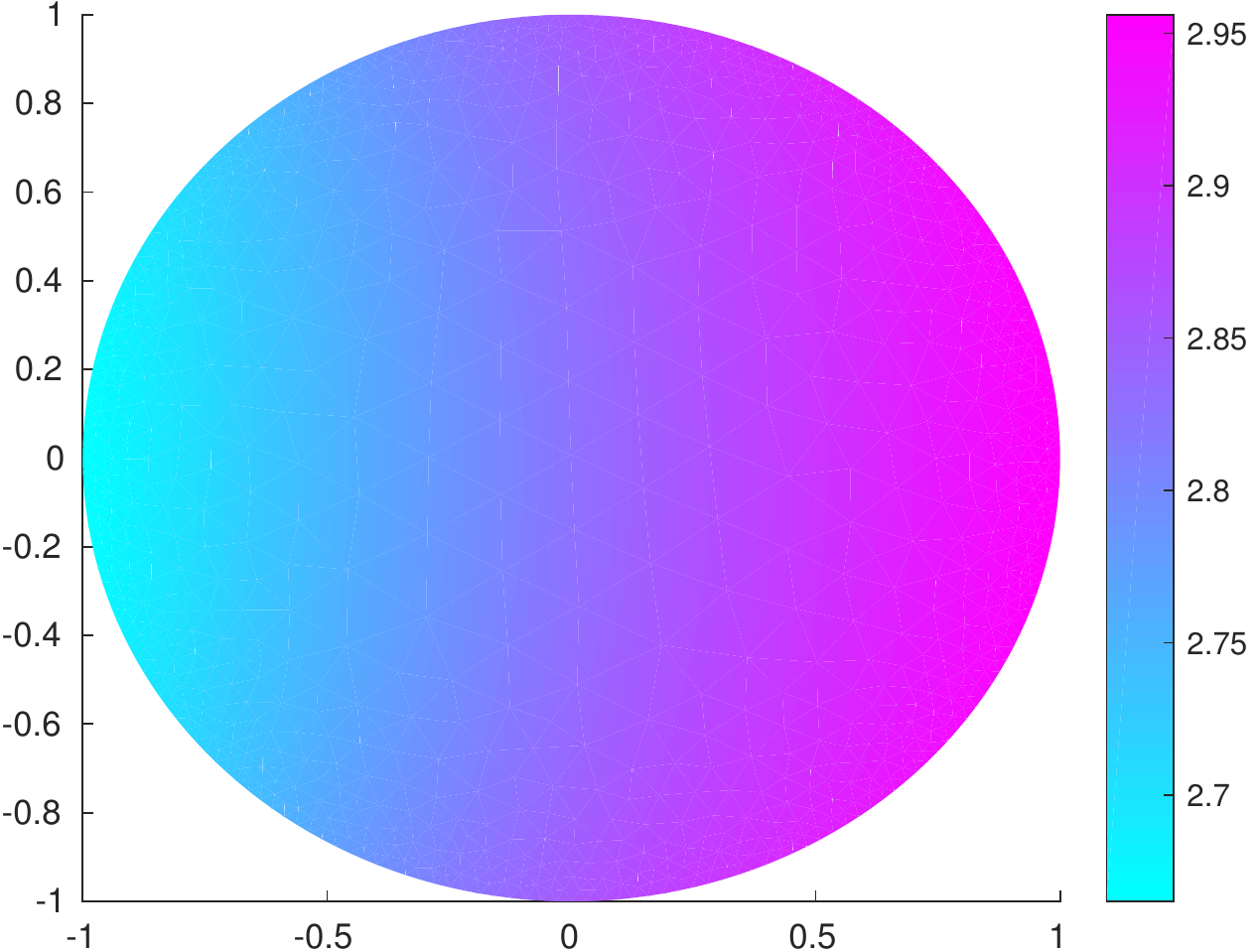}
\end{subfigure}
\begin{subfigure}{0.47\linewidth}
\includegraphics[width=\linewidth,height=5.0cm]{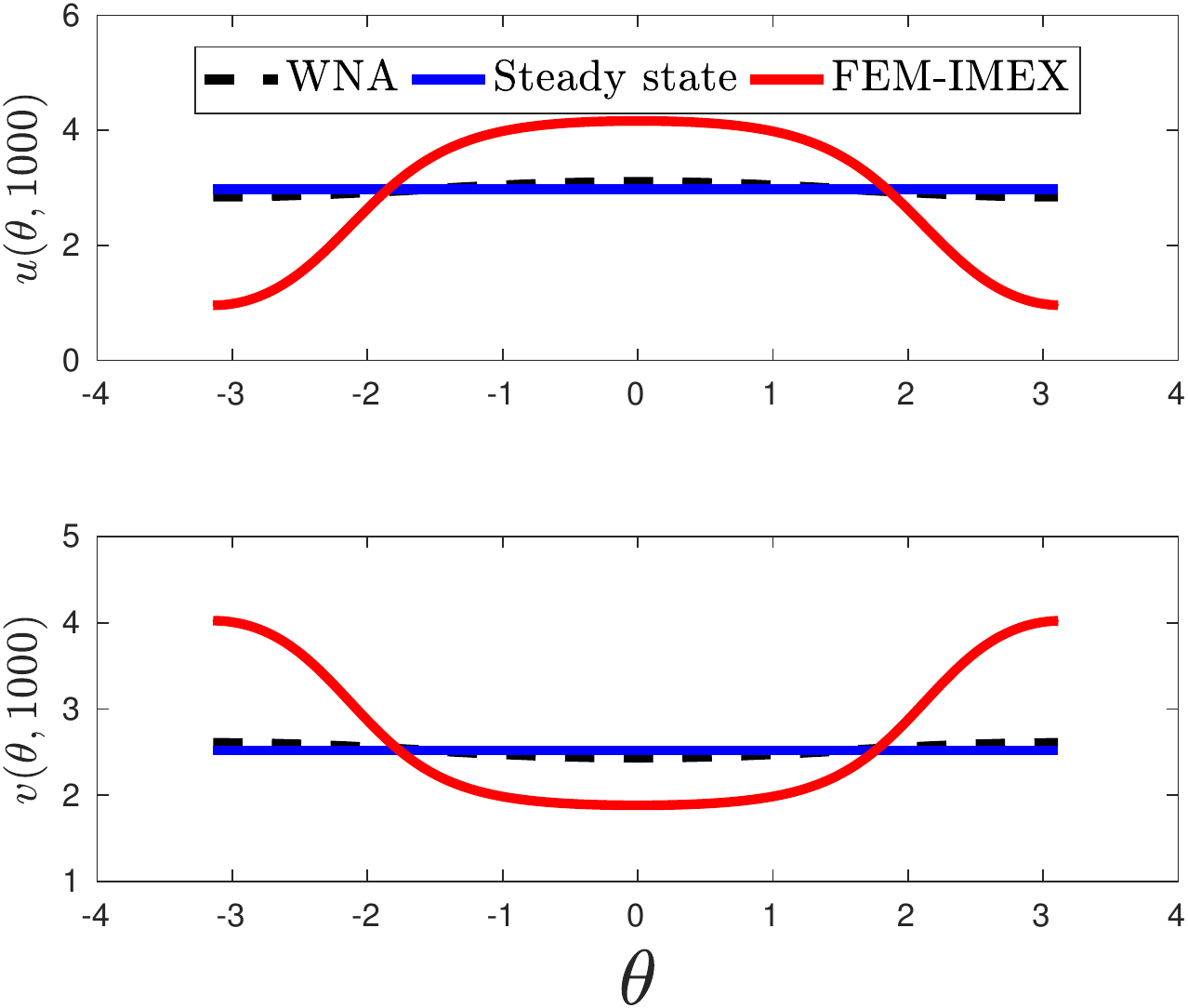}
\end{subfigure}
\caption{\label{fig:pattern_brusselator} A numerically computed stable
  pattern at $t = 1000$ (red curve), near the steady state, as
  evolved from the unstable branch near a subcritical pitchfork
  bifurcation with the Brusselator kinetics and $\eps = 0.1$.
  The left panel shows the corresponding solution in the bulk.}
\end{figure}

{The results of similar experiments using the Brusselator
  kinetics with $D_v=5$ (see the phase diagram in
  Fig.~\ref{fig:stability_diagram_brusselator}) are shown in
  Fig.~\ref{fig:subcritical_pitchfork_brusselator} and
  Fig.~\ref{fig:pattern_brusselator}. Notice here that the unstable
  branch goes farther backward than in the Schnakenberg case before
  reaching a "turning point" at around $K_v \approx 2.66$. The
  pitchfork bifurcation point is at $K_{v0} \approx 3.02$. Again, this
  is an example of a hard loss of stability of the base state.}

\begin{figure}[htbp]
\centering
\begin{subfigure}{0.47\linewidth}
\includegraphics[width=\linewidth,height=5.0cm]{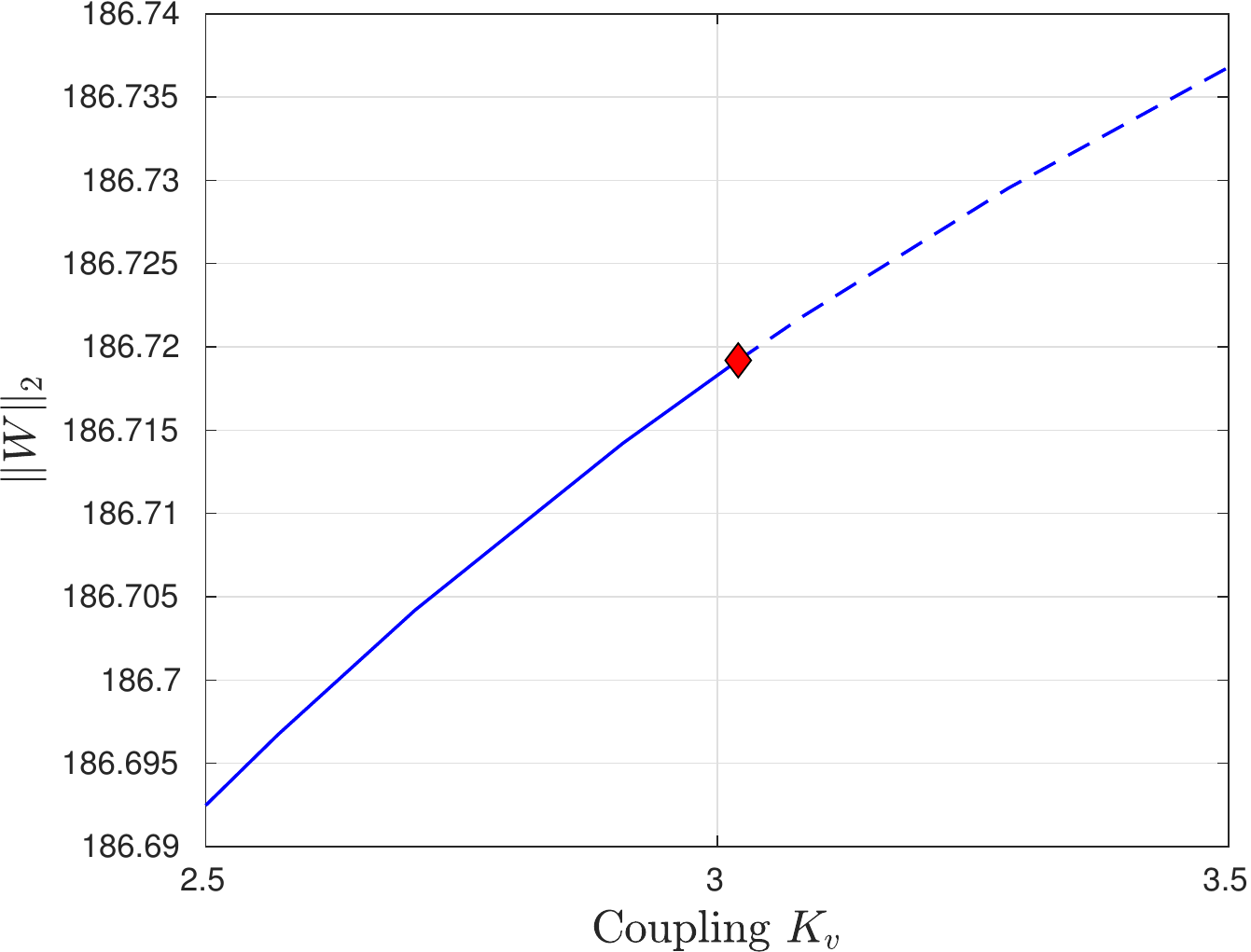}
\end{subfigure}
\begin{subfigure}{0.47\linewidth}
\includegraphics[width=\linewidth,height=5.0cm]{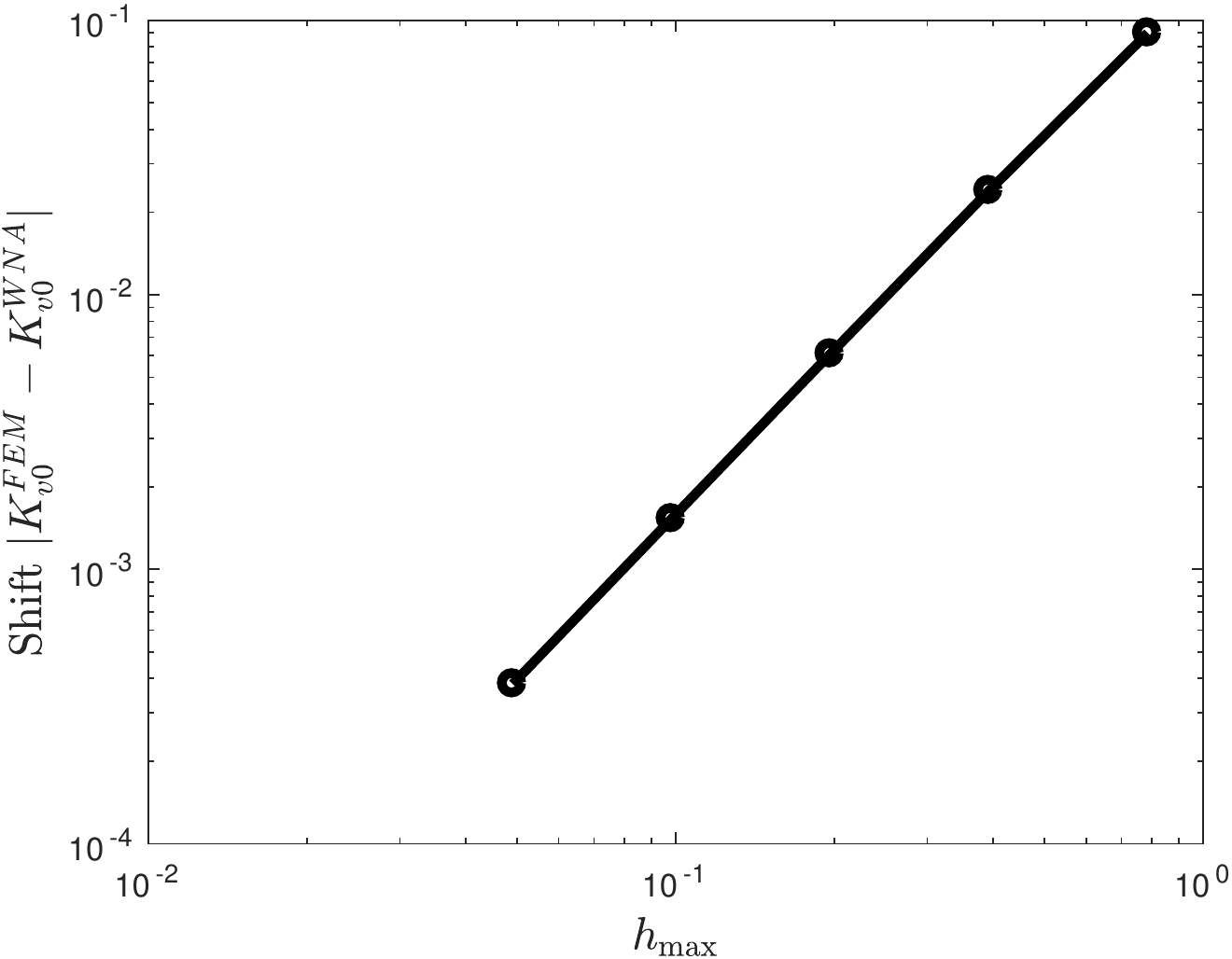}
\end{subfigure}
\caption{\label{fig:coco_FEM_brusselator} Convergence of pitchfork
  bifurcation points between the continuous versus the spatially
  discretized system on the vertical line $D_v = 5$ for the
  Brusselator kinetics (see the plots in
  Fig.~\ref{fig:stability_diagram_brusselator}). {The right panel plots
  on a logarithmic scale the distance} between the
  bifurcation points in the discrete versus continuous systems as
  $h_{\max}$ tends to zero. The left panel shows the standard
  Euclidean norm of the base state solution as the parameter $K_v$
  increases past the pitchfork bifurcation point when $N=128$ nodes
  are used on the boundary, yielding a maximal distance of
  $h_{max} \approx 4.9 \times 10^{-2}$.}
\end{figure}

Despite being unable to compute bifurcating branches using numerical
continuation methods, the {package \textsc{coco}
  (cf.~\cite{danko2013})} can be used to estimate the shift in
bifurcation points between the full model and its finite element
discretization. {In Fig.~\ref{fig:coco_FEM_brusselator} we
  show the results of such a convergence study}, where $h_{\max}$ is
the maximal distance between two nodes on the mesh. Letting $N$ be the
number of equidistant nodes on the circular boundary, then $h_{\max}$
is chosen as $h_{\max} = {2\pi R/N}$. As $h_{\max}$ tends to zero, the
discrepancy between the bifurcation points is expected to converge
like
\begin{equation}
 \|\mu_0^{num} - \mu_0^{wna}\|_2 \leq \mathcal{O}(h_{\max}^{\gamma})\,,
\end{equation}
for some positive power $\gamma$. Estimating the slope of the curve in
the right panel of Fig.~\ref{fig:coco_FEM_brusselator} yields
$\gamma \approx 1.97$. Similar quadratic convergence rate was obtained
for bifurcation points of systems with radial symmetry discretized
with simple finite differences (see \S \ref{subsubsec:subsub_hopf}).

\begin{figure}[htbp]
\centering
\begin{subfigure}{0.47\linewidth}
\includegraphics[width=\linewidth,height=5.0cm]{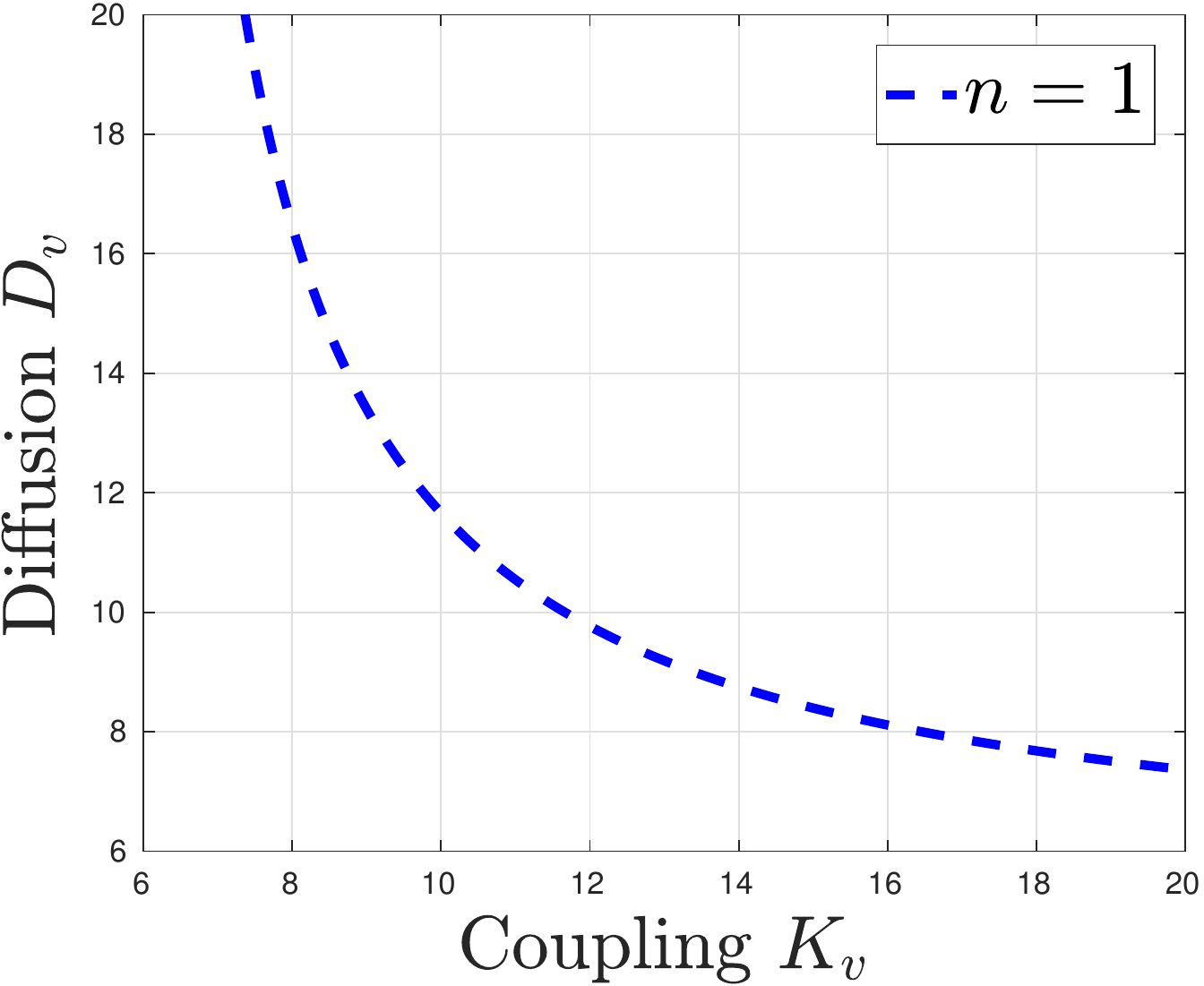}
\end{subfigure}
\begin{subfigure}{0.47\linewidth}
\includegraphics[width=\linewidth,height=5.0cm]{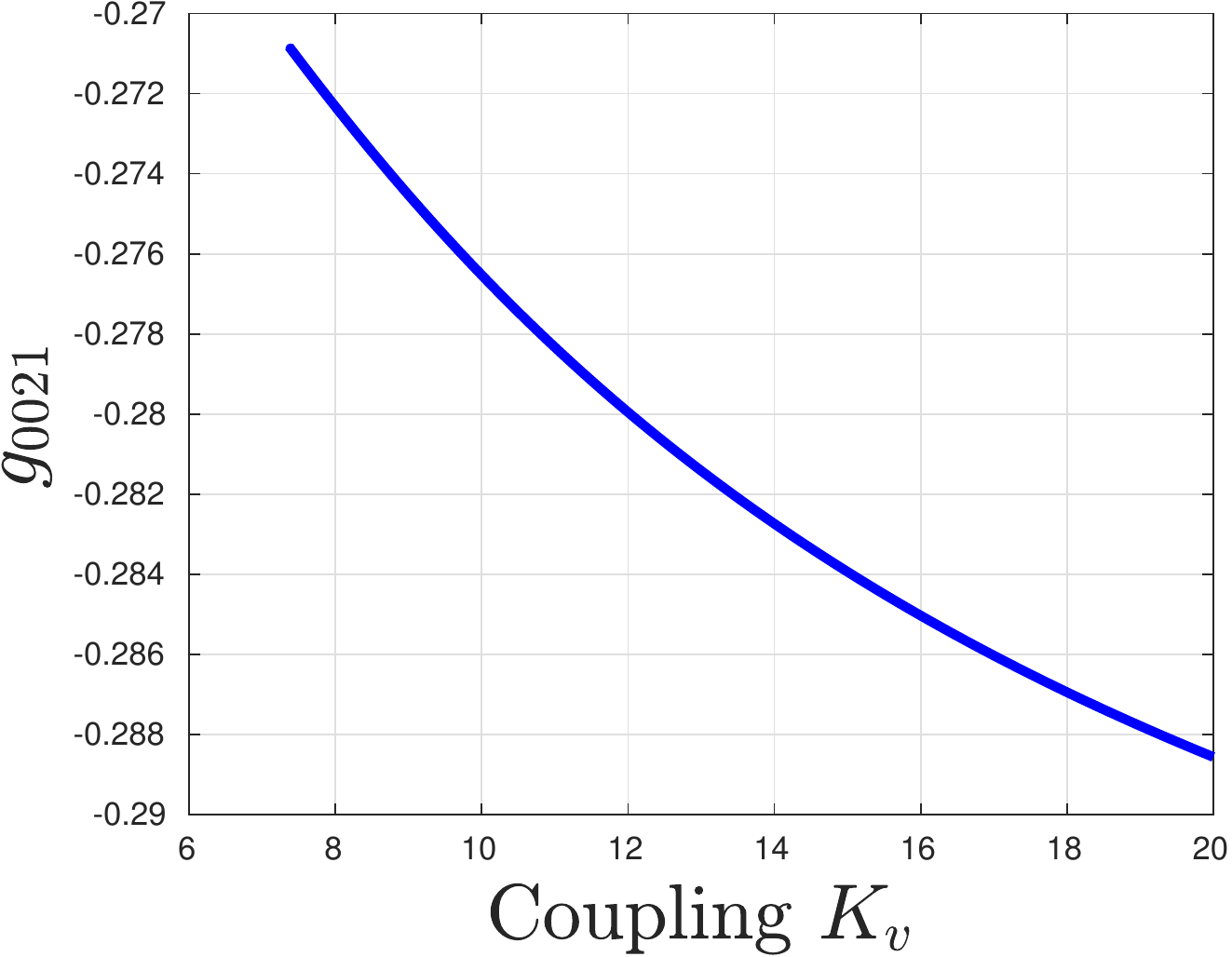}
\end{subfigure}
\caption{\label{fig:stability_diagram_brusselator_EAD_t4}
  {Left panel: the pitchfork bifurcation ($n=1$ mode) curve
    in the $D_v$ versus $K_v$ parameter plane for Brusselator kinetics
    with $a=3$ and $b=5$.  Other parameter values are
    $R=1,\, D_u=1,\, \sigma_u=\sigma_v=0.01,\, K_u=0.1,\, d_u=d_v =
    1$. The region of linear stability is to the left of the
    curve. Right panel: the corresponding coefficient $g_{0021}$ of
    the cubic term in the normal form \eqref{eq:amplitude_pitchfork},
    along the pitchfork bifurcation locus. This coefficient is
    negative, indicating a supercritical pitchfork bifurcation.}}
\end{figure}

{In a different parameter regime, we now show that under
  Brusselator kinetics the branching behavior at the pitchfork point
  can be supercritical instead of subcritical.  For the parameter set
  in Fig.~\ref{fig:stability_diagram_brusselator_EAD_t4} a new
  pitchfork bifurcation locus is plotted in the $D_v$ versus $K_v$
  parameter plane. By numerically evaluating the cubic coefficient of
  the normal form \eqref{eq:amplitude_pitchfork}, the weakly nonlinear
  theory from \S \ref{sec:weakly_nonlinear_theory} now predicts a
  supercritical pitchfork bifurcation (see the right panel of
  Fig.~\ref{fig:stability_diagram_brusselator_EAD_t4}). In comparison
  with the parameter set used for the subcritical case in
  Fig.~\ref{fig:stability_diagram_brusselator}, we took a slightly
  different value for $b$ in the Brusselator kinetics, while the
  surface diffusion coefficients were increased to $d_u = d_v = 1$.}
  
\begin{figure}[htbp]
\centering
\begin{subfigure}{0.47\linewidth}
\includegraphics[width=\linewidth,height=5.0cm]{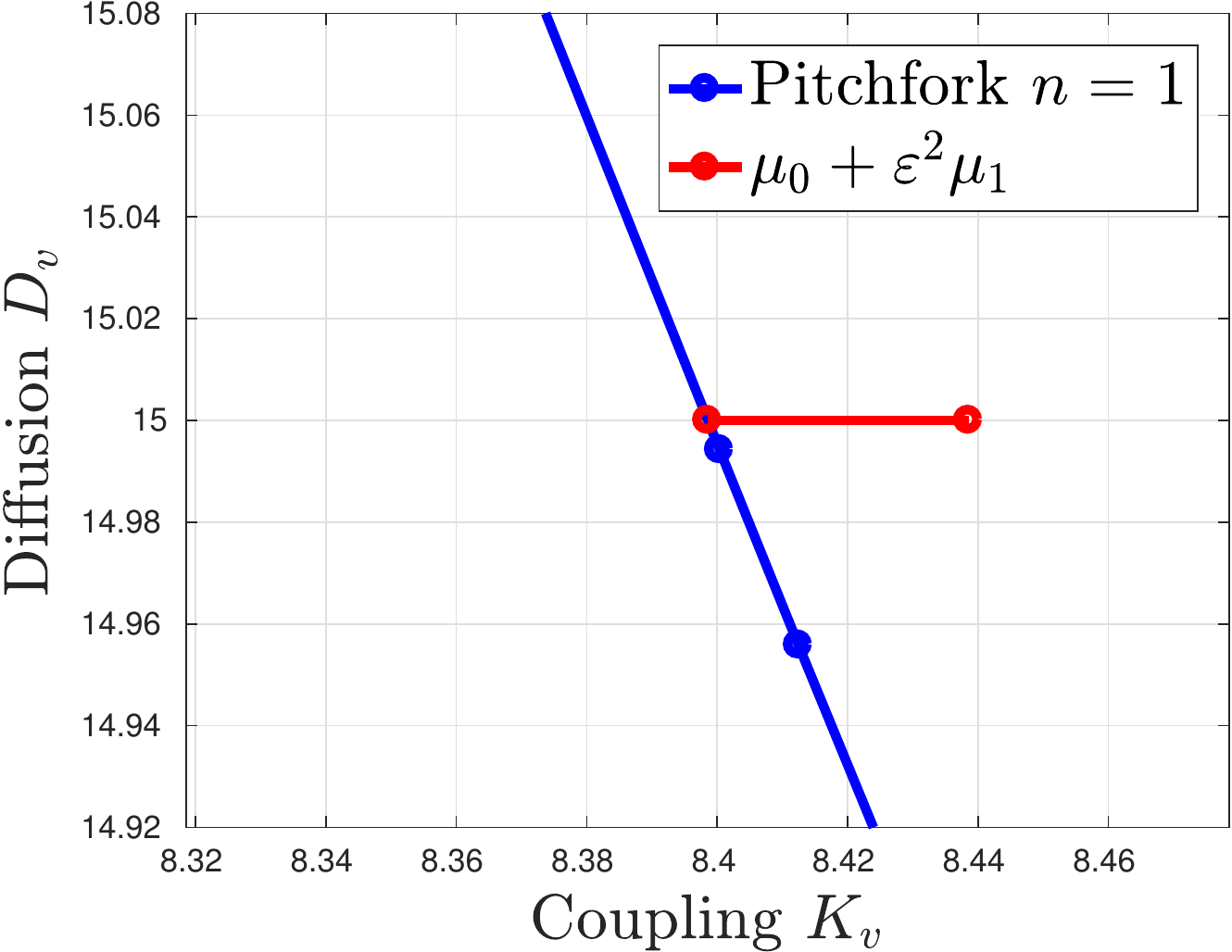}
\end{subfigure}
\begin{subfigure}{0.47\linewidth}
\includegraphics[width=\linewidth,height=5.0cm]{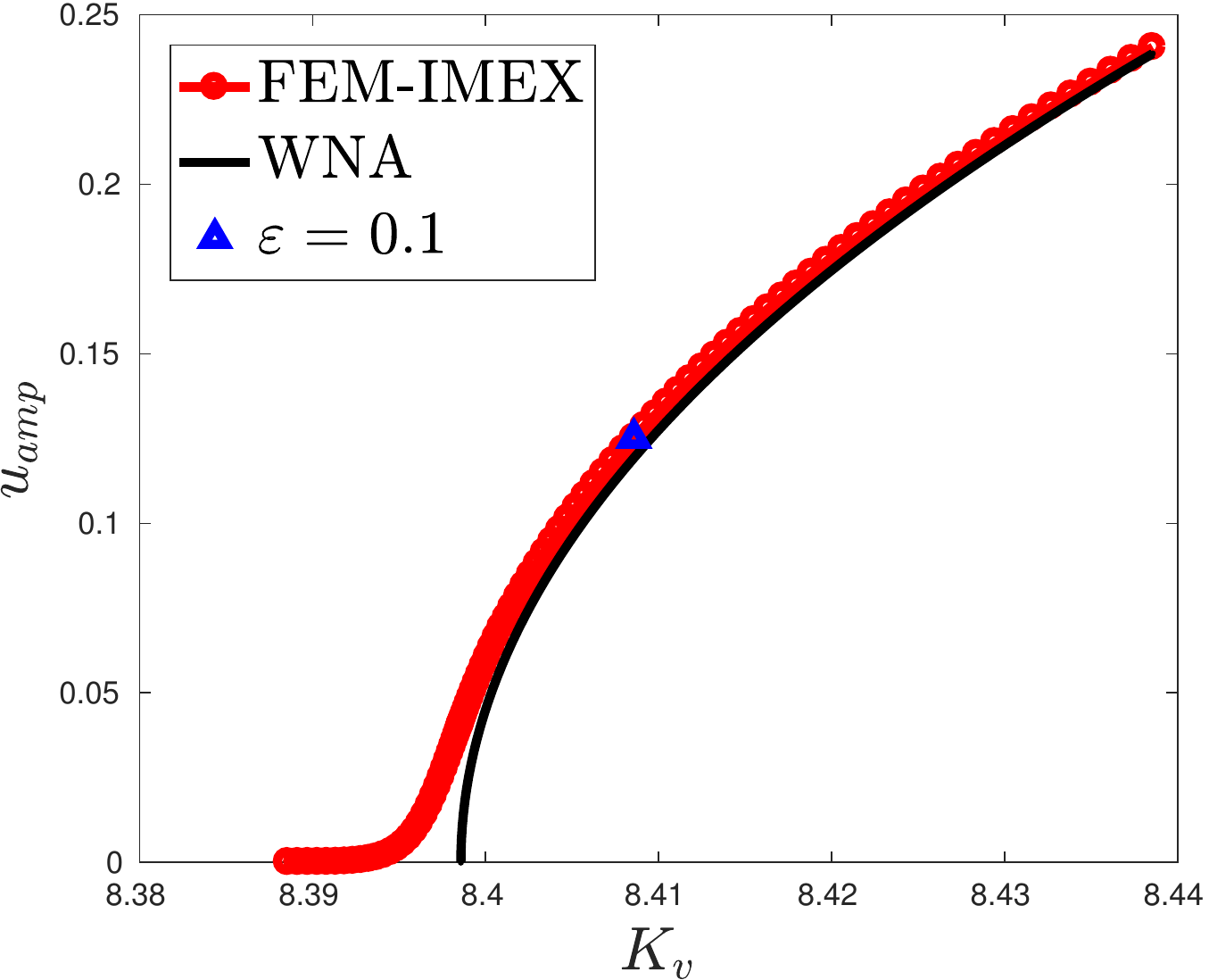}
\end{subfigure}
\caption{\label{fig:supercritical_brusselator} Stable bifurcating
  branch past a supercritical pitchfork bifurcation for the
  Brusselator kinetics, with the rightmost point corresponding to
  $\eps = 0.2$. {The parameters are as given in
  Fig.~\ref{fig:stability_diagram_brusselator_EAD_t4}. The left panel
  shows the parameter path past the bifurcation point. The right panel compares $u_{amp}$, as
  obtained from the weakly nonlinear theory
  \eqref{eq:amplitude_pitchfork}, with numerically computed results
  from PDE simulations.}}
\end{figure}

\begin{figure}[htbp]
\centering
\begin{subfigure}{0.47\linewidth}
\includegraphics[width=\linewidth,height=5.0cm]{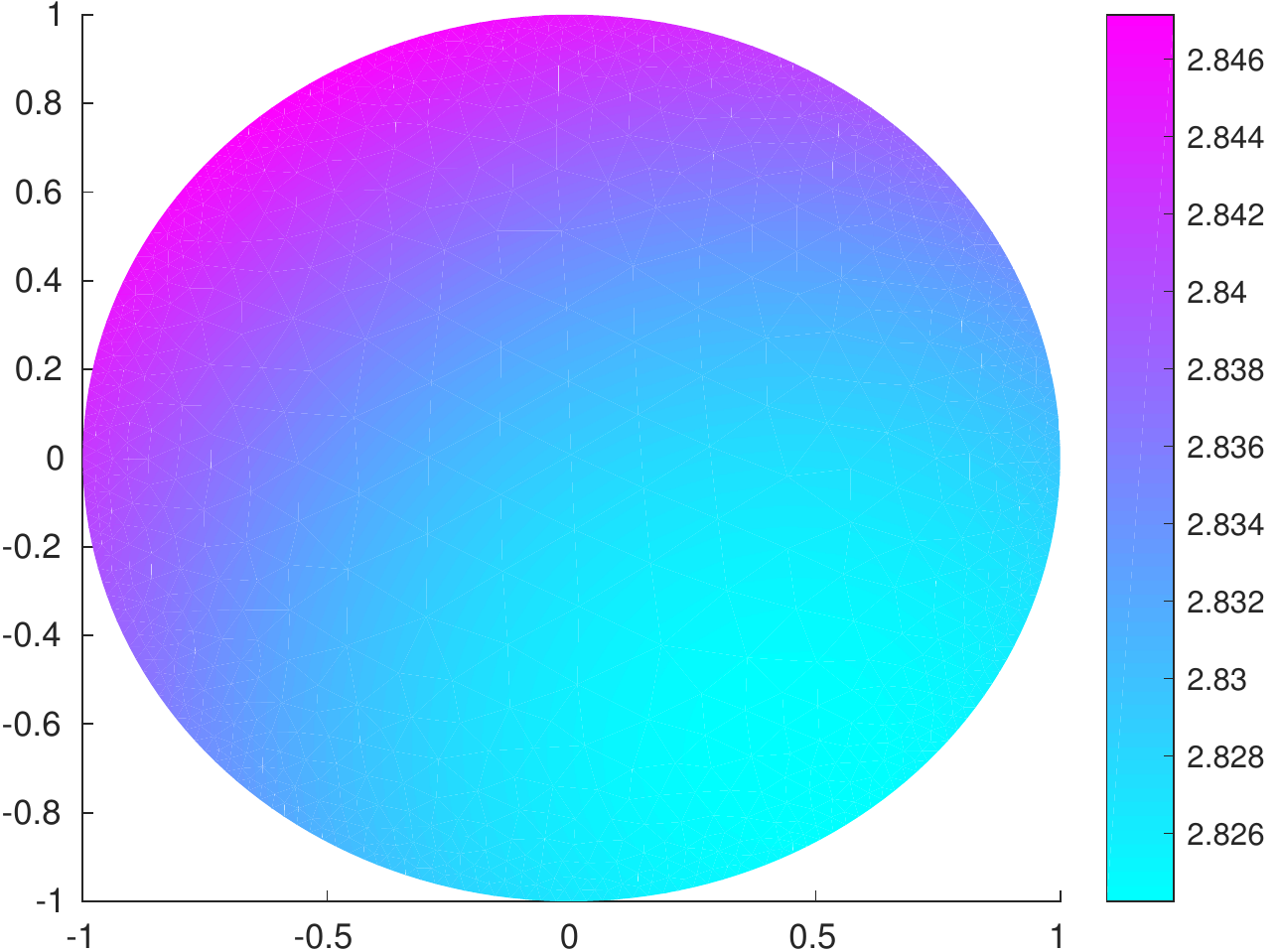}
\end{subfigure}
\begin{subfigure}{0.47\linewidth}
\includegraphics[width=\linewidth,height=5.0cm]{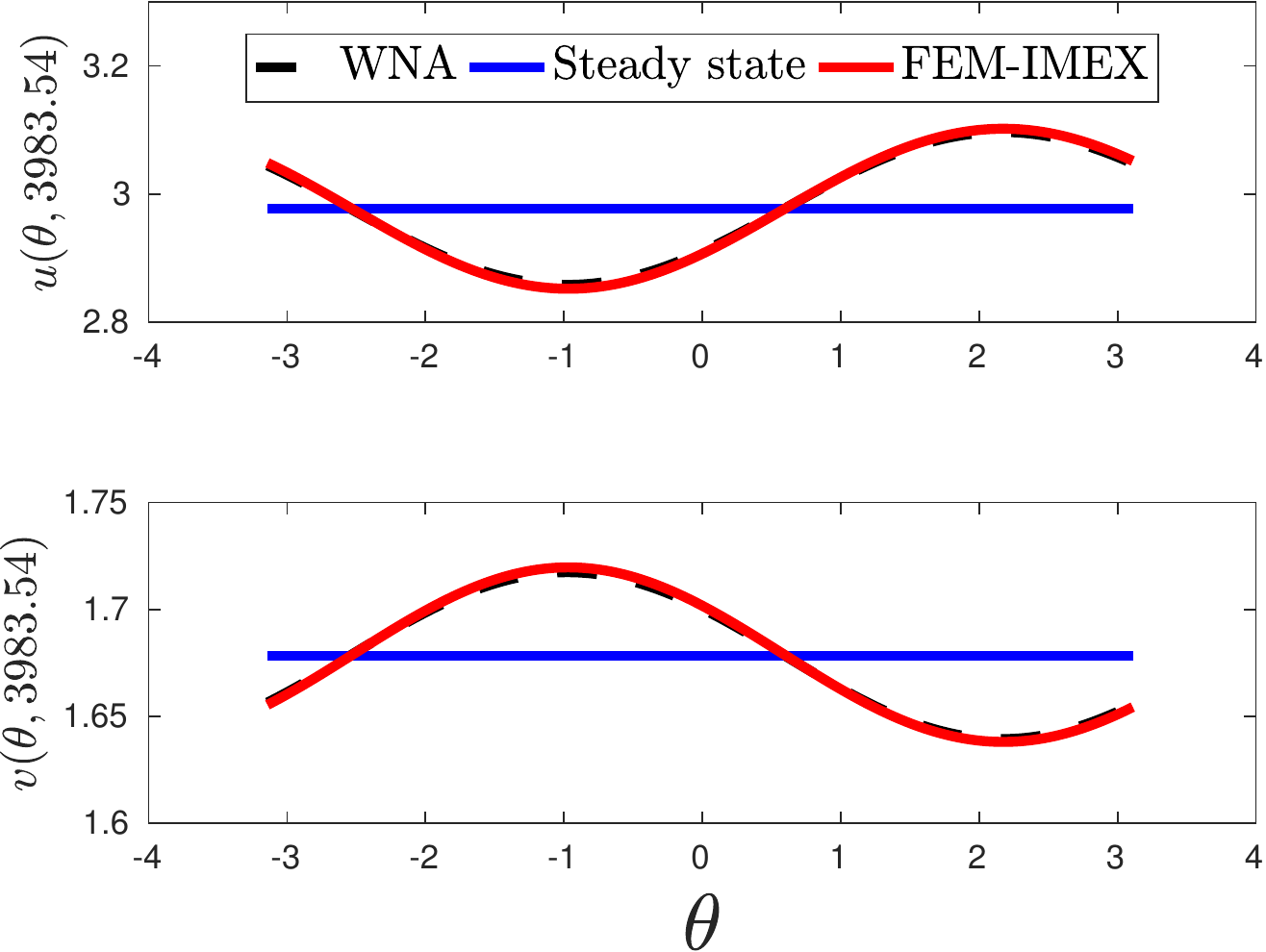}
\end{subfigure}
\caption{\label{fig:supercritical_pattern_brusselator} Stable pattern
  near a supercritical pitchfork bifurcation of the $n=1$ mode with
  $\eps = 0.1$, corresponding to the point indicated by a blue triangle on the
  {bifurcation branch plotted in the right panel of
    Fig.~\ref{fig:supercritical_brusselator}. Left panel: contour plot
    of the activator concentration in the bulk ($U$). Right panel:
    membrane-bound patterns. A close agreement between the weakly
    nonlinear theory predicted by \eqref{eq:patterned_solution} (black
    dashed curve) and the (nearly coinciding) numerical PDE results
    (red curve) is obtained.  Here the nonzero phase $\theta_n$ in
    \eqref{eq:patterned_solution} was calculated from the numerical
    solution, whose initial condition is a linear combination of the
    base state solution with the critical eigenvectors of the
    Jacobian of the spatially discretized system.}}
\end{figure}

{To validate the prediction of supercriticality, numerical
  bifurcation results and full PDE simulations are undertaken near a
  bifurcation point on the stability boundary. In the right panel
  of Fig.~\ref{fig:supercritical_brusselator} the amplitude of the
  patterned state for the membrane-bound activator when $K_v$
  increases on the horizontal line $D_v = 15$ is shown. Here, a rather
  close agreement between \eqref{eq:amplitude_pitchfork} from the
  weakly nonlinear theory and the numerical bifurcation results is
  obtained because the bifurcating branch is stable. Moreover, as
  predicted by the theory, the amplitude of the patterned state scales
  as the square root of the distance from the bifurcation point. The
  corresponding stable pattern computed from full PDE simulations with
  $\eps = 0.1$ is shown in
  Fig.~\ref{fig:supercritical_pattern_brusselator} and favorably
  compared with results from the weakly nonlinear theory.}

{Next, we give a convergence study comparing $W_{FEM}$, the
  numerical solution to the spatially discretized system, with the
  leading-order asymptotic solution \eqref{eq:patterned_solution} in
  the weakly nonlinear regime, and denoted by $W_{WNA}$. Such an
  experiment is only valid when testing a stable patterned
  solution arising from a supercritical pitchfork
  bifurcation. Referring to
  Fig.~\ref{fig:supercritical_pattern_brusselator}}, the stable
pattern {for $\eps = 0.1$} is repeatedly computed while
decreasing $h_{\max}$ on a uniform grid. For each $h_{\max}$ value,
the error is estimated using the weakly nonlinear asymptotic solution
as a substitute for the unknown exact solution.

\begin{figure}[htbp]
\centering
\begin{subfigure}{0.47\linewidth}
\includegraphics[width=\linewidth,height=5.0cm]{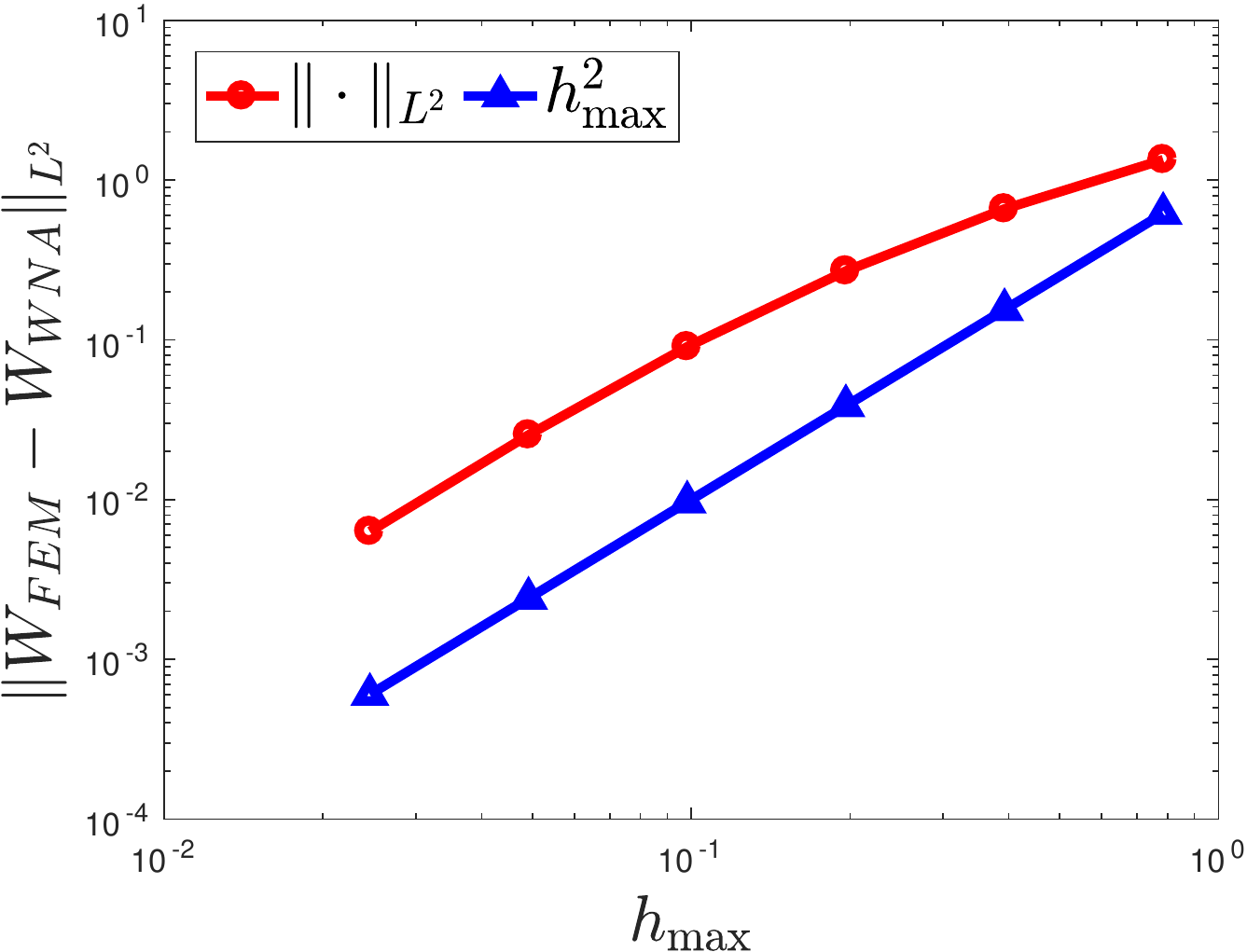}
\end{subfigure}
\begin{subfigure}{0.47\linewidth}
\includegraphics[width=\linewidth,height=5.0cm]{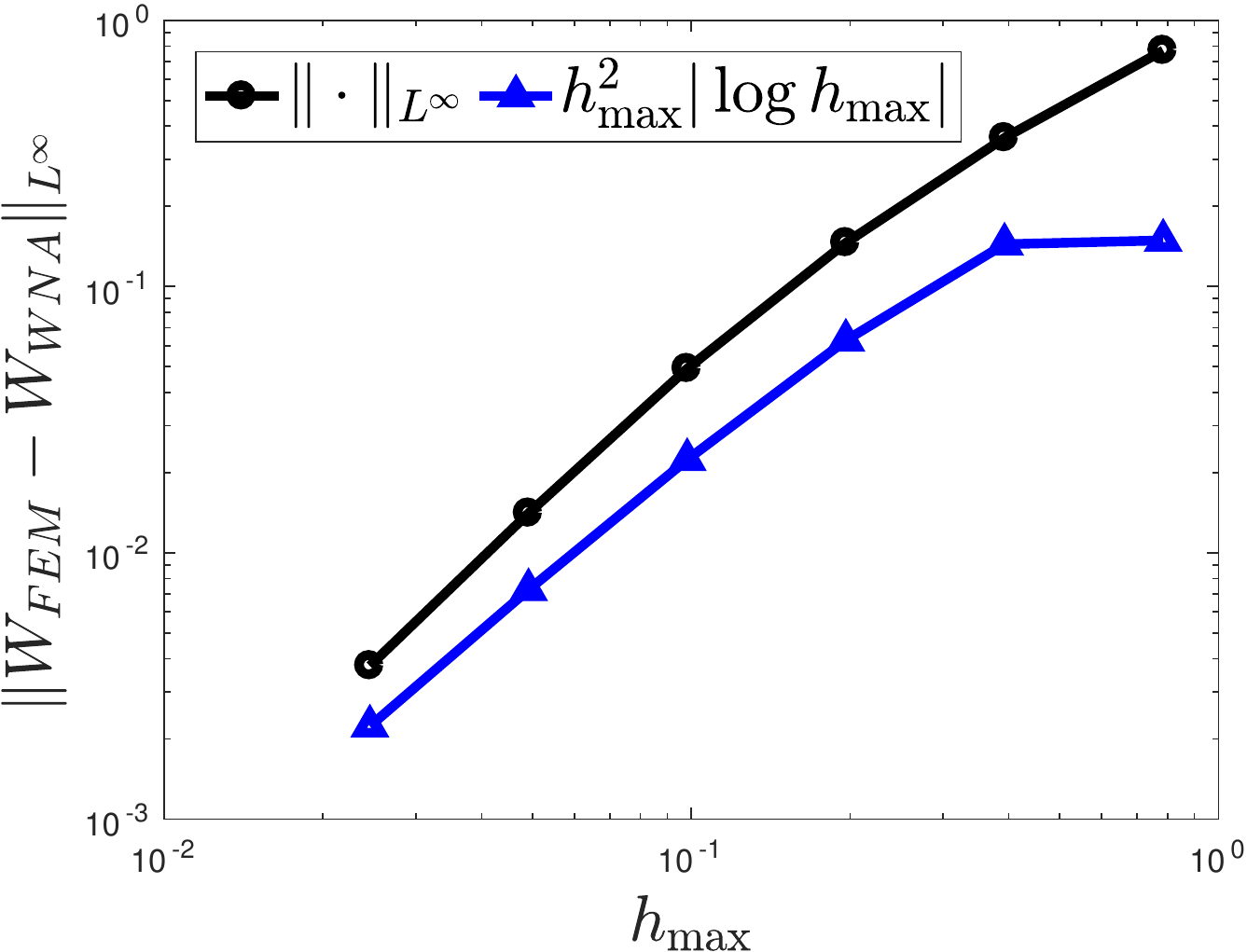}
\end{subfigure}
\caption{\label{fig:convergence_analysis_brusselator} Convergence of
  the numerical solution as $h_{\max}$ approaches zero. {Left panel:
    error measured with the $L^2$ norm (red curve) is compared with
    $h_{\max}^2$. Right panel: the error in the $L^{\infty}$ norm (black curve)
    is compared with $h_{\max}^2|\log h_{\max}|$.} The approximate weakly
    nonlinear solution is obtained from \eqref{eq:patterned_solution} with
  $\eps = 0.1$, $n=1$ and the correct phase $\theta_n$ (so that the
  numerical solution matches the solution in the weakly nonlinear
  regime).}
\end{figure}

A plot of the error as a function of $h_{\max}$ is shown in
Fig.~\ref{fig:convergence_analysis_brusselator}, where two different
solution measures are employed. Assuming sufficient regularity of the
exact solution, and {given that the mesh consists of
linear triangular elements}, the error is expected to behave like
\begin{equation}\label{eq:convergence_Linfty}
  \|W - W_{FEM}\|_{L^\infty} \leq \mathcal{O}\left(h_{\max}^2|\log h_{\max}|\right)
  \,, \quad \mbox{as} \quad h_{\max}\to 0 \,,
\end{equation}
using the $L^\infty$ norm (see remark 4.41 in
\cite{grossmann2007}). For the $L^2$ norm, quadratic convergence rate
is expected (see Theorem 4.34 in \cite{grossmann2007})
\begin{equation}\label{eq:convergence_L2}
  \|W - W_{FEM}\|_{L^2} \leq \mathcal{O}(h_{\max}^2)\,, \qquad
  \quad \mbox{as} \quad h_{\max}\to 0 \,.
\end{equation}
{The plots in Fig.~\ref{fig:convergence_analysis_brusselator}
  confirm the bounds given in \eqref{eq:convergence_Linfty} and
  \eqref{eq:convergence_L2}, with the error as measured with the
  $L^\infty$ norm (right panel) converging slightly faster than
  expected. The choice of a temporal discretization, along with an
  associated stopping criteria for the solver, may influence the
  convergence of the numerical solution. Although it should not affect
  the speed of convergence, the error estimates are also biased since
  we are using the leading-order asymptotic solution
  \eqref{eq:patterned_solution} as a proxy for the exact solution.}

\subsection{Codimension-two bifurcations}\label{subsec:codimension_two_bifurcation}

{As shown in Fig.~\ref{fig:stability_diagram_schnakenberg} and
Fig.~\ref{fig:stability_diagram_brusselator} for the
Schnakenberg and Brusselator kinetics, respectively, the stability
curves associated with the Hopf mode $n=0$ and the pitchfork mode $n=1$
intersect at a unique codimension-two bifurcation point. In this subsection,
we use the weakly nonlinear theory developed in \S \ref{subsec:theory_codim_2}
to explore the dynamics of the full model in the vicinity of such a point.}

{Near a codimension-two bifurcation point, a linear approximation of
the intersecting stability curves is obtained by using the
leading-order result \eqref{eq:amplitude_pitchfork_hopf} from the
weakly nonlinear analysis. This is done by applying Lemma
\ref{lemma:map} to the generic single-mode stability boundary defined
in \eqref{eq:H_1_2}. This yields,}
\begin{equation}\label{eq:beta_1_2}
  \beta_1 = \TT(H_1) = \mu_0 + \RR\left(-\frac{\pi}{2}\right)g_{0010}\,,
  \qquad \beta_2 = \TT(H_2) = \mu_0 + \RR\left(\frac{\pi}{2}\right)\Re(g_{1000})
  \,,
\end{equation}
where $\beta_1$ and $\beta_2$ are, respectively, tangent to the
pitchfork and Hopf stability boundaries.

{Next, the cubic term coefficients in
  \eqref{eq:perturbed_normal_form} are evaluated numerically for the
  Brusselator and Schnakenberg models. These results are given in
  Table \ref{table:p_ij}. The dynamics of the truncated normal form
  \eqref{eq:perturbed_normal_form} are then classified in Table
  \ref{table:GH_hopf_hopf} into two distinct cases. In this table, the
  reader is referred to \eqref{eq:scaling} for the definition of
  $\gamma$, $\eta$ and $d$.}

\begin{table}[htbp]
\centering
\caption{\label{table:p_ij} {Numerical evaluation of the coefficients
  $p_{ij}$ in the normal form \eqref{eq:perturbed_normal_form} for the
  codimension-two bifurcation point for the Schnakenberg and Brusselator
  kinetics. The parameter values are the same as in the caption of
  either Fig.~\ref{fig:stability_diagram_schnakenberg}
  (Schnakenberg) or Fig.~\ref{fig:stability_diagram_brusselator}
  (Brusselator).}}
\begin{tabular}{|c|c|c|}
\hline
& Schnakenberg & Brusselator \\ \hline
$\mu_0$ & $(4.26,3.10)^T$ & $(4.25,3.38)^T$ \\ \hline
$p_{11}$ & $0.19096$ & $1.3146$ \\ \hline
$p_{12}$ & $-1.2752$ & $0.87043$ \\ \hline
$p_{21}$ & $-2.3796$ & $-0.52089$ \\ \hline
$p_{22}$ & $-0.48351$ & $-0.162$ \\ \hline
\end{tabular} 
\end{table}

\begin{table}[htbp]
\centering
\caption{\label{table:GH_hopf_hopf} Two specific pitchfork-Hopf
  unfoldings (see Table 7.5.2 in \cite{guckenheimer1983})}
\begin{tabular}{|c|c|c|}
\hline
& Schnakenberg (case VIII) & Brusselator (case VIa) \\ \hline
$d$ & -1 & -1 \\ \hline
$\gamma$ & $<0$ & $>0$ \\ \hline
$\eta$ & $<0$ & $<0$ \\ \hline
$d - \gamma\eta$ & $<0$ & $>0$ \\ \hline
\end{tabular}
\end{table}

\begin{figure}[htbp]
\centering
\begin{subfigure}{0.47\linewidth}
\includegraphics[width=\linewidth,height=5.0cm]{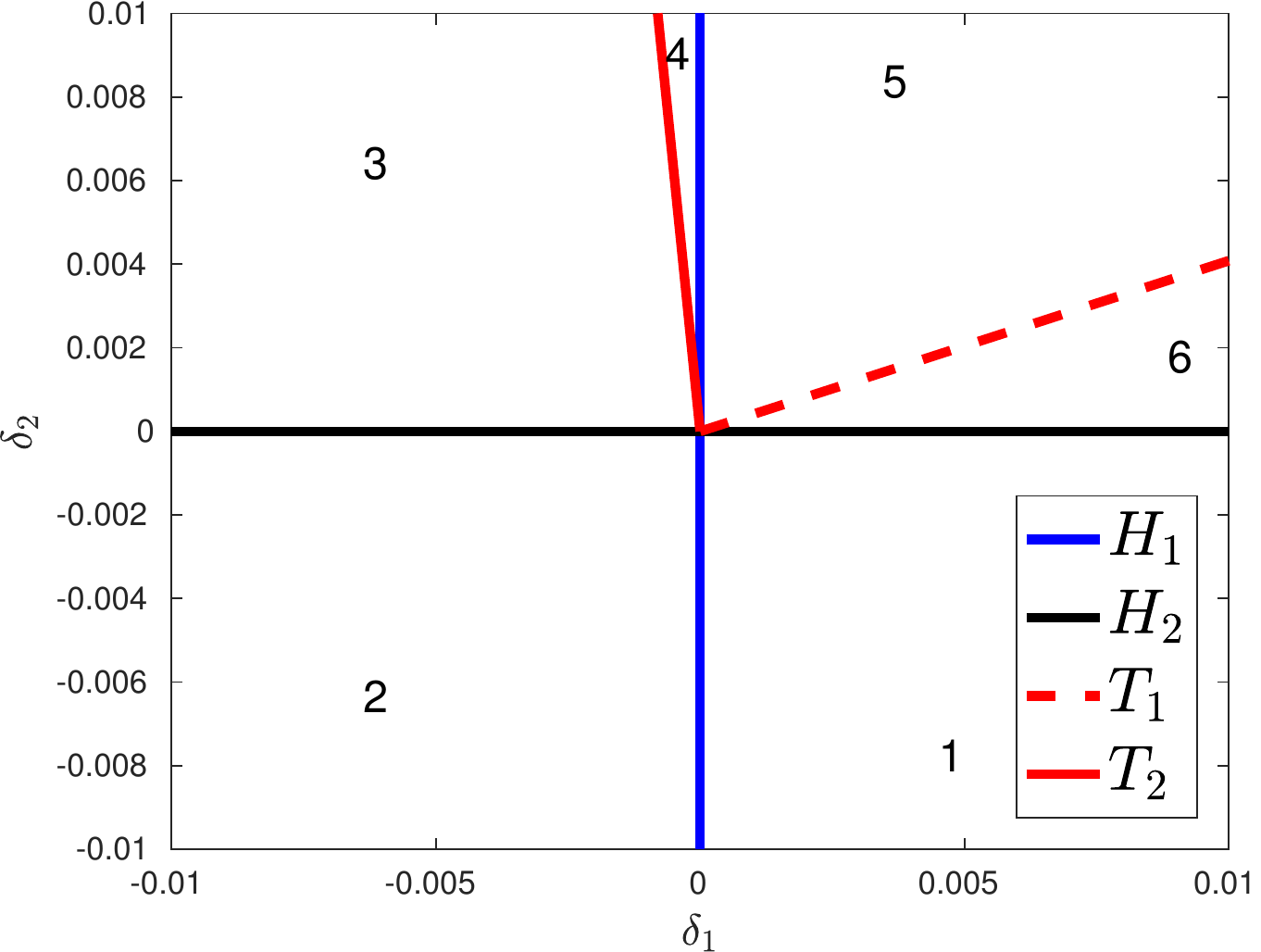}
\end{subfigure}
\begin{subfigure}{0.47\linewidth}
\includegraphics[width=\linewidth,height=5.0cm]{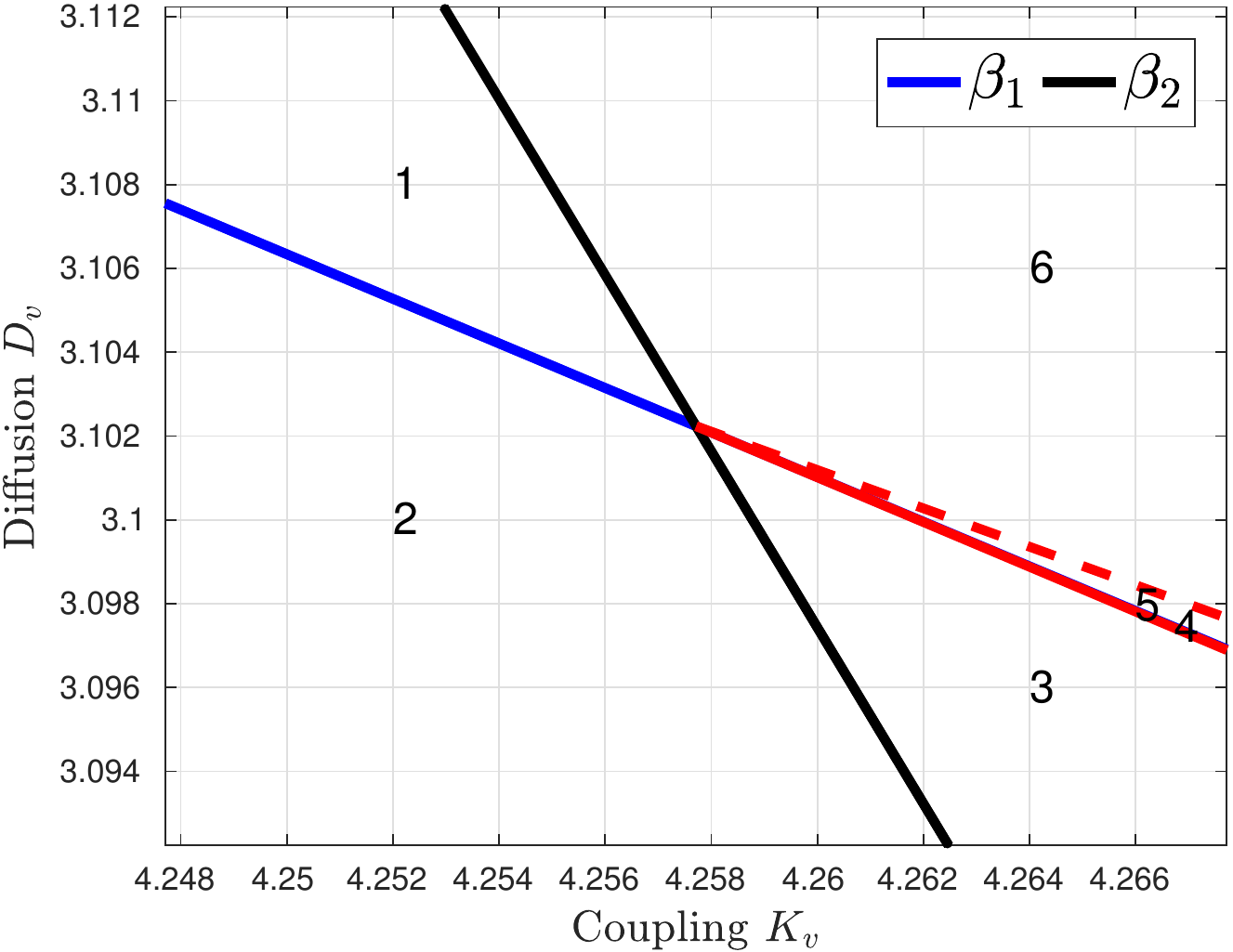}
\end{subfigure}
\caption{\label{fig:pitchfork_hopf_schnakenberg} Parametric portraits
  in the space of generic parameters $(\delta_1,\delta_2)$ (left
  panel) and original bifurcation parameters $(K_v,D_v)$ (right panel)
  {for the Schnakenberg kinetics with parameter values as in
    Fig.~\ref{fig:stability_diagram_schnakenberg}}. In the left panel,
  the line $H_1$ is the vertical $\delta_2$-axis (blue),
  {while the line $H_2$ is the horizontal $\delta_1$-axis
    (black)}. The semi-infinite lines $T_1$ and $T_2$, respectively,
  correspond to the red dashed and full lines. Application of the map
  \eqref{eq:affine_map} to the curves in the left panel yields the
  curves in the right panel. In particular, the lines $H_{1,2}$ are
  mapped to $\beta_{1,2}$ (see \eqref{eq:beta_1_2}), which
  are tangent to the pitchfork and Hopf stability boundaries.}
\end{figure}

\begin{figure}[htbp]
\centering
\begin{subfigure}{0.31\linewidth}
\includegraphics[width=\linewidth,height=4.4cm]{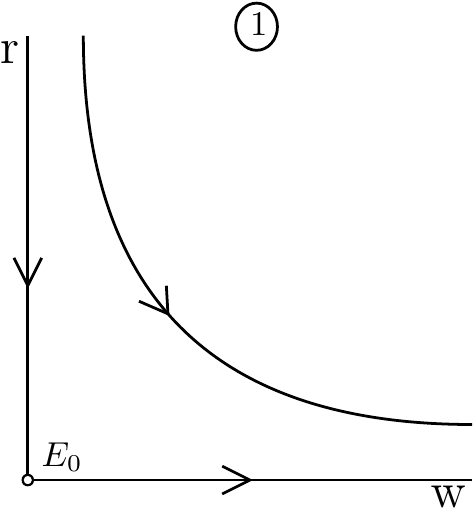}
\end{subfigure}
\begin{subfigure}{0.31\linewidth}
\includegraphics[width=\linewidth,height=4.4cm]{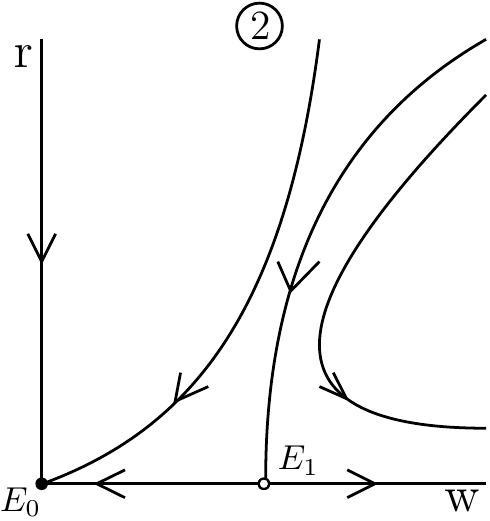}
\end{subfigure}
\begin{subfigure}{0.31\linewidth}
\includegraphics[width=\linewidth,height=4.4cm]{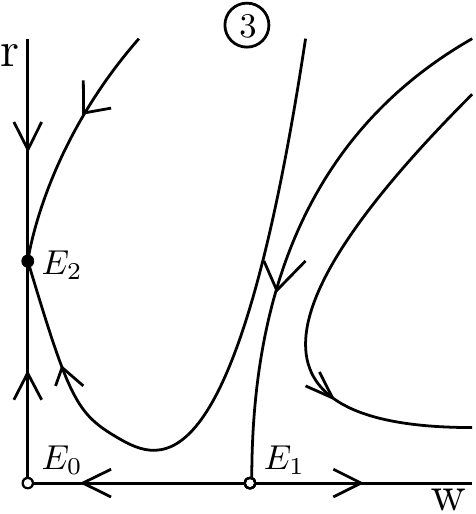}
\end{subfigure}

\begin{subfigure}{0.31\linewidth}
\includegraphics[width=\linewidth,height=4.4cm]{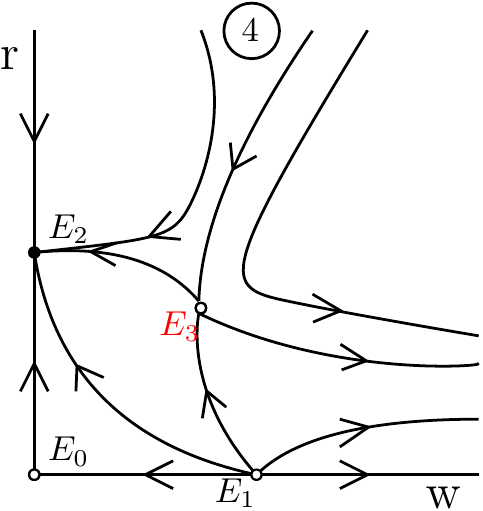}
\end{subfigure}
\begin{subfigure}{0.31\linewidth}
\includegraphics[width=\linewidth,height=4.4cm]{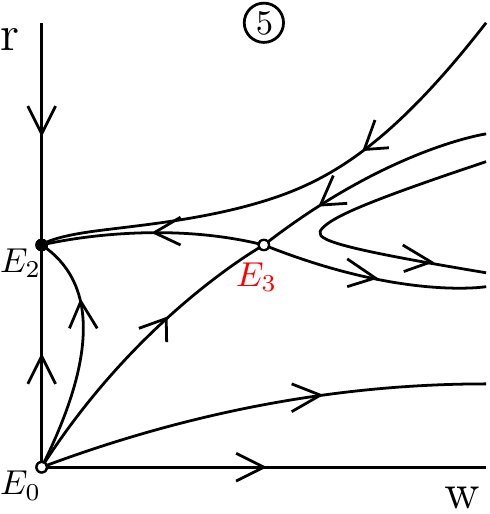}
\end{subfigure}
\begin{subfigure}{0.31\linewidth}
\includegraphics[width=\linewidth,height=4.4cm]{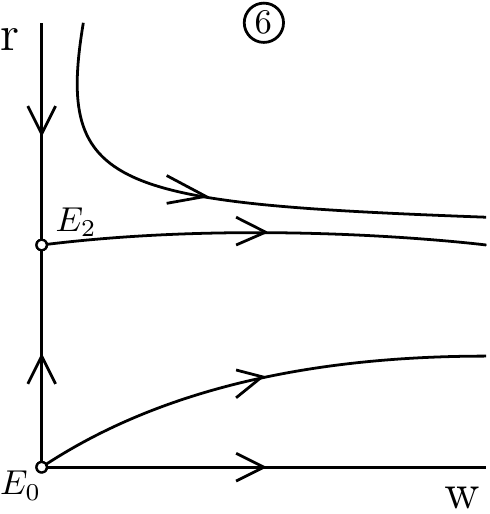}
\end{subfigure}
\caption{\label{fig:phase_diagrams_schnakenberg} Phase diagrams of the
  coupled system of amplitude equations
  \eqref{eq:rescaled_perturbed_normal_form} with unfolding of type
  VIII (see \cite{guckenheimer1983}). The mixed-mode equilibrium $E_3$
  from panels 4 and 5 is a hyperbolic saddle.}
\end{figure}

{From Table \ref{table:GH_hopf_hopf}, and based on \S
  \ref{subsec:codimension_two_bifurcation}, it follows that the case
  involving the Schnakenberg kinetics is simple, whereas for the
  Brusselator kinetics the mixed-mode equilibrium $E_3$ undergoes a
  Hopf bifurcation that is degenerate in the truncated cubic normal
  form. In this more difficult case, it is an open problem to
  calculate the fifth-order term in the normal form which, if nonzero,
  would eliminate this degeneracy.}

Parametric portraits for the simple case are given in
Fig.~\ref{fig:pitchfork_hopf_schnakenberg}. The four lines $H_1, H_2, T_1$,
and $T_2$ divide the $(\delta_1,\delta_2)$ parameter plane into six
open regions, for which the corresponding phase portraits are shown in
Fig.~\ref{fig:phase_diagrams_schnakenberg}. In region 1, there is a
unique unstable equilibrium $E_0$. Stability is gained when crossing
into region 2, which also generates an unstable equilibrium
$E_2$. When entering region 3 from region 2, a stable equilibrium
($E_2$) bifurcates from the origin while $E_0$ becomes unstable again
and $E_1$ remains unstable. From regions 3 to 4, the mixed-mode
equilibrium $E_3$ bifurcates from $E_1$. Next, entering region 5 from
region 4 causes $E_1$ to vanish. When finally crossing the $T_1$ line
into region 6, the mixed-mode equilibrium collapses with the single
mode equilibrium $E_2$, causing $E_2$ to lose stability. Moreover,
when it exists in regions 4 and 5, the mixed-mode $E_3$ is a
hyperbolic saddle whose stable manifold forms the boundary between the
basin of attraction of the stable equilibrium $E_2$ and some unknown
dynamics with large $w$ amplitude.

Restoring the angular variable to \eqref{eq:perturbed_normal_form},
some equilibria must be interpreted differently. In
Fig.~\ref{fig:phase_diagrams_schnakenberg}, the phase portraits may be
viewed with the $r$-axis rotating around the $w$-axis. Hence, both
$E_2$ and $E_3$ now correspond to limit cycles, with their stability
properties remaining the same. The equilibria $E_0$ and $E_1$ each
remain steady states of the system. The line $H_2$ becomes a
supercritical Hopf bifurcation, {$H_1$ remains a subcritical
pitchfork bifurcation, while $T_1$ and $T_2$ each remain
a mixed-mode bifurcation.}

{Since for this simple case the nondegeneracy conditions are
  satisfied, the stability results {associated with} the
  normal form can be interpreted in the context of the bulk-surface
  PDE model. The origin $E_0$ becomes the base state
  \eqref{eq:patternless_solution}, $E_1$ corresponds to an unstable
  Turing-type pattern of the first circular harmonic, and $E_2$ to
  radially symmetric nonlinear oscillations. The mixed-mode $E_3$
  corresponds to nonlinear oscillations around a spatially
  inhomogeneous equilibrium, which is a type of breather
    solution}. When mapped to the parameter space defined by $K_v$ and
  $D_v$, its area of existence becomes fairly narrow (see regions 4
  and 5 in the right panel of
  Fig.~\ref{fig:pitchfork_hopf_schnakenberg}). Because the mixed-mode
  solution possesses the stability property of a saddle, bistability
  between a radially symmetric periodic solution and a large amplitude
  Turing pattern is expected in this region.

{We remark that another equilibrium corresponding to a stable
  Turing pattern state must also exist because of the dissipative
  nature of the system, which prevents the solution from becoming
  unbounded.  In \S \ref{subsec:codim_1_bif}, numerical evidence for
the existence of such a large amplitude stable Turing pattern near a
subcritical pitchfork bifurcation was shown for both the
Schnakenberg and Brusselator kinetics.}

\begin{figure}[htbp]
\centering
\begin{subfigure}{0.47\linewidth}
\includegraphics[width=\linewidth,height=5.0cm]{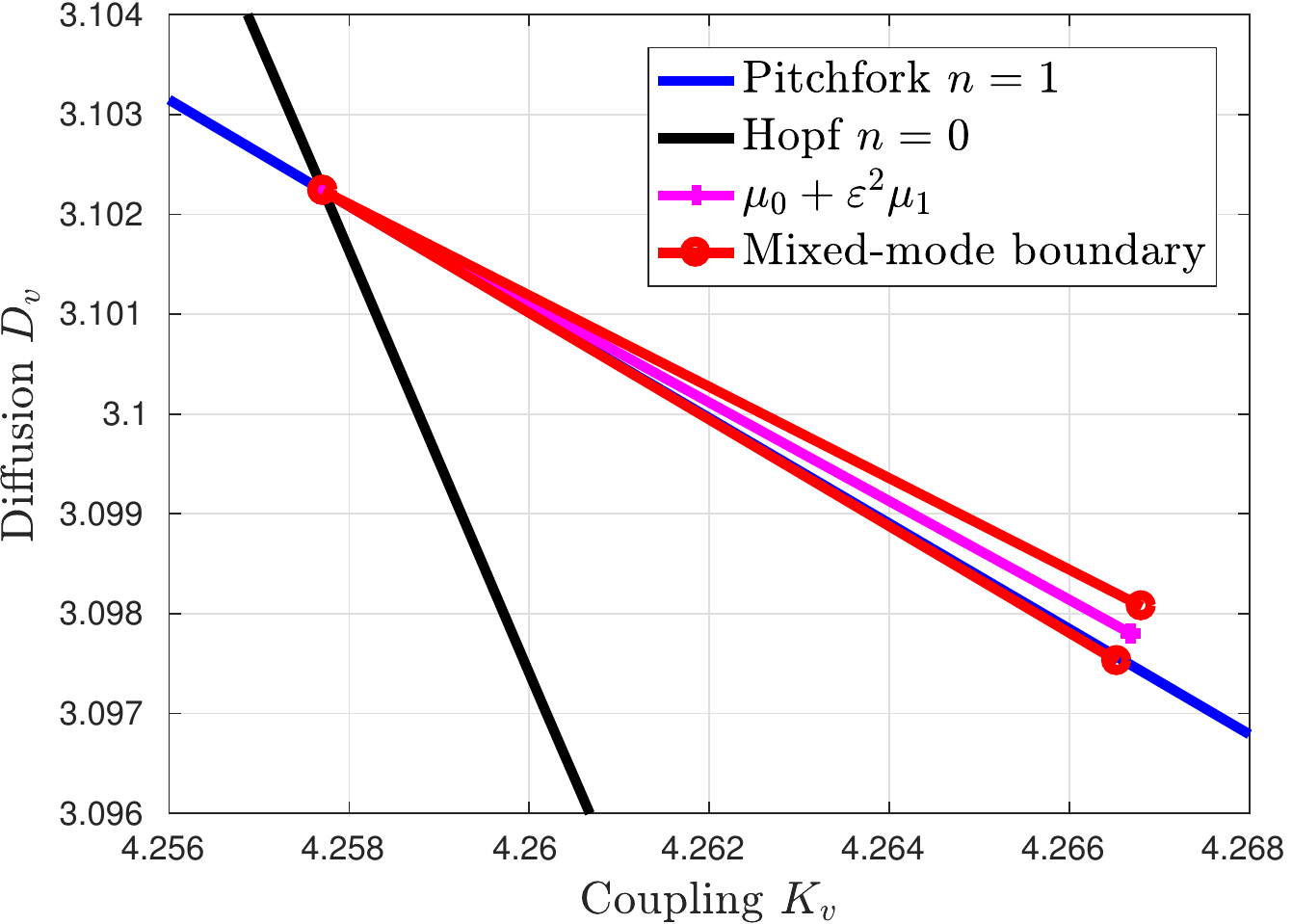}
\end{subfigure}
\begin{subfigure}{0.47\linewidth}
\includegraphics[width=\linewidth,height=5.0cm]{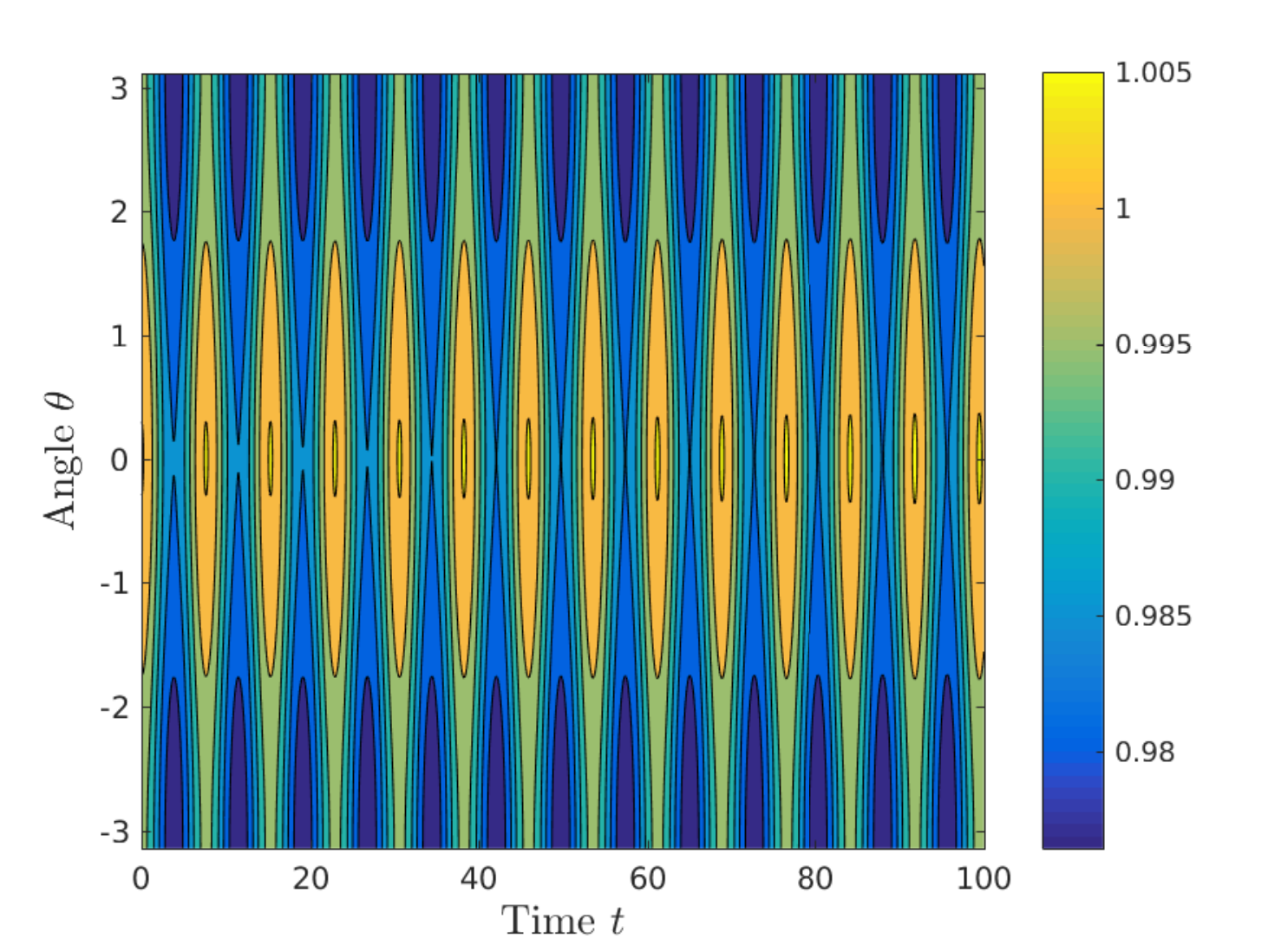}
\end{subfigure}
\caption{\label{fig:numerics_PH_schnakenberg} {Interaction of a
  supercritical Hopf and subcritical pitchfork bifurcations for the
  Schnakenberg kinetics. The simulation corresponds to region 5 (left
  panel). Right panel: a space-time contour plot of the membrane-bound
  activator species $u(\theta,t)$, showing oscillations around a
  spatial pattern. Equation \eqref{eq:spatio_temporal_oscillations}
  with $\eps = 0.1$, $n = 1$ and phases $\theta_0(0) = \theta_n = 0$
  is used as an initial condition.}}
\end{figure}

\begin{figure}[htbp]
\centering
\begin{subfigure}{0.47\linewidth}
\includegraphics[width=\linewidth,height=5.0cm]{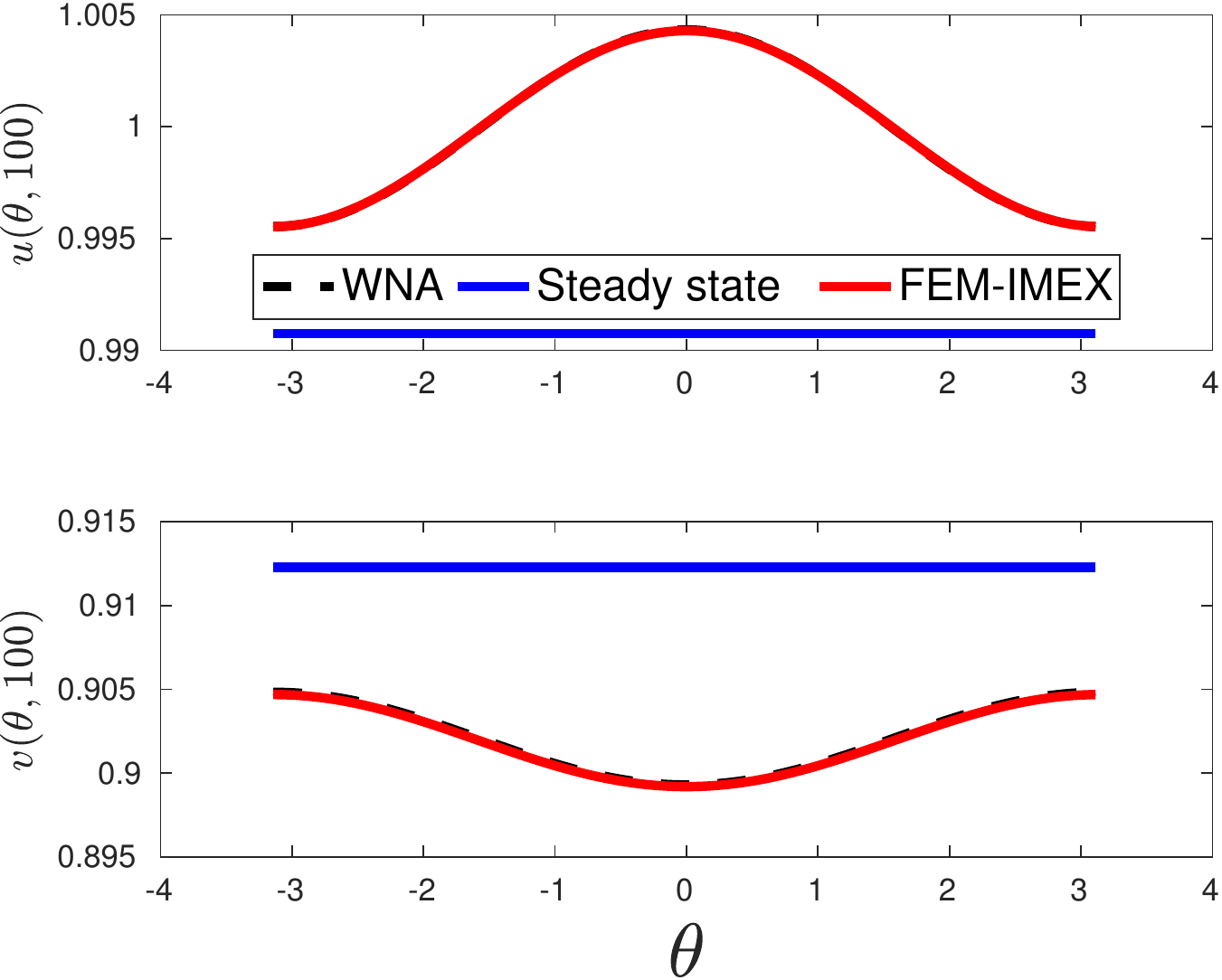}
\end{subfigure}
\begin{subfigure}{0.47\linewidth}
\includegraphics[width=\linewidth,height=5.0cm]{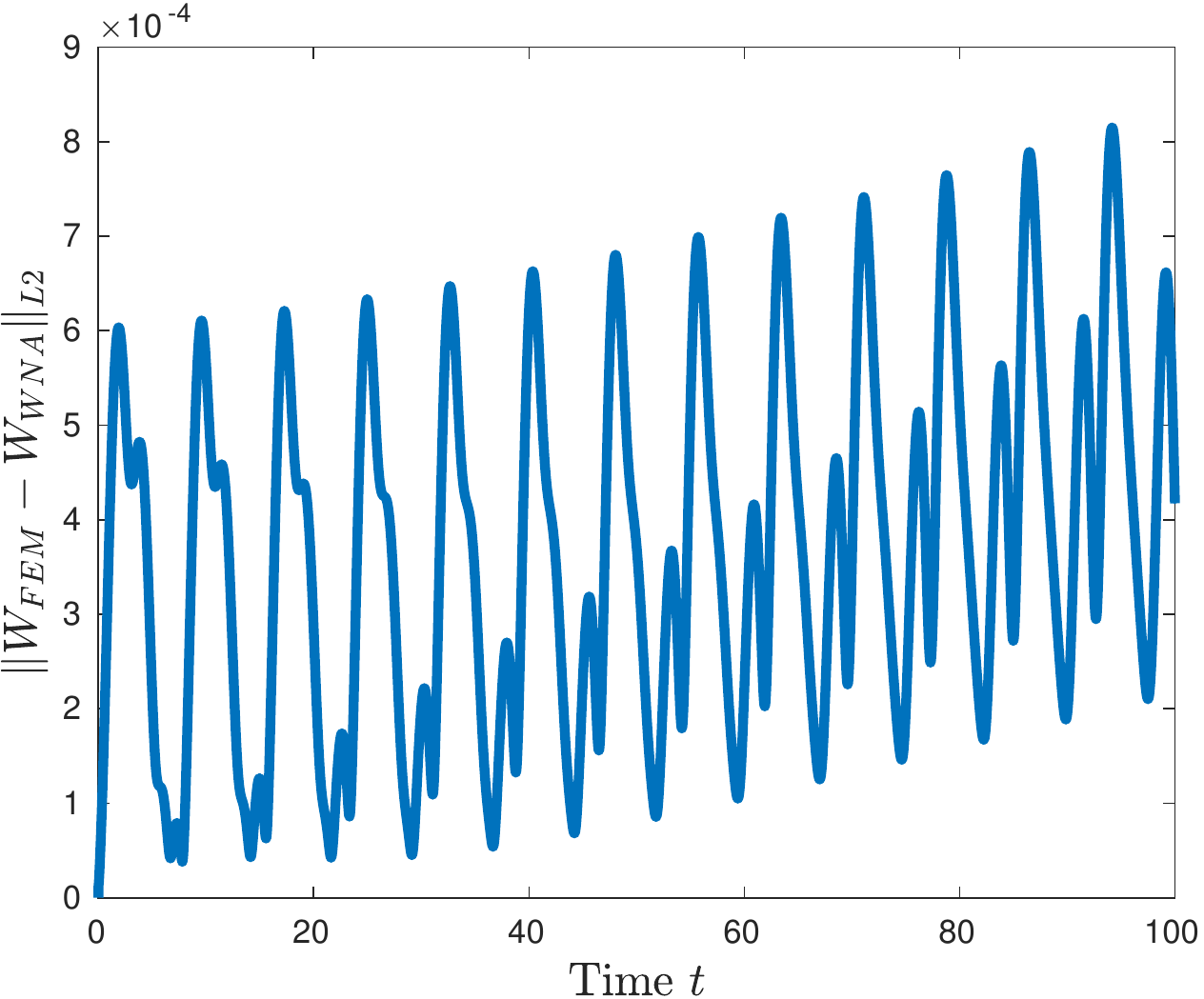}
\end{subfigure}
\caption{\label{fig:WNA_PH_schnakenberg} Same simulation as in
  Fig.~\ref{fig:numerics_PH_schnakenberg}. {Left panel:
    membrane-bound PDE numerically computed solution (red curve) and
    the nearly coinciding weakly nonlinear solution (black dashed
    curve) at time $t = 100$. Right panel: difference between these
    two solutions plotted versus time using the $L^2$ norm.}}
\end{figure}

Although it is expected to be unstable, a numerical simulation
starting very near the mixed-mode solution should stay near it for
some time before drifting away exponentially. {PDE simulation
  results are presented in Fig.~\ref{fig:numerics_PH_schnakenberg} for
  parameter values taken in region 5 and with the initial condition}
corresponding to $E_3$. Good agreement between the weakly nonlinear
and numerically computed mixed-mode PDE membrane-bound patterns is
shown in the left panel of Fig.~\ref{fig:WNA_PH_schnakenberg}. This
agreement is expected for early simulation time. In the right panel of
Fig.~\ref{fig:WNA_PH_schnakenberg}, we {observe that the
  difference between the PDE numerical solution and the asymptotic
  mixed-mode solution in the weakly nonlinear regime grows in time.}

\begin{figure}[htbp]
\centering
\begin{subfigure}{0.47\linewidth}
\includegraphics[width=\linewidth,height=5.0cm]{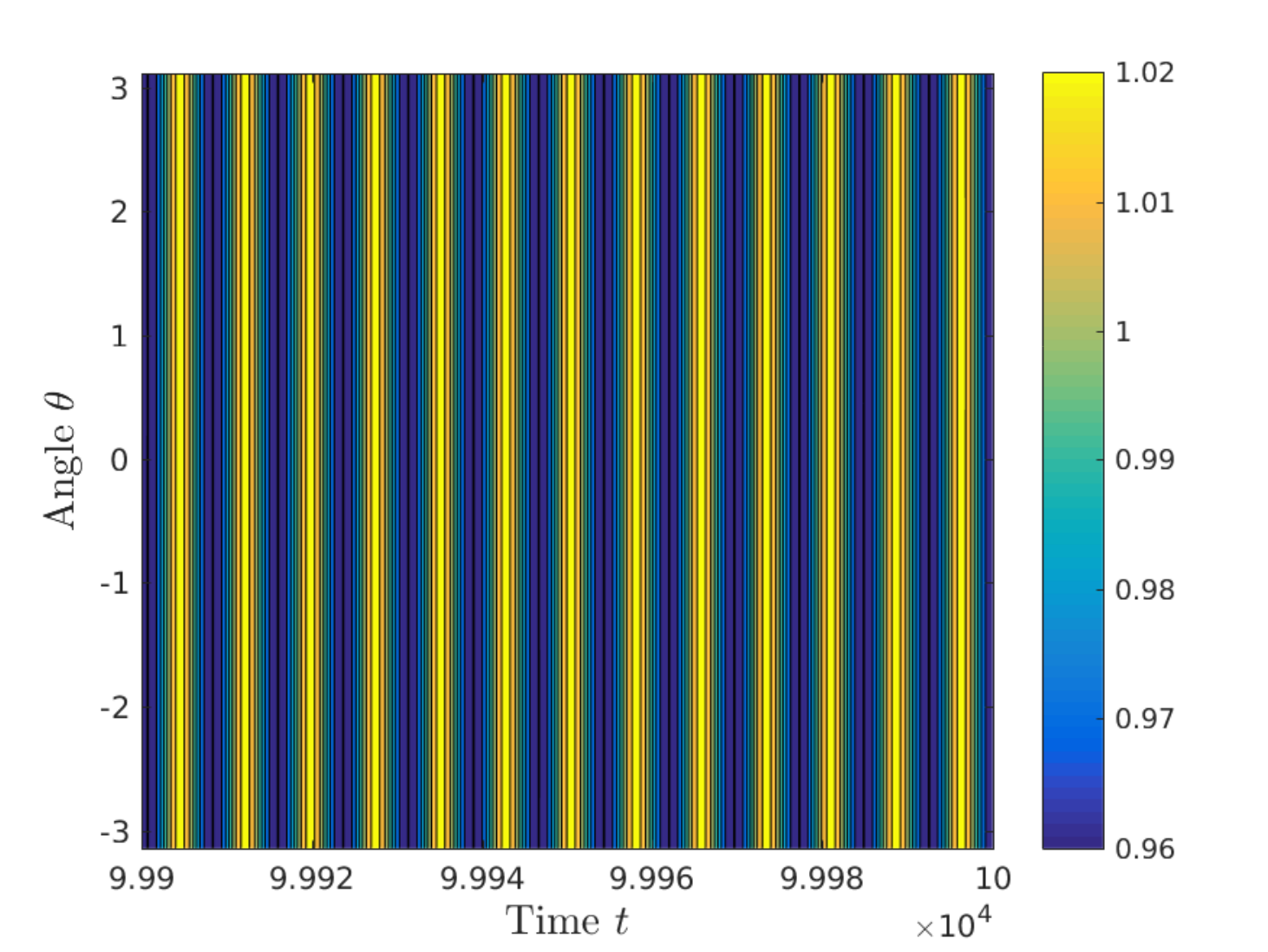}
\end{subfigure}
\begin{subfigure}{0.47\linewidth}
\includegraphics[width=\linewidth,height=5.0cm]{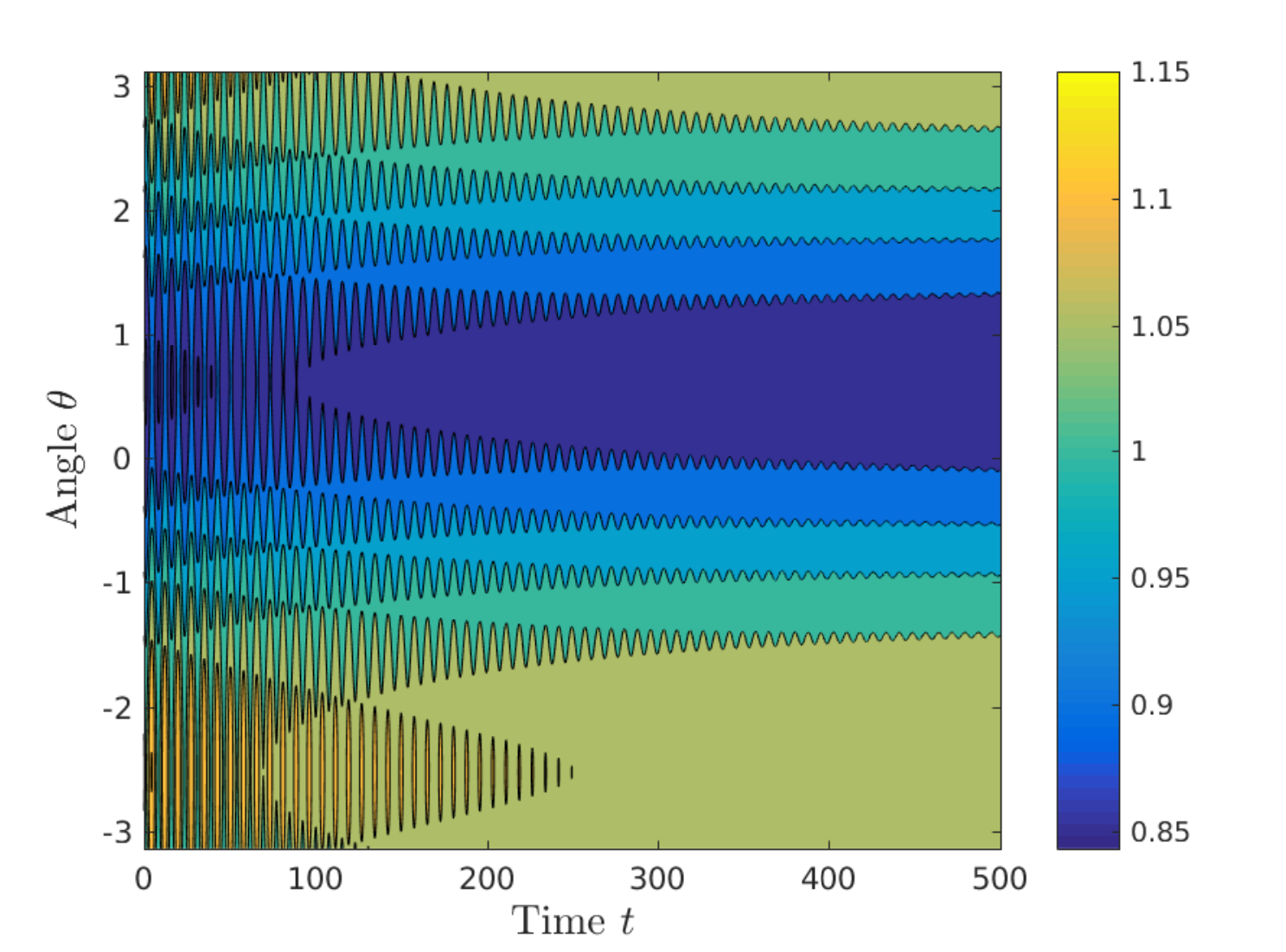}
\end{subfigure}
\caption{\label{fig:numerics_PH_schnakenberg_bistability} Bistability
  between a spatially homogeneous periodic solution (left panel) and a
  stable Turing pattern (right panel) for the Schnakenberg kinetics,
  with bifurcation parameters taken from region 5 with $\eps =
  0.1$. {Left panel: the initial condition is the mixed-mode
    solution in the weakly nonlinear regime, with the space-time
    contour plot showing the long-time oscillatory dynamics of the
    simulation in the right panel of
    Fig.~\ref{fig:numerics_PH_schnakenberg}. Right panel: the initial
    condition is the {base state} solution slightly perturbed
    with the critical eigenvectors of the Jacobian of the spatially
    discretized system. Notice here that the oscillations become
    extinguished as time increases.}}
\end{figure}

{As time increases, our full numerical results show a
  transition toward a spatially homogeneous periodic solution. However, in
this parameter region, bistability is expected and a different initial
condition may lead, instead, to a spatially inhomogeneous equilibrium.}
Fig.~\ref{fig:numerics_PH_schnakenberg_bistability} presents two
simulation outcomes performed in region 5, where different initial
conditions {have led to either} a spatially homogeneous periodic
solution or a stable Turing pattern.

\begin{figure}[htbp]
\centering
\begin{subfigure}{0.47\linewidth}
\includegraphics[width=\linewidth,height=5.0cm]{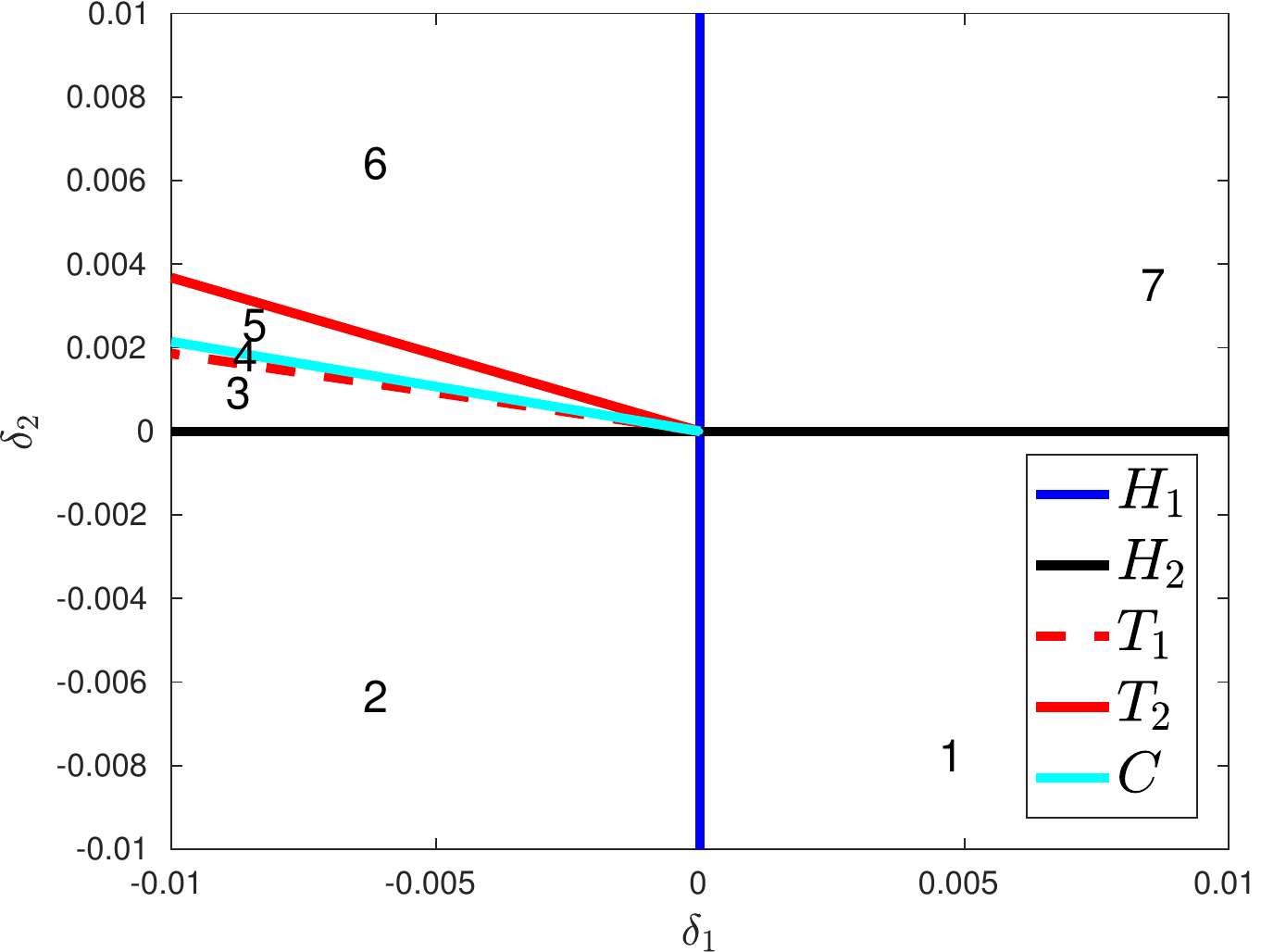}
\end{subfigure}
\begin{subfigure}{0.47\linewidth}
\includegraphics[width=\linewidth,height=5.0cm]{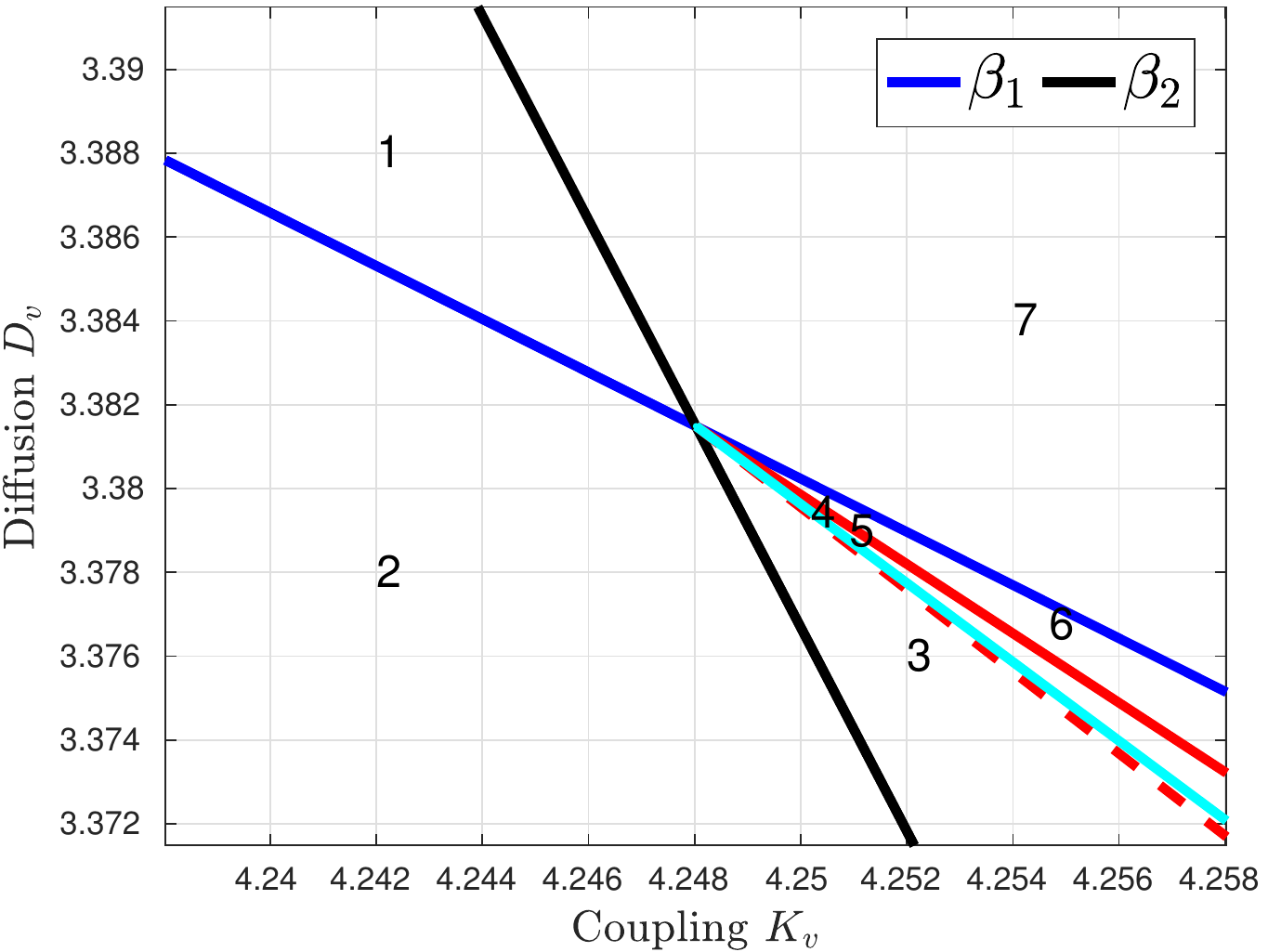}
\end{subfigure}
\caption{\label{fig:pitchfork_hopf_brusselator} Parametric portraits
  in the space of generic parameters $(\delta_1,\delta_2)$ (left
  panel) and original bifurcation parameters $(K_v,D_v)$ (right panel)
  with Brusselator kinetics. The lines $H_{1,2}$ and
  $T_{1,2}$ are described in the caption of
  Fig.~\ref{fig:pitchfork_hopf_schnakenberg}, with the additional line $C$
  (equation \eqref{eq:hopf_line}) in cyan color in the left
  panel. {Applying the affine transformation defined in
  Lemma \ref{lemma:map} yields the plot in the right panel.}}
\end{figure}

\begin{figure}[htbp]
\centering
\begin{subfigure}{0.23\linewidth}
\includegraphics[width=\linewidth]{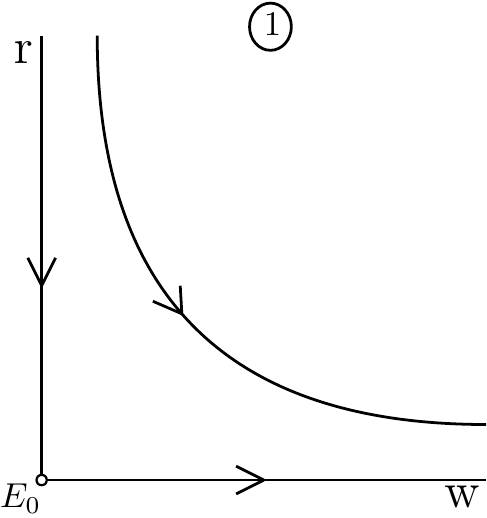}
\end{subfigure}
\begin{subfigure}{0.23\linewidth}
\includegraphics[width=\linewidth]{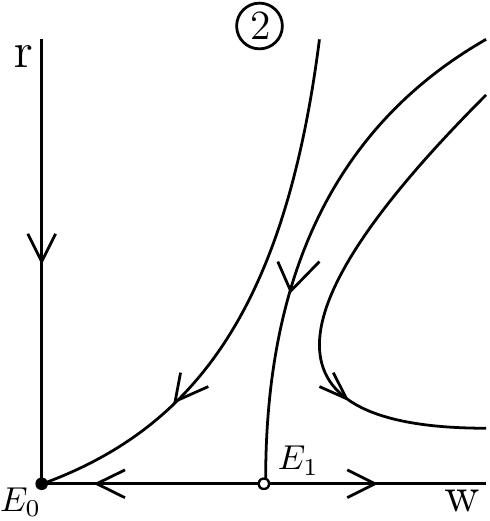}
\end{subfigure}
\begin{subfigure}{0.23\linewidth}
\includegraphics[width=\linewidth]{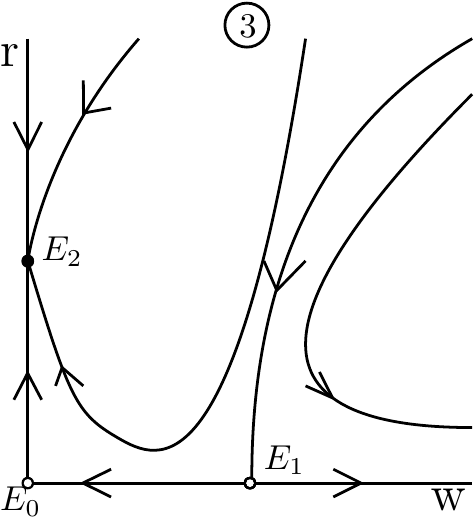}
\end{subfigure}
\begin{subfigure}{0.23\linewidth}
\includegraphics[width=\linewidth]{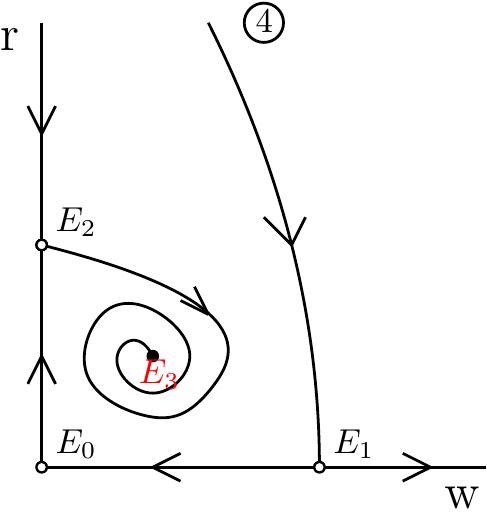}
\end{subfigure}

\begin{subfigure}{0.23\linewidth}
\includegraphics[width=\linewidth]{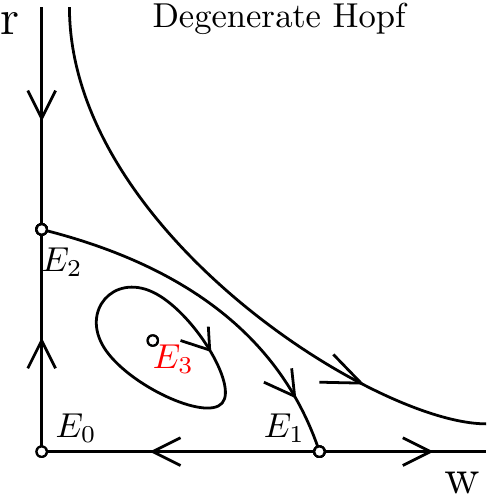}
\end{subfigure}
\begin{subfigure}{0.23\linewidth}
\includegraphics[width=\linewidth]{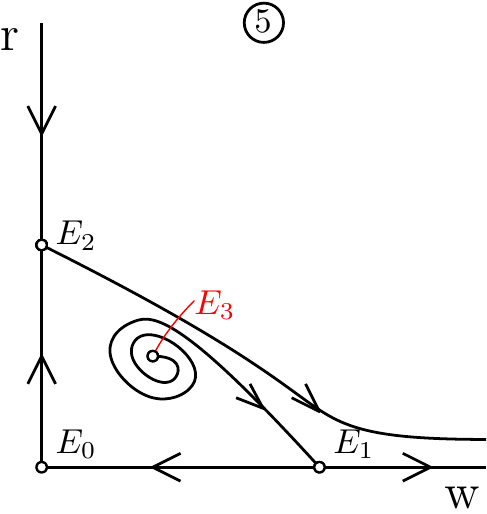}
\end{subfigure}
\begin{subfigure}{0.23\linewidth}
\includegraphics[width=\linewidth]{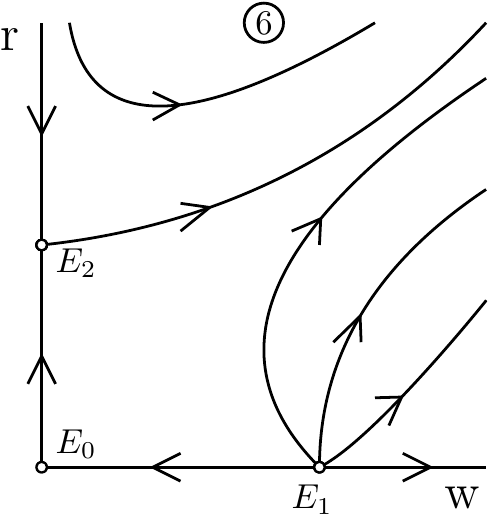}
\end{subfigure}
\begin{subfigure}{0.23\linewidth}
\includegraphics[width=\linewidth]{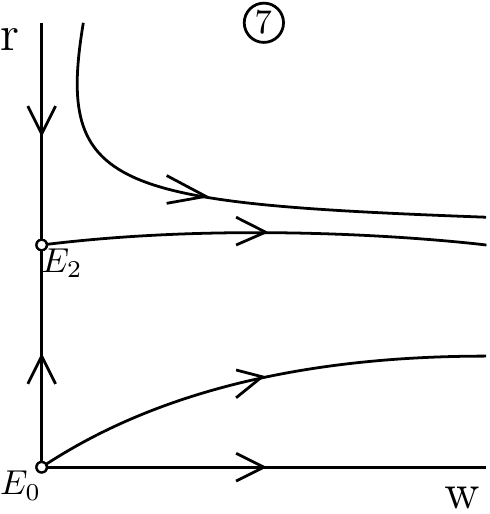}
\end{subfigure}
\caption{\label{fig:phase_diagrams_brusselator} Phase diagrams of the
  truncated system of amplitude equations
  \eqref{eq:rescaled_perturbed_normal_form} with unfolding of type VIa
  (see Fig.~7.5.5 of \cite{guckenheimer1983}). The phase diagram
  {corresponding to the line $C$ in
    Fig.~\ref{fig:pitchfork_hopf_brusselator} is structurally
  unstable, i.e. the retention of generic higher-order terms will
  remove the degeneracy of the Hopf bifurcation and introduce a
  heteroclinic orbit (see \cite{guckenheimer1983} and
  \cite{kuznetsov2004} for more details).}}
\end{figure}

{Next, we discuss the more intricate case that results from
  the Brusselator kinetics (see Table \ref{table:GH_hopf_hopf}).
Parametric portraits are given in
Fig.~\ref{fig:pitchfork_hopf_brusselator}, with corresponding phase
diagrams provided in Fig.~\ref{fig:phase_diagrams_brusselator}.} The
regions 1, 2, 3, and 7 yield the same phase diagrams as for the simple
case analyzed above. Here, region 7 of
Fig.~\ref{fig:pitchfork_hopf_brusselator} corresponds to region 6 of
Fig.~\ref{fig:pitchfork_hopf_schnakenberg}. When crossing the line
$T_1$ from regions 3 to 4, $E_3$ bifurcates from $E_2$, with $E_2$
losing stability. In region 4, $E_3$ is a stable focus while $E_0$,
$E_1$, and $E_2$ are all unstable. On the line $C$, the equilibrium
$E_3$ undergoes a Hopf bifurcation within the truncated system of
amplitude equations \eqref{eq:rescaled_perturbed_normal_form}. Because
only cubic terms are included, the bifurcation is degenerate and the
family of limit cycles persist only on the line itself. Also, for this
threshold value there is a heteroclinic connection between the two
single mode equilibria. In region 5, the four equilibria are
unstable. Finally, between regions 5 and 7, the successive crossing of
the lines $T_2$ and $H_2$ causes the mixed-mode equilibrium to
collapse on $E_1$, after which $E_1$ collapses at the origin.

Restoring the angular variable to the truncated system of amplitude
equations, we expect torus (Neimarck-Sacker) bifurcations for
parameter values on a curve tangent to $C$ and an exponentially thin
(as $(\delta_1,\delta_2) \to 0$) region of parameters near $C$
corresponding to some kind of chaotic behaviour (see
\cite{kuznetsov2004} and \cite{wiggins2003}). The possibility of
{such intricate} dynamics is interesting, but it seems likely
to be confined to an extremely small region of parameter space that
would be virtually undetectable in PDE simulations.

\begin{figure}[htbp]
\centering
\begin{subfigure}{0.47\linewidth}
\includegraphics[width=\linewidth,height=5.0cm]{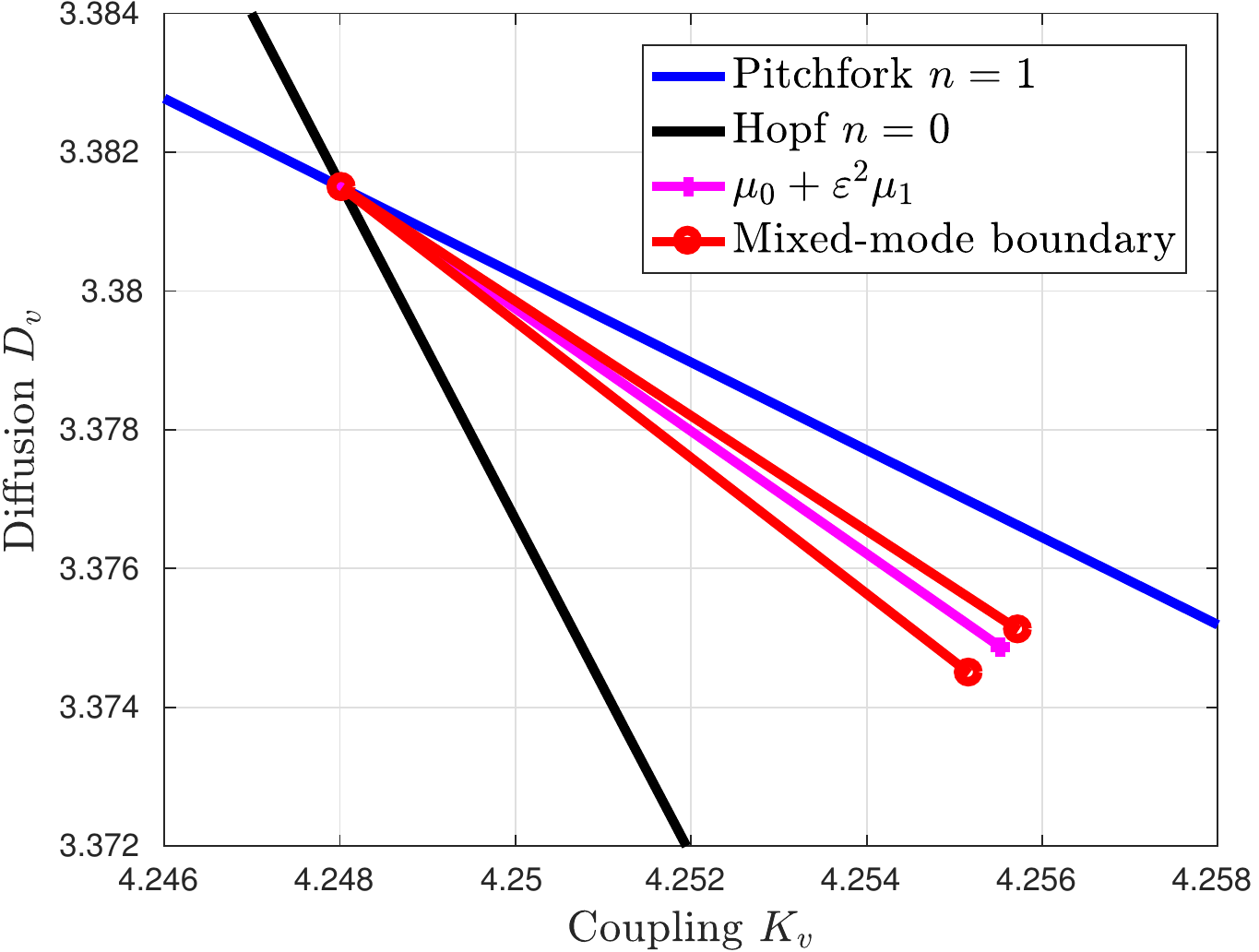}
\end{subfigure}
\begin{subfigure}{0.47\linewidth}
\includegraphics[width=\linewidth,height=5.0cm]{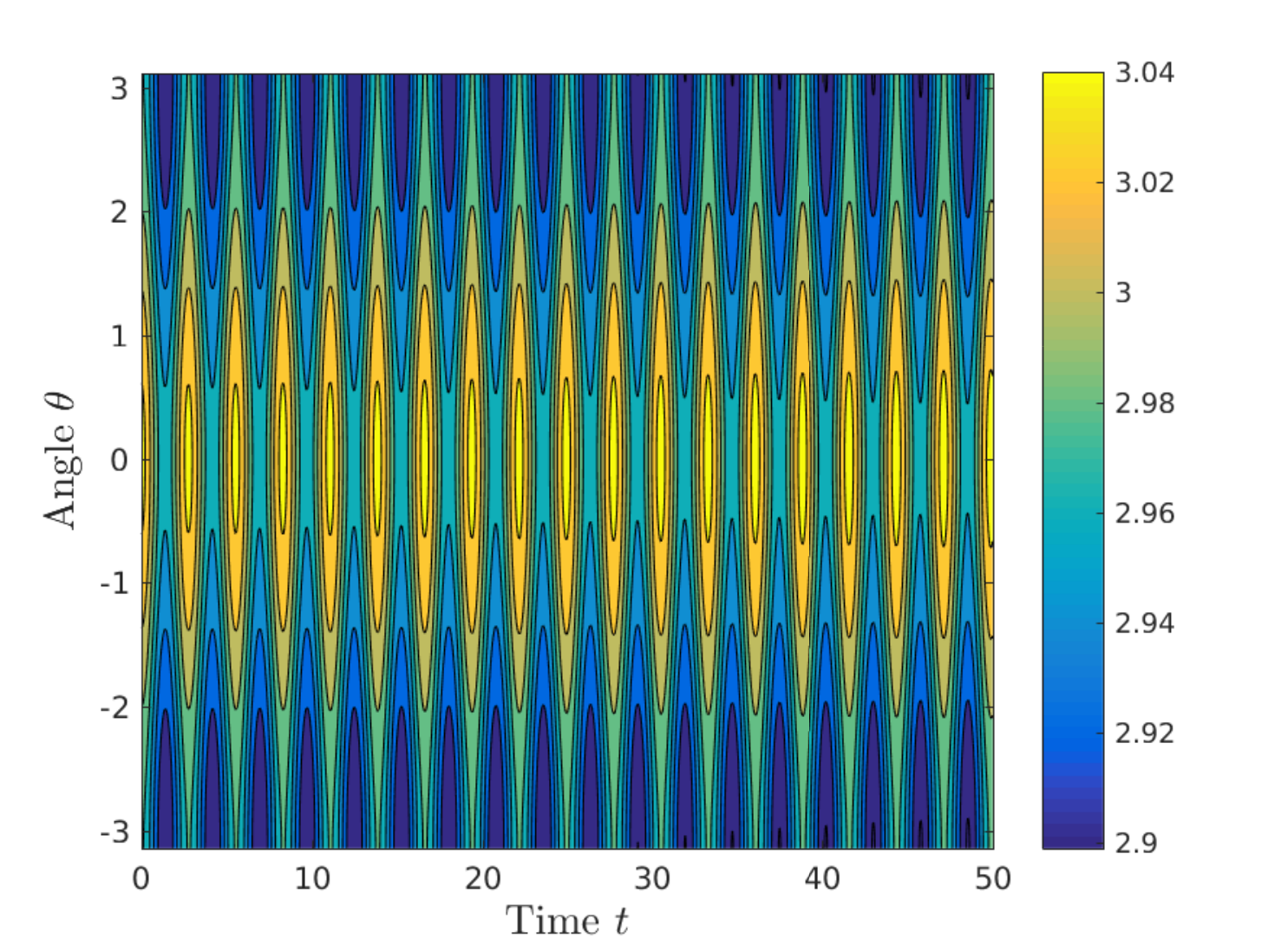}
\end{subfigure}
\caption{\label{fig:numerics_PH_brusselator} Interaction of a
  supercritical Hopf and subcritical pitchfork bifurcations
  {for Brusselator kinetics with $\eps = 0.1$.} {The
    simulation corresponds to region 5 (see left panel). Right panel:
    a space-time contour plot of the membrane-bound activator species
    $u(\theta,t)$, which exhibits oscillations around some spatial
    pattern. The initial condition is the mixed-mode solution in the
    weakly nonlinear regime.}}
\end{figure}

In order to remove the degeneracy of the Hopf bifurcation,
higher-order terms should be added to the normal form. Since this
challenging computation, starting from our coupled bulk-surface PDE
model, {is left as an open problem}, some bifurcation results
from regions 4, 5 and {on the line $C$ in
  Fig.~\ref{fig:pitchfork_hopf_brusselator}} cannot be transferred to
the original system. It is nevertheless possible to investigate
numerically the breather-type solutions within this narrow parameter
regime (when considering the space defined by the original bifurcation
parameters). Fig.~\ref{fig:numerics_PH_brusselator} shows simulation
results for short times, and for parameter values taken in region 5,
where the mixed-mode solution is expected to be unstable.

\begin{figure}[htbp]
\centering
\begin{subfigure}{0.47\linewidth}
\includegraphics[width=\linewidth,height=5.0cm]{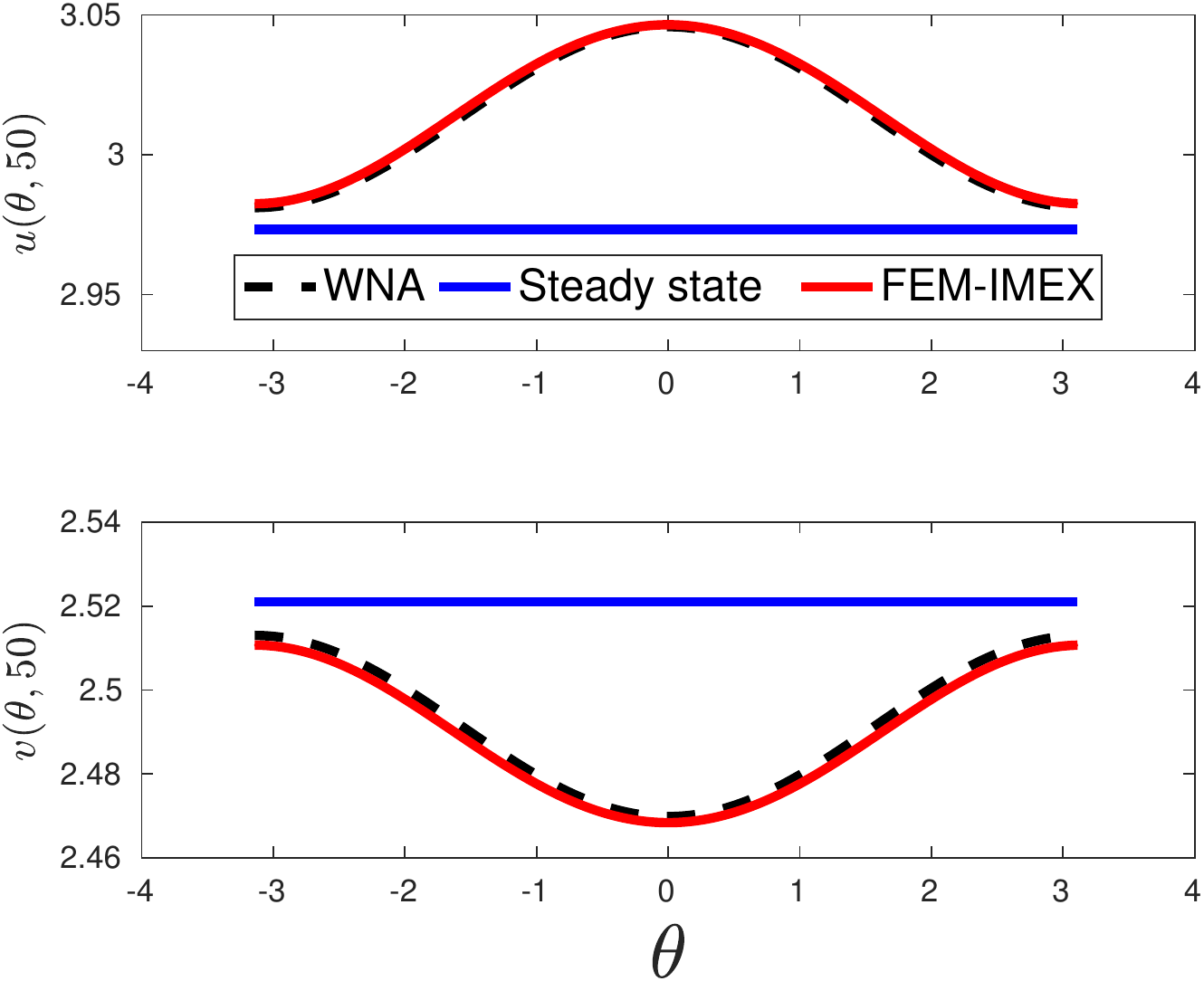}
\end{subfigure}
\begin{subfigure}{0.47\linewidth}
\includegraphics[width=\linewidth,height=5.0cm]{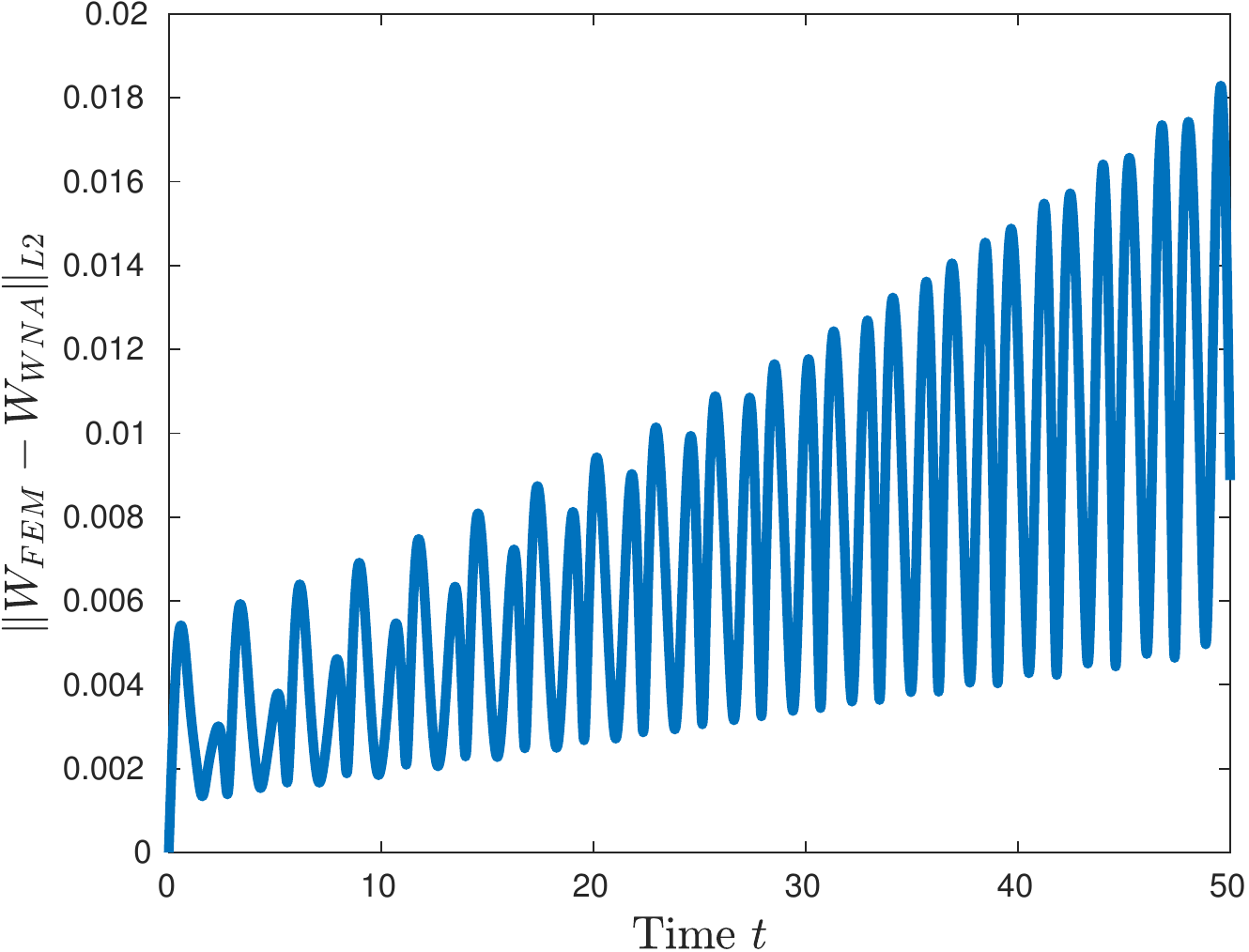}
\end{subfigure}
\caption{\label{fig:WNA_PH_brusselator} Same simulation as in
  Fig.~\ref{fig:numerics_PH_brusselator} for the Brusselator kinetics,
  with $\eps = 0.1$. Left panel: the membrane-bound numerical (red
  curve) and the {nearly coinciding} weakly nonlinear (black
  dashed curve) solutions at time $t = 50$.  {Right panel:
    difference between these two solutions plotted over time using the
    $L^2$ norm.}}
\end{figure}

\begin{figure}[htbp]
\centering
\begin{subfigure}{0.47\linewidth}
\includegraphics[width=\linewidth,height=5.0cm]{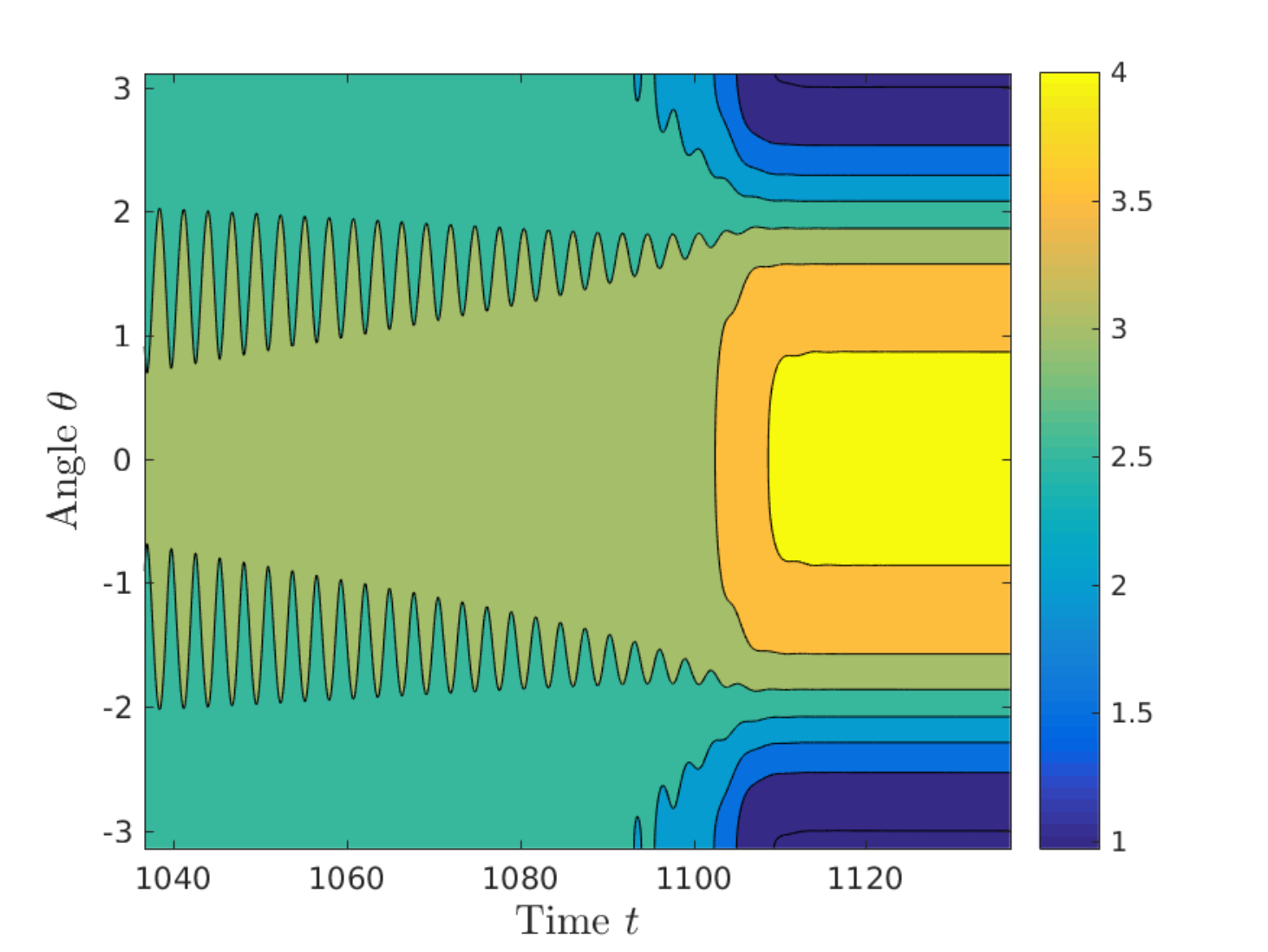}
\end{subfigure}
\begin{subfigure}{0.47\linewidth}
\includegraphics[width=\linewidth,height=5.0cm]{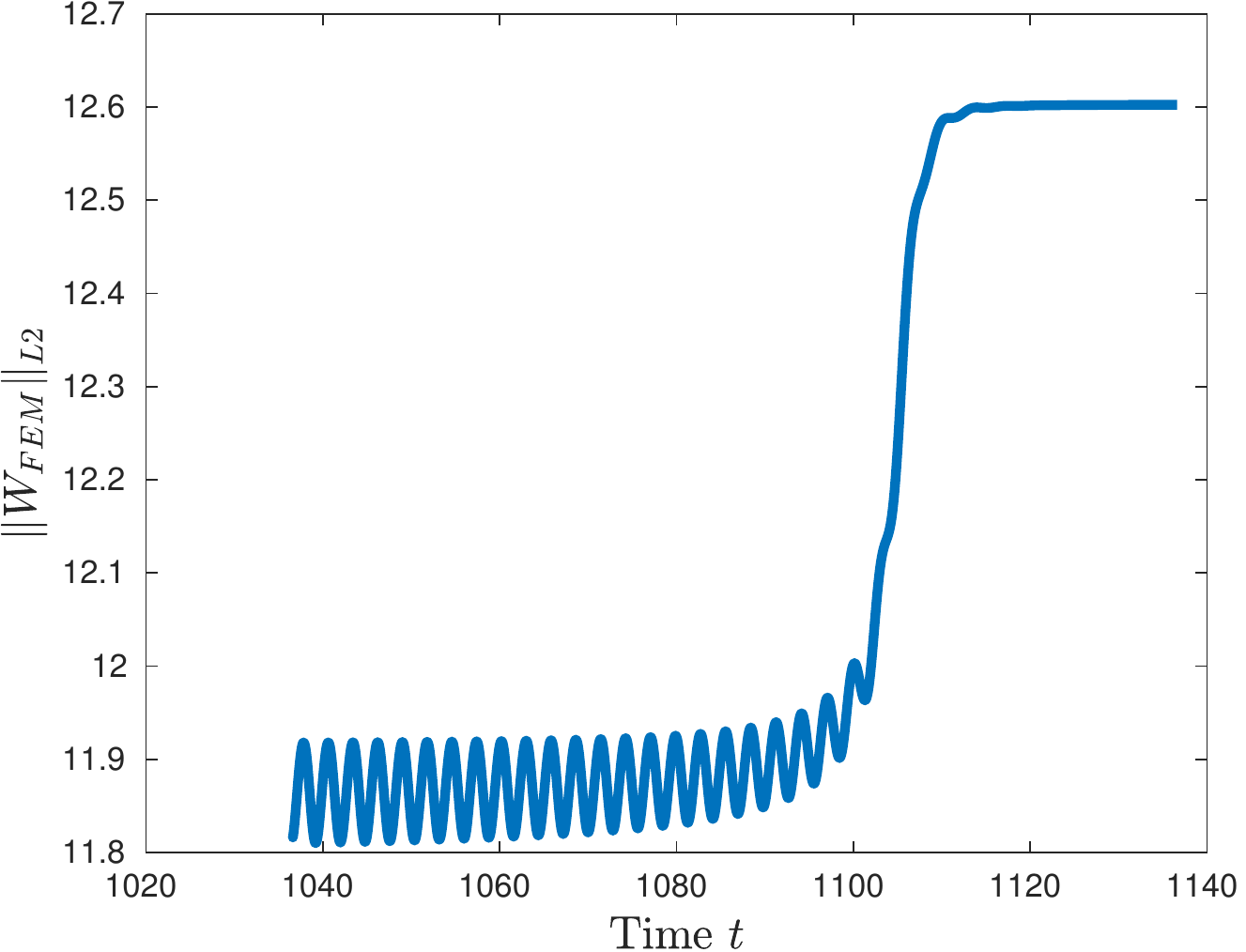}
\end{subfigure}
\caption{\label{fig:LT_PH_brusselator} {Transition between
    the mixed-mode solution and a stable Turing pattern for the
    Brusselator kinetics for bifurcation parameters taken from region
    5 of Fig.~\ref{fig:phase_diagrams_brusselator} with $\eps = 0.1$.
    Left panel: a space-time contour plot. Right panel: plot of the
    $L^2$ norm of the numerical solution. The initial transient has
    been removed for both plots.}}
\end{figure}

{The right panel of Fig.~\ref{fig:WNA_PH_brusselator} shows
  that the distance between the numerical solution and the solution in
  the weakly nonlinear regime grows over time. However, in contrast to
  the previous case with the Schnakenberg kinetics, the long time
  integration in Fig.~\ref{fig:LT_PH_brusselator} clearly reveals a
  transition towards a spatially inhomogeneous steady state and the
  absence of bistability with the spatially homogeneous periodic
  solution. This is consistent with the phase diagram for parameters
  in region 5 in Fig.~\ref{fig:pitchfork_hopf_brusselator} and
  Fig.~\ref{fig:phase_diagrams_brusselator}.}

{Finally, we remark that when applied to systems of PDEs, the
  normal form analysis of codimension-two bifurcations must be
  interpreted with care, especially when degenerate local bifurcations
  occur in the system of amplitude equations
  \cite{witten1997}. Moreover, the parameter regime and phase space
  ranges where the conclusions hold can be fairly narrow, making it
  very difficult for PDE direct numerical simulations to reproduce
  delicate dynamical behaviors that occur in the ODE amplitude
  equations.}

\section{Global dynamics via full numerics: rotating waves}\label{sec:global_dynamics}

In this section, full numerical simulations are used to briefly explore novel dynamical behaviors in the highly nonlinear regime, away from bifurcation points, that are due to the bulk-surface coupling. For the Brusselator reaction kinetics, we study the formation of rotating waves and show that they arise when a nontrivial spatial mode undergoes a Hopf bifurcation. Allowing for different adsorption and desorption rates for each species seems to be a key condition behind the formation of such waves.

We consider here a {modified coupled bulk-surface model} for
which the rates of adsorption and desorption are different for each
species. Let $r_a$ and $p_a$ be the activator and inhibitor
rates of adsorption. Similarly, let $r_d$ and $p_d$ be the
desorption rates. {Then, the boundary conditions in \eqref{eq:BC}
are reformulated as
\begin{equation}
 D_u \partial_r U|_{r=R} = r_d u - r_a U|_{r=R} \,, \qquad
 D_v \partial_r V|_{r=R} = p_d v - p_a V|_{r=R}\,.
\end{equation}
Similar boundary conditions are considered in \cite{levine2005} and
\cite{madzvamuse2015}. These new boundary conditions modify the
dynamics on the surface, so that \eqref{eq:surface_RD} is replaced by
\begin{equation}
  \partial_t u = \frac{d_u}{R^2}\partial_{\theta\theta}u - r_d u +
  r_a U|_{r = R} + f(u,v)\,, \qquad
  \partial_t v = \frac{d_v}{R^2}\partial_{\theta\theta}v - p_d v +
  p_a V|_{r = R} + g(u,v)\,.
\end{equation}
After calculating the radially symmetric base state for this modified
bulk-surface model, a linear stability analysis readily provides a
transcendental equation for the growth rate $\lambda$ associated with the
circular harmonic of mode $n$. In place of \eqref{eq:transcendental}, the
growth rates are roots of $F_n(\lambda)=0$, where
\begin{equation}\label{eq:transcendental_RW}
  F_n(\lambda) = \left( \lambda - f_u^e + \frac{r_d}
    {1 + \frac{r_a I_n(\Omega_u R)}
      {D_u \Omega_u I_n^{\prime}(\Omega_u R)}} + \frac{n^2 d_u}{R^2}\right)
  \left(\lambda - g_v^e + \frac{p_d}{1 + \frac{p_a I_n(\Omega_v R)}
      {D_v \Omega_v I_n^{\prime}(\Omega_v R)}} + \frac{n^2 d_v}{R^2} \right) - f_v^e g_u^e\,.
\end{equation}
}

\begin{figure}[htbp]
\centering
\begin{subfigure}{0.47\linewidth}
\includegraphics[width=\linewidth,height=5.0cm]{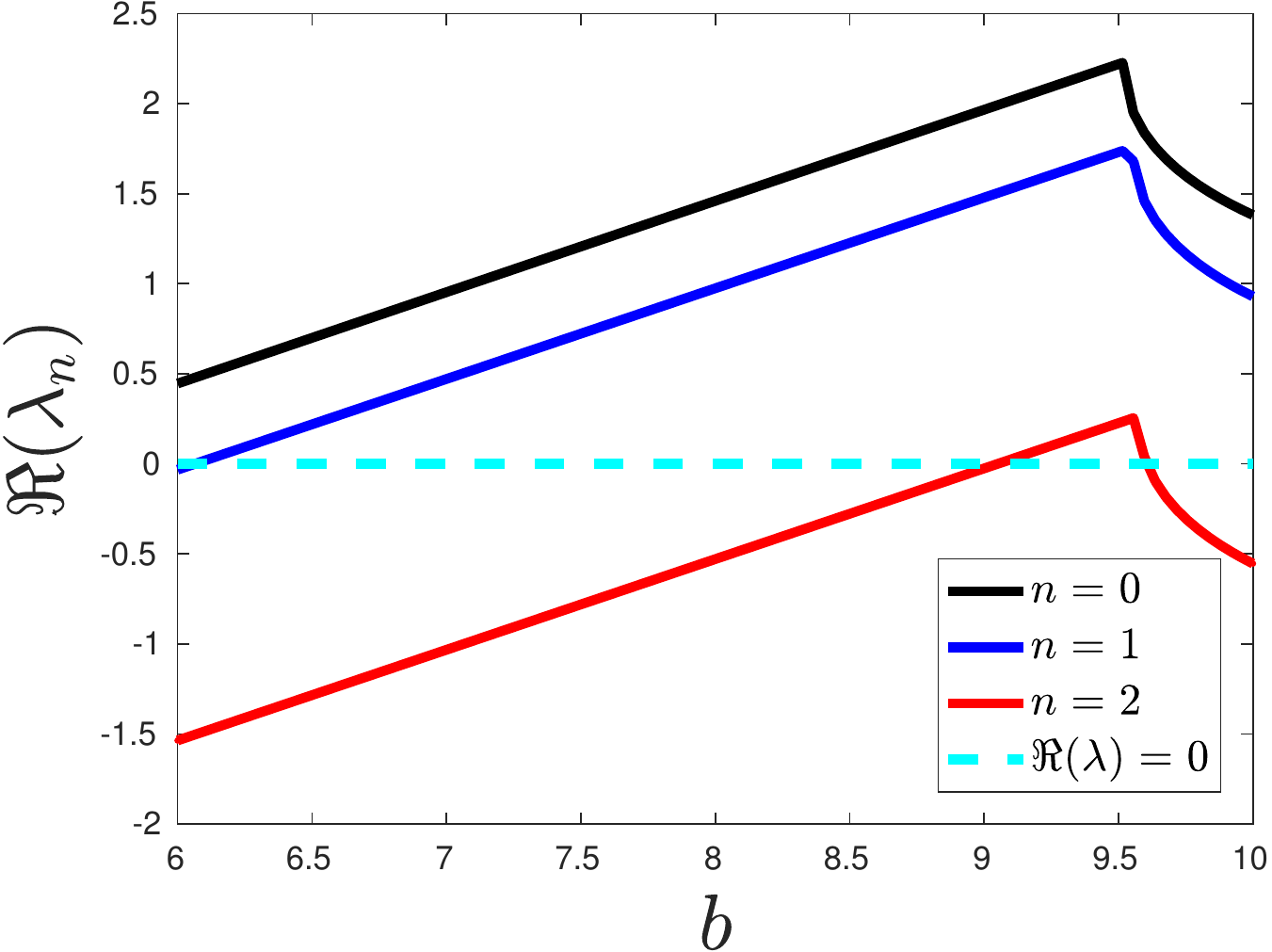}
\end{subfigure}
\begin{subfigure}{0.47\linewidth}
\includegraphics[width=\linewidth,height=5.0cm]{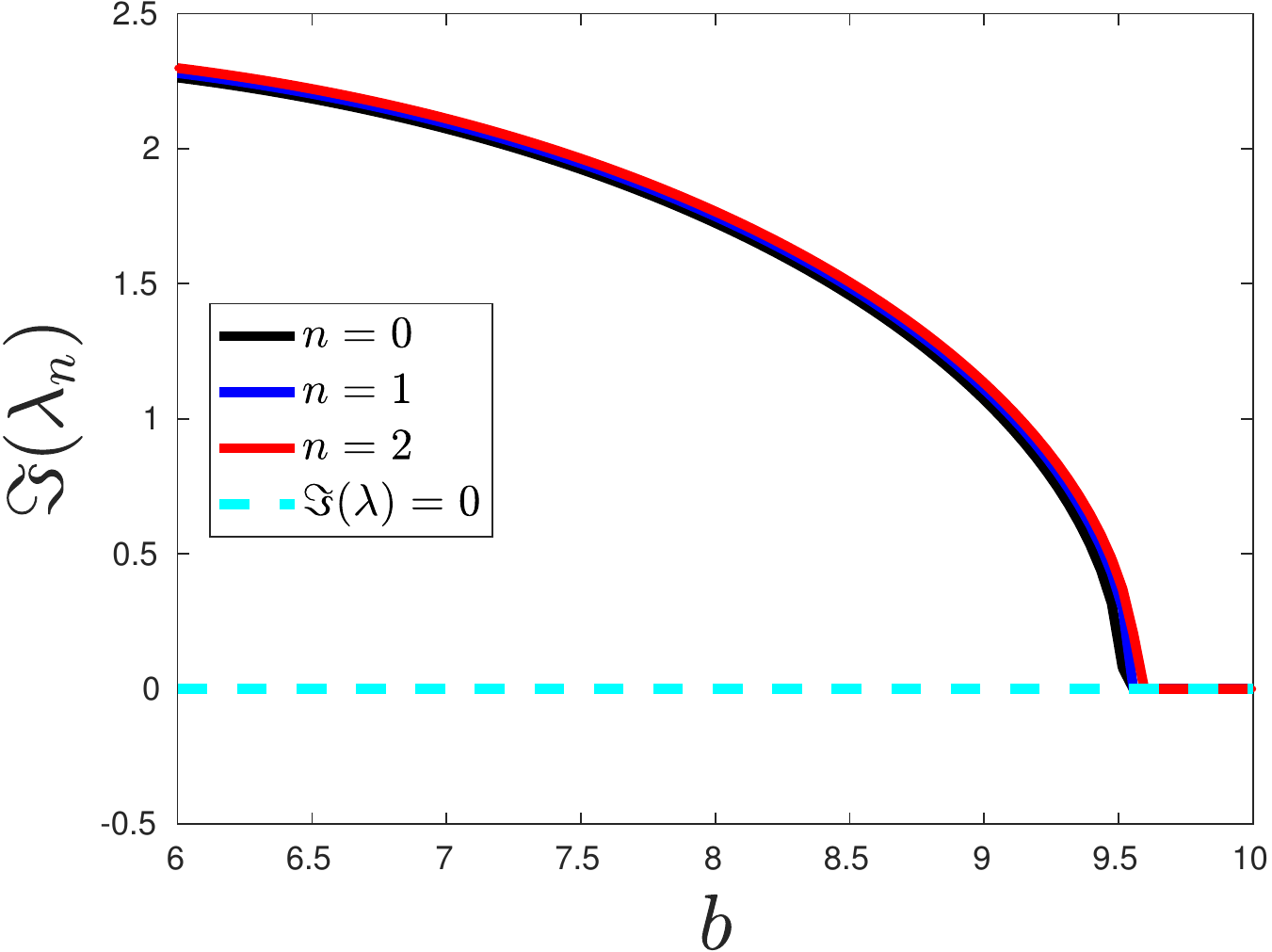}
\end{subfigure}
\caption{\label{fig:sweep_b} Real (left panel) and imaginary (right
  panel) parts of the most unstable eigenvalues, {computed
    from \eqref{eq:transcendental_RW}, for the mode $n=0,\, 1,\, 2$ as
    the kinetic parameter $b$ increases. The parameters are given in
    \eqref{eq:parameters}.}}
\end{figure}

Following a remark from \cite{levine2005} on the conditions
{underlying} the emergence of traveling waves, we restrict
the parameter space by setting the diffusion coefficients to be equal
for both species. More specifically, the following set of parameters
is considered:
\begin{equation}\label{eq:parameters}
  R = 1,\,\, D_u = D_v = 1,\,\, \sigma_u = \sigma_v = 0.5,\,\,
  d_u = d_v = 0.5,\,\, r_a = 0.1,\,\, r_d = 1,\,\,
  p_a = 1,\,\, p_d = 0.1,\,\, a = 3\,.
\end{equation}
By allowing the {Brusselator} kinetic parameter $b$ in
(\ref{eq:brusselator}) to be free in equation
\eqref{eq:transcendental_RW}, {in Fig.~\ref{fig:sweep_b} we show
  numerically that the system undergoes a series of Hopf bifurcations,}
  each of which is associated with a spatial mode $n$. Notice in the
left panel of Fig.~\ref{fig:sweep_b} that the trivial mode is the
first to lose stability. Hence, we expect the waves to coexist with
radially symmetric oscillations in the fully nonlinear regime.

\begin{figure}[htbp]
\centering
\begin{subfigure}{0.47\linewidth}
\includegraphics[width=\linewidth,height=5.0cm]{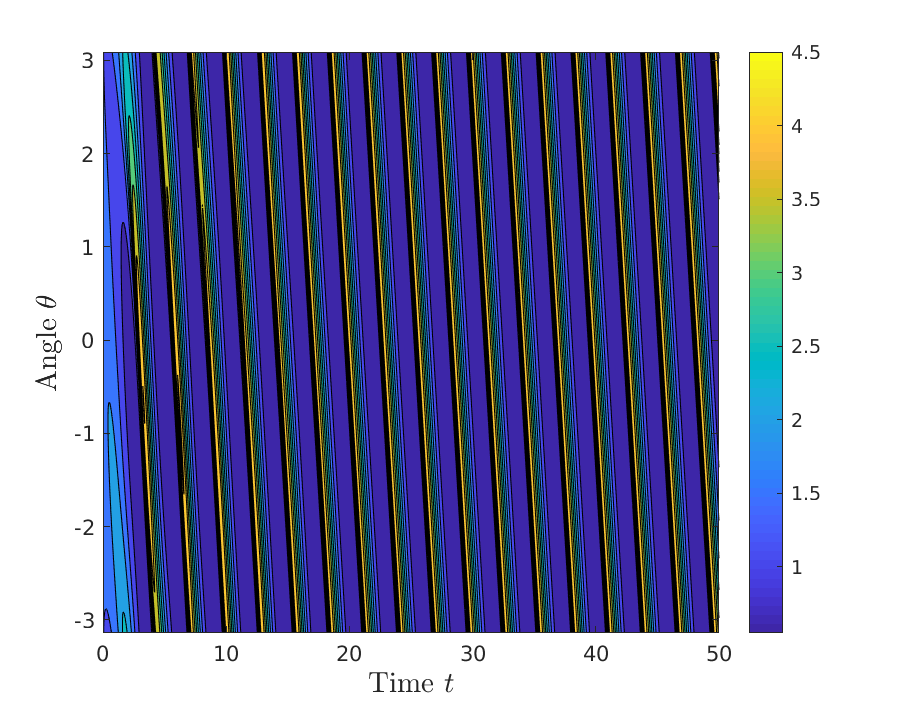}
\end{subfigure}
\begin{subfigure}{0.47\linewidth}
\includegraphics[width=\linewidth,height=5.0cm]{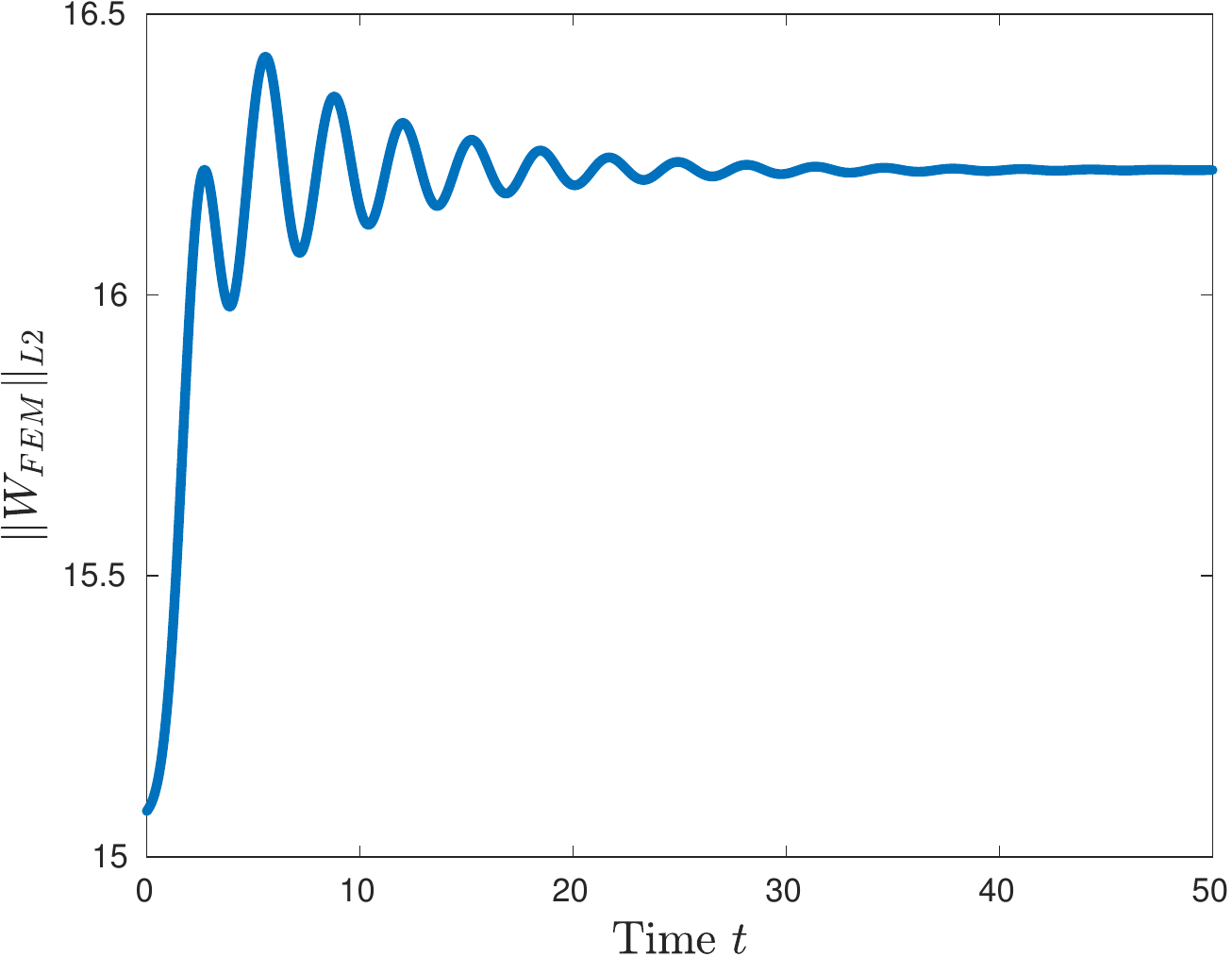}
\end{subfigure}
\caption{\label{fig:CW_wave} {Clockwise rotating waves for
    the Brusselator kinetics (\ref{eq:brusselator}) with $b=8$ and for
    the parameter set \eqref{eq:parameters}.} The initial condition
  corresponds to a perturbation of the {base state} solution
  favoring the mode $n=1$. The left panel shows a space-time contour
  plot of the membrane-bound activator species. In the right panel,
  the $L^2$ norm of the solution converges to some equilibrium values
  after an initial oscillatory transient.}
\end{figure}

\begin{figure}[htbp]
\begin{subfigure}{0.47\linewidth}
\includegraphics[width=\linewidth,height=5.0cm]{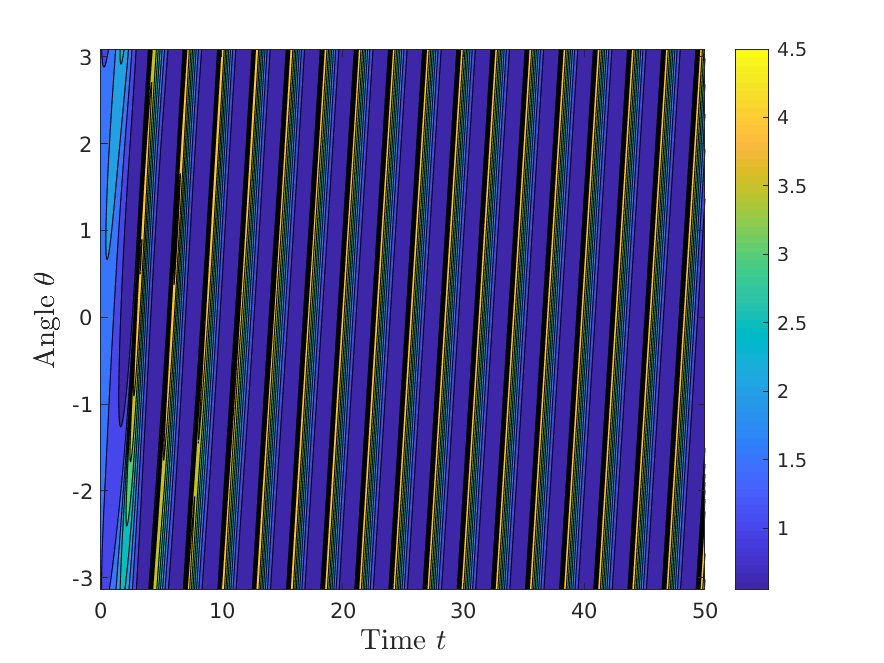}
\end{subfigure}
\begin{subfigure}{0.47\linewidth}
\includegraphics[width=\linewidth,height=5.0cm]{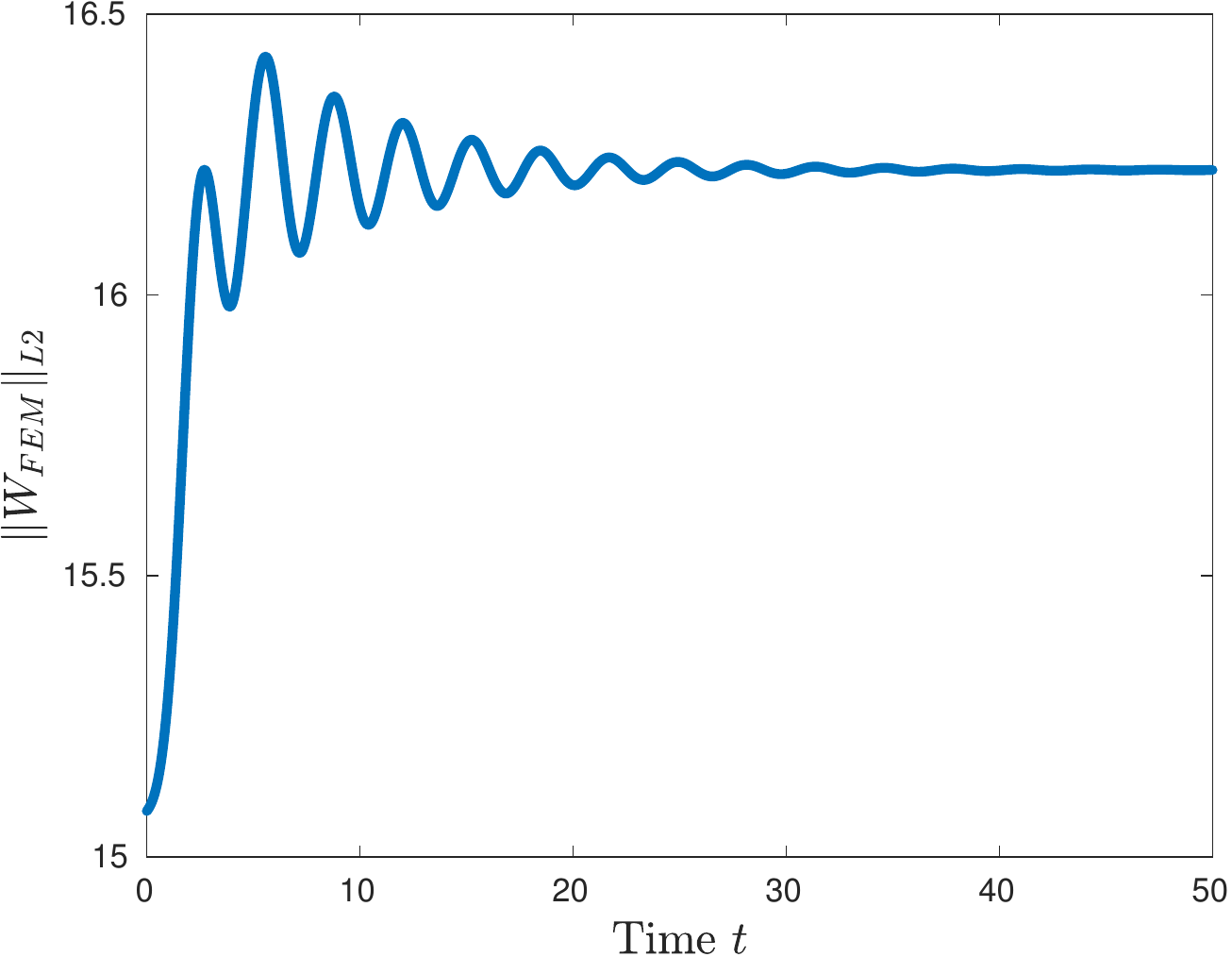}
\end{subfigure}
\caption{\label{fig:ACW_wave} Anti-clockwise rotating waves with the
  Brusselator kinetics \eqref{eq:brusselator} and the same parameter
  values as in Fig.~\ref{fig:CW_wave}. The initial condition
  corresponds to a perturbation of the {base state} solution
  favoring the mode $n=-1$. The left panel shows a space-time contour
  plot of the membrane-bound activator species. In the right panel, we
  see the $L^2$ norm of the solution converging to some equilibrium
  values in a similar fashion as for the clockwise waves.}
\end{figure}

\begin{figure}[htbp]
\begin{subfigure}{0.47\linewidth}
\includegraphics[width=\linewidth,height=5.0cm]{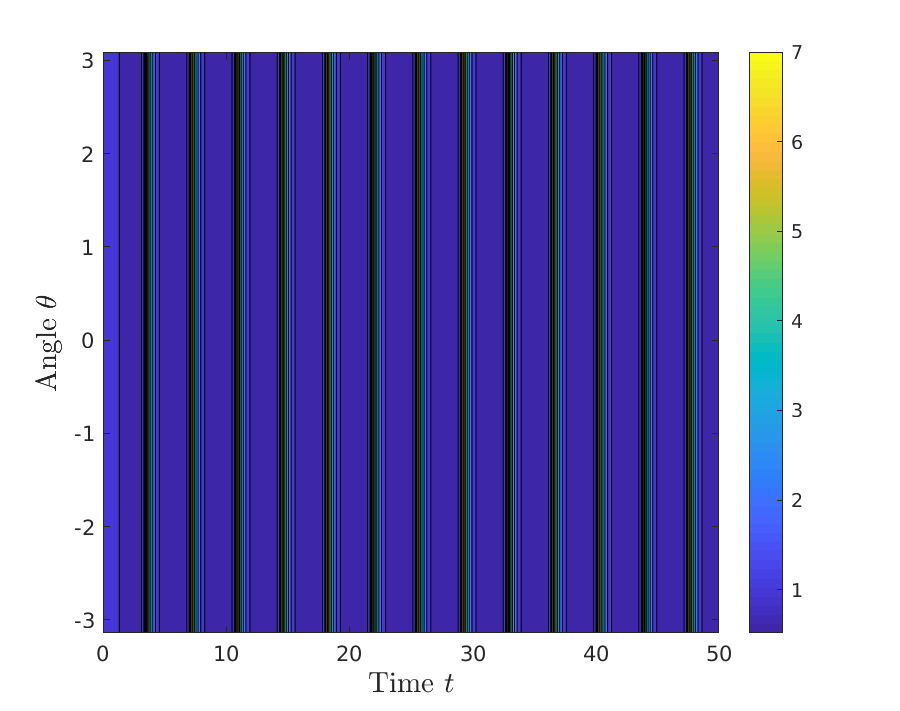}
\end{subfigure}
\begin{subfigure}{0.47\linewidth}
\includegraphics[width=\linewidth,height=5.0cm]{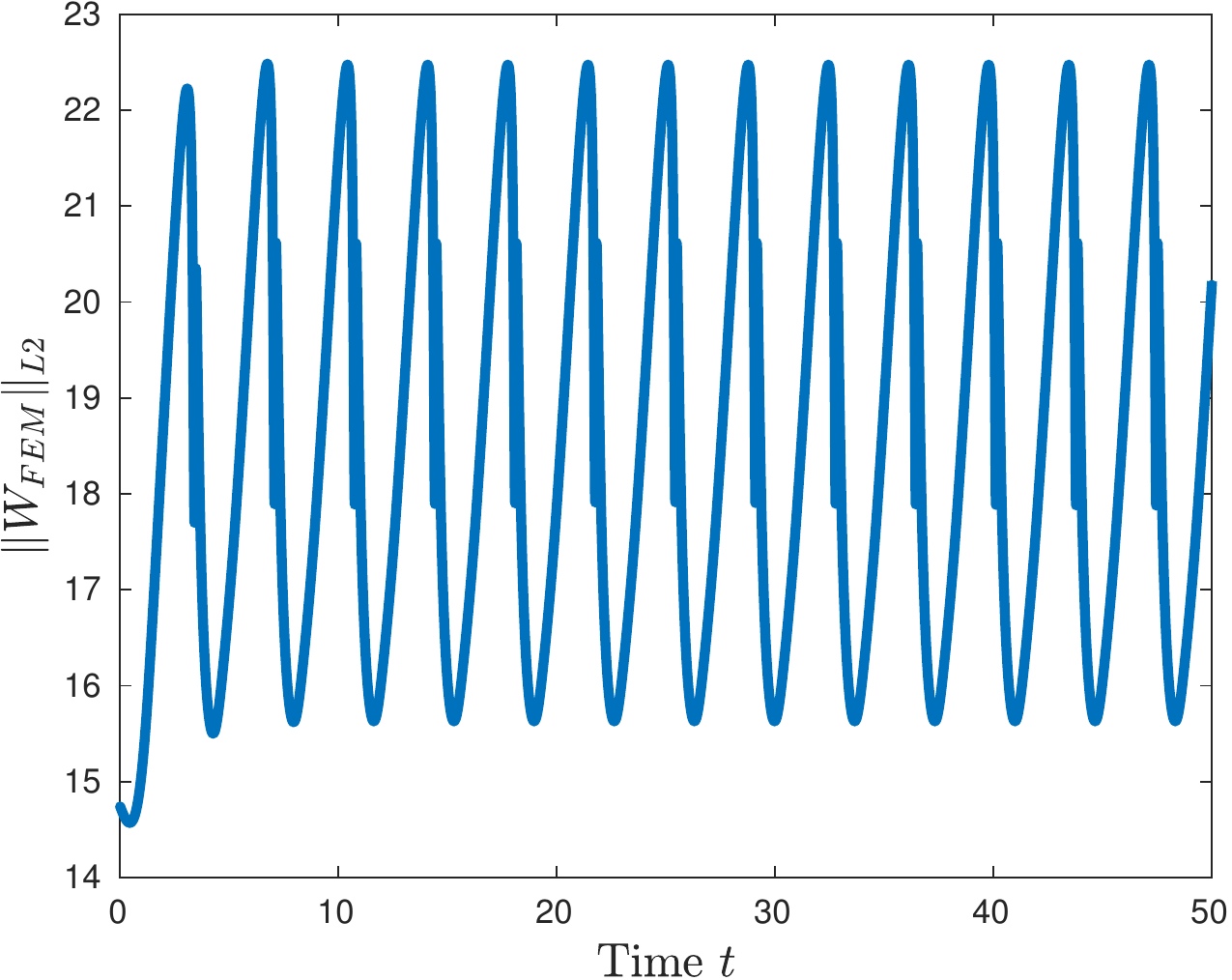}
\end{subfigure}
\caption{\label{fig:uniform_oscillations} Radially symmetric
  oscillations with the Brusselator kinetics and the same parameter
  values as in Figs.~\ref{fig:CW_wave}--\ref{fig:ACW_wave}. The
  initial condition corresponds to a perturbation of the {base state}
  solution favoring the trivial mode ($n=0$). The left panel shows a
  space-time contour plot of the membrane-bound activator species,
  which clearly exhibits spatially uniform oscillations. In the right
  panel, the solution in $L^2$ norm undergoes sustained oscillations.}
\end{figure}

{For $b=8$, full PDE numerical computations of this modified
  bulk-surface model reveal three distinct types of temporally
  oscillatory solutions depending on the initial data. A clockwise
  rotating wave is shown in Fig.~\ref{fig:CW_wave}, an anti-clockwise
  rotating wave is shown in Fig.~\ref{fig:ACW_wave}, and finally a
  radially symmetric oscillatory solution is shown in
  Fig.~\ref{fig:uniform_oscillations}. For each case, appropriate
  initial conditions favoring a particular mode have led to the
  desired dynamics. We have also tried to compute a standing wave by
  stimulating the modes $n=\pm 1$, but our numerical results suggest
  such a solution to be unstable.}

{These numerical results give only a glimpse of novel global
  solution structures for coupled bulk-surface models that can occur
  away from bifurcation points. A rigorous analysis of the existence
  of such rotating traveling waves, including a precise determination
  of the parameter space involving the adsorption and desorption rates
  where they occur, is beyond the scope of this paper.}

\section{Discussion}\label{sec:discussion}

On a two-dimensional circular domain, we have introduced and analyzed
a class of coupled bulk-surface reaction-diffusion models for which a
passive diffusion process occurring in the interior bulk domain {is
  linearly coupled to a nonlinear reaction-diffusion process on the
  domain boundary.} In \S \ref{sec:weakly_nonlinear_theory}, a
multiple time-scale approach was employed to systematically derive
amplitude equations near three different instabilities: the Hopf, the
pitchfork (or Turing), and the pitchfork-Hopf
bifurcations. {An interesting feature of our development of the weakly nonlinear theory of pattern formation for coupled bulk-surface PDE models was the analysis of the spectral problem for the linearization, which involved the eigenvalue parameter appearing in the boundary conditions.} In \S \ref{sec:bifurcation_analysis}, we used the normal form equations to determine the stability of bifurcating branches in the weakly nonlinear regime. The theory was illustrated using the classical Schnakenberg and Brusselator kinetics in \S \ref{sec:validation_weakly_nonlinear}, where good agreement between numerical and analytical solutions was observed. Our hybrid analytical-numerical approach has shown that the linear coupling of a diffusive bulk to an active membrane can lead to either oscillatory dynamics or pattern formation. Finally, the formation of rotating waves is explored through numerical simulations in \S \ref{sec:global_dynamics}.

Several open problems related to coupled bulk-surface
reaction-diffusion systems warrant further investigation. One
challenge concerns the computation of global bifurcating branches, a
task amenable to numerical bifurcation analysis. {The classical software AUTO \cite{doedel2007} has been successfully applied to the reduced 1-D radially symmetric model with angular invariance (in the context of Hopf bifurcations), but it cannot handle implicit systems of differential equations such as those obtained after discretizing the full 2-D model using finite elements (see Appendix \ref{sec:numerical_methods} for details).} A promising
alternative to AUTO that has been explored is the software package
\textsc{coco} (cf.~\cite{danko2013}). The \textit{Equilibrium Point} toolbox
from \textsc{coco}, combined with the PDE toolbox from MATLAB
\cite{matlab2016}, has been used to compute base state solution
families of the full model. Successive mesh refinement has revealed
quadratic convergence between the bifurcation points predicted by the
weakly nonlinear theory and as detected by \textsc{coco}.  However,
because of rotational symmetries, we have been unable to numerically
branch off at a pitchfork bifurcation point. These bifurcations are
characterized by the crossing of two nearly identical eigenvalues
(spatial discretization causes some loss of symmetry) through the
origin, and thus there are two critical eigenvectors at the branch
point. Further work in this direction is needed in order to
numerically resolve the bifurcating branch arising from this rather
{degenerate bifurcation point.}

The weakly nonlinear analysis carried out in this work has revealed a
rich bifurcation structure {consisting of both subcritical
  and supercritical} codimension-one bifurcations, as well as
codimension-two pitchfork-Hopf bifurcations. On two occasions, the
cubic normal forms derived in \S \ref{sec:weakly_nonlinear_theory}
were not sufficient to capture the dynamical behavior of the original
system up to topological equivalence. The first situation, discussed
in \S \ref{subsubsec:subsub_hopf} for the {Brusselator
  membrane kinetics}, concerns the transition from a supercritical to
a subcritical Hopf bifurcation. There, parameter values at which the
bifurcation becomes a degenerate Hopf (Bautin) bifurcation were
found. For the same kinetics, the classification of codimension-two
bifurcations in \S \ref{subsec:codimension_two_bifurcation} has also
revealed some degeneracy in the phase portraits of the truncated
system of amplitude equations, which resulted from the mixed-mode
equilibrium undergoing a Hopf bifurcation. For those two cases, the
computation of an additional (nonzero) term is needed to fully resolve
the degeneracy in the normal forms. Further details on this lengthy
computation {for simpler ODE models} can be found in
\cite{kuznetsov2004} and \cite{guckenheimer1983}.

Through numerical simulations, our work has revealed the existence of
clockwise and anti-clockwise rotating waves coexisting with radially
symmetric oscillations. Hence, an open question amenable to a more
rigorous PDE theory approach consists of proving the existence and the
stability of the waves. Key to this problem is the appropriate
reformulation of the model into a moving coordinate frame. {Also, it would be
worthwhile to precisely delineate the region in the
adsorption-desorption parameter space where rotating traveling waves
can occur.}

With their confined geometry, and because of the clear distinction
between the dynamics in the domain and on its boundary, coupled
bulk-surface reaction-diffusion models are ideal for investigating
intracellular pattern-forming systems. For instance, a bulk-surface
model for the spatio-temporal Min protein patterning within
\textit{E. Coli} was formulated in \cite{halatek2012} in a
two-dimensional elliptical geometry. To our knowledge, prior studies
are often limited to linear stability analysis and full numerical
simulations (cf.~\cite{levine2005}, \cite{madzvamuse2015}). An open
problem is to extend the weakly nonlinear theory developed in \S
\ref{sec:weakly_nonlinear_theory} to some biologically relevant
bulk-surface models in more general classes of domains (cylinders,
spheres, ellipses).

\begin{appendices}

\section{Numerical methods}\label{sec:numerical_methods}

In this appendix, the various numerical techniques employed in this
paper are briefly explained. We first focus on the finite differences
discretization of the model with radial symmetry. Then, we present the
finite element discretization of the full bulk-surface
reaction-diffusion system. Finally, the specific Implicit-Explicit
time-stepping method used for most numerical simulations is discussed.

\subsection{Finite differences for the radially symmetric case}

Assuming angular invariance of the coupled bulk-surface system, \eqref{eq:bulk_RD} and \eqref{eq:surface_RD} become
\begin{align}
  \frac{\partial U}{\partial t} &= \frac{D_u}{r}\frac{\partial}{\partial r}
  \left( r\frac{\partial U}{\partial r} \right) - \sigma_u U\,,\qquad
\frac{\partial V}{\partial t} = \frac{D_v}{r}\frac{\partial}{\partial r}
\left( r\frac{\partial V}{\partial r} \right) - \sigma_v V\,, \qquad
                                  0 < r < R\,, \\
  \frac{du}{dt} &= - K_u\left(u - U|_{r=R}\right) + f(u,v)\,, \qquad
                  \frac{dv}{dt} = - K_v\left(v - V|_{r=R}\right) + g(u,v)\,.
\end{align}
The coupling between the PDEs in the bulk and the ODEs on the boundary occurs through the same linear Robin-type boundary conditions as given in \eqref{eq:BC}.

We let $h = {R/(N-1)}$ be the step size, where $N$ is the number of
mesh points. We then approximate $U_j(t) \approx U(h(j-1),t)$ and
$V_j(t) \approx V(h(j-1),t)$, for $j = 1,\ldots,N$. Next, employing
the method of lines yields the following system of ODEs for the vector
$\bW = \left(U_1,\ldots,U_N,V_1,\ldots,V_n,u,v\right)^T \in
\R^{2N+2}$:
\begin{equation}\label{eq:radial_symmetry_FD}
 \dot{\bW} = \A \bW + \bF(\bW)\,.
\end{equation}
Here, $\A \in \R^{(2N+2) \times (2N+2)}$ is the block diagonal matrix defined by
\begin{equation}
 \A = \begin{pmatrix}
   D_u \Laplacian - \sigma_u \I - K_u (\frac{2}{h} + \frac{1}{R})
   e_Ne_N^T & \zero & \zero \\
   \zero & D_v \Laplacian - \sigma_v \I - K_v (\frac{2}{h} +
   \frac{1}{R}) e_Ne_N^T & \zero \\
       \zero & \zero & \zero
      \end{pmatrix},
\end{equation}
where $\I \in \R^{N \times N}$ is the identity matrix,
$e_N = (0,\ldots,1)^T \in R^N$ and each instance of $\zero$ is an
appropriate matrix of zeros. Also, $\Laplacian \in \R^{N \times N}$
corresponds to the discrete radially symmetric Laplacian, defined by
\begin{equation}
\Laplacian = \frac{1}{h^2}\begin{pmatrix}
         -4 & 4 & 0 & \ldots & 0 \\
\frac{1}{2} & -2 & \frac{3}{2} & \ldots & 0\\
     \vdots & \ddots & \ddots & \ddots & \vdots\\
          0 & \ldots & 1 - \frac{1}{2(N-2)} & -2 & 1 + \frac{1}{2(N-2)} \\
                  0 & \ldots & 0 & 2 & -2
                 \end{pmatrix}.
\end{equation}
Finally, the nonlinear function $\bF(\bW): \R^{2N+2} \to \R^{2N+2}$ is
defined by
\begin{equation}
\bF(\bW) = \begin{pmatrix}
            K_u \left( \frac{2}{h} + \frac{1}{R} \right)ue_N \\
            K_v \left( \frac{2}{h} + \frac{1}{R} \right)ve_N \\
            -K_u \left(u - U_N \right) + f(u,v) \\
            -K_v \left(v - V_N \right) + g(u,v)
           \end{pmatrix}.
\end{equation}

\subsection{Finite element discretization}

Let $\Omega$ be the two-dimensional circular bulk domain of radius
$R$. In order to derive the weak formulation for equation
\eqref{eq:bulk_RD} at each $t>0$, we multiply it by
$\phi \in H^1(\Omega)$ and integrate by parts using the boundary
conditions \eqref{eq:BC}. This yields that
\begin{subequations}\label{eq:weak}
\begin{align}
  \int_{\Omega} \phi U_t &= K_u\int_{\partial \Omega} \phi (u - U) -
 D_u \int_{\Omega} \nabla \phi\cdot\nabla U - \sigma_u \int_{\Omega} \phi U\,,
    \qquad \forall \phi \in H^1(\Omega), \\
  \int_{\Omega} \phi V_t &= K_v\int_{\partial \Omega} \phi (v - V) -
 D_v \int_{\Omega} \nabla \phi\cdot\nabla V - \sigma_v \int_{\Omega} \phi V\,,
   \qquad \forall \phi \in H^1(\Omega)\,.
\end{align} 
\end{subequations}

\begin{figure}[htbp]
\begin{center}
\begin{subfigure}{0.45\linewidth}
\includegraphics[width=0.95\linewidth]{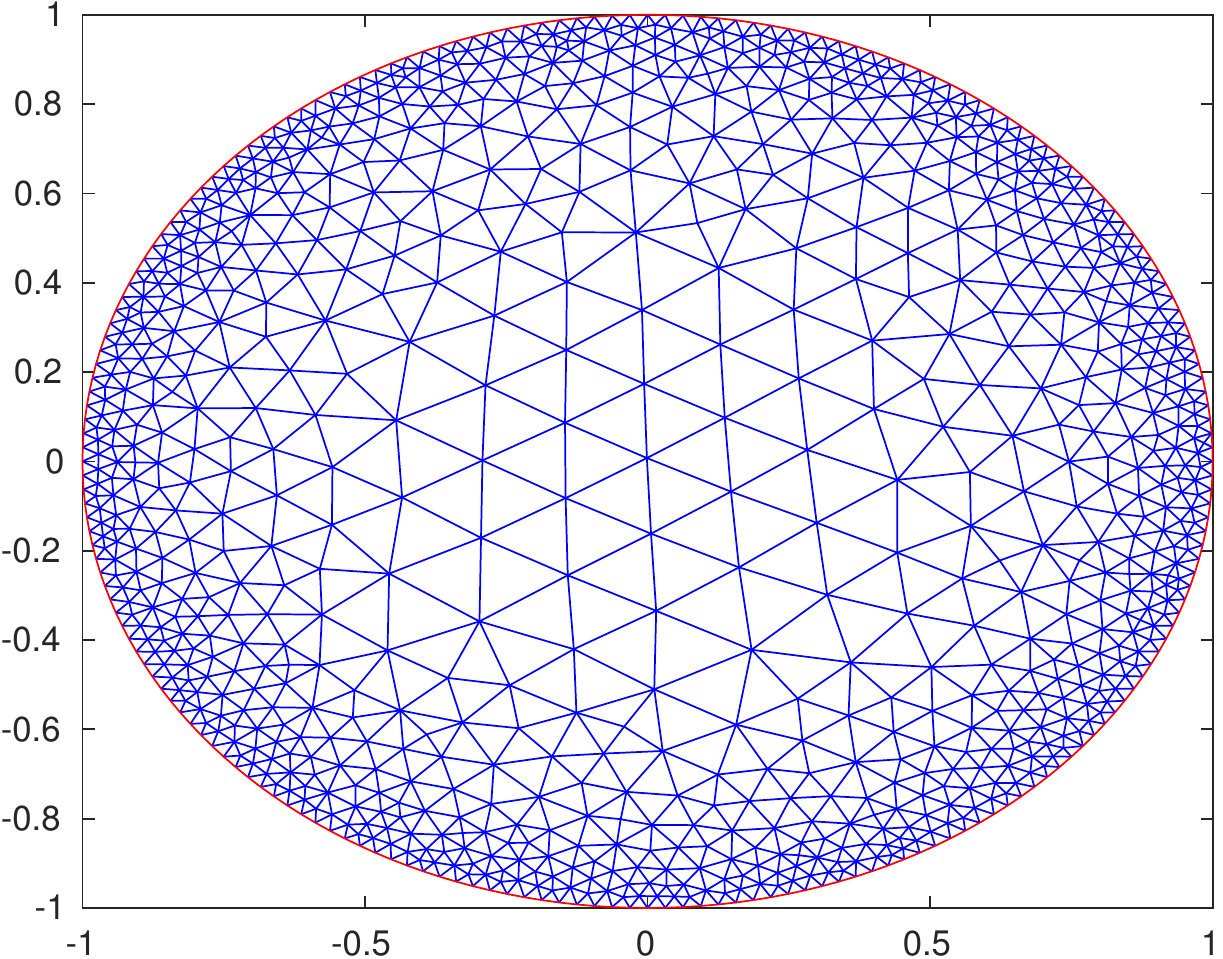}
\end{subfigure}
\begin{subfigure}{0.45\linewidth}
\includegraphics[width=0.95\linewidth]{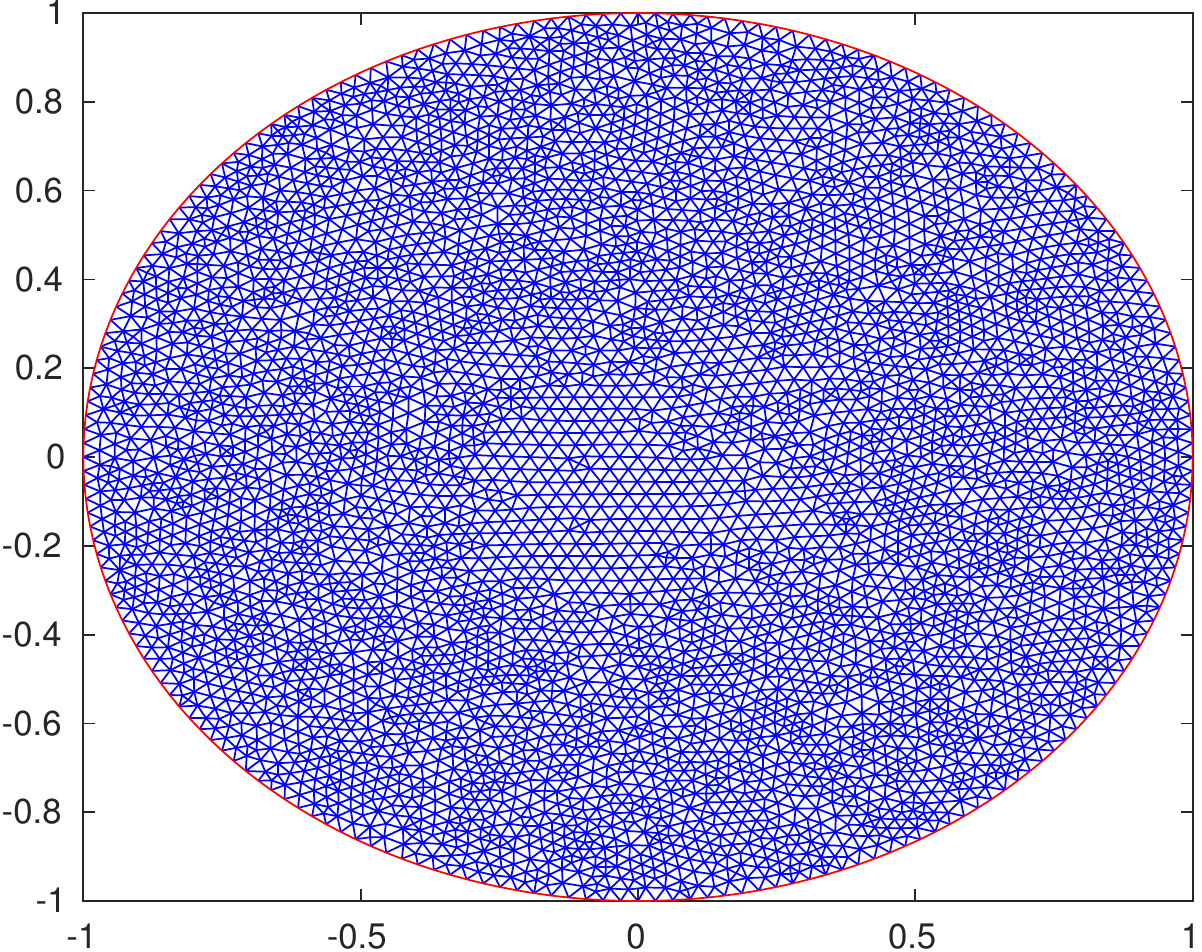}
\end{subfigure}
\end{center}
\caption{\label{fig:mesh} Two different meshes approximating the unit
  disk for $N=200$, obtained with the PDE Toolbox of MATLAB
  \cite{matlab2016}. In the left panel, the mesh is finer near the
  boundary than in the center of the bulk domain. {In
    contrast, in the right panel, we have set
  $h_{\max} \equiv d\sigma$.}}
\end{figure}

We then define an appropriate mesh on $\Omega$. First, we can
parametrize the boundary $\partial \Omega$ by the arc-length as
\begin{equation}
  \partial \Omega = \left\{\bm{X}(\sigma) \in \R^2 \, | \,\,  0 \leq \sigma <
    2\pi R \right\}\,.
\end{equation}
For simplicity, the nodes on the boundary are chosen to be evenly
{spaced by an arc-length step size of $d\sigma ={2\pi R/N}$,
  where $N$ is the number of nodes} on the boundary. Let $N_{total}$
denote the total number of mesh points in $\Omega$, which includes
those on the boundary. A partition can then be defined {as
  follows:}
\begin{equation}
  \Delta_{h_{\max}} = \left\{ x_i = \bm{X}((i-1)d\sigma) \,
    |\, i = 1,\dots,N\right\}\cup \left\{ x_i\, |\, \|x_i\| < R
    \text{  for  } i = N+1,\ldots, N_{total} \right\}\,,
\end{equation}
where $h_{\max}$ is the maximal distance between two adjacent nodes,
defined by
\begin{equation}
 h_{\max} = \max_{j}\min_{i \neq j} \| x_i - x_j \|.
\end{equation}
{In Fig.~\ref{fig:mesh}, we plot two different meshes
  approximating the unit disk given $N = 200$ boundary nodes.}

Now, let $S_{h_{\max}}(\Omega)$ be the space of piecewise linear
functions defined on the mesh $\Delta_{h_{\max}}$ and define a basis
$\{\phi_i\}$ such that any element
$U_{h_{\max}} \in S_{h_{\max}}(\Omega)$ can be uniquely written as
\begin{equation}
 U_{h_{\max}} = \sum_{i=1}^{N_{total}} U_i(x)\phi(x)\,.
\end{equation}
{Hence, the weak formulation \eqref{eq:weak} is approximated as}
\begin{subequations}\label{eq:weak_approx}
  \begin{align}
    \int_{\Omega} \phi_i U_t &= K_u\int_{\partial \Omega} \phi_i (u - U) -
                               D_u \int_{\Omega} \nabla \phi_i\cdot\nabla U -
                               \sigma_u \int_{\Omega} \phi_i U\,, \quad
                               \text{ for } i=1,\ldots,N_{total} \,, \\
    \int_{\Omega} \phi_i V_t &= K_v\int_{\partial \Omega} \phi_i (v - V)
                               - D_v \int_{\Omega} \nabla \phi_i\cdot\nabla V -
                               \sigma_v \int_{\Omega} \phi_i V\,, \quad
                               \text{ for } i=1,\ldots,N_{total}\,.
\end{align} 
\end{subequations}
Assuming that the first basis functions $\phi_i(x)$ for $i=1,\ldots,N$
form a piecewise linear basis for the polygonal approximation of the
boundary, we can approximate the bulk $U,\, V$ and surface $u,\, v$
{concentrations by
\begin{equation}\label{eq:basis_form}
  U_{h_{\max}} = \sum_{i=1}^{N_{total}} U_i(t) \phi_i(x)\,, \quad
  V_{h_{\max}} = \sum_{i=1}^{N_{total}} V_i(t) \phi_i(x)\,, \quad
  u_{h_{\max}} = \sum_{i=1}^N u_i(t) \phi_i(x)\,, \quad
  v_{h_{\max}} = \sum_{i=1}^N v_i(t) \phi_i(x)\,.
\end{equation}
}Then, substituting \eqref{eq:basis_form} into \eqref{eq:weak_approx}
we obtain the following linear system of ODEs:
\begin{subequations}\label{eq:bulk_RD_FEM}
 \begin{align}
   \M \dot{\bU} &= -\left(K_u\Q + D_u \K + \sigma_u \M \right)\bU +
                  K_u\Q\B^T\U, \\
   \M \dot{\bV} &= -\left(K_v\Q + D_v \K + \sigma_v \M \right)\bV +
                  K_v\Q\B^T\V,
 \end{align}
\end{subequations}
where the vectors $\bU,\, \bV\, \in \R^{N_{total}}$ and
$\U,\, \V\, \in \R^{N}$ are defined by
\begin{equation}
 \bU = \begin{pmatrix} U_1(t) \\ \vdots \\ U_{N_{total}}(t) \end{pmatrix}, \quad
 \bV = \begin{pmatrix} V_1(t) \\ \vdots \\ V_{N_{total}}(t) \end{pmatrix}, \quad
 \U = \begin{pmatrix} u_1(t) \\ \vdots \\ u_N(t) \end{pmatrix}, \quad
 \V = \begin{pmatrix} v_1(t) \\ \vdots \\ v_N(t) \end{pmatrix},
\end{equation}
and the matrices $\M$, $\K$, and $\Q$, all in $\R^{N_{total}\times N_{total}}$, are
\begin{equation}\label{eq:FEM_matrices}
  \M_{i,j} = \int_{\Omega} \phi_i\phi_j\,, \qquad \K_{i,j} = \int_{\Omega} \nabla
  \phi_i \nabla \phi_j\,, \qquad \Q_{i,j} = \int_{\partial\Omega} \phi_i \phi_j\,,
  \qquad i,j = 1,\ldots,N_{total}\,.
\end{equation}
Finally, the rectangular matrix $\B \in \R^{N \times N_{total}}$ is defined by
\begin{equation}
 \B = [\I_{N\times N}|\zero_{N \times (N_{total} - N)}]\,,
\end{equation}
where $\I_{N\times N}$ is the identity matrix and
$\zero_{N \times (N_{total} - N)}$ is the appropriate matrix of zeros.

{Simple finite differences are used to approximate the
reaction-diffusion process on the boundary. Using the same notation,
\eqref{eq:surface_RD} is approximated by
\begin{equation}\label{eq:surface_RD_FD}
  \dot{\U} = d_u \D_2 \U - K_u\left(\U - \B\bU\right) +
  \sum_{i=1}^N f(u_i,v_i)e_i\,, \qquad
  \dot{\V} = d_v \D_2 \V - K_v\left(\V - \B\bV\right) +
  \sum_{i=1}^N g(u_i,v_i)e_i\,,
\end{equation}
}where the vectors $e_i$ for $i = 1,\dots,N$ form the standard
Euclidean basis in $\R^N$. Also, $\D_2\, \in \R^{N\times N}$ is the
discrete one-dimensional Laplacian with periodic boundary conditions,
defined as
\begin{equation}
 \D_2 = \frac{1}{(d\sigma)^2}\begin{pmatrix}
                  -2 & 1 & 0 & \ldots & 1 \\
                  1 & -2 & 1 & \ldots & 0\\
                  \vdots & \ddots & \ddots & \ddots & \vdots\\
                  0 & \ldots & 1 & -2 & 1 \\
                  1 & \ldots & 0 & 1 & -2 
                 \end{pmatrix}.
\end{equation}

Now, let $\bW(t)\, \in \R^{2N_{total} + 2N}$ be the time-dependent
solution of the spatially discretized bulk-surface system, defined as
\begin{equation}
 \bW = \begin{pmatrix} \bU \\ \bV \\ \U \\ \V \end{pmatrix}.
\end{equation}
Combining the equations in \eqref{eq:bulk_RD_FEM} with
\eqref{eq:surface_RD_FD}, we obtain an implicit system of differential
equations for $\dot{\bW}$, given by
\begin{equation}\label{eq:implicit_ODE}
\C \dot{\bW} = \A\bW + \bF(\bW),
\end{equation}
where $\C,\, \A\, \in \R^{(2N_{total} + 2N) \times (2N_{total} + 2N)}$ are
block diagonal matrices defined by
\begin{equation}
\begin{split}
 \C &= \begin{pmatrix}
       \M & \zero & \zero & \zero \\
       \zero & \M & \zero & \zero \\
       \zero & \zero & \I_{N\times N} & \zero \\
       \zero & \zero & \zero & \I_{N\times N}
      \end{pmatrix}, \\
 \A &= \begin{pmatrix}
       -\left(K_u\Q + D_u \K + \sigma_u \M \right) & \zero & \zero & \zero \\
       \zero & -\left(K_v\Q + D_v \K + \sigma_v \M \right) & \zero & \zero \\
       \zero & \zero & d_u \D_2 & \zero \\
       \zero & \zero & \zero & d_u \D_2
      \end{pmatrix},
\end{split}
\end{equation}
where once again each instance of $\zero$ is an appropriate matrix of
zeros. Finally, the nonlinear terms in \eqref{eq:implicit_ODE} are
defined as
\begin{equation}
 \bF(\bW) = \begin{pmatrix}
             K_u \Q \B^T \U \\
             K_v \Q \B^T \V \\
             - K_u\left( \U- \B \bU \right) + \sum_{i=1}^N f(u_i,v_i)e_i \\
             - K_v\left(\V - \B\bV\right) + \sum_{i=1}^N g(u_i,v_i)e_i
            \end{pmatrix}.
\end{equation}

{We conclude this appendix} with the definition of two
different solution measures used in this paper. First, we define the
infinity norm as
\begin{equation}
 \|\bW\|_{\infty} = \max_{i=1,\ldots,N_{total}} |e_i^T\bW|\,,
\end{equation}
where the set $\{e_i\}$ forms the standard Euclidean basis in
$\R^{2N_{total}+2N}$. Then, we can approximate the $L^2(\bm{\WW})$
{norm using quadratures as}
\begin{equation}
 \|\bW\|_{L^2(\bm{\WW})} = \sqrt{\bW^T\tilde{\C}\bW}\,,
\end{equation}
where $\tilde{\C}$ is a mass matrix, {defined in terms of $\M$ (see
\eqref{eq:FEM_matrices}), by}
\begin{equation}
\tilde{C} = \begin{pmatrix}
       \M & \zero & \zero & \zero \\
       \zero & \M & \zero & \zero \\
       \zero & \zero & \M_{N \times N} & \zero \\
       \zero & \zero & \zero & \M_{N \times N} 
       \end{pmatrix}, \quad \text{with} 
 \quad \M_{N \times N} = \frac{d\sigma}{6}\begin{pmatrix}
                  4 & 1 & 0 & \ldots & 1 \\
                  1 & 4 & 1 & \ldots & 0\\
                  \vdots & \ddots & \ddots & \ddots & \vdots\\
                  0 & \ldots & 1 & 4 & 1 \\
                  1 & \ldots & 0 & 1 & 4 
                 \end{pmatrix}\,.                
\end{equation}
 
\subsection{Implicit-explicit time-stepping}
 
Two different implicit-explicit time-stepping schemes have been used
{in our PDE numerical simulations:} 1-SBDF and 2-SBDF, where the
acronym SBDF stands for Semi-Implicit Backward Difference Formula
\cite{ruuth1995}. The single-step method 1-SBDF is employed to obtain
the appropriate initial condition for the multi-step method
2-SBDF. When applied to the system \eqref{eq:implicit_ODE}, the two
methods yield
\begin{align}
  & \text{1-SBDF : } (\C - \Delta t \A)\bW^{n+1} = \C \bW^n +
    \Delta t \bF(\bW^n)\,, \\
  & \text{2-SBDF : } (3\C - 2\Delta t \A)\bW^{n+1} =
    4\C \bW^n + 4\Delta t \bF(\bm{W}^n) - \C \bW^{n-1} -
    2\Delta t \bF(\bW^{n-1})\,,
\end{align}
where $\Delta t$ is the time step and $\bW^n \approx \bW(n\Delta t)$
is the approximate solution. Hence, given an initial condition
$\bW^0$, we can compute the solution at the next time step $\bW^1$
using 1-SBDF, after which both $\bW^0$ and $\bW^1$ are used as initial
conditions in 2-SBDF.

The same time-stepping can also be applied to the spatially discretized radially symmetric system \eqref{eq:radial_symmetry_FD}, where one simply needs to replace the matrix $\C$ with the appropriate identity matrix.

Finally, we remark that the time step $\Delta t$ used in our simulations never exceeded $10^{-2}$.

\section{Nondimensionalization}\label{sec:dimensionless}

{We derive here the dimensionless coupled bulk-surface
  reaction-diffusion system defined in equations
  \eqref{eq:bulk_RD}--\eqref{eq:surface_RD}. Let us first consider an
  arbitrary two-dimensional bounded bulk domain $\Omega_\xi$, along
  with its one-dimensional boundary $\partial \Omega_\xi$. If
  $\UU(\xi,T)$ and $\VV(\xi,T)$ denote the bulk variables undergoing
  diffusion and linear decay, then they must satisfy 
\begin{equation}\label{eq:d_bulk_RD}
  \frac{\partial \UU}{\partial T} = D_\UU \Delta \UU - \sigma_\UU \UU\,,
  \qquad \frac{\partial \VV}{\partial T} = D_\VV \Delta \VV - \sigma_\VV \VV\,,
  \qquad \xi \in \Omega_\xi\,, \quad T>0 \,.
\end{equation}
Here $D_\UU$ and $D_\VV$ are bulk diffusion coefficients, while
$\sigma_\UU$ and $\sigma_\VV$ are constant decay rates. Next,
equations \eqref{eq:d_bulk_RD} are supplemented with linear Robin-type
boundary conditions given by
\begin{equation}\label{eq:d_BC}
  D_\UU (\partial_{n_\xi} \UU) = K_\mathfrak{u} \mathfrak{u} - K_\UU \UU\,,
  \qquad D_\VV (\partial_{n_\xi} \VV) = K_\mathfrak{v} \mathfrak{v} - K_\VV \VV
  \,, \qquad \xi \in \partial \Omega_\xi,
\end{equation}
where $n_\xi$ is the outward normal unit vector to the domain
$\Omega_\xi$ while $\mathfrak{u}$, $\mathfrak{v}$ are membrane-bound
variables. The four coupling parameters
$K_\mathfrak{u}\,, K_\mathfrak{v}\,, K_\UU$ and $K_\VV$ in equation
\eqref{eq:d_BC} model local exchanges between membrane-bound and bulk
variables. On the boundary $\partial \Omega_\xi$, the dynamics of
$\mathfrak{u}(\xi,T)$ and $\mathfrak{v(\xi,T)}$ are governed by a
system of reaction-diffusion equations of the form
\begin{equation}\label{eq:d_surface_RD}
  \frac{\partial \mathfrak{u}}{\partial T} = d_\mathfrak{u}\Delta_s \mathfrak{u}
  - K_\mathfrak{u} \mathfrak{u} + K_\UU \UU + \gamma_\mathfrak{u}
  \frac{\mu}{L}f\left(\frac{L\mathfrak{u}}{\mu},
    \frac{L\mathfrak{v}}{\mu}\right)\,, \quad
  \frac{\partial \mathfrak{v}}{\partial T} =
  d_\mathfrak{v}\Delta_s \mathfrak{v} - K_\mathfrak{v} v + K_\VV \VV +
  \gamma_\mathfrak{v}\frac{\mu}{L}g\left(\frac{L\mathfrak{u}}{\mu},
    \frac{L\mathfrak{v}}{\mu}\right)\,, \quad \xi \in \partial \Omega_\xi\,,
\end{equation}
where $d_\mathfrak{u}$ and $d_\mathfrak{v}$ are surface diffusion
coefficients, while $f(\cdot,\cdot)$ and $g(\cdot,\cdot)$ are
arbitrary dimensionless nonlinear reaction kinetics. Here, $\mu$, $L$,
$\gamma_\mathfrak{u}$ and $\gamma_\mathfrak{v}$ are respectively
typical mass, dimension and time-scale measures, needed for the units
to balance in each equation.

Next, we explicitly state the units of the variables,
\begin{equation}
  [\xi]: [\rm length], \quad [T]: [\rm time], \quad [\mathfrak{u}]\,,
  [\mathfrak{v}]: \frac{[\rm mass]}{[\rm length]}, \quad [\UU]\,, [\VV]:
  \frac{[\rm mass]}{[\rm length]^2},
\end{equation}
and the units of the parameters,
\begin{equation}
\begin{split}
  & [L]: [\rm length]\,, \quad [\mu]: [\rm mass]\,, \quad K_\mathfrak{u},
  K_\mathfrak{v}: \frac{1}{[\rm time]}\,, \quad [K_\UU], [K_\VV]:
  \frac{[\rm length]}{[\rm time]}\,, \\
  & [D_\UU]\,, [D_\VV]\,, [d_\mathfrak{u}]\,, [d_\mathfrak{v}]:
  \frac{[\rm length]^2}{[\rm time]}\,, \quad [\sigma_\UU]\,,
  [\sigma_\VV]\,, [\gamma_\mathfrak{u}]\,, [\gamma_\mathfrak{v}]:
  \frac{1}{[\rm time]}\,.
\end{split}
\end{equation}
In this way, new dimensionless variables can be defined as
\begin{equation}\label{eq:nd_var}
\begin{split}
  & t = \gamma_\mathfrak{u}T\,, \quad x = \frac{\xi}{L}\,, \quad U(x,t) =
  \frac{L^2}{\mu} \UU\left(Lx,\frac{t}{\gamma_\mathfrak{u}}\right)\,,
  \quad V(x,t) = \frac{L^2}{\mu} \VV\left(Lx,\frac{t}{\gamma_\mathfrak{u}}
  \right)\,, \\
  & u(x,t) = \frac{L}{\mu} \mathfrak{u}\left(Lx,\frac{t}{\gamma_\mathfrak{u}}
  \right)\,, \quad v(x,t) = \frac{L}{\mu} \mathfrak{v}\left(Lx,\frac{t}
    {\gamma_\mathfrak{u}}\right)\,.
\end{split}
\end{equation}
We remark that the scaled bulk domain, denoted as $\Omega_x$, has a
typical dimension of $\mathcal{O}(1)$. For the case of a disk, the
typical dimension corresponds to the radius.

Upon substituting the variables defined in \eqref{eq:nd_var} within
the system defined by equations \eqref{eq:d_bulk_RD}, \eqref{eq:d_BC}
and \eqref{eq:d_surface_RD}, we readily obtain a dimensionless coupled
bulk-surface reaction-diffusion system given by
\begin{equation}
\begin{split}
  & \frac{\partial U}{\partial t} = D_u \Delta U - \sigma_u U\,, \quad
  \frac{\partial V}{\partial t} = D_v \Delta V - \sigma_v V\,, \quad x \in
  \Omega_x\,, \quad t>0 \,, \\
  & D_u (\partial_{n_x} U) = r_d u - r_a U\,, \quad D_v (\partial_{n_x} V) = p_d v
  - p_a V \,, \quad x \in \partial \Omega_x\,, \\
  & \frac{\partial u}{\partial t} = d_u \Delta_s u - r_d u + r_a U + f(u,v)\,,
  \quad \frac{\partial v}{\partial t} = d_v \Delta_s v - p_d v + p_a V +
  \gamma g(u,v)\,, \quad x \in \partial \Omega_x\,,
\end{split}
\end{equation}
where the new dimensionless parameters are each defined as
\begin{equation}
\begin{split}
  D_u &\equiv \frac{D_\UU}{L^2 \gamma_\mathfrak{u}}\,, \quad D_v \equiv
  \frac{D_\VV}{L^2\gamma_\mathfrak{u}}\,, \quad d_u \equiv
  \frac{d_\mathfrak{u}}{L^2 \gamma_\mathfrak{u}}\,, \quad d_v \equiv
  \frac{d_\mathfrak{v}}{L^2 \gamma_\mathfrak{u}}\,, \quad \sigma_u \equiv
  \frac{\sigma_\UU}{\gamma_\mathfrak{u}}\,, \quad \sigma_v \equiv
  \frac{\sigma_\VV}{\gamma_\mathfrak{u}}\,,\\
r_d &\equiv \frac{K_\mathfrak{u}}{\gamma_\mathfrak{u}}\,, \quad r_a \equiv
\frac{K_\UU}{L\gamma_\mathfrak{u}}\,, \quad p_d \equiv \frac{K_\mathfrak{v}}
{\gamma_\mathfrak{u}}\,, \quad p_a \equiv \frac{K_\VV}{L\gamma_\mathfrak{u}}\,,
\quad \gamma \equiv \frac{\gamma_\mathfrak{v}}{\gamma_\mathfrak{u}}\,.
\end{split}
\end{equation}

Our first assumption is that the time-scales of the nonlinear reaction
kinetics are the same, which yields $\gamma = 1$. Next, to reduce the size of the parameter space, we will assume equal adsorption and desorption rates for each variable. Letting $K_u$ and $K_v$ be new coupling rate constants, defined as
\begin{equation}\label{eq:assumption}
 K_u \equiv r_a = r_d\,, \quad K_v \equiv p_a = p_d\,,
\end{equation}
we obtain the dimensionless coupled bulk-surface
reaction-diffusion system given in \S \ref{sec:introduction}.
In terms of the parameters of (\ref{eq:d_bulk_RD})--(\ref{eq:d_surface_RD}), we are assuming by \eqref{eq:assumption} that
\[ {K_\mathfrak{u}}=\frac{K_\UU}{L},\quad {K_\mathfrak{v}}=\frac{K_\VV}{L}. \]
However in \S \ref{sec:global_dynamics}, we relax assumption \eqref{eq:assumption} and allow for distinct adsorption and desorption rates in order to explore the formation of rotating waves in a circular bulk domain.}

\end{appendices}


\section*{Acknowledgments}
F.~Paquin-Lefebvre was partially supported by a NSERC Doctoral Award
and a UBC Four-Year Doctoral Fellowship. W.~Nagata and M.~J.~Ward
gratefully acknowledge the support of the NSERC Discovery Grant
Program. We would also like to thank Daniel Gomez for discussions on
implementing the finite-element solver.

\bibliographystyle{plainnat}
\bibliography{references}

\end{document}